%% file: Thesis.tex
\documentclass[11pt, openany]{book}

\def\thechapter{\Roman{chapter}} 

\usepackage{titlesec}
\usepackage{appendix}
\usepackage[usenames,dvipsnames]{color}
\usepackage{amsthm,amsmath,amsfonts,graphicx,bm,verbatim,bbold}
\usepackage{dsfont}
\usepackage{trsym}
\usepackage{pifont}
\usepackage{fullpage}
\usepackage[latin1]{inputenc}
\usepackage{sectsty}
\usepackage{soul}

\usepackage{mathtools,amssymb}

\usepackage{xcolor}

\usepackage{dcolumn,booktabs}

\usepackage[colorlinks=true,bookmarks=true]{hyperref}

\usepackage{fancyhdr}
\fancypagestyle{plain}{
\fancyhf{} 
\fancyfoot[C]{\bfseries \thepage} 

}

\usepackage{blindtext}
\usepackage{enumerate}
\usepackage{setspace}

\usepackage{lmodern}

\newlength\longest

\fancypagestyle{plain}

\newcommand{\myauthor}{THEAUTHOR}
\newcommand{\mytitle}{THETITLE}

\definecolor{col}{rgb}{0.49,0.24,0.10}

\newcommand{\Id}{\mathbb{1}}
\newcommand{\be}{\begin{equation}}
\newcommand{\ee}{\end{equation}}
\newcommand{\bea}{\begin{eqnarray}}
\newcommand{\eea}{\end{eqnarray}}

\newtheorem{thm}{Theorem}[section]

\newtheorem{lem}[thm]{Lemma}

\newtheorem{ex}[thm]{Example}

\newcommand{\ket}[1]{\vert#1\rangle}
\newcommand{\bra}[1]{\langle#1\vert}
\newcommand{\tr}{\mathrm{tr}}
\newcommand{\mc}[1]{\mathcal{#1}}

\newcommand{\mr}{\mathrm}
\newcommand{\dg}{\dagger}
\newcommand{\e}{\mathcal{E}}
\newcommand{\mb}{\mathbf}
\renewcommand{\oe}{\overline{\mathcal{E}}}

\renewcommand{\O}{\mathcal{O}}
\renewcommand{\S}{\mathcal{S}}
\renewcommand{\sp}{\mathrm{span}}
\renewcommand{\P}{\mathbb{P}}
\renewcommand{\d}{\mathrm{d}}

\begin{document}

\hypersetup{
linkcolor=col,          
citecolor=blue,        
filecolor=blue,      
urlcolor=blue 
}

\def\thechapter{\arabic{chapter}}

\renewcommand{\mytitle}{Tensor network states for the description of quantum many-body systems}
\renewcommand{\myauthor}{Thorsten Bernd Wahl}


\thispagestyle{empty}
\begin{center} \huge{
Fakult\"{a}t f\"{u}r Physik, Technische Universit\"{a}t M\"{u}nchen \\
Max-Planck-Institut f\"{u}r Quantenoptik \par}
\vspace{54pt}

\Huge{\bfseries  \mytitle }\\
\vspace{20pt}
\huge{\myauthor\\}
\vspace{40pt}

{\flushleft \Large Vollst\"{a}ndiger Abdruck der von der Fakult\"{a}t f\"{u}r Physik der Technischen Universit\"{a}t M\"{u}nchen 
zur Erlangung des akademischen Grades eines \\ 
\vspace{4pt}
{\centering Doktors der Naturwissenschaften \\ } 
\vspace{4pt}
genehmigten Dissertation.\\}
\vspace{34pt}

{\Large
\begin{flushleft}
Vorsitzender: \hspace{3.49cm} Univ.-Prof. Dr. Rudolf Gross \\
\vspace{12pt}
Pr\"{u}fer der Dissertation: \\
\hspace{6cm}
1. Hon.-Prof. J. Ignacio Cirac, Ph.D. \\
\hspace{6cm} 2. Univ.-Prof. Dr. Michael Knap \\
\hspace{6cm} 3. Univ.-Prof. Dr. Ulrich Schollw\"{o}ck, \\
\hspace{6cm} \ \ \ \ Ludwig-Maximilians-Universit\"{a}t M\"{u}nchen \\
\hspace{6cm} \ \ \ \ (nur schriftliche Beurteilung)

\vspace{30pt}
Die Dissertation wurde am 11. Mai 2015 bei der Technischen Universit\"{a}t M\"{u}nchen eingereicht und durch die Fakult\"{a}t f\"{u}r Physik am 3. August 2015 angenommen.

\end{flushleft}
}
\end{center}

\newpage
\thispagestyle{empty}
\mbox{}

\newpage

\pagenumbering{roman}

\include{0}

\tableofcontents

\newpage
{

\clearpage

\thispagestyle{empty}
\null\vfill

\settowidth\longest{\huge\itshape is the joy of understanding.}
\centering
\parbox{\longest}{%
  \raggedright{\huge\itshape%
   The noblest pleasure \\ 
   is the joy of understanding.\par\bigskip
  }   
  \raggedleft\Large\MakeUppercase{Leonardo da Vinci}\par%
}

\vfill\vfill

\clearpage
}

\newpage

\pagenumbering{arabic}

\include{1} 
\include{2.1}

\include{2.2} 
\include{2.3}

\include{3.1}

\include{3.2}

\include{3.3} 
\include{3.4} 
\include{4}


\include{A}

\phantomsection
\addcontentsline{toc}{chapter}{List of Figures}
\listoffigures
\clearpage

\phantomsection
\addcontentsline{toc}{chapter}{List of Tables}
\listoftables
\clearpage

\phantomsection
\addcontentsline{toc}{chapter}{Bibliography}
\bibliographystyle{ieeetr}
\bibliography{References}
\clearpage

\phantomsection
\addcontentsline{toc}{chapter}{Acknowledgements}
\include{B}

\clearpage

\end{document}

%% file: 0.tex
\newpage

\section*{Abstract}

This thesis is divided into two mainly independent parts: The first one elaborates on the classification of all Matrix Product States (MPS) with long range localizable entanglement, where the second is devoted to free fermionic and interacting chiral Projected Entangled Pair States (PEPS).

In the first part, we derive a criterion to determine when a translationally invariant MPS has long range localizable entanglement, which indicates that the corresponding state has some kind of non-local hidden order. We give examples fulfilling this criterion and eventually use it to obtain all such MPS with bond dimension 2 and 3 (which is the dimension of the virtual indices of the MPS tensor). 

In the second part, we show that PEPS in two spatial dimensions can describe chiral topological states by explicitly
constructing a family of such states with a non-trivial Chern number, thus defying previous arguments which seemed to rule out their very existence. They turn out to have power law decaying correlations, i.e., the local parent Hamiltonians for which they are ground states, are gapless. We show that such parent Hamiltonians lie at quantum phase transition points between different topological phases. 

Furthermore, we also construct long range Hamiltonians with a flat energy spectrum for which those PEPS are unique ground states. The gap is robust against local perturbations, which allows us to define a Chern number for the PEPS. We demonstrate that free fermionic PEPS (GFPEPS) must necessarily be non-injective and have gapless parent Hamiltonians. Moreover, we provide numerical evidence that they can nevertheless approximate well the physical properties of Chern insulators with local Hamiltonians at arbitrary temperatures.

As for non-chiral topological PEPS, the non-trivial, topological properties can be traced down to the existence of a symmetry on the virtual level of the PEPS tensor that is used to build the state. Based on that symmetry, we construct string-like operators acting on the virtual modes that can be continuously deformed without changing the state. On the torus, the symmetry implies that the ground state space of the local parent Hamiltonian is two-fold degenerate. By adding a string wrapping around the torus one can change one of the ground states into the other. We use the special properties of PEPS to build the boundary theory and show how the symmetry results in the appearance of chiral modes and in a universal correction to the area law for the zero R\'{e}nyi entropy. 

Finally, we show that PEPS can also describe chiral topologically ordered phases. For that, we construct a simple PEPS for spin-1/2 particles in a two-dimensional lattice. We reveal a symmetry in the local projector of the PEPS that gives rise to the global topological character. We also extract characteristic quantities of the edge Conformal Field Theory using the bulk-boundary correspondence.

\newpage

\section*{Zusammenfassung}

Diese Dissertation ist in zwei weitgehend unabh\"{a}ngige Teile gegliedert: Der erste beschreibt die Klassifizierung aller Matrixproduktzust\"{a}nde (MPS) mit langreichweitiger lokalisierbarer Verschr\"{a}n-kung, w\"{a}hrend der zweite frei fermionischen und wechselwirkenden chiralen projizierten verschr\"{a}nk-ten Paarzust\"{a}nden (PEPS) gewidmet ist.

Im ersten Teil leiten wir ein Kriterium her, mit dem sich bestimmen l\"{a}sst, ob ein MPS lang-reichweitige lokalisierbare Verschr\"{a}nkung besitzt, was eine Art verborgene nichtlokale Ordnung des entsprechenden Zustandes bedeutet. Wir liefern Beispiele, die dieses Kriterium erf\"{u}llen, und verwenden es letztendlich um alle solchen MPS mit Bindungsdimension 2 und 3 (die Dimension der virtuellen Indizes des MPS-Tensors) zu erhalten.

Im zweiten Teil zeigen wir, dass in zwei Dimensionen PEPS chirale topologische Zust\"{a}nde beschreiben k\"{o}nnen, indem wir explizit eine Familie solcher Zust\"{a}nde mit nichttrivialer Chernzahl konstruieren, was bisherige Argumente widerlegt, welche deren Existenz auszuschlie{\ss}en schienen. Es stellt sich heraus, dass deren Korrelationen einem inversen Potenzgesetz folgend abfallen, d.h. die lokalen Parent Hamiltonians, f\"{u}r welche sie Grundzust\"{a}nde sind, haben keine Energiel\"{u}cke. Wir zeigen, dass solche Parent Hamiltonians auf Quantenphasen\"{u}bergangspunkten zwischen verschiedenen topologischen Phasen liegen. 

Au{\ss}erdem konstruieren wir langreichweitige Hamiltonians mit einem flachen Energiespektrum, f\"{u}r welche diese PEPS nichtentartete Grundzust\"{a}nde sind. Die Energiel\"{u}cke ist stabil gegen lokale St\"{o}rungen, was es uns erlaubt eine Chernzahl f\"{u}r diese PEPS zu definieren. Wir demonstrieren, dass frei fermionische PEPS (GFPEPS) notwendigerweise nichtinjektiv sein m\"{u}ssen und Parent Hamiltonians mit einer verschwindenden Energiel\"{u}cke besitzen.  Dar\"{u}berhinaus liefern wir numerische Belege daf\"{u}r, dass sie dennoch die physikalischen Eigenschaften von Chernisolatoren mit lokalen Hamiltonians bei beliebigen Temperaturen hinreichend genau approximieren k\"{o}nnen. 

Wie bei nichtchiralen topologischen PEPS k\"{o}nnen die nichttrivialen, topologischen Eigenschaften auf die Existenz einer Symmetrie im virtuellen Raum des PEPS-Tensors zur\"{u}ckgef\"{u}hrt werden, aus dem der Zustand aufgebaut wird. Auf dieser Symmetrie basierend konstruieren wir stringartige Operatoren, die auf die virtuellen Moden wirken und kontinuierlich verformt werden k\"{o}nnen ohne dass sich der Zustand dabei \"{a}ndert. Auf dem Torus hat die Symmetrie eine zweifache Entartung des Grundzustandsraumes des lokalen Parent Hamiltonians zur Folge. Wenn man einen String einf\"{u}gt, der den Torus umwickelt, kann man zwischen den beiden Grundzust\"{a}nden wechseln. Wir verwenden die speziellen Eigenschaften von PEPS um die Randtheorie abzuleiten und weisen auf, wie die Symmetrie zu chiralen Moden und einer universellen Korrektur zum Fl\"{a}chengesetz der nullten R\'{e}nyi-Entropie  f\"{u}hrt

Schlie{\ss}lich zeigen wir, dass PEPS auch chirale topologisch geordnete Systeme beschreiben k\"{o}nnen. Dazu konstruieren wir einfache PEPS von Spin-1/2-Teilchen auf einem zweidimensionalen Gitter. Wir enth\"{u}llen eine Symmetrie des lokalen PEPS-Projektors, die den globalen \mbox{topologischen} Charakter bedingt. Au{\ss}erdem leiten wir durch Verwendung der Volumen-Rand-Entsprechung be-zeichnende Gr\"{o}{\ss}en der Konformen Feldtheorie des Randes ab.

\newpage

\section*{Publications}

\begin{enumerate} 

\item Shuo Yang, Thorsten B. Wahl, Hong-Hao Tu, Norbert Schuch, and J. Ignacio Cirac, ``Chiral projected entangled-pair states with topological order'', 	Phys. Rev. Lett. \textbf{114}, 106803 (2015). 

\item Thorsten B. Wahl, Stefan T. Ha{\ss}ler, Hong-Hao Tu, J. Ignacio Cirac, and Norbert
Schuch, ``Symmetries and boundary theories for chiral projected entangled pair 	states'', Phys. Rev. B \textbf{90}, 115133 (2014).

\item Thorsten B. Wahl, Hong-Hao Tu, Norbert Schuch, and J. Ignacio Cirac, ``Projected entangled-pair states can describe chiral topological states'', Phys. 	Rev. Lett. \textbf{111}, 236805 (2013).

\item Thorsten B. Wahl, David P\'{e}rez-Garc\'{i}a, and J. Ignacio Cirac, ``Matrix Product 	States with long-range Localizable Entanglement'', Phys. Rev. A \textbf{86}, 062314 (2012). \\

\end{enumerate}

\noindent All publications have been incorporated into this thesis.

\newpage

%% file: 1.tex
\chapter{Introduction}

The state of a system of classical particles is completely characterized by the particles' positions and velocities. However, at the atomic length scale, the motion of particles can no longer be described by classical mechanics, but instead the rules of quantum mechanics need to be taken into consideration. In quantum mechanics, the motion of particles is dictated by the Schr\"odinger equation, and the particles' positions and momenta no longer have fixed values, but follow a certain probability distribution. Since for many particles joint probabilities need to be taken into consideration, the complexity of the Schr\"odinger equation grows exponentially with the number of particles. This makes the exact description of quantum many-body systems of very few particles already in general intractable. However, many fascinating phenomena in Condensed Matter Physics emerge entirely due to the interplay of many quantum particles and cannot be explained without the laws of quantum mechanics. Examples include superconductivity, Bose-Einstein condensation~\cite{BEC_Bose,BEC_Einstein} and the Quantum Hall Effect~\cite{QHE_ex}. That is why efficient approximation methods to the many-particle problem have been devised.

Many of the above phenomena can be explained by assuming the particles to be effectively non-interacting, such that the Schr\"odinger equation can be decoupled into  equations for the individual particles. The solution to the latter yields an approximate description of the quantum many body system under consideration. However, apart from obtaining inaccuracies via this method, it simply fails to predict the properties induced by strong interactions between the particles, like the Fractional Quantum Hall Effect~\cite{FQHE_ex,FQHE_th}. For this reason, other methods had to be conceived, which do not rely on the non-interacting picture, such as Quantum Monte Carlo simulations~\cite{QMC_Rev}. 

\begin{figure}
\centering\includegraphics[width=0.8\columnwidth]{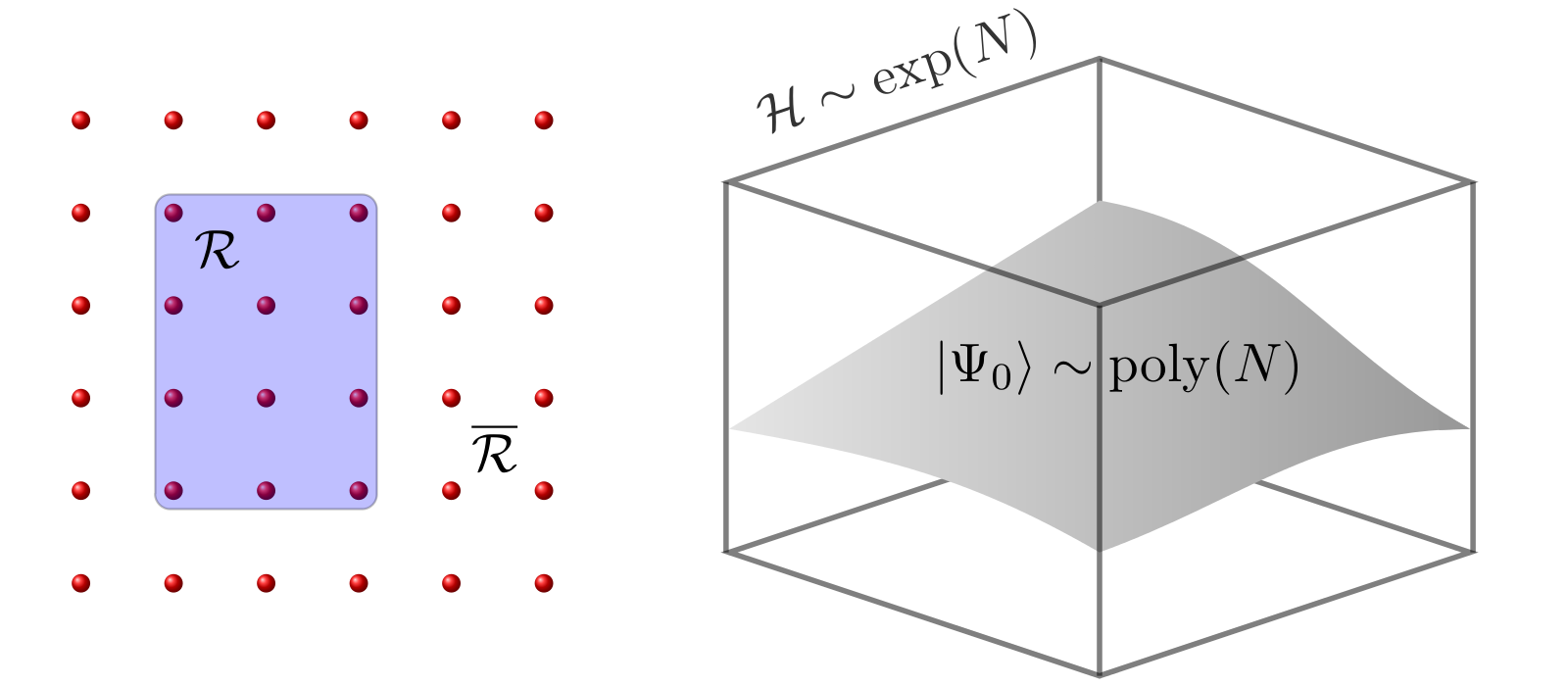}
\caption{Left: For ground states of local gapped Hamiltonians, the entanglement entropy of the reduced density matrix of a region $\mc R$ grows only like the ``area'' (which in two dimensions is the perimeter) of region $\mc R$ and not as its volume. Right: Because of that, the ground states $|\Psi_0\rangle$ of such Hamiltonians live only on a subspace of the full Hilbert space $\mc{H}$, which grows polynomially with the number of particles $N$.}
\label{fig:area}
\end{figure}

Throughout this thesis, we will consider systems defined on a lattice, where on the lattice sites either immobile spins reside or electrons that are allowed to hop. Although this is already a strong simplification, such systems are believed to accurately describe certain materials, and, most importantly, also host the above phenomena observed in continuous Condensed Matter systems. In second quantization, such systems are described by diagonalizing the many-body Hamiltonian, which is again exponentially hard in the number of particles. Therefore, the system sizes amenable to exact diagonalization are often too small in order to investigate the thermodynamic limit. The exponentially huge Hilbert space (the space the many-body Hamiltonian acts on) seems to pose strong restrictions on the system sizes that can be tackled precisely. However, it turns out that the ground states of most relevant Hamiltonians are not randomly distributed in the exponentially large Hilbert space, but instead live on a subspace that only grows polynomially with the system size. This seems to be in particular the case for Hamiltonians with a gap to the first excited state in the thermodynamic limit and local interactions, that is, interactions whose amplitudes decay at least exponentially with distance. The reason is that such Hamiltonians fulfill the so-called \textit{area law}~\cite{area_Rev,PEPS_eff,Area_exp}, which states that for their ground states ($\rho$), the entanglement entropy of the reduced density matrix $\rho_\mc{R} = \tr_{\overline{\mc{R}}}(\rho)$ of some simply connected region $\mc{R}$ grows only as its surface and not as its volume if the size of the region is increased. These points are illustrated in Fig.~\ref{fig:area}. 

\begin{figure}
\centering\includegraphics[width=0.8\columnwidth]{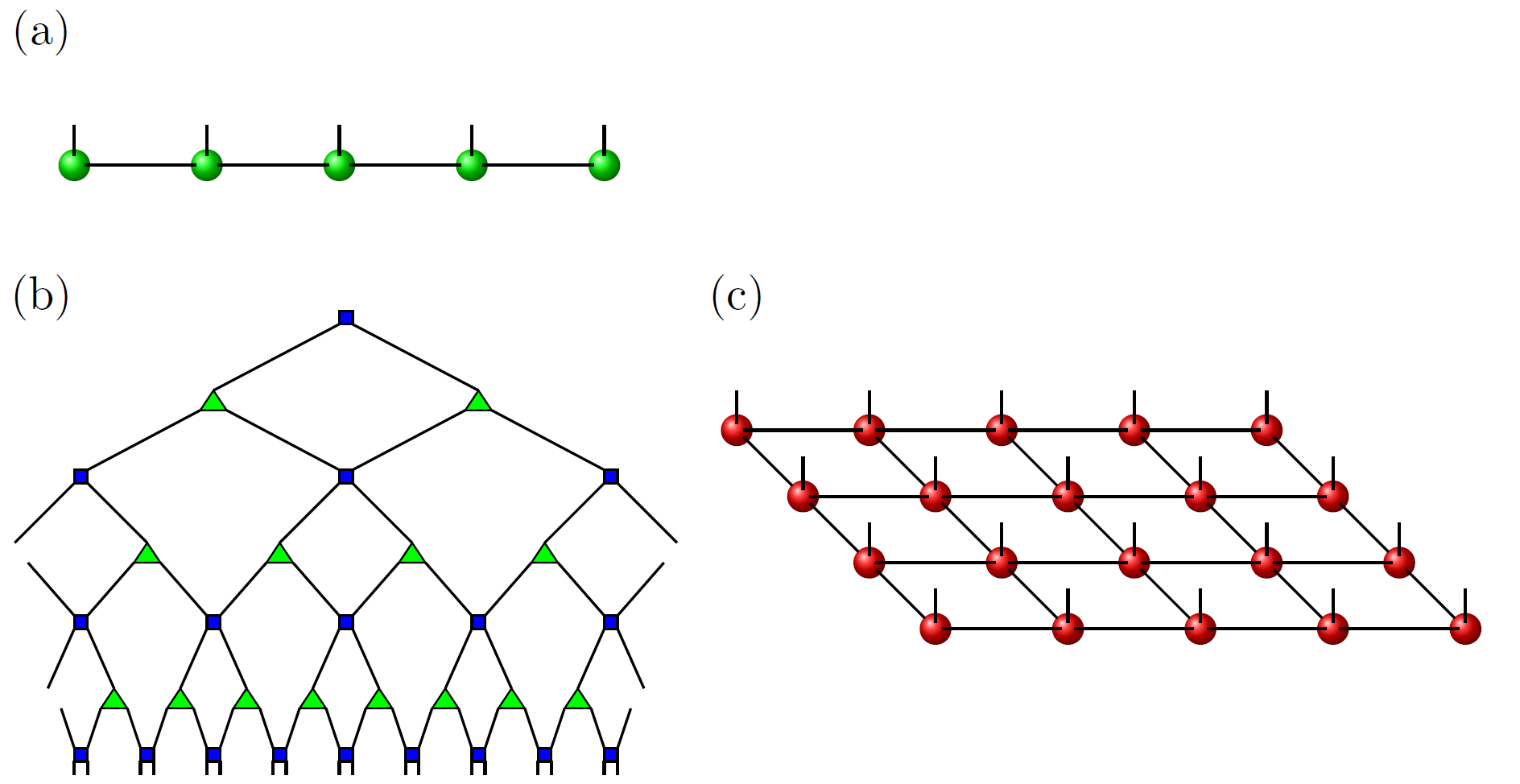}
\caption{Examples of Tensor Network States. Each ball or box denotes a tensor. The number of legs is the number of indices. Connected legs are the ones which are contracted, whereas open legs are the physical indices. (a) Matrix Product State (MPS), (b) Multiscale Entanglement Renormalization Ansatz (MERA), (c) Projected Entangled Pair State (PEPS).}
\label{fig:TNS}
\end{figure}

Even more excitingly, the above-mentioned relevant subspace of the Hilbert space can be parameterized by \textit{Tensor Network States}~\cite{MPS_math, MERA_org, TTN_org, PEPS_org, bMERA_org,bMERA_org2} (see Fig.~\ref{fig:TNS}). They can be constructed in such a way that they automatically fulfill the area law.  Each component of the wave function can in principle be obtained by carrying out a contraction of a tensor network. The highest dimension of the contracted tensor indices is referred to as the \textit{bond dimension}. By increasing the bond dimension, the desired ground state can be approximated with higher accuracy. It is believed that in arbitrary dimensions, the bond dimension and thus the number of parameters grows only polynomially with the system size if the TNS is required to have a certain overlap with the ground state wave function of a local Hamiltonian. This has been proven rigorously in one dimension~\cite{MPS_faithful}. Apparently, nature does not occupy the exponentially large Hilbert space she has at her disposal, but contents herself with a polynomially large subspace that is in principle describable for us!

Another way to grasp this huge reduction in the complexity of the ground states of realistic Hamiltonians is to count their number of parameters: Consider for instance a Hamiltonian defined for a system of $N$ particles with spin $1/2$. The dimension of the Hilbert space is $2^N$. However, any physically relevant Hamiltonian has at most four-particle interactions. The number of possible four-particle interactions scales as $\O({N \choose 4}) = \O(N^4)$. Hence, the number of possible Hamiltonians, and therefore also of possible ground states is polynomial in the system size and not exponential - even without the assumption of local interactions. However, to the best of the author's knowledge, only for the ground states of \textit{local} Hamiltonians a suitable parameterization of the corresponding polynomial subspace has been provided.

Let us now summarize important classes of TNS: In one dimension, \textit{Matrix Product States} (MPS)~\cite{MPS_math} have been shown to provide an \textit{efficient} approximation of realistic local Hamiltonians~\cite{PEPS_eff,Area_exp,MPS_faithful}. By efficient we mean that the number of parameters required to obtain a fixed accuracy of the approximation of any observable grows at most polynomially with the systems size. MPS can be obtained by sequential generation~\cite{MPS_sequential} and are the variational class underlying the extremely successful \textit{Density Matrix Renormalization Group} (DMRG)~\cite{DMRG_org, DMRG_org2, DMRG_Rev} algorithms. Note that MPS not only allow for an efficient representation of the ground states of local Hamiltonians, but it has also been shown that the best MPS with a given bond dimension can be found in a time that is polynomial in the system size~\cite{poly_time_org, poly_time}. Moreover, once the best MPS representation has been found, local observables and few-particle correlations can be computed efficiently. MPS can also be applied to critical systems (that is, gapless Hamiltonians); however, their polynomial scaling is worse in this case and other tensor network ans\"{a}tze have been devised to remedy this problem. The first one was the \textit{Tree Tensor Network}~\cite{TTN_org} (TTN), which has a real space renormalization group structure. However, the tensors on the highest level of an optimized TTN contain information about the local entanglement on its lowest level. The resulting reduction in computational efficiency could be overcome by the  \textit{Multiscale Entanglement Renormalization Ansatz} (MERA)~\cite{MERA_org}, which removes short range entanglement. Critical systems in one dimension display a logarithmic violation of the area law, i.e., the entanglement entropy does not saturate as a function of the length of region $\mc R$ (as is the case for gapped local Hamiltonians), but scales like the logarithm of its length. The TTN and the MERA also possess this logarithmic violation~\cite{TTN_violation, MERA_org}. 

The generalization of MPS to two and higher dimensions is denoted as \textit{Projected Entangled Pair States} (PEPS)~\cite{PEPS_org,PEPS_Rev,PEPS_Rev2}. As opposed to Quantum Monte Carlo methods, they do not suffer from the so-called sign problem~\cite{sign_prob}, i.e., they also allow for the description of frustrated and fermionic systems~\cite{fPEPS_Kraus}. However, PEPS still compete with other methods which are not that well suited for the thermodynamic limit of two dimensional systems, such as DMRG in two dimensions~\cite{DMRG_2D, DMRG_Rev}. The reason is that the scaling of the number of parameters required for a given overlap with the true ground state, although polynomial in the system size, is much worse then for MPS. Furthermore, there are some technical problems, which arise due to the fact that PEPS do not allow for an efficient calculation of observables (or even of the norm of the state)~\cite{PEPS_complexity}. This implies that approximate methods need to be applied in the contraction of PEPS, which are, however, usually well controlled~\cite{PEPS_2Dbosons,Lubi_comparison,Lubis_algorithm}. In two dimensions, critical systems do not necessarily display logarithmic corrections to the area law. If they do, it would still be sensible to use PEPS to describe them, as PEPS probably also allow for an efficient approximation of critical local Hamiltonians as is the case for MPS in one dimension. Note that higher dimensional MERA~\cite{MERA_org} obey the area law and can in fact be represented efficiently by PEPS~\cite{MERA_PEPS}. However, for two dimensional critical systems it is possible to obtain a better (polynomial) scaling by employing the branching MERA~\cite{bMERA_org,bMERA_org2}, which displays, depending on its branching parameter, logarithmic violations to the area law or even a volume law~\cite{bMERA_area}.

At finite temperature, the area law of the entanglement entropy for gapped local Hamiltonians is replaced by an area law of the mutual information~\cite{area_finiteT}.
Tensor Network States can easily be generalized to mixed states. An obvious way is to define a purification of the mixed state one intends to represent~\cite{MPO_puri}. This can be achieved by doubling the physical dimension at each site (such that there is an additional physical index) and obtaining the mixed state by tracing out the extra degrees of freedom. Another possibility is to again double the physical dimension, but to consider the additional physical index as defined in the dual space (ket and bra). Then, one has a representation of mixed states by the same contraction scheme as for pure states. However, some care has to be taken to ensure that the constructed state is positive semidefinite. The simplest cases are Matrix Product Operators (MPO)~\cite{MPO_puri,MPO_org2} and Projected Entangled Pair Operators (PEPO)~\cite{PEPO_org} in one and higher dimensions, respectively. The two approaches have been compared in Ref.~\cite{puri_vs_MPO}, where it has been shown that mixed states obtained by purifying MPS can be efficiently represented by MPOs, but not the other way around.

Most TNS can be defined such that they allow to read off properties of the wave function from the constituent tensors: Symmetries and topological properties of MPS and PEPS can be encoded locally in the corresponding tensors. Moreover, there is a
standard procedure of constructing a \textit{parent Hamiltonian}~\cite{MPS_math,TNS_intersection} for any MPS or PEPS as a sum of local projectors which each annihilate the state. Ground states of this kind are called \textit{frustration free}. For MPS it can be easily checked whether it is the unique ground state of its parent Hamiltonian. In one dimension, several such MPS with exact parent Hamiltonians of a simple form, such as the Majumdar-Ghosh state~\cite{Majumdar_Ghosh} and the AKLT model~\cite{AKLT_org} (a special point of the spin-1 bilinear biquadratic Heisenberg model) have been found. For two dimensional systems determining whether the parent Hamiltonian has a unique ground state is a rather intricate issue. 

The first part of this work is devoted to MPS with long range \textit{localizable entanglement} (LRLE)~\cite{LE_org,MPS_LRLE} and to chiral topological PEPS~\cite{cPEPS_org,cPEPS_Read,cPEPS_long,interacting_cPEPS}. Localizable entanglement plays an essential role in the field of quantum communication. It quantifies the amount of entanglement that can on average be localized between the spins at the ends of a spin chain by measuring all the remaining ones. If it is non-vanishing in the thermodynamic limit, the state is said to have LRLE. Such states can be used as ideal quantum repeaters~\cite{quantum_repeater}, since they allow for the creation of entangled pairs over arbitrary long distances. Moreover, they appear at certain topological quantum phase transitions~\cite{LRLE_QPT}, i.e., LRLE can be used for their detection. A criterion will be presented which classifies all translationally invariant MPS with a fixed bond dimension, which possess LRLE. This is very relevant as it basically also classifies all ground states of local gapped Hamiltonians with LRLE, since the bond dimension of the corresponding MPS saturates as a function of the size of the system. 

\begin{figure}
\centering\includegraphics[width=0.4\columnwidth]{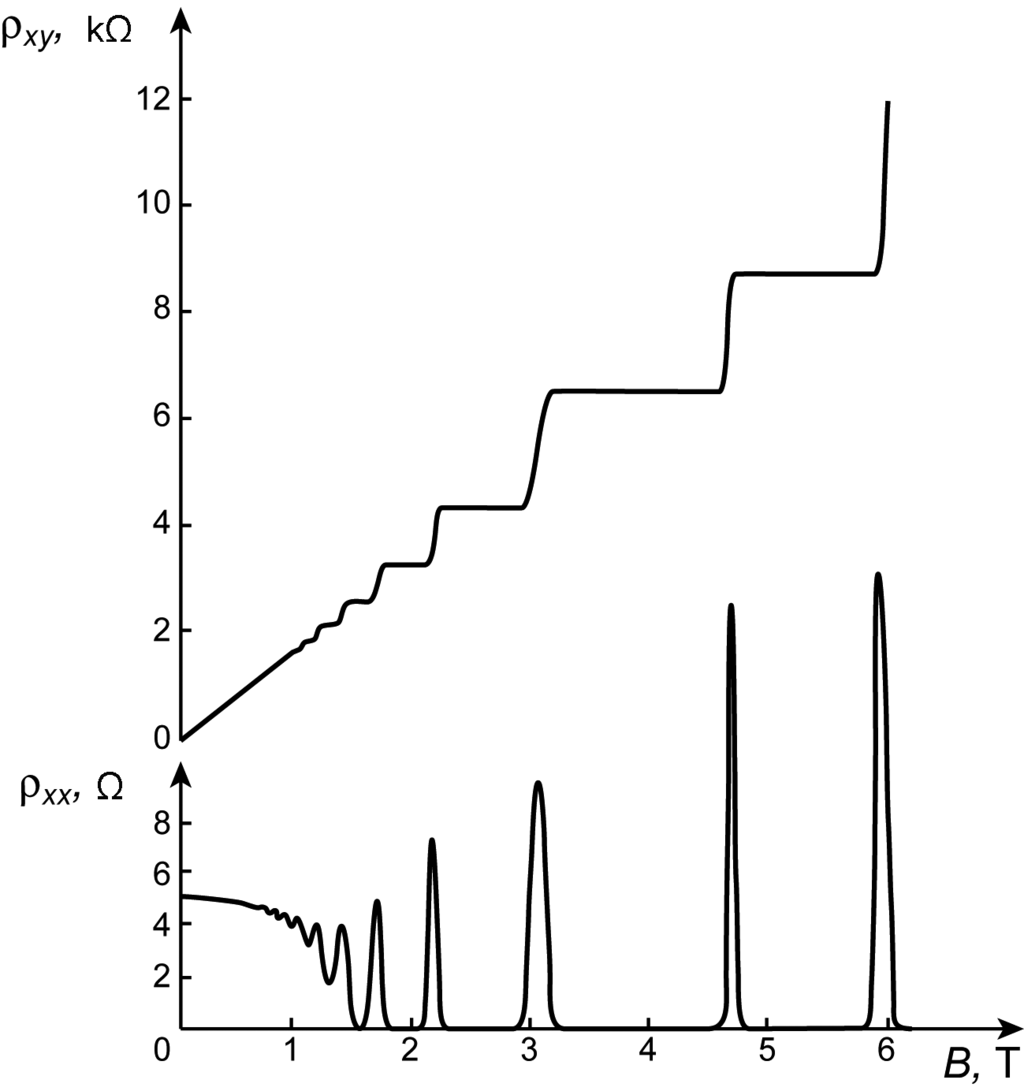}
\caption{Integer Quantum Hall Effect. Top: The transverse resistance (Hall resistance) displays sharp plateaus as a function of the magnetic field. Bottom: The longitudinal resistance peaks at transition points between the plateaus owing to the Schubnikov-de Haas effect~\cite{SdH_org,SdH_org2}. (Reproduced from \cite{QHE_Wiki} under the terms of the GNU Free Documentation License.)}
\label{fig:QHE}
\end{figure}

The major part of this thesis contains our contribution to the discovery and description of chiral topological PEPS~\cite{cPEPS_org,cPEPS_long,interacting_cPEPS}. In many cases, the interplay between the quantum particles of a topological state gives rise to macroscopic properties which are quantized and protected to disorder and perturbations: The field of topological systems emerged with the discovery of the Quantum Hall Effect in 1980 by von Klitzing~\cite{QHE_ex}, where it has been observed that for a sufficiently pure metallic sample at high magnetic fields, the transverse resistance (Hall resistance) displays sharp quantized plateaus if the magnetic field is varied, cf., Fig.~\ref{fig:QHE}. The Hall resistance is an integer multiple of $h/e^2$, which is why the effect is also denoted as the Integer Quantum Hall Effect. In the meantime, the accuracy of the quantization of the Hall conductance has been measured to be around one in a billion~\cite{QHE_precision}. Moreover, a closer investigation for sufficiently pure samples revealed several smaller plateaus with conductances that are fractional multiples of $h/e^2$, which is denoted as the Fractional Quantum Hall Effect discovered in 1982~\cite{FQHE_ex}. Due to the topological protection against perturbations and disorder, the following definition of a topological phase has been put forward: Two Hamiltonians are in the same topological phase if and only if they can be connected via a gapped path of Hamiltonians. The phase which contains the Hamiltonian in the atomic limit (i.e., where all sites of the lattice are decoupled) is also called the topologically trivial phase. The topological phases containing the Integer Quantum Hall states are labeled by the \textit{Chern number}~\cite{TKNN_Chern}, which is up to a sign the Hall conductance expressed in units of $h/e^2$.

Apart from systems that are intrinsically topological, i.e., that are stable to any small perturbation, the notion of \textit{symmetry protected topological phases}~\cite{SPT_org} has been coined: Some states that possess a certain symmetry are topologically protected against any small perturbation that does not break the symmetry. However, an arbitrarily small perturbation that does break the symmetry can destroy the topological phase. More mathematically, a symmetry protected topological phase can be considered as the equivalence class of Hamiltonians that possess a certain symmetry and can be connected by a gapped path of Hamiltonians that obey the same symmetry.

Coming back to ``genuine'' topological systems, another distinction is usually made: They are divided into short range entangled and long range entangled topological systems~\cite{SRE_Wen}. The latter are said to have \textit{topological order}~\cite{TO_org,TO_org2}. Such states are characterized by a constant sub-leading correction to the area law, i.e., the entanglement entropy of a region $\mc{R}$ grows like the size of its boundary minus a constant term, known as the \textit{topological entanglement entropy}~\cite{TEE_org,TEE_Levin}. A characteristic property of such systems is that the ground state degeneracy of the corresponding Hamiltonian depends on the topology of the manifold they are defined on. All other topological states are referred to as being short range entangled. They in particular include symmetry protected topological states, since, as elaborated above, they are connected to the trivial phase by small perturbations that break the corresponding symmetry. Another class of short range entangled states is the one of free fermionic systems, i.e., those fermionic systems that are described by a Hamiltonian that is quadratic in creation and annihilation operators. They can be diagonalized exactly by a computational cost that is polynomial in the system size, since there exist simple methods to write the Hamiltonian as a sum of terms that act only on single particle modes~\cite{Gaussian_Bravyi}. Due to this great simplification, a complete characterization of free fermionic phases has been achieved~\cite{Phases_Schnyder,Phases_Kitaev}. All free fermionic examples of ``genuine'' topological systems possess chiral edge modes at the boundary of the system, i.e., one or several quasi-particle modes that propagate in one direction along the boundary of the system and thus break time reversal symmetry. Those quasi-particle modes can be either fermionic modes, such as for the Integer Quantum Hall effect, or Majorana modes, such as in the case of topological superconductors. 

\begin{figure}
\centering{\includegraphics[width=0.25\columnwidth]{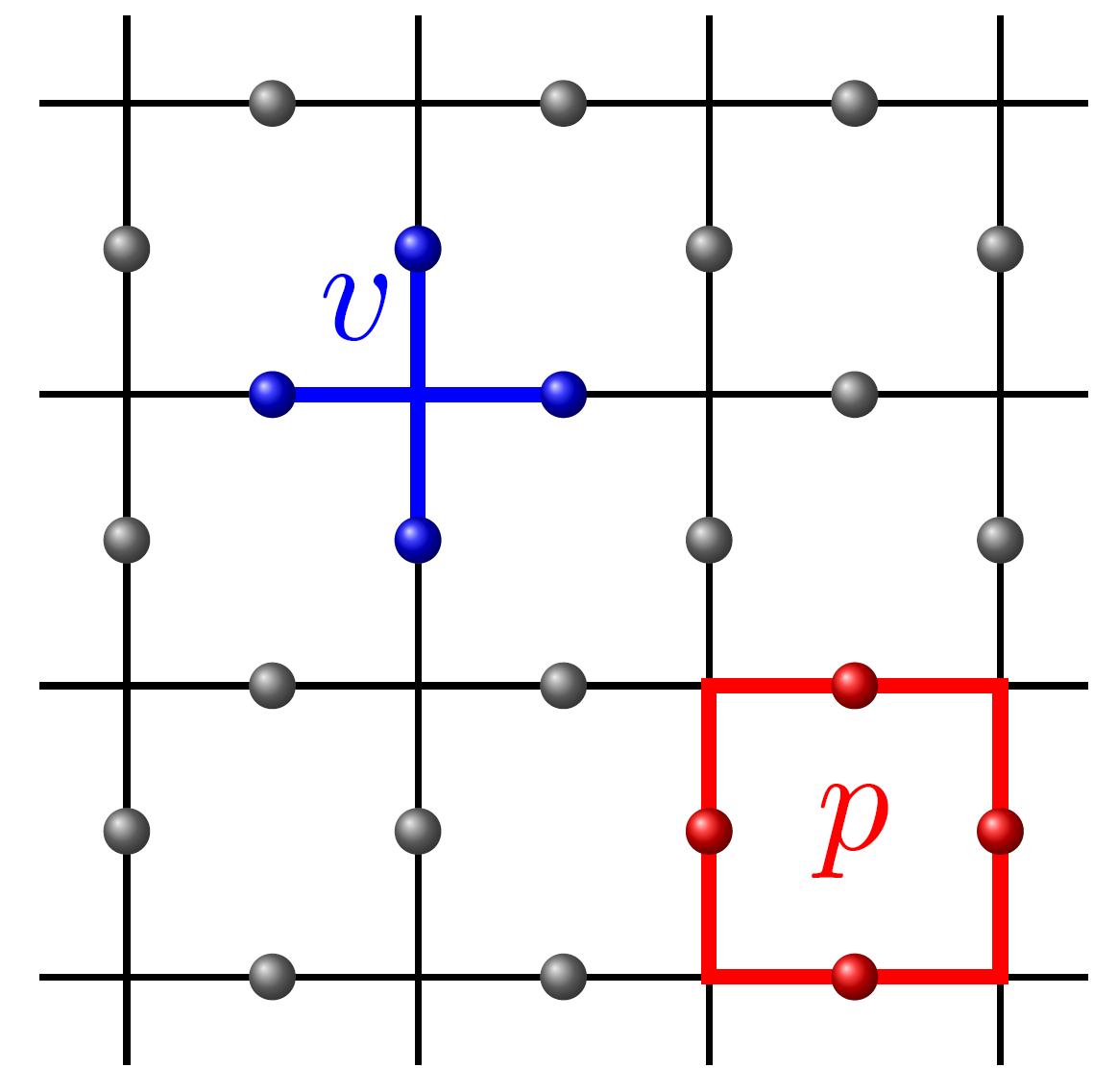}
\hspace{2cm} 
\includegraphics[width=0.27\columnwidth]{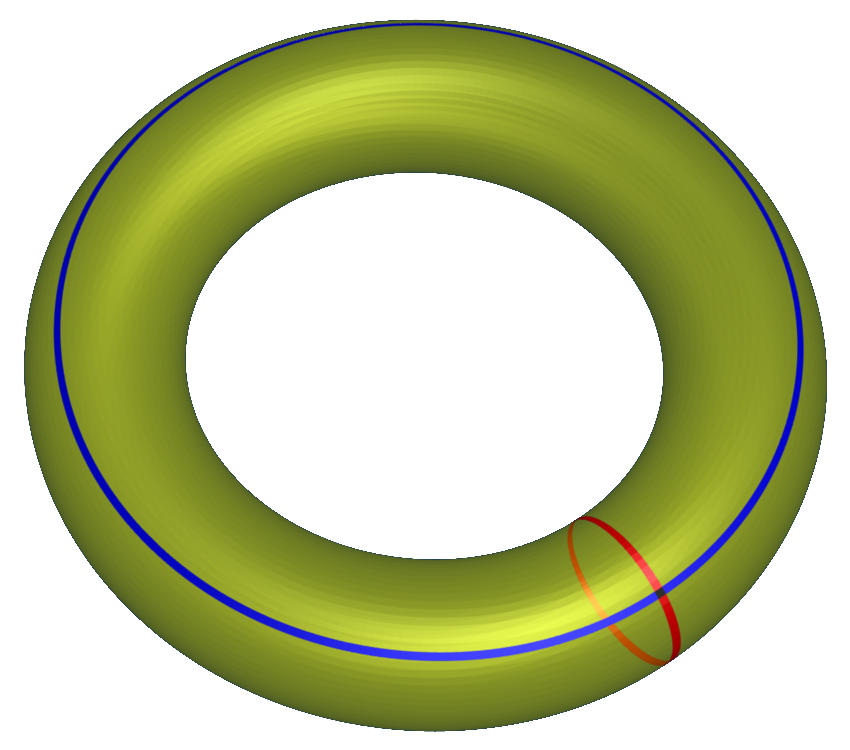}}
\caption{The toric code model. Left: The Hamiltonian of the toric code model is a sum of terms acting on plaquettes (red) and vertices (blue) of qubits on a lattice. Right: It has topological order and a ground state degeneracy of four if it is defined on a torus. The origin of the degeneracy is the existence two inequivalent non-contractible loops on the torus.}
\label{fig:toric}
\end{figure}

Moving on to the interacting regime, which includes interacting electrons and spin systems, the picture is less clear and a variety of ``genuine'' topological systems has been found theoretically, and in the case of the Fractional Quantum Hall Effect also experimentally. They no longer have to possess chiral edge modes (or even gapless edge states). Prominent examples are Resonating Valence Bond states~\cite{RVB_org} proposed by Anderson for the description of high-temperature superconductivity, Levin-Wen~\cite{Levin_Wen} models and the toric code~\cite{toric_code} (see Fig.~\ref{fig:toric}). Being all topologically ordered, these examples have been studied intensively, since the topological degeneracy of the ground state subspace is stable to local perturbations and can thus be used as a quantum memory. Moreover, several systems with topological order have non-Abelian anyonic excitations, whose braiding might be used in order to carry out topologically protected quantum computations. It is thus a formidable task to find efficient numerical and analytic tools to describe those systems. 

While for the above non-chiral examples of topologically ordered systems exact PEPS representations with low bond dimensions exist~\cite{RVB_PEPS, Levin_PEPS, toric_PEPS}, no example of a PEPS has been known previously that represents a chiral topological system, not even in the free fermionic limit. This circumstance invoked arguments that seemed to rule out the existence of chiral PEPS. One of them is that at least free fermionic chiral PEPS could not exist, since the fact that they are generically unique frustration free ground states of local Hamiltonians allows for the construction of a complete set of localized Wannier wave functions~\cite{Wannier_org,Wannier_org2} - something that is incompatible with a non-vanishing Chern number~\cite{exp_Wannier}. However, this work presents the discovery of the first family of free fermionic chiral PEPS~\cite{cPEPS_org}, which was obtained independently from a similar example provided by Dubail and Read~\cite{cPEPS_Read} at the same time. Hence, PEPS \textit{can} be used for the description of chiral systems, as will be demonstrated by concrete numerical examples. Moreover, the properties of such chiral PEPS will be thoroughly investigated and eventually an example of an interacting chiral PEPS will be presented, whose construction is based on two copies of free fermionic chiral PEPS. This opens the doors to the description of general chiral topological systems by PEPS, from the numerical side, but also from the analytic side, via the construction of new interacting chiral models with exact PEPS ground states. This will enhance our understanding of interacting chiral topological systems and might eventually even help to classify them.

This thesis is organized as follows:
Chapter 2 presents the description of states with long range localizable entanglement by MPS. Its first section, MPS will be introduced via the local projection of maximally entangled pairs that are located between neighboring sites. The graphical representation of MPS and the connection to the area law will be explained. Furthermore, it will be shown how symmetries on the physical level can be locally incorporated into the MPS tensor and also how it can be used to decide whether the corresponding parent Hamiltonian has a gap in the thermodynamic limit. 
In section~\ref{sec:LE_intro}, the notion of localizable entanglement will be introduced and its significance for quantum repeaters and the detection of quantum phase transitions. Our new findings are contained in section~\ref{sec:LRLE}, where it will be shown how to determine if a given translationally invariant MPS has LRLE. Based on that condition, a criterion for all such MPS will be presented, which will in turn be used to parameterize all MPS with LRLE for bond dimension $2$ and $3$ explicitly. 

Chapter 3 is devoted to chiral PEPS. In section~\ref{sec:PEPS_intro} the construction of section~\ref{sec:MPS_intro} will be extended to two dimensional systems. This construction will be extended both to general fermionic systems and explicitly to free fermionic systems. Furthermore, a special emphasis will be made on how physical symmetries and topological properties can be encoded in the PEPS tensor.
Section~\ref{sec:chiral_intro} contains a general introduction to chiral topological systems, first focusing on free fermionic topological systems and later providing tools from Conformal Field Theory~\cite{CFT_Ginsparg,yellow_book} for the description of interacting chiral fermionic phases. 
After establishing these concepts, the examples of chiral free fermionic PEPS we obtained will be presented in section~\ref{sec:cGFPEPS}. It will be shown that there are strong restrictions on free fermionic PEPS that support chiral edge modes - which in all generality lead to algebraically decaying correlations. However, they are nonetheless unique ground states of long-ranged gapped Hamiltonians, which will be demonstrated to be topologically protected against perturbations. 
The correspondence between the bulk energy spectrum and the spectrum of the boundary reduced density matrix will be conveyed (\textit{bulk-boundary correspondence}~\cite{bulk_boundary}).
A detailed treatment of the symmetries of chiral free fermionic PEPS will be given, which will eventually be used to fully classify all such PEPS with the smallest non-trivial bond dimension. 
Section~\ref{sec:cPEPS} includes our results on chiral PEPS with topological order. Such a PEPS is obtained by applying a Gutzwiller projector on two copies of a free fermionic chiral PEPS. The resulting symmetries of the PEPS tensor will be presented and used to construct all ground states of the local frustration-free parent Hamiltonian. Furthermore, it will be explained how the bulk-boundary correspondence can be used to calculate the entanglement spectra~\cite{ES_org} and how the later can be used to confirm the expected Conformal Field Theory of the model~\cite{ES_extract} (which characterizes the topological state). Our calculations showing that the gap of the transfer operator vanishes in the thermodynamic limit are presented. This implies that, again, correlations decay algebraically in real space.  

Finally, chapter 4 concludes our results and gives an outlook on future research directions.

\noindent Chapter 2 and 3 are two mostly self-contained parts and can be read independently.

%% file: 2.1.tex
\chapter{Matrix Product States with Long Range Localizable Entanglement}
\chaptermark{MPS with Long Range Localizable Entanglement}

\section{Introduction to Matrix Product States}\label{sec:MPS_intro}

\subsection{Background}

As illustrated above, Matrix Product States (MPS) are well suited for the description of ground states of local gapped Hamiltonians in one dimension, as both fulfill the area law (this argument becomes rigorous if one refers to the area law of the R\'{e}nyi entropy with parameter $\alpha < 1$~\cite{area_efficient}). Historically, the first non-trivial example of an MPS was provided by Affleck, Kennedy, Lieb and Tasaki (AKLT)~\cite{AKLT_org} in 1987 as the ground state of a certain parameter value of the bilinear biquadratic spin-1 Heisenberg Hamiltonian. However, they were formally introduced only in 1992 under the name of Finitely Correlated States~\cite{FCS_org}, which are translationally invariant MPS. They captured broad interest only from 2007 on when it was realized~\cite{MPS_math} that their non-translationally invariant version is the variational class underlying the DMRG algorithm~\cite{DMRG_org}. Furthermore, it was observed that many statements about the whole MPS can be made (e.g., regarding the decay of correlations~\cite{MPS_math}, symmetries~\cite{MPS_sym}, etc.) by studying the rank-3 tensors the MPS is composed of. In the meantime, MPS have been generalized to higher dimensional systems, resulting in Projected Entangled Pair States (PEPS), which are the subject of chapter 3, to critical systems violating the area law (Multiscale Entanglement Renormalization Ansatz~\cite{MERA_org}) and to continuous systems~\cite{cMPS, cMPS_calculus} (both not discussed in this work).

This section is organized as follows: In the next subsection, the construction of MPS via pairs of maximally entangled spins is presented. The final form of MPS will be given, along with their graphical representation, which will be very useful to illustrate the contraction scheme of PEPS. 
Subsection~\ref{sec:area} elaborates on the area law and the resulting efficient approximability of ground states of local Hamiltonians by MPS. In subsection~\ref{sec:symmetries} symmetries and, finally, in subsection~\ref{sec:MPS_parent} parent Hamiltonians for MPS will be considered.

\subsection{Construction}\label{sec:MPS_construction}

We consider a chain of $N$ spin-$S$ particles. In order to construct an MPS for this system, we attach two virtual particles of spin $S'$ to each site, one on the left and on one the right. Furthermore, we assume that they are in a product state of maximally entangled pairs connecting neighboring sites. E.g., the right virtual particle of site 1 and the left virtual particle of site 2 are in a maximally entangled state $|\omega_{1,2} ) = \sum_{\alpha = 1}^D |r_1 = \alpha, l_2 = \alpha)$, where $D = 1 + 2 S'$ is the \textit{bond dimension} (dimension of the Hilbert space of each virtual particle) and round brackets are used for vectors defined in the virtual Hilbert space. 

\begin{figure}
\centering\includegraphics[width=\columnwidth]{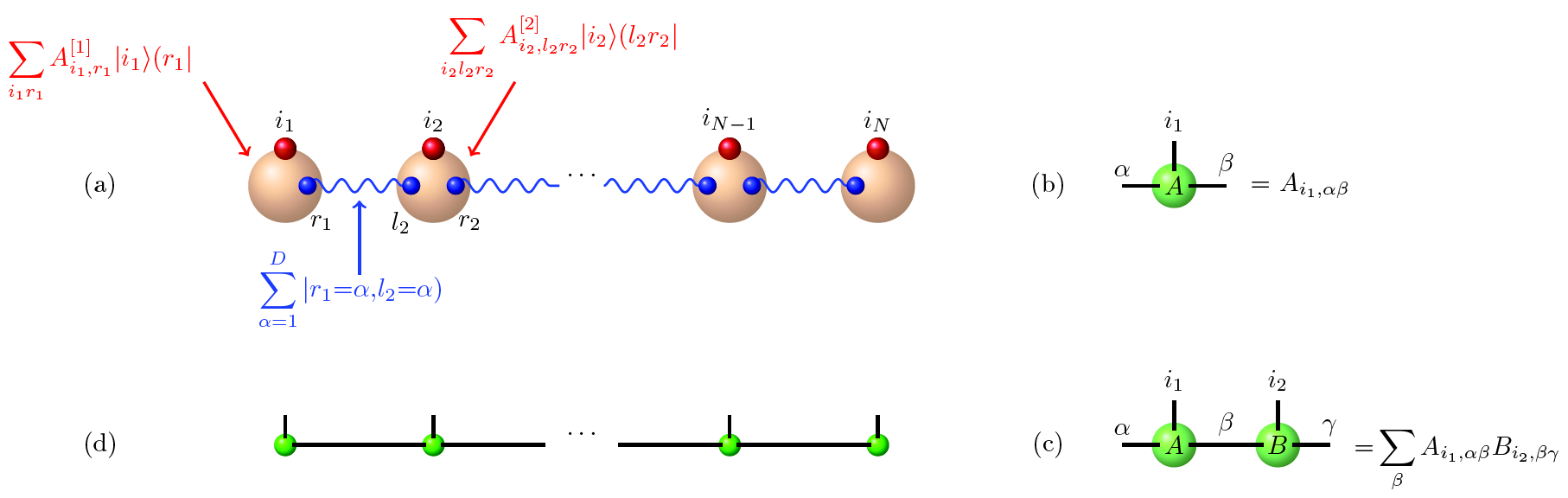}
\caption{Construction of an MPS with open boundary conditions. (a) Maximally entangled virtual pairs are placed between neighboring sites (blue balls). Afterwards, linear maps (red transparent spheres) are carried out on each site from the local Hilbert space of virtual particles to the local Hilbert space of the physical particle (red balls), see text for details. (b) Graphical representation of a rank-3 tensor. Each index of the tensor is represented by one leg. (c) Graphical representation of a contraction of two rank-3 tensors. Connected legs correspond to contractions (in this case over the index $\beta$. (d) Pictorial representation of an MPS with open boundary conditions, as constructed in (a). Open legs are physical indices, which one needs to multiply with the corresponding configuration $|i_1 i_2 \ldots i_N\rangle$ of the overall physical Hilbert space and to subsequently sum over, as in Eq.~\eqref{eq:MPS_OBC}.}
\label{fig:MPS_OBC}
\end{figure}

\begin{figure}
\centering\includegraphics[width=0.8\columnwidth]{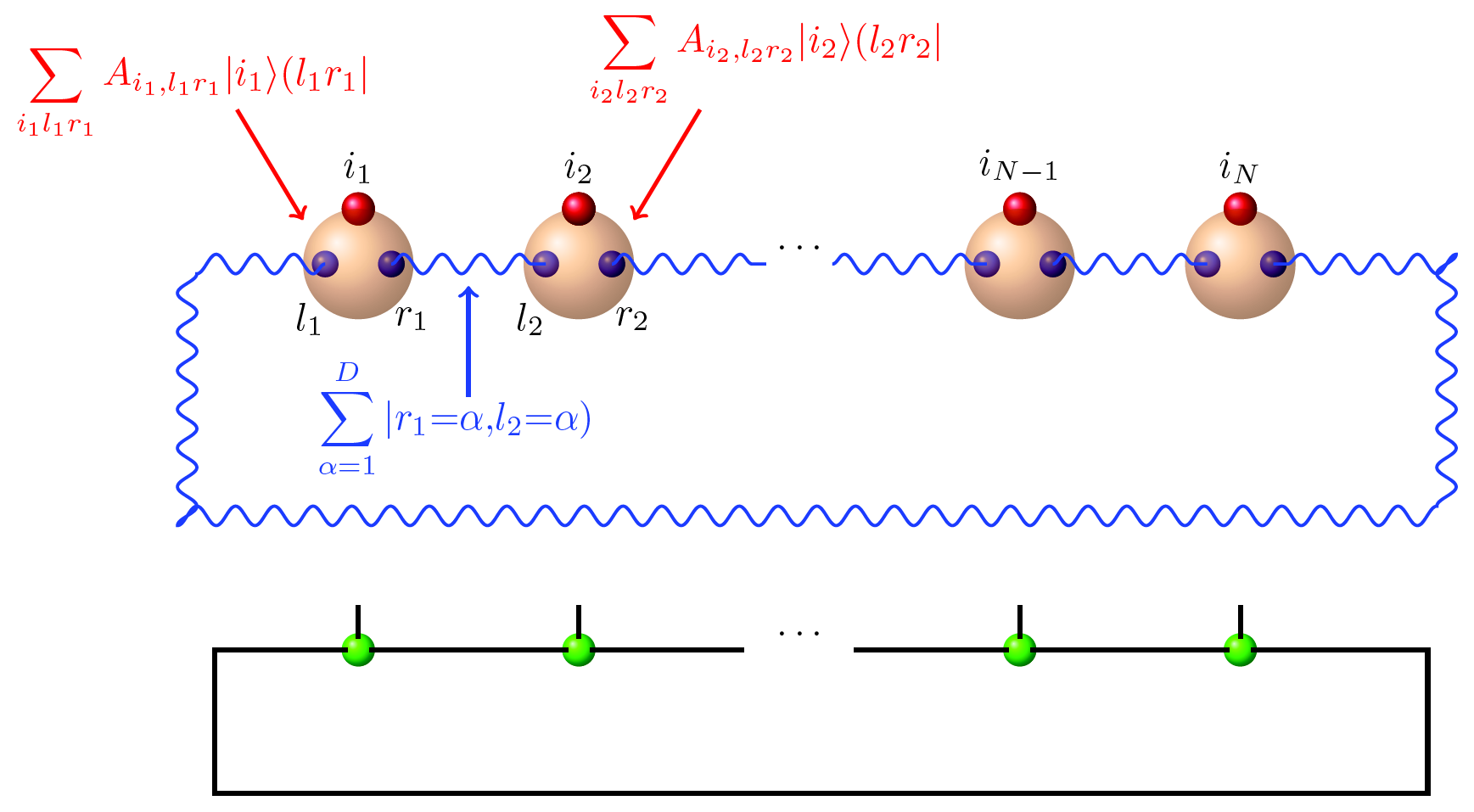}
\caption{Construction of an MPS with periodic boundary conditions. Top: Maximally entangled virtual pairs (blue balls) are placed between neighboring sites and also connect the first and last site. Bottom: Graphical representation as in Fig.~\ref{fig:MPS_OBC}(b-d) of the obtained MPS, Eq.~\eqref{eq:MPS_PBC}.}
\label{fig:MPS_PBC}
\end{figure}

In the case of open boundary conditions, the first site has only a right virtual particle and the $N$-th site only a left virtual particle, see Fig.~\ref{fig:MPS_OBC}. For periodic boundary conditions, each site has two virtual particles, where the left one of the first site is maximally entangled with the right one of the $N$-th, cf., Fig.~\ref{fig:MPS_PBC}. In both cases, we perform an arbitrary linear map $\mc{M}_n$ at each site $n = 1, \ldots, N$ from the Hilbert space of the virtual particles to the Hilbert space of the corresponding physical particle. The map $\mc{M}_n$ can be written in terms of a rank-3 tensor $A_{i,l r}^{[n]}$
\be
\mc{M}_n = \sum_{i = 1}^d \sum_{l,r = 1}^D A_{i,l r}^{[n]} |i\rangle(l,r|,
\ee
where $d = 1 + 2 S$ is the dimension of the physical Hilbert space for each site. In the case of open boundary conditions, the maps for $n = 1$ and $n = N$ are different: For them, it is sufficient to use rank-2 tensors (i.e., matrices),
\be
\mc{M}_1 = \sum_{i = 1}^d \sum_{r = 1}^D A_{i,r}^{[1]} |i\rangle(r|, \ \mc{M}_N = \sum_{i = 1}^d \sum_{l = 1}^D A_{i,l}^{[N]} |i\rangle(l|.
\ee
Hence, the resulting state for open boundary conditions is
\begin{align}
|\Psi_\mr{OBC}\rangle = \left(\bigotimes_{n=1}^N \mc{M}_n \right) \left(\bigotimes_{n=1}^{N-1} |\omega_{n,n+1}) \right)=\sum_{i_1, i_2, \ldots, i_N} A^{[1]}_{i_1} A^{[2]}_{i_2} \ldots A^{[N]}_{i_N} |i_1 i_2 \ldots i_N\rangle,
\label{eq:MPS_OBC}
\end{align}
where $A^{[n]}_{i_n}$ represents the matrix with entries $A^{[n]}_{i_n,l_n r_n}$. Hence, Eq.~\eqref{eq:MPS_OBC} is a product of matrices for each configuration $i_1, \ldots i_N$, thus the name Matrix Product State. Note that $A^{[1]}_{i_1}$ is a row vector and $A^{[N]}_{i_N}$ a column vector. The other matrices can also be chosen to have other dimensions ($A^{[n]}_{i_n}$ being a $D_{n-1} \times D_n$ matrix) by using virtual particles of different spins $S'$. This case is included in the above construction if one sets $D = \max_n D_n$.
The graphical representation of Eq.~\eqref{eq:MPS_OBC} is given in Fig.~\ref{fig:MPS_OBC}. 

On the other hand, for periodic boundary conditions, one can easily derive
\begin{align}
|\Psi_\mr{PBC}\rangle = \left(\bigotimes_{n=1}^N \mc{M}_n \right) \left(\bigotimes_{n=1}^{N} |\omega_{n,n+1}) \right) =\sum_{i_1, i_2, \ldots, i_N} \tr(A^{[1]}_{i_1} A^{[2]}_{i_2} \ldots A^{[N]}_{i_N} ) |i_1 i_2 \ldots i_N\rangle
\label{eq:MPS_PBC}
\end{align}
with $|\omega_{N,N+1}) \equiv |\omega_{N,1})$, as shown in Fig.~\ref{fig:MPS_PBC}. In Ref.~\cite{MPS_math} it has been proven that MPS with periodic boundary conditions can be represented by the same rank-3 tensor $A_{l r}^i$ for each site, so the site index $[n]$ will be dropped in equations relating to periodic boundary conditions from now on. 

It is obvious that an MPS with open boundary conditions is invariant under the replacement $A^{[n]}_i \rightarrow (Y^{[n-1]})^{-1} A^{[n]}_i Y^{[n]}$ for a set of invertible matrices $Y^{[n]}$. For periodic boundary conditions, the MPS is invariant under $A_{i} \rightarrow Y^{-1} A^{[n]}_i Y$ for any invertible $D \times D$ matrix $Y$.
This \textit{gauge freedom} can be used to derive canonical forms for MPS~\cite{MPS_math}: Any MPS with open boundary conditions and bond dimension $D$ can be expressed as in Eq.~\eqref{eq:MPS_OBC}, where the rank-3 tensors $A^{[n]}_i$ fulfill
\begin{align}
\sum_{i=1}^d A^{[n]}_i (A^{[n]}_i)^\dg &= \Id \\
\sum_{i=1}^d (A^{[n]}_i)^\dg \Lambda^{[n-1]} A^{[n]}_i &= \Lambda^{[n]},
\end{align}
where $\Lambda^{[0]} = \Lambda^{[N]} = 1$ and $\Lambda^{[n]}$ is a diagonal strictly positive $D \times D$ matrix with $\tr(\Lambda^{[n]}) = 1$. For translationally invariant MPS, the rank-3 tensors $A_i$ can be gauged to a block form
\begin{align}
A_i &= \left(\begin{array}{cccc}
\lambda^{(1)} A^{(1)}_i&0&&0\\
0&\lambda^{(2)} A^{(2)}_i&&0\\
&&\ldots\\
0&0&&\lambda^{(r)} A^{(r)}_i
\end{array}\right), \label{eq:A_blocks}\\
\sum_{i=1}^d A_i^{(b)} (A_i^{(b)})^\dg &= \Id, \label{eq:FP_e}\\
\sum_{i=1}^d (A_i^{(b)})^\dg \Lambda^{(b)} A^{(b)}_i &= \Lambda^{(b)}. \label{eq:FP_oe}
\end{align}
The gauge given by Eqs.~\eqref{eq:A_blocks} to~\eqref{eq:FP_oe} is fixed in such a way that the \textit{completely positive map} $\mc{E}^{(b)} (X) = \sum_{i=1}^d A_i^{(b)} X (A^{(b)}_i)^\dg$ with $b = 1, \ldots, r$ (studied in more detail in subsection~\ref{sec:MPS_parent}) has a unique fixed point of maximum absolute eigenvalue, which is $\Id$. Note that an MPS with tensors as in Eq.~\eqref{eq:A_blocks} is a sum of $r$ MPS whose tensors are the diagonal blocks. In the limit of $N \rightarrow \infty$ only those terms corresponding to the $\lambda^{(b)}$'s of maximum magnitude contribute to the sum.

Using the completely positive map, one can easily show that the correlations of any MPS decay exponentially in real space~\cite{FCS_org}: Let us consider the completely positive map of one block (we drop the index $(b)$ for clarity). It is equivalent to the so-called \textit{double tensor} $\mathbb T = \sum_{i=1}^d A_i \otimes A_i^\dg$. It has $1$ as its highest eigenvalue and we denote by $|R\rangle$ and  $\langle L|$ the corresponding right and left eigenvector. For simplicity, let us assume that the eigenvalue of second highest magnitude, $\zeta_1$, is non-degenerate (the proof for the degenerate case is completely analogous). We call the respective right and left eigenvectors $|R_1 \rangle$ and $\langle L_1|$. For large $m$, the leading terms of $\mathbb T^m$ are thus $|R\rangle\langle L| + \zeta_1^m |R_1\rangle \langle L_1| + \mc{O}\left(\left|{\zeta_2}\right|^m\right)$, where $|\zeta_2| < |\zeta_1| < 1$. The correlation of two single-site operators $O_1$ and $O_2$  is given by
\be
C(O_1^n,O_2^{n+x}) = \langle O_1^n O_2^{n+x} \rangle - \langle O_1^n \rangle \langle O_2^{n+x} \rangle,
\ee
where $O_1$ acts on site $n$ and $O_2$ on site $n + x$. In the case of our MPS described by the tensor $A$, we obtain for $x \ll N$ (with $\mathbb{O}_{1,2} := \sum_{i,j=1}^d  [O_{1,2}]_{ij} A_i \otimes A_j^\dg$)
\begin{align}
C(O_1^n,O_2^{n+x}) &= \tr\left(\mathbb{T}^{n-1} \mathbb{O}_1 \mathbb{T}^{x-1} \mathbb{O}_2 \mathbb{T}^{N-n-x}\right) - \tr\left(\mathbb{T}^{n-1} \mathbb{O}_1 \mathbb{T}^{N-n}\right) \tr\left(\mathbb{T}^{n+x-1} \mathbb{O}_2 \mathbb{T}^{N-n-x}\right) \notag \\
&= \tr\left(\mathbb{T}^{N-x-1} \mathbb{O}_1 \mathbb{T}^{x-1} \mathbb{O}_2\right) - \tr\left(\mathbb{T}^{N-1} \mathbb{O}_1 \right) \tr\left(\mathbb{T}^{N-1} \mathbb{O}_2 \right) \notag \\
&= \langle L| \mathbb{O}_1 \mathbb{T}^{x-1} \mathbb{O}_2 | R \rangle - \langle L | \mathbb{O}_1 |R\rangle \langle L| \mathbb{O_2}|R\rangle + \mc{O}(|\zeta_1|^{N-x-1}) \notag \\
&= \zeta_1^{x-1} \langle L|\mathbb{O}_1 |R_1\rangle \langle L_1| \mathbb{O}_2 |R\rangle\left(1 + \mc{O}(\left|{\zeta_2}/{\zeta_1}\right|^{x-1})\right) + \mc{O}(|\zeta_1|^{N-x-1}). \label{eq:exp_corr}
\end{align}
This demonstrates the exponential decay of all two-point correlations for the MPS corresponding to a single block $(b)$ in Eq.~\eqref{eq:A_blocks} and therefore also for the full MPS, which is a superposition of the MPSs corresponding to individual blocks. 

In subsection~\ref{sec:PEPS_transfer}, we will come back to the above approach in the context of PEPS, where cases with algebraically decaying correlation functions exist.

\subsection{The area law and efficient approximability}\label{sec:area}

Why are MPS useful for the description of one-dimensional spin systems? The reason is that ground states of local gapped Hamiltonians obey the \textit{area law}~\cite{area_Rev}, which constrains such ground states to occupy only a subspace of the full Hilbert space that grows polynomially with the system size $N$. As MPS (and also PEPS) by construction fulfill the area law, their cost to describe such ground states grows only polynomially with the system size, too. Let us settle these statements more concretely for the one dimensional case (MPS): In one dimension, the area law demands that for the reduced density matrix $\rho_L$ of a local gapped Hamiltonian the \textit{entanglement entropy} saturates as a function of $L$ (for $N \gg L$). The entanglement entropy $S_{\mr{v N}}$~\cite{vN_ent} (also known as the \textit{von Neumann entropy}) of a density matrix $\rho$ is defined as 
\be
S_{\mr{vN}}(\rho) = - \tr(\rho \log(\rho)).
\ee
It is the limiting case $\alpha \rightarrow 1$ of the R\'enyi entropy~\cite{Renyi_ent}
\be
S_\alpha(\rho) = \frac{1}{1-\alpha} \log(\tr(\rho^\alpha)).
\ee
Hence, the area law states $S_\mr{vN}(\rho_L) \leq c$ for arbitrary $L$. That MPS fulfill the area law can be seen by considering the Schmidt decomposition of a region $\mc R$ of length $L$ with the rest
\be
|\Psi \rangle = \sum_{\kappa} \mu_\kappa |\psi_{\mc R,\kappa}\rangle \otimes | \psi_{\overline{\mc R},\kappa} \rangle,
\label{eq:SD_1D}
\ee
where $\psi_{\mc R, \kappa}$ lives in region $\mc R$, $\psi_{\overline{\mc R},\kappa}$ refers to the complementary region $\overline{\mc R}$ and $\mu_\kappa > 0$. If one decomposes the wave functions described by Eqs.~\eqref{eq:MPS_OBC} and~\eqref{eq:MPS_PBC} into a connected region of $L$ sites and its complement, i.e., a basis for $\psi_{\mc R}$ is $|i_{x+1}, i_{x+2}, \ldots, i_{x+L}\rangle$, it becomes obvious that there can be at most $D^2$ Schmidt coefficients $\mu_\kappa$. Realizing that the eigenvalues of $\rho_L$ are the Schmidt coefficients squared, the von Neumann entropy of~\eqref{eq:SD_1D} reads
\be
S_{\mr{vN}}(\rho_L) = - \tr(\mu_\kappa^2 \log(\mu_\kappa^2)) \leq 2 \log(D),
\ee
since the entropy is maximized for $\mu_\kappa^2 = \frac{1}{D^2}, \ \kappa = 1, \ldots, D^2$. Hence, the von Neumann entropy is bounded by $\log(D)$ times the number of bonds that are cut when tracing out all but $L$ consecutive sites (cf. Figs.~\ref{fig:MPS_OBC} and~\ref{fig:MPS_PBC}). By the same token, this can be shown for PEPS.

The argument that both gapped ground states of local Hamiltonians and MPS fulfill the area law is of course by far not sufficient to prove that they can be used to efficiently approximate the ground states of those Hamiltonians. The rigorous proof can be found in Ref.~\cite{MPS_faithful} and shows that for any local one-dimensional Hamiltonian acting on $N$ sites whose ground state $|\Psi_0(N)\rangle$ violates the area law at most logarithmically, an MPS approximation $|\Psi_\mr{MPS}(D)\rangle$ can be found with overlap $|\langle \Psi_\mr{MPS}(D)|\Psi_0(N)\rangle| \geq 1 - \epsilon$ with $D$ polynomial in $N$ for fixed $\epsilon > 0$. Therefore, also critical systems can be efficiently approximated by MPS, where MERA, however, scales better as a function of the system length.

Let us add that in Ref.~\cite{poly_time} it has been shown that MPS do not only allow for an efficient representation of ground states of gapped local Hamiltonians, but that this representation can also be found numerically in a time that is polynomial in the system size.

\subsection{Symmetries}\label{sec:symmetries}

The results of Refs.~\cite{MPS_string,MPS_sym} shall be restated here (for a related approach, see Ref.~\cite{MPS_renormalization}): Let us consider translationally invariant states with local symmetries with a symmetry group $\mc G$, which is represented by the unitary matrices $u^g$ acting on the physical degree of freedom, that is,
\be
(u^g)^{\otimes N} |\Psi\rangle = e^{i N \theta^g} |\Psi \rangle, \label{eq:def_symmetry}
\ee
where the phases $\theta^g$ form a one dimensional representation of $\mc G$. 

Let us now consider separately the two cases of discrete and continuous $\mc G$. 
If $\mc G$ is discrete and $|\Psi\rangle$ is an MPS obeying that symmetry, its tensor $A$ fulfills
\be
\sum_{i j} u_{ij}^g A_j = W U A_i U^{\dg},
\ee
with $U = P(\oplus_{b=1}^r V^{(b)})$ acting only on the virtual degrees of freedom and $P$ being a matrix permuting the blocks with index $b$ as in Eq.~\eqref{eq:A_blocks}. $W$ is given by $W = \oplus_{b=1}^r e^{i \vartheta_b} \Id^{(b)}$ (with arbitrary phases $\vartheta_b$). 

For a compact connected Lie group $\mc G$ representing a continuous symmetry (such as rotation symmetry), the permutation matrix $P$ is trivial~\cite{MPS_sym}, which can be used to show that  
\be
\sum_{i j} u_{ij}^g A_j^{(b)} = e^{i \theta^g} V^{(b) g} A_i^{(b)} (V^{(b) g})^\dg,
\ee
with $b = 1, \ldots r$ and the phases $\theta^g$ as in Eq.~\eqref{eq:def_symmetry}. The matrices $V^{(b)g}$ form a projective representation of the symmetry group $\mc G$. 

\subsection{Injectivity and parent Hamiltonians}\label{sec:MPS_parent}

For each translational invariant MPS a Hamiltonian can be found that acts on $L$ sites and has the MPS as a ground state. The question is whether this ground state is unique. To tackle this problem, the concept of \textit{injectivity} of MPS has been introduced~\cite{FCS_org,MPS_math}: Consider the map
\be
\Gamma_L: X \rightarrow \sum_{i_1, ..., i_L = 1}^d \tr(X A_{i_1} ... A_{i_L}) |i_1 ... i_L\rangle
\label{eq:def_inj}
\ee
with $A$ being the tensor of a translationally invariant MPS in the canonical form~\eqref{eq:A_blocks}.
An MPS for which the map $\Gamma_{L_0}$ is injective, but not for any $L < L_0$ is said to be $L_0$-injective or simply injective. In this case, $L_0$ is denoted as the \textit{injectivity length} of the MPS. If $\sum_{i=1}^d A_i A_{i}^\dg = \Id$ and one has only one block in Eq.~\eqref{eq:A_blocks}, $L_0$-injectivity implies that $\Gamma_L$ is injective for any $L \geq L_0$ (e.g., for $L = L_0 + 1$ this is shown by inserting $X' = \sum_{i_{L_0+1}} c_{i_{L_0+1}}  A_{i_{L_0+1}} X$  such that $\sum_{i_{L_0+1}} c_{i_{L_0+1}} A_{i_{L_0+1}}$ is full rank into Eq.~\eqref{eq:def_inj}). In words, an $L_0$-injective MPS has the property that by blocking $L_0$ sites the whole virtual space is accessible by acting on the physical space of the block (or, conversely, the virtual space is fully mapped to the physical space).

Injectivity turns out to be a generic property of MPS, i.e., the tensors $A$ which correspond to non-injective MPS have Haar measure zero. For injective MPS it has been demonstrated~\cite{MPS_math} that there is only one block in the canonical form Eq.~\eqref{eq:A_blocks} and that the reduced density operator $\rho_L$ has rank $D^2$ for $L \geq L_0$. For injective MPS, $L_0 \leq (D^2 - d + 1) D^2$

Quite related to injectivity is the \textit{completely positive map} $\e$ defined via
\be
\e(X) = \sum_{i=1}^d A_i X A_{i}^\dg.
\label{eq:CP_map}
\ee
If and only if $\e$ has a unique eigenvalue of magnitude 1 (the eigenvector is $\Id$), \eqref{eq:def_inj} is injective~\cite{inj_length}. 

Let us consider how to construct the so-called \textit{parent Hamiltonian} of an MPS, which is frustration free and has the MPS as its ground state. By frustration free it is meant that the Hamiltonian is a sum of projectors (possibly with prefactors) that each act only on a finite range of sites. E.g., a parent Hamiltonian with sums of projectors $h$ acting non-trivially on $L$ sites can be obtained by calculating the reduced density matrix $\rho_L$ of the MPS and setting
\begin{align}
\mc H &= \sum_{n = 1}^N h_n, \label{eq:parent_Ham} \\
\ker(h) &= \mr{supp}(\rho_L).
\end{align}
$\ker(h)$ denotes the kernel (the null space) of a matrix $X$ and $\mr{supp}(X)$ represents its support, i.e., the span of its column vectors. The reduced density matrix $\rho_L$ can be calculated via
\be
\rho_L = \sum_{\substack{i_1, \ldots, i_L \\ j_1, \ldots, j_L}} \tr\left(A_{i_1} \ldots A_{i_L} A_{j_L}^\dg \ldots A_{j_1}^\dg\right) |i_1 \ldots i_4 \rangle \langle j_1 \ldots j_L|,
\ee
assuming that the MPS is normalized according to Eqs.~\eqref{eq:A_blocks} and~\eqref{eq:FP_e}. That is, the reduced density matrix can be obtained by the contraction of a block of $L$ sites.

For any MPS with injectivity length $L_0$, there is a gapped parent Hamiltonian which is a sum of projectors acting on $L_0+1$ sites and has the MPS as its \textit{unique} ground state. Note that for such an MPS it is also possible to construct a Hamiltonian with a continuous spectrum and the MPS as its unique ground state, denoted as the \textit{uncle Hamiltonian}~\cite{uncle_Ham}.

%% file: 2.2.tex
\section{Introduction to Localizable Entanglement}\label{sec:LE_intro}

\subsection{Background}

Soon after a theoretical framework for MPS had been established~\cite{MPS_math}, they were used to detect certain kinds symmetries and orders that may be present in a system. One of them is string order~\cite{SO_org}, which is a non-local quantity defined as the expectation value of a string of unitaries multiplied with local operators acting on the end sites of the string. Within the set of MPS it was shown to be related to the existence of a \textit{local} symmetry in the system. Another such quantity, which appears to be truly non-local in general, is the \textit{localizable entanglement} (LE)~\cite{LE_org}. It quantifies the amount of entanglement that can on average be generated between the two spins at the ends of a spin chain by measuring all the remaining ones. It can be lower bounded by two-point connected correlation functions (this has been shown rigorously for qubits in Ref.~\cite{LE_corr} and extended to spin-1 systems in Ref.~\cite{LE_org}). 

LE is not only of importance as a non-local order parameter, but also as a resource for quantum communication, because if it is close to 1 for a long spin chain, one can create with high probability highly entangled states over large distances. The limiting case is known as \textit{long range localizable entanglement} (LRLE), which means that the LE is finite even for an infinitely long spin chain. This scenario corresponds to an ideal \textit{quantum repeater}~\cite{quantum_repeater}, which is a device to create maximally entangled pairs over large distances. LRLE is also relevant for Condensed Matter Physics, as, like string order, it appears at certain quantum phase transitions~\cite{LRLE_QPT} and can therefore be used for their detection. 

We will proceed by defining LE and revealing its long range behavior at certain quantum phase transitions.

\subsection[Definition and relation to quantum phase transitions]{Definition and relation to quantum phase transitions\footnote{The remainder of this chapter is a slight modification of the content of Ref.~\cite{MPS_LRLE}, copyright American Physical Society.}}

The LE is defined as the maximum average of the entanglement that can be generated between two spins of a spin chain by measuring the remaining ones~\cite{LE_org}. 
Let $\rho$ denote the density matrix of the original state. With probability $p_{\mb i}$ the outcome of a measurement $\mathcal{M}$ will be $\mb i$ and the system will be in the corresponding two-particle state $\rho_{\mb i}$. Hence, the LE is given by
\begin{align}
L^{\mathcal{C},E}(\rho) = \sup_{\mathcal{M} \in \mathcal{C}} \sum_{\mb i} p^\mathcal{M}_{\mb i} E(\rho^\mathcal{M}_{\mb i}),
\end{align}
where $\mathcal{C}$ is the class of allowed measurements and $E(\cdot)$ an entanglement measure.


Let us consider the slightly more general case of a chain of $N$ spin-$S$ particles along with two auxiliary particles of spin $S'$ at each of the boundaries. The $N$ particles of the actual chain are the ones to be measured (measurement outcomes $\mb i = (i_1, \ ..., \ i_N)$), and the class of allowed measurements $\mathcal{C}$ is the set of local projective von Neumann measurements, where the same measurement is carried out on each party (in particular, we exclude adaptive strategies). Therefore, the maximization of the average entanglement is performed by choosing the optimal physical basis $\{| i \rangle \}_{i=1}^{2S+1}$.

An example of a state with LRLE is the ground state of the AKLT model~\cite{AKLT_org} defined for a spin-1 chain with periodic boundary conditions; its Hamiltonian is given by
\be
\mc{H}_\mr{AKLT} = \sum_{n = 1}^N \left( \mb S_n \cdot \mb S_{n+1} + \frac{1}{3} (\mb S_n \cdot \mb S_{n+1})^2 + \frac{2}{3}\right).
\ee
As mentioned above, it has a unique ground state that is an MPS given by the matrices 
$A^0 = \frac{1}{\sqrt{3}} \left(\begin{smallmatrix} 
1&0\\
0&-1
\end{smallmatrix}\right)$, 
$A^1 = \sqrt{\frac{2}{3}} \left(\begin{smallmatrix} 
0&1\\
0&0
\end{smallmatrix}\right)$ and
$A^{-1} = \sqrt{\frac{2}{3}} \left(\begin{smallmatrix} 
0&0\\
1&0
\end{smallmatrix}\right)$. If one measures the spins in the basis $\{|0\rangle, \tfrac{1}{\sqrt{2}}(|1\rangle \pm |-1\rangle\}$, the LE is non-vanishing in the thermodynamic limit. In Ref.~\cite{LRLE_QPT} a deformation has been introduced into the AKLT Hamiltonian via a parameter $\phi \in \mathbb{R}$, such that for $\phi = 0$ the AKLT Hamiltonian is recovered. If one varies $\phi$, LRLE is observed only at $\phi = 0$, signaling a quantum phase transition at this point. What is even more important is that the correlation length is finite for any $\phi$, i.e., it does not allow for the detection of this quantum phase transition (note that correlations provide only a \textit{lower} bound on the LE~\cite{LE_corr,LE_corr2}). However, it turns out that this particular quantum phase transition can also be detected by the emergence of string order at $\phi = 0$~\cite{SO_org}. An example of a state that has LRLE but no string order will be provided below. 

%% file: 2.3.tex
\section{A criterion for Matrix Product States with Long Range Localizable Entanglement}
\label{sec:LRLE}

In this section a necessary and sufficient criterion will be derived for all translationally invariant MPS with LRLE.
In subsection~\ref{sec:LRLE_MPS}, we will derive the LE of an MPS and show how it can be simplified to the case of the spins at the ends of the spin chain being qubits. Thereafter, in subsection~\ref{sec:criterion}, a criterion will be derived which allows to check for an MPS of arbitrary bond dimension based on a single tensor (i.e., without having to contract them), whether it possesses LRLE. This criterion will be used in subsection~\ref{sec:parameterize_MPS} to parameterize all MPS with bond dimension 2 and 3 that display LRLE.

\subsection{Localizable entanglement of an MPS}\label{sec:LRLE_MPS}

\begin{figure}[b!]
\centering\includegraphics[scale=.14]{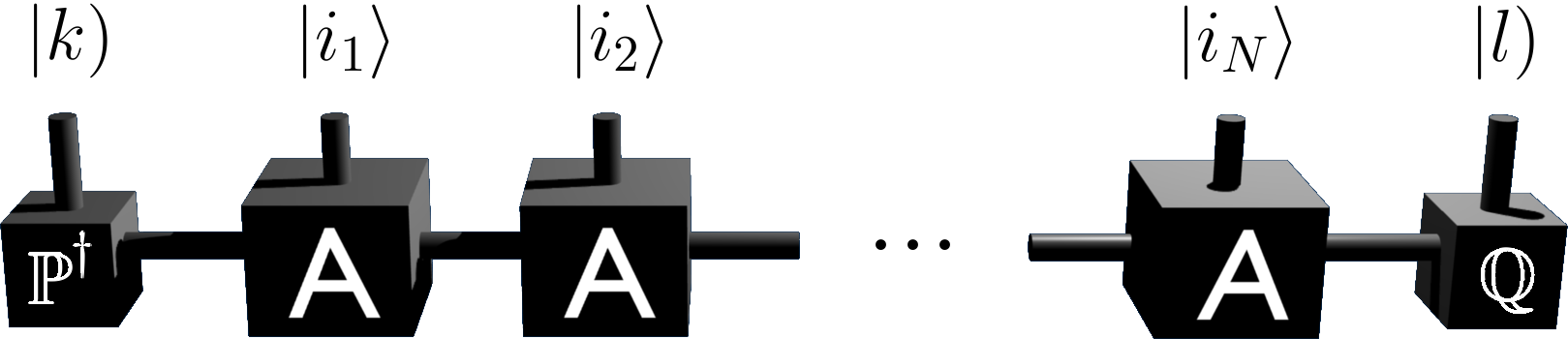}
\caption{Spin chain with $N$ spin-$S$ particles and one auxiliary particle of spin $S'$ at each of the borders. The matrices of the real particles are $A_i$ and those of the auxiliary particles $\P^\dg$ and $\mathbb{Q}$, respectively. This figure was reproduced from Ref.~\cite{MPS_LRLE}.}
\label{fig}
\end{figure}

The question to be answered in this section is for which translationally invariant MPS a finite amount of entanglement can be localized between the two ancillas in the limit $N \rightarrow \infty$. We assume the state of the system to be translationally invariant apart from boundary effects; for this reason, the rank-three tensors corresponding to the spin-$S$ particles are taken equal. Those consist of complex $D \times D$ matrices $A_i$ ($i = 1, \ ..., \ d \equiv 2S + 1$) and can be assumed to be in the canonical form~\eqref{eq:A_blocks}. As mentioned at the end of section~\ref{sec:MPS_construction}, the MPS decomposes into a sum of $r$ MPS, where $r$ is the number of blocks in Eq.~\eqref{eq:A_blocks}. Thus, we will only consider one block, which we simply call $A_i$, in the following. The results to be derived will apply to the other blocks, too. The maps $\e(X) = \sum_{i=1}^d A_i X A_i^\dg$ and $\oe (X) = \sum_{i=1}^d A_i^\dg X A_i$ satisfy (cf. Eqs.~\eqref{eq:FP_e} and~\eqref{eq:FP_oe})
\begin{align}
\e(\Id) = \Id, \  \  \  \oe(\Lambda) = \Lambda, \label{gauge}
\end{align}
with a diagonal matrix $\Lambda > 0$, cf. Eqs.~\eqref{eq:FP_e} and \eqref{eq:FP_oe}. Our goal is to find necessary and sufficient conditions on the matrices $\{A_i\}_{i=1}^d$ to give rise to LRLE for some matrices of the auxiliary particles. The latter can be chosen at will and are denoted by $\P,\mathbb{Q}: \mathbb{C}^{D'} \rightarrow  \mathbb{C}^D$, where $D' = 2S' + 1 \leq D$ is the Hilbert space dimension of the individual auxiliary spins. The initial MPS is therefore
\begin{align}
| \psi \rangle = \sum_{k,l=1}^{D'} \sum_{i_1, ..., i_N = 1}^d (k|\P^\dg A_{i_1} ... A_{i_N} \mathbb{Q}|l) | i_1 ... i_N \rangle \otimes |k,l), \label{eq:initial_MPS}
\end{align}
where the Hilbert space vectors of the auxiliary particles are denoted by round brackets, cf. Fig. \ref{fig}.
Subsequent to a measurement $\mathcal{M}$, the initial state of the system $\rho = | \psi \rangle \langle \psi |$ reduces with probability 
\be
p_{\mb i} = \tr(\P^\dg A_{i_1} ... A_{i_N} \mathbb{Q} \mathbb{Q}^\dg A_{i_N}^\dg ... A_{i_1}^\dg \P) \label{eq:measurement_prob}
\ee 
to $\rho^\mathcal{M}_{\mb i} = | \psi^\mathcal{M}_{\mb i} \rangle \langle \psi^\mathcal{M}_{\mb i}|$, where (excluding cases with $p_{\mathbf i} = 0$)
\begin{align}
| \psi^\mathcal{M}_{\mb i} \rangle = \frac{1}{\sqrt{p_{\mb i}}} \sum_{k,l=1}^{D'} (k|\P^\dg A_{i_1} ... A_{i_N} \mathbb{Q}|l)|kl) \label{eq:MPS_proj}
\end{align}
is the normalized two-particle state after the measurement.
In appendix \ref{app_show_qubits} it is shown that any MPS with LRLE for $D' > 2$ also has LRLE for $D' = 2$ (the converse is obvious), and also that w.l.o.g. we can take $\P$ and $\mathbb{Q}$ as isometries. We can thus choose $D' = 2$ and the concurrence \cite{concurrence} as the measure of entanglement, $E(\rho_{\mb i}^\mathcal{M}) = 2|\det ({\Psi}_{\mb i})|$, where ${\Psi}_{\mb i}$ is the matrix with coefficients $(\Psi_{\mb i})_{kl} = \frac{1}{\sqrt{p_{\mb i}}} (k|\P^\dg A_{i_1} ... A_{i_N} \mathbb{Q}|l)$.  Dropping the superscripts $\mathcal{C}$ and $E$, the LE reads
\begin{align}
L(\rho) = 2 \sup_{\{|i\rangle\}} \sum_{i_1, ..., i_N} \left| \det(\P^\dg A_{i_1} ... A_{i_N} \mathbb{Q}) \right|. \label{LE_det}
\end{align}
This optimization problem is in general hard, since the sum needs to be evaluated for large $N$ before optimizing over the physical basis $\{|i\rangle\}_{i=1}^d$. Its maximization succeeded only in special cases, like
%
for a chain of $N$ spins without auxiliary ones and measurement on all spins but those at the borders \cite{LRLE_QPT}. In this case, $\prod_{j=2}^{N-1} \sum_{i_j=1}^d | \det (A_{i_j}) |^{2/D}=1$, which implies that there is LRLE for $A_i = \alpha_i U_i$. In the following, we slightly generalize this family of states and adapt them to our system that includes the auxiliary spins at the boundary:

\begin{ex}Block structure of unitaries.
\end{ex}

Consider an MPS described by the matrices
\begin{align}
A_i = (P_i \otimes \Id_{n \times n})\,  \bigoplus_{k=1}^{q} \, \alpha^k_i U^k_i,
\end{align}
where $P_i$ is a $q \times q$ permutation matrix. The $U^k_i$ are $n \times n$ unitaries ($D= n q$) and $\alpha^k_i \geq 0 \ \forall \ k = 1, \ ..., \ q, \ i = 1, \ ..., \ d$.
We can realize that this MPS has LRLE for $n = 2$ by choosing $\P^\dg = |\uparrow)(1|$ + $|\downarrow)(2|$ (where $|\uparrow), \ |\downarrow)$ are basis vectors of the auxiliary qubits and $|1), \ldots, |D)$ the basis vectors of the basis in which the matrices are given) and $\mathbb{Q} = D^{-1/2} \sum_{j=1}^D \left(|j)(\uparrow| + (-1)^j |j)(\downarrow| \right)$. Then, due to $\mathcal{E}(\Id) = \Id$ and therefore $\sum_{i=1}^d |\alpha_i^k|^2 = 1 \ \forall \ k = 1, \ ..., \ D/2$, Eq. \eqref{LE_det} is finite in the thermodynamic limit. The fact that there is also LRLE for $n > 2$ will become clear below. 

In the above example, the matrices $A_i$ exhibit a block structure of unitaries. One may think that all states with LRLE need to have this property in some basis $\{|i\rangle\}_{i=1}^d$.
Interestingly, this is not the case as shown by the following counterexample for $D = 3$:

\begin{ex}Non-unitary matrices.
\end{ex} 

For the MPS with
\begin{align}
A_{1} = \frac{1}{2}\left(\begin{array}{ccc}
1&0&0\\
0&1&0\\
1&0&0
\end{array}\right), \
A_{2} = \frac{1}{\sqrt{2}}\left(\begin{array}{ccc}
0&0&1\\
0&1&0\\
0&0&-1
\end{array}\right), \
A_{3} =\frac{1}{2}\left(\begin{array}{ccc}
0&1&0\\
1&0&0\\
0&1&0
\end{array}\right),
\end{align}
$\P^\dg = |\uparrow)(1| + |\downarrow)(2|$ and $\mathbb{Q} = \frac{1}{\sqrt{3}} \left[ |1) + |2) + |3)\right] (\uparrow|$  $+ \frac{1}{\sqrt{2}}\left[|2) - |3)\right](\downarrow|$ it is a simple exercise to show that \eqref{LE_det} remains finite in the limit $N \rightarrow \infty$. The matrices $A_i$ fulfill $\e(\Id) = \Id$. However, $\oe(\Id) \ne \Id$, and thus they are not of the form $A_i = \alpha_i U_i$ in any basis $\{|i\rangle\}_{i=1}^d$. Note that the MPS can numerically be verified to be injective \cite{MPS_math}, and thus it is the unique ground state of a local translationally invariant gapped Hamiltonian. 
Moreover, it can be easily shown (analytically) that it is invariant under
the $\mathbb Z_2$-symmetry generated by the transformation $|1\rangle \rightarrow |1\rangle$, $|2\rangle \rightarrow |2\rangle$, $|3\rangle \rightarrow - |3\rangle$. 


\subsection{Derivation of the criterion}\label{sec:criterion}

The question arises of how one can check whether a given MPS has LRLE without having to resort to evaluating Eq.~\eqref{LE_det} for large $N$ numerically. 
In the following, a necessary and sufficient criterion will be derived, which allows to decide this based on the matrices $\{A_i\}_{i=1}^d$ directly. 
We first rewrite \eqref{LE_det} by inserting to the right of $A_{i_j}$ a projector on the subspace spanned by the row vectors of $\P^\dg A_{i_1} ... A_{i_j}$, which can be written as $\P^{j}_{i_1 ... i_j} \P^{j \, \dg}_{i_1 ... i_j}$, where $\P^j_{i_1 ... i_j} \in \mc{P}$ is an isometry with $\mc{P} := \{\P: \mathbb{C}^2 \rightarrow \mathbb{C}^D \ \mathrm{s. t.} \ \P^\dg \P = \Id_{2 \times 2} \}$. 
Thus, Eq.~\eqref{LE_det} reads now
\begin{align}
L(\rho) = 2 \sup_{\{|i\rangle\}} \sum_{i_1} \left| \det (\P^\dagger A_{i_1} \P^1_{i_1})\right|
\sum_{i_2} \left| \mathrm{det} (\P^{1 \, \dagger}_{i_1} A_{i_2} \P^2_{i_1,i_2})\right| 
... \sum_{i_N} \left| \mathrm{det}(\P^{N-1 \ \dagger}_{i_1, ..., i_{N-1}} A_{i_N} \mathbb{Q}) \right|.
\label{projectors}
\end{align}
Using the SVD of $\P^{j-1 \dg}_{i_1,...,i_{j-1}} A_{i_j} \P^j_{i_1,..., i_{j}}$, which is $U_{i_j} \mathrm{diag}(\sigma_{i_j}^{(1)}, \sigma_{i_j}^{(2)}) V_{i_j}^\dg$ (neglecting the indices $i_1, \ ..., \\ i_{j-1}$), along with the inequality of arithmetic and geometric means, one obtains for the factors of Eq.~\eqref{projectors} 
\begin{align}
\sum_{i_j} \left| \det (\P^{j-1 \dg}_{i_1,...,i_{j-1}} A_{i_j} \P^j_{i_1,..., i_{j}}) \right| \leq \frac{1}{2} \sum_{i_j} \tr \left(\P^{j-1 \,\dg}_{i_1, ..., i_{j-1}} A_{i_j} \P^{j}_{i_1, ..., i_j} \P^{j \dg}_{i_1, ..., i_j} A^\dg_{i_j} \P^{j-1}_{i_1, ..., i_{j-1}} \right)  = 1, \label{eq:factors_leq1}
\end{align}
since $\sum_{i_j} A_{i_j} A_{i_j}^\dg = \Id$. The inequality becomes an equality if and only if $\sigma_{i_j}^{(1)} = \sigma_{i_j}^{(2)}$, i.e., \\ $\P^{j-1 \dg}_{i_1,...,i_{j-1}} A_{i_j} \P^j_{i_1,..., i_{j}}$ is proportional to a unitary. The intuitive fact that all factors but the last one of Eq.~\eqref{projectors} have to be exactly 1 to get LRLE turns out to be correct:

\begin{lem} In the limit $N \rightarrow \infty$ the sum~\eqref{projectors} 
can be non-zero only if one can choose $\P^\dg$ such that for all well-defined $\P^{j-1 \,\dg}_{i_1, ..., i_{j-1}}$ (i.e. those for which $\P^\dg A_{i_1} ... A_{i_{j-1}} \neq 0$) the inequality~\eqref{eq:factors_leq1}
is an equality.
\end{lem}

\noindent\textbf{Proof} \\
Formally, this lemma reads
\begin{align}
\mathrm{LRLE} \ \Rightarrow \ \exists \ \P^\dg \ \mathrm{s.t.}\ \forall \ i_1, \ ..., \ i_{j-1}  \in  \{1, \ldots,\ d \} \notag \\
\sum_{i_j} \left| \det\left( \P^{j-1 \,\dg}_{i_1, ..., i_{j-1}} A_{i_j} \P^{j}_{i_1, ..., i_j} \right)\right| = 1,
\end{align}
or equivalently,
\begin{align}
\neg \, \mathrm{LRLE} \ &\Leftarrow \ \forall \ \P^\dg \ \exists \ (i_1, \ ..., \ i_{j-1}) \ \mathrm{s.t.} \notag \\ &\sum_{i_j} \left| \det\left( \P^{j-1 \,\dg}_{i_1, ..., i_{j-1}} A_{i_j} \P^{j}_{i_1, ..., i_j} \right)\right| \neq 1,
\label{lemma2}
\end{align}
which is what we want to show in the following. Thus, we assume that for the sum
\begin{align}
\sum_{i_1, ..., i_s} \left| \det \left( \P^\dg A_{i_1} ... A_{i_s} \P^{s}_{i_1 ... i_s} \right)\right| := 1 - \Delta^{(s)}(\P^\dg)
\label{sum_Delta}
\end{align}
there is a minimum integer $s_*$ such that
\begin{align}
\min_{\P^\dg} \Delta^{(s_*)}(\P^\dg) := \Delta_* > 0.
\label{min_Delta}
\end{align}
After building blocks of $s$ terms in Eq.~\eqref{projectors}, each of them can be upper bounded by
\begin{align}
1 - \Delta(\P_{i_1, ..., i_{j}}^{j \, \dg}) \leq 1 - \Delta_*,
\label{star}
\end{align}
which shows that the LE must fulfill
\begin{align}
L(\rho) \leq (1 -\Delta_*)^{\frac{N}{s_*}-1} \xrightarrow{N \rightarrow \infty} \ 0,
\end{align}
i.e., there is no LRLE. \qed

Therefore, a necessary condition for any MPS to give rise to LRLE is that there exists an isometry $\P^\dg$, such that for a certain basis $\{ |i \rangle \}_{i=1}^d$ 
\begin{align}
\P^{j-1 \dg}_{i_1,...,i_{j-1}} A_{i_j} \P^j_{i_1,..., i_{j}} \propto U_{i_1 ... i_{j}} \ \forall \ i_1, \ ..., \ i_j  \in  \{1, \ldots, d \},
\label{prop_unitary}
\end{align}
$U_{i_1 ... i_{j}}$ denoting some $2 \times 2$ unitary. Redefining $\P^j_{i_1,..., i_{j}} \ \rightarrow \ \P^j_{i_1,..., i_{j}} U_{i_1 ... i_{j}}$ shows that one can require the RHS of \eqref{prop_unitary} to be the identity. After multiplying this from the right by $\P^{j \, \dg}_{i_1,..., i_{j}}$, one obtains
\begin{align}
\P^{j-1 \dg}_{i_1,...,i_{j-1}} A_{i_j} \P^j_{i_1,..., i_{j}} \P^{j \, \dg}_{i_1,..., i_{j}}
= \P^{j-1 \dg}_{i_1,...,i_{j-1}} A_{i_j} \propto \P^{j \, \dg}_{i_1,..., i_{j}},
\label{new_isometry}
\end{align}
which can also be written as
\begin{align}
\P^\dg A_{i_1} ... A_{i_j} \propto \P^{j \, \dg}_{i_1, ..., i_j}.
\label{basic_rel}
\end{align}
It follows that Eq.~\eqref{basic_rel} is a necessary criterion to have LRLE. Whether it is also sufficient depends on if $\mathbb{Q}$ can always been chosen such that \eqref{projectors} is finite, which is confirmed by the following lemma:

\begin{lem} If criterion~\eqref{basic_rel} holds, there exists $\mathbb{Q}$ such that \eqref{projectors} is finite in the limit $N \rightarrow \infty$.
\end{lem}

\noindent\textbf{Proof}\\
We denote the proportionality factor in \eqref{basic_rel} by $\gamma_{i_1, ..., i_j}$, for which $\mathcal{E}(\Id) = \sum_{i=1}^d A_i A_i^\dg = \Id$ implies $\sum_{i_j} | \gamma_{i_1, ..., i_j}|^2 = | \gamma_{i_1, ..., i_{j-1}}|^2$. \eqref{LE_det} thus reads
\begin{align}
L(\rho) = 2 \sum_{i_1, ..., i_N} \left| \det(\P^{N \, \dg}_{i_1 ...i_N} \mathbb{Q})\right| |\gamma_{i_1 ... i_N}|^2.
\end{align}
Following the approach in appendix \ref{app_show_qubits} we introduce an $\epsilon$-net $\mathcal{N}_\epsilon$ in the isometries of $\mathcal{P}$ and obtain (the remaining equations of this proof are understood up to order $\mc{O}(\epsilon)$)
\begin{align}
L(\rho) = 2 \sum_{\mathbb{V} \in \mathcal{N}_\epsilon} \sigma(\mathbb{V}) | \det(\mathbb{V}^\dg \mathbb{Q})|, \label{sum_V}
\end{align}
where we defined
\begin{align}
\sigma(\mathbb{V}) = \sum_{i_1 ... i_N} |\gamma_{i_1 ... i_N}|^2 \delta_{\P^{N}_{i_1 ... i_N}, \mathbb{V}}.
\end{align}
$\delta_{\P^{N}_{i_1 ... i_N}, \mathbb{V}}$ is 1 if $\P^{N}_{i_1 ... i_N}$ and $\mathbb{V}$ lie in the same $\epsilon$-hypercube and 0 otherwise. 
Consider now the sum
\begin{align}
\sum_{\mathbb{V} \in \mathcal{N}_\epsilon} \sigma(\mathbb{V}) = \sum_{i_1 ... i_N} |\gamma_{i_1, ..., i_N}|^2 = 1.
\end{align}
Since it takes the value 1, for any $r > \epsilon$ there must exist a $\mathbb{V}'$ such that for $\mc K_r(\mathbb{V}') := \{\mathbb{V} \in \mc N_\epsilon \ \mr{s.t.} \ \| \mathbb{V} - \mathbb{V}'\|_\mr{F} \leq r\}$ ($\| \cdot \|_\mr{F}$ denoting the Frobenius norm) we have
\begin{align}
\sum_{\mathbb{V} \in \mathcal{K}_r} \sigma(\mathbb{V}) \geq \frac{|\mathcal{K}_r(\mathbb{V}')|}{|\mathcal{N}_\epsilon|},
\end{align}
where $|\mathcal{K}_r(\mathbb{V}')|$ and $|\mathcal{N}_\epsilon|$ are the cardinalities of the sets $\mc K_r(\mathbb{V}')$ and $\mc N_\epsilon$, respectively. In words, there has to be a spherical region $\mc K_r(\mathbb{V}')$ around $\mathbb{V}'$ whose points have at least average total weight.
For sufficiently small $r$, we consider only the following part of the sum in \eqref{sum_V}
\begin{align}
2 \sum_{\mathbb{V} \in \mathcal{R}_r} \sigma(\mathbb{V}) | \det(\mathbb{V}^\dg \mathbb{Q})| \notag = 2 \sum_{\mathbb{V} \in \mathcal{R}_r} \sigma(\mathbb{V}) | \det({\mathbb{V}'}^\dg \mathbb{Q})| \left(1 + \mathcal{O}(r)\right).
\end{align}
If we set $\mathbb{Q} = \mathbb{V}'$, this is lower bounded by $2(1 + \mathcal{O}(r))|\mathcal{K}_r(\mathbb{V}')|/|\mathcal{N}_\epsilon|$, i.e.,
\begin{align}
L(\rho) \geq 2(1 + \mathcal{O}(r))\frac{|\mathcal{K}_r(\mathbb{V}')|}{|\mathcal{N}_\epsilon|},
\end{align}
which is small but positive for sufficiently small $r$. Taking the limit $\epsilon \rightarrow 0$ and afterwards $N \rightarrow \infty$ completes the proof (the ratio of the cardinalities converges to an $r$-dependent finite value). 
\qed

Criterion \eqref{basic_rel} is thus necessary and sufficient for the emergence of LRLE. Note that for $D = 2$ all isometries are unitaries, i.e., according to \eqref{new_isometry} all matrices $A_i$ have to be proportional to unitaries themselves.


If we now define $\P^\dg := |\uparrow)(x| + |\downarrow)(y|$, i.e., $(x|x) = (y|y) = 1$ and $(y|x) = 0$, \eqref{basic_rel} is equivalent to
\begin{align}
(x| A_{i_1} ... A_{i_j} A_{i_j}^\dg ... A_{i_1}^\dg | x) &= (y | A_{i_1} ... A_{i_j} A_{i_j}^\dg ... A_{i_1}^\dg | y), \label{norm} \\
(y| A_{i_1} ... A_{i_j} A_{i_j}^\dg ... A_{i_1}^\dg | x) &= 0. \label{orth}
\end{align}
After further setting $V := |x)(x| - |y)(y|$, $W := |x)(y|$, the map $\oe_i (X) := A_i^\dg X A_i$ and
\begin{align}
V_{i_1, ... ,i_j} &:= A_{i_j}^\dg ... A_{i_1}^\dg |x)(x| A_{i_1} ... A_{i_j} - A_{i_j}^\dg ... A_{i_1}^\dg |y) (y| A_{i_1} ... A_{i_j} \notag \\
&= \oe_{i_j} \circ ... \circ \oe_{i_1} (V), \ \label{norm2} \\
W_{i_1, ... ,i_j} &:= A_{i_j}^\dg ... A_{i_1}^\dg |x)(y| A_{i_1} ... A_{i_j} = \oe_{i_j} \circ ... \circ \oe_{i_1} (W), \ \label{orth2}
\end{align}
we see that Eqs.~\eqref{norm} and \eqref{orth} are equivalent to
\begin{align}
\tr(V_{i_1, ... ,i_j}) &= \tr(\oe_{i_j} \circ ... \circ \oe_{i_1} (V)) = 0, \label{norm3} \\
\tr(W_{i_1, ... ,i_j}) &= \tr(\oe_{i_j} \circ ... \circ \oe_{i_1} (W)) = 0, \label{orth3}
\end{align}
respectively. Thus, if we define the subspace $\S := \mathrm{span} \{V, \ W,  \ ..., \ V_{i_1, ..., i_j}, \ W_{i_1, ..., i_j}, \ ... \}_{i_1, ..., i_j = 1}^d$, Eqs.~\eqref{norm3} and \eqref{orth3} indicate that the occurrence of LRLE is equivalent to $\tr(S) = 0 \ \forall \ S \in \S$. By definition $\S$ is closed under the application of any $\oe_i$, 
which leads to the following criterion characterizing translationally invariant MPS with LRLE:

\begin{thm}
For the MPS considered here, there is LRLE if and only if for a certain basis $\{|i\rangle\}_{i=1}^d$ there exists a subspace $\S$ of the vector space of $D \times D$ matrices satisfying the following conditions
\begin{enumerate}[i]
\item $\S$ \textit{is closed under all linear maps $\oe_i$ defined as $\oe_i(X) = A_i^{\dg} X A_i$, i.e., $\oe_i(\S) \subseteq \S \ \forall \ i = 1, \ ..., \ d$}, \label{i}
\item $\tr(S) = 0 \ \forall \ S \in \S$, \label{ii}
\item $\exists \ |x), \ |y) \ \in \mathbb{C}^D$ \textit{s.t.} $V = |x) (x| - |y)(y|$, $W = |x)(y|\in \S$.
\label{iii}
\end{enumerate}
\end{thm}

Note that if the theorem is fulfilled, $\oe_{i_j} \circ ... \circ \oe_{i_1} (W^\dg) = W_{i_1, ..., i_j}^\dg$ is also traceless, i.e., $\S$ could additionally be required to contain $W^\dg = |y)(x|$ and correspondingly to be equal to its adjoint, $\forall \ S \in \S \Rightarrow S^\dg \in \S$. 

The above theorem can be used numerically to determine whether a given MPS has LRLE, 
since it imposes conditions on the matrices $\{A_i\}_{i=1}^d$, which can be represented by a set of polynomial equations: 
The entries of $\{A_i\}_{i=1}^d$ will give rise to the coefficients of those equations, whereas all other quantities introduced below will constitute their variables to be determined numerically. The first variables to be introduced are orthonormal basis vectors $\{S^k\}_{k=1}^n$ of $\S$ , where $n \leq D^2 - 1$ is the dimension of $\S$. Condition~\ref{i} is equivalent to requiring that any basis vector $S^k$ of $\S$ is mapped by any $\oe_i$ into $\S$, i.e., for all $k = 1, \ ..., \  n$
\begin{align}
\mathrm{I.}& \ \ \tilde A_i^\dg \, S^{k} \tilde A_i = \sum_{l=1}^n a_i^{k,l} S^l \ \forall \ i = 1, \ ..., \ d, \label{Groebner_first}\\
& \ \ \tr(S^{k \dg} S^l) = \delta_{k,l} \ \forall \ l = 1, \ ..., \ n, \\
\intertext{where $\{a_i^{k,l}\}_{i=1, ..., d}^{k,  l = 1,  ...,  n}$ are scalar complex variables of the set of equations to be solved numerically. The $\{\tilde{A_i}\}_{i=1}^d$ are the matrices in a possibly different physical basis $\{|\tilde i\rangle\}_{i=1}^d$. 
condition~\ref{ii} can be stated as}
\mathrm{II.}& \ \ \tr(S^k) = 0 \ \forall \ k = 1, \ ..., \ n. \\
\intertext{Furthermore, since according to condition~\ref{iii} $V = |x) (x| - |y)(y|$ and $W = |x)(y|$ have to be also contained in $\S$,}
\mathrm{III.}& \ \ \sum_{k=1}^n v^k S^k = |x)(x| - |y)(y|, \\
& \ \ \sum_{k=1}^n w^k S^k = |x)(y| \\
\intertext{with $\{v^k\}_{k=1}^n$, $\{w^k\}_{k=1}^n$ and the coefficients of $|x)$ and $|y)$ as other scalar complex variables. Last, a rotation in the basis of measurement $\{|i\rangle\}_{i=1}^d$ is implemented by}
\mathrm{IV.}& \ \ \tilde A_i = \sum_{j=1}^d U_{ij} A_j, 
\ \ U U^\dg = \Id_{d \times d}, \label{Groebner_last}
\end{align}
constituting the last of the set of equations \eqref{Groebner_first} to \eqref{Groebner_last} to be solved. Generally, a set \\ $p_\ell(t_1,  \ldots,  t_m) = 0$ of $s$ polynomial equations ($\ell = 1, \ldots,  s$) with $m$ complex variables $\{t_\ell\}_{\ell=1}^m$ can be solved by means of a Gr\"obner basis \cite{Groebner,Buchberger}. A Gr\"obner basis $\{g_\ell(t_1, \ldots ,  t_m)\}_{\ell=1}^s$ is a special basis in the vector space of functions involving the variables $\{t_\ell\}_{\ell=1}^m$: It has the property that the set of equations to be solved is equivalent to the set $g_\ell(t_1, \ldots ,  t_m) = 0$ ($\ell = 1,  \ldots,  s$), which, in contrast, can be solved by back-substitution while having to deal with the solution of polynomials involving only one variable at a time. For instance, one of the new equations might involve only $t_1$. After solving it numerically the result can be inserted into another equation involving, e.g., only $t_1$ and $t_2$ etc. (cf. Gaussian elimination). A Gr\"obner basis can be found systematically by use of Buchberger's algorithm \cite{Buchberger}, which is doubly exponential in the complexity of the set of equations to be solved. In our case, this implies a computational cost that is doubly exponential in the square of the bond dimension. However, $D$ is a constant and in particular independent of the length of the spin chain. If a simultaneous solution to Eqs.~\eqref{Groebner_first} to \eqref{Groebner_last} is found for some $n < D^2$, the MPS has LRLE. If even for $n = D^2 - 1$ no solution is found, it does not.


\subsection{Parameterization of MPS with LRLE of low bond dimension}\label{sec:parameterize_MPS} 

We now employ Theorem 2.3.5. analytically to determine the complete sets of MPS with LRLE for $D = 2$ and $D = 3$. In the former case, we reproduce the finding that the matrices need to be proportional to unitaries. In the latter, we obtain the result that matrices of the type of Example 2.3.2. are the only non-trivial additional matrices 
which give rise to LRLE. 

In both cases we take $\S$ to be equal to its adjoint, i.e., it has to contain the matrices $V = |x)(x| - |y)(y|$, $W = |x)(y|$ and $W^\dg$.
For $D = 2$ this implies that $\S$ is the full subspace of traceless $2 \times 2$ matrices. 
If we take any element $S \in \mathcal{S}$, $\tr (A_i^\dg S A_i) = 0$ shows that $A_i A_i^\dg$ is orthogonal to $\mathcal{S}$ with respect to the Hilbert-Schmidt scalar product. However, the orthogonal subspace of $\mathcal{S}$ is spanned by the identity, i.e., $A_i A_i^\dg \propto \Id$ for any MPS with LRLE. 

For an MPS with LRLE and matrices $\{A_i\}^d_{i=1}$ of bond dimension $D = 3$, the dimension of the smallest self-adjoint subspace $\mc S$ is denoted by $n_{\mathrm{min}}$. In the following, we will determine all possible values for $n_{\mathrm{min}}$ and the associated types of matrices $A_i$.

The simplest case is $\S = \sp\{V,\ W,\ W^\dg\} \subset \mathbb{C}^{3 \times 3}$, i.e., $n_{\mathrm{min}} = 3$. We define the two dimensional subspace $\tau = \sp\{|x), \ |y)\} \subset \mathbb{C}^3$, where tracelessness of $W$ and $V$ implies $(x|y) = 0$ and $(x|x) = (y|y)$, respectively (the latter will be taken equal to 1 in the following). It follows that $\S \subset \tau \times \tau$. If w.l.o.g. we set $|x) = |1)$ and $|y) = |2)$, we observe that we retrieve the case of $D = 2$, as 
\begin{align}
A_i^{D=3} = \left(\begin{array}{cc}
A_i^{D=2}&0\\
B_i&c_i
\end{array}\right), \label{3D_2D}
\end{align}
$\forall \ i = 1, \ ..., \ N$, where $A_i^{D=2}$ is proportional to a unitary. $B_i \in \mathbb{C}^{1 \times 2}$ and $c_i \in \mathbb{C}$ are arbitrary given the normalization~\eqref{gauge}.

Consider now $n_{\mathrm{min}} > 3$. Since $V_{i_1, ... ,i_j} = \oe_{i_j} \circ ... \circ \oe_{i_1} (V) := |x_{i_1, ..., i_j}) (x_{i_1, ..., i_j}| - |y_{i_1, ..., i_j}) (y_{i_1, ..., i_j}|$ and $W_{i_1, ..., i_j} = |x_{i_1, ..., i_j})(y_{i_1,...,i_j}|$ with $|x_{i_1, ..., i_j}) := A_{i_j}^\dg ... A_{i_1}^\dg |x)$ and $|y_{i_1, ..., i_j}) := A_{i_j}^\dg ... A_{i_1}^\dg |y)$, the smallest self-adjoint subspace $\S$ associated with $\{A_i\}_{i=1}^d$ can be written as the span of triples of matrices, which are of the type $V' = |x')(x'| - |y')(y'|$, $W' = |x')(y'|$, $W'^\dg = |y')(x'|$ each. Those triples are associated with the two vectors $|x')$ and $|y')$.
Thus, if $d_{\mathrm{min}} > 3$, there are vectors $|x'), \ |y')$, which do not both lie in the two-dimensional subspace $\tau = \sp\{|x), |y)\}$. We define $\tau' := \sp\{|x'), |y')\} \neq \tau$ and take $|x')$ and $|y')$ as orthonormal (which is possible due to $\tr(V') = \tr(W') = 0$). Since $\tau$ and $\tau'$ have to intersect, we can assume the intersection to be w.l.o.g. along $|y)$.  As $V', \ W'$ and $W'^\dg$ span the full space of traceless matrices contained in $\tau' \times \tau'$, we are allowed to choose the basis $\{|x'), \ |y')\}$ for $\tau'$ such that $|y) = |y')$.  In this case, $\sp\{V, \ W, \ W^\dg\}$ and $\sp\{V', \ W', \ W'^\dg\}$ could only intersect in $|y)(y|$, which is, however, not traceless. Consequently, if $n_{\mathrm{min}} > 3$, $\S$ is at least of dimension 6. 

In the case of $n_{\mathrm{min}} = 6$, that is, $\S = \sp\{V, \ W, \ W^\dg, \ V', \ W', \ W'^\dg\}$ it is simple to construct the orthogonal complement of $\S$, which is 
$\overline{\S} = \sp\{\Id, \ |n)(n'|, \ |n')(n|\}$, where $|n)$ and $|n')$ are the normal vectors of $\tau$ and $\tau'$, respectively. This may be easily verified by taking the trace of the products of $\Id, \ |n)(n'|$ or $|n')(n|$ with $V, \ W, \ W^\dg, \ V', \ W'$ or $W'^\dg$. This structure of $\overline{\S}$ uniquely defines $\tau$ and $\tau'$ (and thus also $|n)$ and $|n')$) and vice versa. Therefore, another $\tau'' \neq \tau, \ \tau'$ would not be consistent with this $\overline{\S}$, and the only possibility would be $\overline{\S} = \sp\{\Id\}$ corresponding to $A_i A_i^\dg \propto \Id$ (i.e., $n_{\mathrm{min}} = 8$). 
Thus, $n_{\mathrm{min}} = 6$ is the only remaining case which might lead to a new type of MPS with LRLE.  
Then, it is obvious that for some $\oe_i$ $V_i$ must happen not to be in $\sp\{V, \ W, \ W^\dg\}$, otherwise $\S$ would be of the form~\eqref{3D_2D}. We note that in the former case, $A_i^\dg|x)$ and $A_i^\dg |y)$ are new orthogonal vectors with equal norms (due to $\tr(V_i) = \tr(W_i) = 0$). From them we obtain the orthonormal vectors $[(x|A_i A_i^\dg |x)]^{-1/2} A_i^\dg |x)$ and $[(y|A_i A_i^\dg |y)]^{-1/2} A_i^\dg |y)$ contained in $\tau'$. Thus, $A_i$ applied from the right maps all vectors contained in $\tau = \sp\{|x), \ |y)\}$ to vectors in $\tau' = \sp\{|x'), \ |y')\}$, while preserving their relative lengths, or equivalently, the angles between them. 
As noted above, the existence of another $\tau'' \neq \tau, \ \tau'$ (supporting $V'', \ W''$ and $W''^\dg$) is excluded for $n_{\mathrm{min}} < 8$. Consequently, any 
$A_i$ applied from the right has to map $\tau$ to $\tau$ or $\tau'$ and $\tau'$ to $\tau$ or $\tau'$ in such a way that the relative lengths of all vectors lying in one of those subspaces are preserved. 
After choosing w.l.o.g. $|x) := |1)$ and $|y) = |y') := |2)$, we can make the ansatz
\begin{align}
A_i = \gamma_i \left(\begin{array}{ccc}
1&0&0\\
0&1&0\\
a_i&b_i&c_i
\end{array}
\right)U_i,
\label{Ansatz}
\end{align}
$\gamma_i, \ a_i, \ b_i, \ c_i \in \mathbb{C}$, and $U_i$ is a unitary. We see that indeed $\tr(A_i^\dg V A_i) = 0$, i.e., $(1|A_i A_i^\dg |1) = (2| A_i A_i^\dg |2) = |\gamma_i|^2$ and $\tr(A_i^\dg W A_i) = 0$, i.e., $(2|A_i A_i^\dg |1) = 0$. 
The normal vector of $\tau$ is $|n) = |3)$, and the one of $\tau'$ is of the form $|n') = r |1) + s|3)$, since $\tau$ and $\tau'$ intersect in the $|2)$-axis. Because of $\tr(A_i^\dg S A_i) = 0 \ \forall \ S \in \S$, it follows that $A_i A_i^\dg \in \overline{\S} = \sp\{\Id, |n)(n'|, |n')(n|\}$. Therefore, 
\begin{align}
A_i A_i^\dg = \alpha_i \Id + \beta_i \big(|3)[r(1| + s(3|] + [r^*|1) + s^* |3)](3|\big)
\end{align}
(the asterisk denotes complex conjugation), and comparison with \eqref{Ansatz} results in $b_i = 0 \ \forall \ i = 1, \ ..., \ d$. Now, the matrix
\begin{align}
M_i = \left(\begin{array}{ccc}
1&0&0\\
0&1&0\\
a_i&0&c_i
\end{array}\right)
\end{align}
applied from the right maps $\tau$ to $\tau$ and $\tau'$ to, say, $\tilde \tau$ preserving the relative lengths of vectors lying in one of the subspaces $\tau$ or $\tau'$. As a result, there are four possible cases for the action of $U_i$
\begin{align}
a) \ \ U_i:& \ (\tau, \ \tilde \tau) \rightarrow (\tau, \ \tau), \\
b) \ \ U_i:& \ (\tau, \ \tilde \tau) \rightarrow (\tau, \ \tau'), \\
c) \ \ U_i:& \ (\tau, \ \tilde \tau) \rightarrow (\tau', \ \tau), \\
d) \ \ U_i:& \ (\tau, \ \tilde \tau) \rightarrow (\tau', \ \tau'). 
\end{align}
a) and d) imply $\tilde \tau = \tau$ and hence $|a_i| = 1$ and $c_i = 0$, whereas b) and c) mean $\tilde \tau \neq \tau$. Hence, in the latter case, the intersection of $\tau$ and $\tilde \tau$ is the $|2)$-axis. Since it has to be mapped by $U_i$ to the intersection of $\tau$ and $\tau'$, which is likewise the $|2)$-axis, $|2)$ must be an eigenvector of $U_i$. Then, it follows that $U_i$ has a unitary action in the $|1)$-$|3)$-plane. Consequently, b) corresponds to $\tilde \tau = \tau'$, and therefore $a_i = 0$, \ $|c_i| = 1$ (otherwise $(x'| M_i$ would not be equal to $(x'|$ up to a phase and thus $\tilde \tau \neq \tau'$). Last, in c) double application of $U_i$ would map $\tilde \tau \xrightarrow{U_i} \tau \xrightarrow{U_i} \tau'$, wherefore $M_i$ has to carry out a reflection of $\tau'$ on the $\tau$-plane, i.e., $a_i = 0$ and $c_i = -1$, and hence, $\tilde \tau = \sp\{|2),|3)\} = \tau'$. 
We thus obtain either $a_i = 0$ and $|c_i| = 1$ or $|a_i| = 1$ and $c_i = 0 \ \forall \ i = 1, \ ..., \ d$ and conclude that $\tau' = \sp\{|2), |3)\}$.

The requirement that $A_i$ maps $\tau$ to $\tau$ or $\tau'$ and $\tau'$ to $\tau$ or $\tau'$ preserving relative lengths of vectors contained in them also holds for $A_i^2$, from which we deduce that $U_i$ must contain only one non-vanishing entry per row and column (with magnitude 1). All results combined together yield
\begin{align}
A_i \propto [e^{i \phi_i} |1) + e^{i \phi_i'} |3)](l| + e^{i \phi_i''} |2)(m|, 
\end{align}
with $l, m = 1, \ 2, \ 3$, $l \neq m$.

\subsection{Conclusions}

Within the framework of MPS we have specified the localizable entanglement of a spin chain with one auxiliary spin at each of the boundaries. We have shown that LRLE can be detected by placing qubits at the ends of the chain. Based on that we were able to derive a theorem according to which it can be checked directly from the matrices of the MPS, whether it possesses LRLE. How this can been done in practice has been indicated by Eqs. \eqref{Groebner_first} to \eqref{Groebner_last}, which is a set of polynomial equations that can be solved numerically in a systematic manner. Furthermore,  we provided non-trivial examples of MPS for which those equations have a simultaneous solution, determining the full sets of MPS with LRLE for $D = 2$ and $D = 3$.

%% file: 3.1.tex
\chapter{Chiral Topological Projected Entangled Pair States}

\section{Introduction to Projected Entangled Pair States}\label{sec:PEPS_intro}

\subsection{Background}

After it had been understood that Matrix Product States (MPS) can be constructed by applying local projections on maximally entangled pairs (see subsection~\ref{sec:MPS_construction}), the generalization to higher dimensions, Projected Entangled Pair States (PEPS)~\cite{PEPS_org,PEPS}, was obvious. Given this analogy, PEPS share many properties with MPS, e.g., they also obey the area law~\cite{area_Rev} and, as a consequence, might provide a set of states that can efficiently approximate the ground state of any gapped local Hamiltonian (note however, that there are ground states of non-local Hamiltonians which do fulfill the area law, but cannot be approximated efficiently by PEPS~\cite{area_not_efficient}). At finite temperatures, they have already been shown to allow for an efficient approximation of the thermal state~\cite{PEPS_eff,PEPS_eff2} (i.e., a constant accuracy is obtained by increasing the number of parameters only polynomially with the system size). However, PEPS are much harder to handle computationally, since the exponents of the polynomial scaling are much higher than for MPS. Furthermore, given a certain PEPS, calculating local observables, or even its norm, is in general not efficient in the system size and the computation time grows exponentially with all but one spatial direction. However, this obstacle can be overcome by approximately contracting the PEPS with usually well controlled errors. Due to their worse scaling, variational calculations with PEPS in two dimensions achieve far worse accuracies than Density Matrix Renormalization Group (DMRG)~\cite{DMRG_org} calculations in one dimension (using MPS as variational wave functions), which is why reasonable system sizes for PEPS are still roughly the same as those that can be tackled with two dimensional DMRG algorithms~\cite{DMRG_2D}. Using PEPS is of high interest in frustrated systems and fermionic systems, where Quantum Monte Carlo Methods~\cite{QMC_Rev} fail due to the sign problem~\cite{sign_prob}. For this reason, fermionic PEPS have been introduced, where maximally entangled spins are replaced by a maximally entangled state in fermionic Fock space. 

The structure of this section is the following: In the next subsection, the construction of PEPS via a projection of maximally entangled pairs, as in subsection~\ref{sec:MPS_construction} for MPS, will be presented. Furthermore, the extension to fermionic PEPS  will be given. In subsection~\ref{sec:PEPS_parent} it will be shown how parent Hamiltonians for PEPS can be constructed and in subsection~\ref{sec:PEPS_sym} how to implement physical symmetries locally in the PEPS tensor. Thereafter, in subsection~\ref{sec:top_PEPS}, it will be discussed how symmetries that merely act on the virtual indices of the PEPS tensor give rise to topological states. 
Finally, in subsection~\ref{sec:PEPS_cylinder} PEPS on a cylinder geometry with open virtual legs at its boundaries will be considered. The concept of a transfer operator will be introduced, which allows to investigate whether the PEPS is has algebraically decaying correlations or not. Furthermore, the bulk-boundary correspondence in PEPS will be presented which is a holographic principle that states that there is a one to one map between the PEPS reduced density matrix of a certain region and a state defined entirely on its virtual boundary.

\subsection{Construction}\label{sec:PEPS_construction}

\subsubsection{Spin-PEPS}

\begin{figure}
\centering\includegraphics[width=0.8\textwidth]{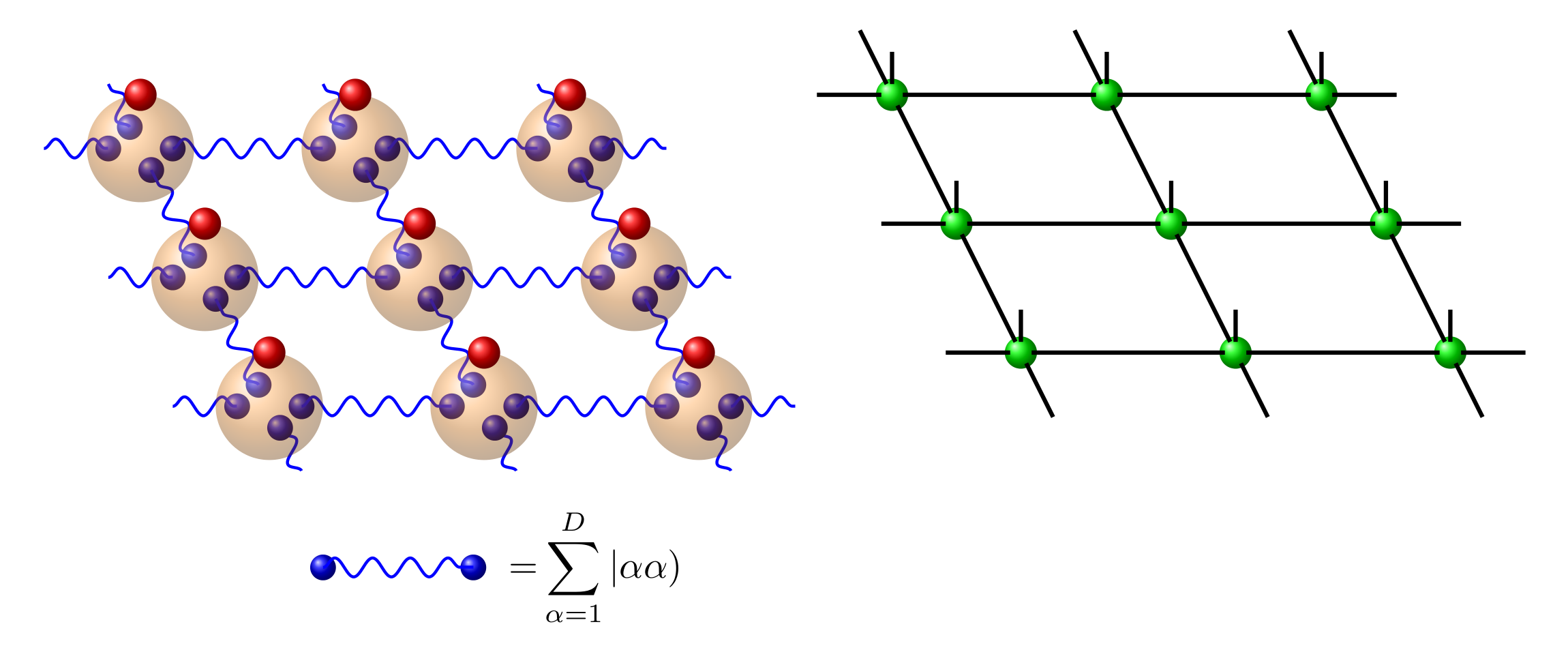}
\caption{Left: Construction of PEPS via local projections applied on maximally entangled pairs. Blue balls denote virtual spins and red particles physical spins. Right: Resulting pictorial representation in terms of rank-5 tensors (green balls). For an explanation of the pictorial representation of tensor networks, see Fig.~\ref{fig:MPS_OBC}.}
\label{fig:PEPS}
\end{figure}

The recipe for constructing spin-PEPS will be the same as the one for obtaining MPS out of maximally entangled pairs: Recall (subsection~\ref{sec:MPS_construction}) that an MPS can be acquired by placing pairs of maximally entangled virtual spins between neighboring sites and applying on each site an arbitrary linear map from the space of virtual particles to the physical Hilbert space. Hence, in order to obtain a PEPS we place maximally entangled pairs between neighboring sites and apply an arbitrary local linear map on each site from the space of $z$ virtual particles (where $z$ is the coordination number) to the physical Hilbert space of each site. For concreteness, let us consider a square two dimensional lattice of $N_v \times N_h$ sites with periodic boundary conditions. We assign to each site four virtual particles. 
We label the spin-z component of the left, right, up, down virtual particle at site $\mb r$ by $l_{\mb r}, r_{\mb r}, u_{\mb r}, d_{\mb r}$, respectively. The maximally entangled pairs between two sites that are horizontally displaced by one unit (with vector $\hat {\mb x}$) form the state 
\be
|\omega_{\mb r, \mb r + \hat{\mb x}}) = \sum_{\alpha=1}^D | r_{\mb r} = \alpha, l_{\mb r + \hat{\mb x}} = \alpha).
\ee
For vertically displaced sites (by one unit vector $\hat{\mb y}$) the maximally entangled state is 
\be
|\omega_{\mb r, \mb r + \hat{\mb y}}) = \sum_{\alpha=1}^D | d_{\mb r} = \alpha, u_{\mb r + \hat{\mb y}} = \alpha). 
\ee
Afterwards, local linear maps $\mc{M}_{\mb r}$ from the Hilbert space of the four virtual particles at site $\mb r$ to the Hilbert space of the physical particle (spin-z projection $i_\mb{r} = 1, \ldots, d$) are applied, where
\be
\mc{M}_\mb{r} = \sum_{i=1}^{d_\mr{ph}} \sum_{l_\mb{r},r_\mb{r},u_\mb{r},d_\mb{r}=1}^D A^{[\mb r]}_{i_\mb{r},l_\mb{r},r_\mb{r},u_\mb{r},d_\mb{r}} |i_\mb{r}\rangle(l_\mb{r},r_\mb{r},u_\mb{r},d_\mb{r}|,
\label{eq:PEPS_map}
\ee
 such that the final PEPS is given by 
\be
|\Psi\rangle = \left(\bigotimes_\mb{r} \mc{M}_\mb{r}\right) \left(\bigotimes_{\mb r} |\omega_{\mb r, \mb r + \hat{\mb x}}) \otimes |\omega_{\mb r, \mb r + \hat{\mb y}})\right).
\label{eq:psi_PEPS}
\ee
The graphical representation of this tensor network state is given in Fig.~\ref{fig:PEPS}. The construction for open boundary conditions is analogous; one just has to modify the map~\eqref{eq:PEPS_map} at the edges, i.e., they will be given by rank-4 tensors at the edges and by rank-3 tensors at the corners of the $N_v \times N_h$ region.

\subsubsection{Fermionic PEPS}\label{sec:fPEPS}

For simplicity, we consider again a two dimensional square lattice, but now with one fermionic mode per site (annihilation operator $a_{\mb r}$).
One could easily write down a fermionic PEPS by defining a spin wave function via Eq.~\eqref{eq:psi_PEPS} for $d=2$ and then setting $|i_{(1,1)},i_{(1,2)}, \ldots, i_{(N_h,N_v)}\rangle = (a_{(1,1)}^\dg)^{i_{(1,1)}} (a_{(1,2)}^\dg)^{i_{(1,2)}} \ldots (a_{(N_h,N_v)}^\dg)^{i_{(N_h,N_v)}} |\Omega\rangle$ ($i_\mb{r} = 0,1$), where $| \Omega\rangle$ is the vacuum of the physical modes. This corresponds to applying a Jordan-Wigner transformation on the state given by Eq.~\eqref{eq:psi_PEPS}. As for MPS, there is a systematic way of constructing (local) parent Hamiltonians for PEPS (see subsection~\ref{sec:PEPS_parent}). However, the Jordan-Wigner transformation of a spin-PEPS to a fermionic PEPS would transform the corresponding local spin-parent Hamiltonian to a non-local fermionic Hamiltonian (the resulting fermionic PEPS would also violate the area law). Hence, this strategy is not amenable of producing good ansatz states for the ground states of local fermionic Hamiltonians. In order to overcome this problem, one replaces the maximally entangled spins $|\omega_{\mb r, \mb r'})$ by operators $\omega_{\mb r, \mb r'}$ that create maximally entangled states of fermions in Fock space~\cite{fPEPS_Kraus}, 
\begin{align}
\omega_{\mb r, \mb r + \hat{\mb x}} &= 1 + \beta^\dg_\mb{r} \alpha^\dg_{\mb r + \hat x}, \\
\omega_{\mb r, \mb r + \hat{\mb y}} &= 1 + \delta^\dg_\mb{r} \gamma^\dg_{\mb r + \hat y},
\end{align}
where $\alpha^\dg_\mb{r}$, $\beta^\dg_\mb{r}$, $\gamma^\dg_\mb{r}$, $\delta^\dg_\mb{r}$ creates a virtual fermion at the left, right, top, bottom of site $\mb r$, respectively. This corresponds to bond dimension $D = 2$, but can be easily generalized to higher bond dimensions by placing several virtual fermions along each direction. The projection $\mc{M}_\mr{r}$ is now written as
\be
\mc{M}_\mb{r} =  \sum_{\substack{i_\mb{r}, l_\mb{r},r_\mb{r},u_\mb{r},d_\mb{r}=0 \\ (i_\mb{r} + l_\mb{r} + r_\mb{r} + u_\mb{r} + d_\mb{r}) \mr{mod} \, 2 = c_\mb{r}}}^1 A^{[\mb r]}_{i_\mb{r},l_\mb{r} r_\mb{r} u_\mb{r} d_\mb{r}} (a_\mb{r}^\dg)^{i_\mb{r}} \alpha_\mb{r}^{l_\mb{r}} \beta_\mb{r}^{r_\mb{r}} \gamma_\mb{r}^{u_\mb{r}} \delta_\mb{r}^{d_\mb{r}}, 
\label{eq:fPEPS_map}
\ee
such that the final state takes the form
\be
|\Psi\rangle = ( \Omega_v| \left(\prod_\mb{r} \mc{M}_\mb{r}\right) \left(\prod_{\mb r} \omega_{\mb r, \mb r + \hat{\mb x}} \, \omega_{\mb r, \mb r + \hat{\mb y}}\right) |\Omega_{v,p}\rangle.
\label{eq:psi_final}
\ee
Note that the sum in Eq.~\eqref{eq:fPEPS_map} is restricted to $(i_\mb{r} + l_\mb{r} + r_\mb{r} + u_\mb{r} + d_\mb{r}) \mr{mod} \, 2 = c_\mb{r} \in \{0, 1\}$ in order to obtain a state with fixed fermionic parity. $|\Omega_v )$ refers to the vacuum of virtual modes and $|\Omega_{v,p}\rangle$ to the vacuum of virtual and physical fermionic modes. The generalization to several virtual and physical fermionic modes per site is also straightforward.

\subsection{Injectivity and Parent Hamiltonians}\label{sec:PEPS_parent}

\begin{figure}
\centering\includegraphics[width=0.8\textwidth]{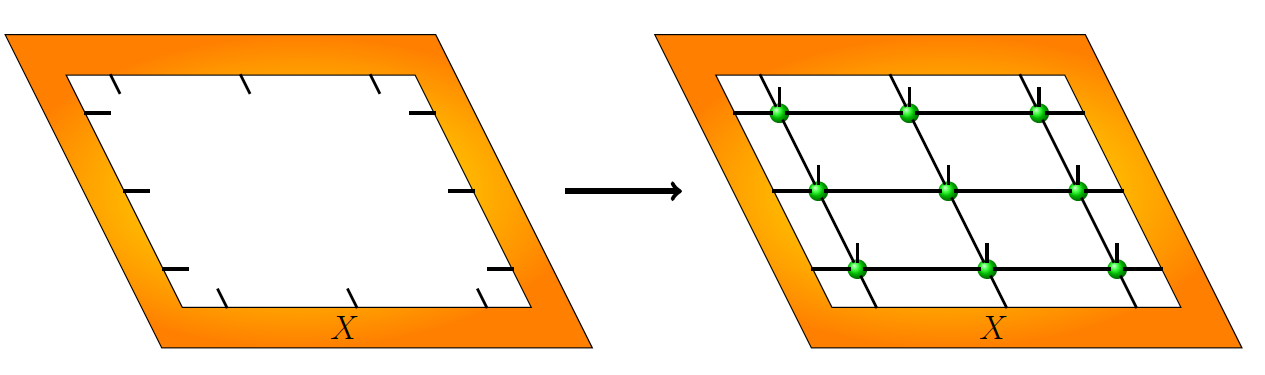}
\caption{We say a PEPS is injective if the shown map is injective for sufficiently large plaquette sizes. For the shown example of a $3 \times 3$ plaquette, this means that the tensor $X$ has twelve virtual indices and that the linear space spanned by such tensors is fully mapped (i.e., without decreasing its dimension) to the local physical space of the $3 \times 3$ plaquette.}
\label{fig:PEPS_map}
\end{figure}

As in the case of MPS, by \textit{injective PEPS}~\cite{PEPS_inj} we mean that by blocking sites, the full virtual space is accessible by acting on the local physical space. We continue to focus on the case of a square lattice. Then, we say a PEPS is injective if for sufficiently large plaquettes the linear map
\be
\Gamma_{L_y \times L_x}: X \rightarrow \sum_{i_{1,1}, ..., i_{L_x,L_y} = 1}^d \tr_v(X A_{i_{1,1}} ... A_{i_{L_x,L_y}}) |i_{1,1} ... i_{L_x,L_y}\rangle
\label{eq:PEPS_inj}
\ee
is injective, where $L_y \times L_x$ is the plaquette size and $\tr_v$ denotes the (implicit) summation over all virtual degrees of freedom. (For fermionic PEPS the anticommutation of the virtual fermionic modes needs to be taken into account when calculating this trace). A graphical illustration is presented in Fig.~\ref{fig:PEPS_map}. Again, injectivity for an $L_{y,0} \times L_{x,0}$ plaquette implies injectivity for plaquettes $L_y \times L_x$ with $L_x \geq L_{x,0}$, $L_y \geq L_{y,0}$. 

As in one dimension, the parent Hamiltonian is defined as the sum of translated projector on the kernel of the reduced density matrix of a certain region ($L_y \times L_x$), cf. Eq.~\eqref{eq:parent_Ham}. The reduced density matrix $\rho_{L_y \times L_x}$ can be calculated by first contracting the PEPS on the $L_y \times L_x$ plaquette with \textit{open} virtual indices $\alpha_1, \alpha_2, \ldots, \alpha_{2 L_x + 2 L_y}$ at its boundary, wherefore the obtained physical state depends on those indices $|\Psi_{\alpha_1, \alpha_2, \ldots, \alpha_{2 L_x + 2 L_y}}\rangle$. For a PEPS with norm 1 the desired reduced density matrix is 
\be
\rho_{L_x \times L_y} = \sum_{\alpha_1, \alpha_2, \ldots, \alpha_{2 L_x + 2 L_y}=1}^D |\Psi_{\alpha_1, \alpha_2, \ldots, \alpha_{2 L_x + 2 L_y}}\rangle \langle \Psi_{\alpha_1, \alpha_2, \ldots, \alpha_{2 L_x + 2 L_y}}|. \label{eq:rho_plaq}
\ee

The parent Hamiltonian is defined as a sum of projectors $h_{\mb n}$ acting non-trivially on $L_y \times L_x$ plaquettes,
\begin{align}
\mc H &= \sum_{\mb n} h_\mb{n}, \label{eq:parent_PEPS}\\
\mr{ker}(h) &= \mr{supp}(\rho_{L_x \times L_y}). \label{eq:kersupp_PEPS}
\end{align}
We set $L_x > L_{x,0}, L_y > L_{y,0}$, which renders the PEPS the unique ground state (with energy zero).

\subsection{Symmetries}\label{sec:PEPS_sym}

\begin{figure}
\centering\includegraphics[width=0.6\textwidth]{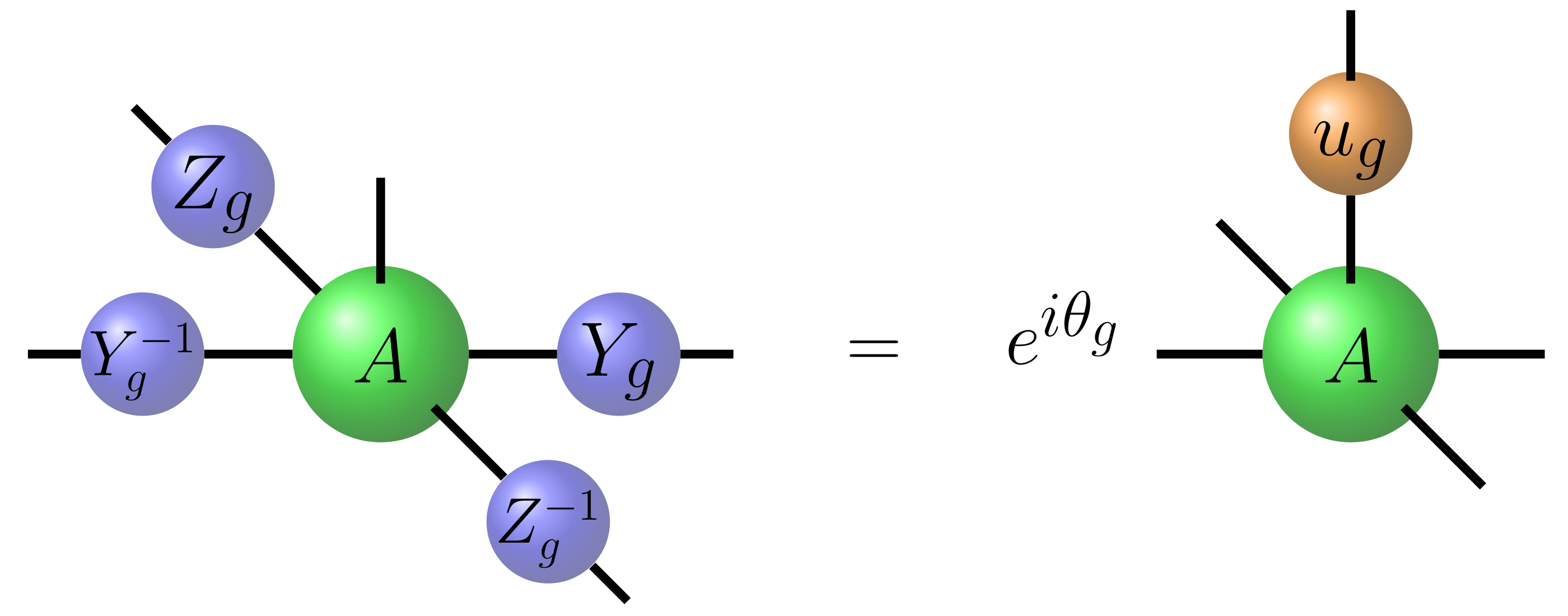}
\caption{PEPS tensor $A_{i,lrud}$ with a continuous symmetry. The matrices $Y_g$ and $Z_g$ represent the symmetry group $\mc G$ (whose elements are denoted by $g$). $u_g$ is a unitary representation of $G$ that acts on the physical level and $\theta_g$ is a one dimensional representation.}
\label{fig:PEPS_sym}
\end{figure}

The results on symmetries of two dimensional PEPS on a square lattice stated here can be found in Ref.~\cite{PEPS_sym}. We assume that it is injective and that its size $N_v \times N_h$ is sufficiently large. Say the symmetry group under consideration $\mc G$ is represented by unitary matrices $u_g$ acting on the local physical degree of freedom. The state $|\Psi\rangle$ is invariant under the symmetry if 
\be
(u_g)^{\otimes N_h N_v} |\Psi\rangle = e^{i N_h N_v \theta_g} |\Psi\rangle,
\ee
with the phases $\theta_g$ being a one dimensional representation of $\mc G$. If $\Psi$ describes a PEPS, the PEPS tensor $A_{i,lrud}$ can be brought into a form such that the relation shown in Fig.~\ref{fig:PEPS_sym} holds. 

A similar relation holds for spatial symmetries, such as inversion symmetry or reflection symmetry; the interested reader is referred to Ref.~\cite{PEPS_sym}.

\subsection{$\mc G$-injectivity and non-chiral topological PEPS}\label{sec:top_PEPS}

Topological phases are special in that they possess global (``topological'') order that is not amenable to Landau's characterization of symmetry broken phases. The ground states of two gapped Hamiltonians are said to be in different topological phases if they cannot be adiabatically connected without closing the energy gap. Hence, we call a state topological whenever it does not lie in the topologically trivial phase, that is, the phase which can be connected to a product state (for details, see next section).  

\begin{figure}[t]
\centering\includegraphics[width=\textwidth]{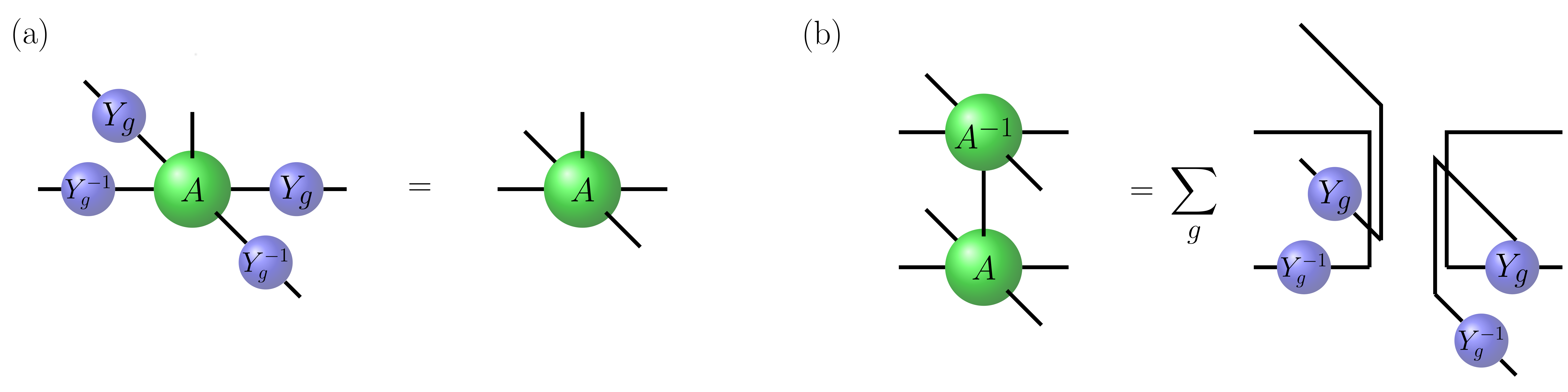}
\caption{Sufficient conditions on a PEPS tensor $A_{i,lrud}$ to give rise to a topological state. (a) The PEPS tensor possesses a symmetry that is only present on the virtual level. The matrices $Y_g$ represent the symmetry group $\mc G$ (whose elements are denoted by $g$). (b) The PEPS tensor - at least after blocking - is invertible within each invariant subspace of the symmetry group $\mc G$. (Crossing lines correspond to independent indices.)}
\label{fig:PEPS_Ginj}
\end{figure}

Many of the known examples of topological PEPS satisfy a condition that is very similar to the one for PEPS with a physical symmetry shown in Fig.~\ref{fig:PEPS_sym}. The only difference is that the tensors of such topological PEPS possess a symmetry that exists only on the virtual level, which corresponds to replacing $u_g$ by $\Id$ (and setting $\theta_g = 0$, $Z_g = Y_g$) in Fig.~\ref{fig:PEPS_sym}, as shown explicitly in Fig.~\ref{fig:PEPS_Ginj}(a). However, this condition alone is not sufficient for a topological PEPS. It becomes sufficient if we add the relation shown in Fig.~\ref{fig:PEPS_Ginj}(b), which together define the set of \textit{$\mc G$-injective} PEPS~\cite{TNS_intersection}. 

\begin{figure}[b]
\centering\includegraphics[width=0.8\textwidth]{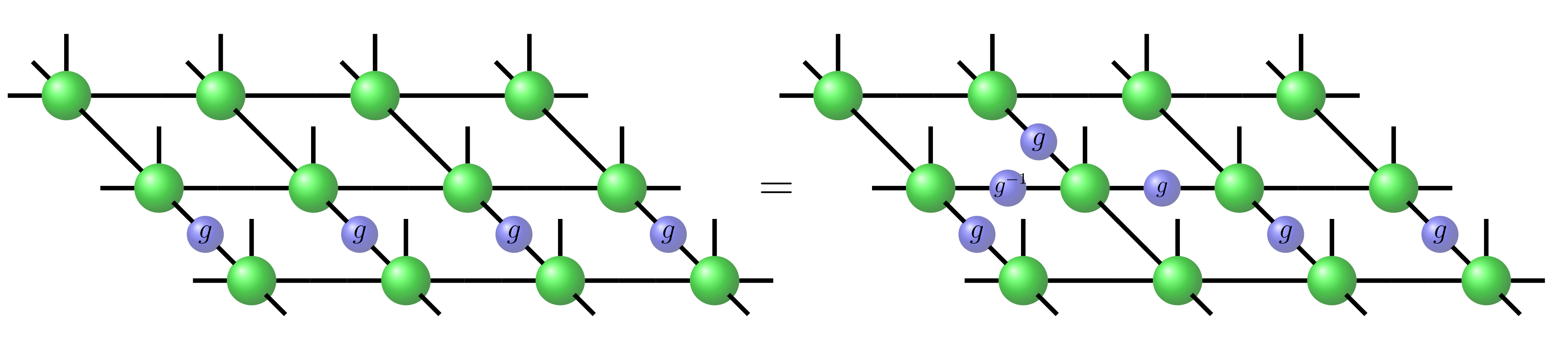}
\caption{Moving strings without changing the PEPS. Strings are tensor products of the matrices $Y_g$ and their inverses $Y_g^{-1}$ denoted by $g$ and $g^{-1}$ in the figure, respectively. The equation shows how a string can be moved upward by consecutively using the relation of Fig.~\ref{fig:PEPS_Ginj}(a). The PEPS is assumed to be defined on a torus, i.e., with periodic boundary conditions.}
\label{fig:PEPS_strings}
\end{figure}

The property in Fig.~\ref{fig:PEPS_Ginj}(a) allows for the insertion of strings (tensor products) of $Y_g$ matrices on the virtual level along non-contractible loops, which can be moved without changing the resulting state on the torus (i.e., with periodic boundary conditions), see Fig.~\ref{fig:PEPS_strings}. Note however, that the insertion of the string changed the state as compared to the original PEPS. Since the strings cannot be detected by local measurements, the states with strings are also ground states of the local Hamiltonian defined by Eqs.~\eqref{eq:parent_PEPS} and~\eqref{eq:kersupp_PEPS}. (The insertion of strings does not change the reduced density matrix $\rho_{L_y \times L_x}$, so the PEPS with strings have the same parent Hamiltonian.) Let us call the state with one horizontal string of group element $g$ and one vertical string of group element $h$ $|\Psi(g,h)\rangle$. For this case it has been shown in Ref.~\cite{TNS_intersection} that the ground state subspace is spanned by the states $|\Psi(g,h)\rangle$ with $gh = hg$. Furthermore, anyonic excitations can be understood as the extreme points of open strings and the braiding properties related to the group $\mc G$.

Examples of $\mc G$-injective topological PEPS are the toric code~\cite{toric_code,toric_PEPS}, where $Y_g = \{\Id, \sigma_z\}$, leading to a fourfold degenerate ground state subspace on the torus, RVB states~\cite{RVB_org,RVB_PEPS} and the quantum double models~\cite{toric_code,TNS_intersection}.

A general characterization of topological PEPS has recently been carried out in Ref.~\cite{Top_PEPS}, but will not be considered here.

\subsection{PEPS on a cylinder}\label{sec:PEPS_cylinder}

\begin{figure}[b]
\centering\includegraphics[width=0.8\textwidth]{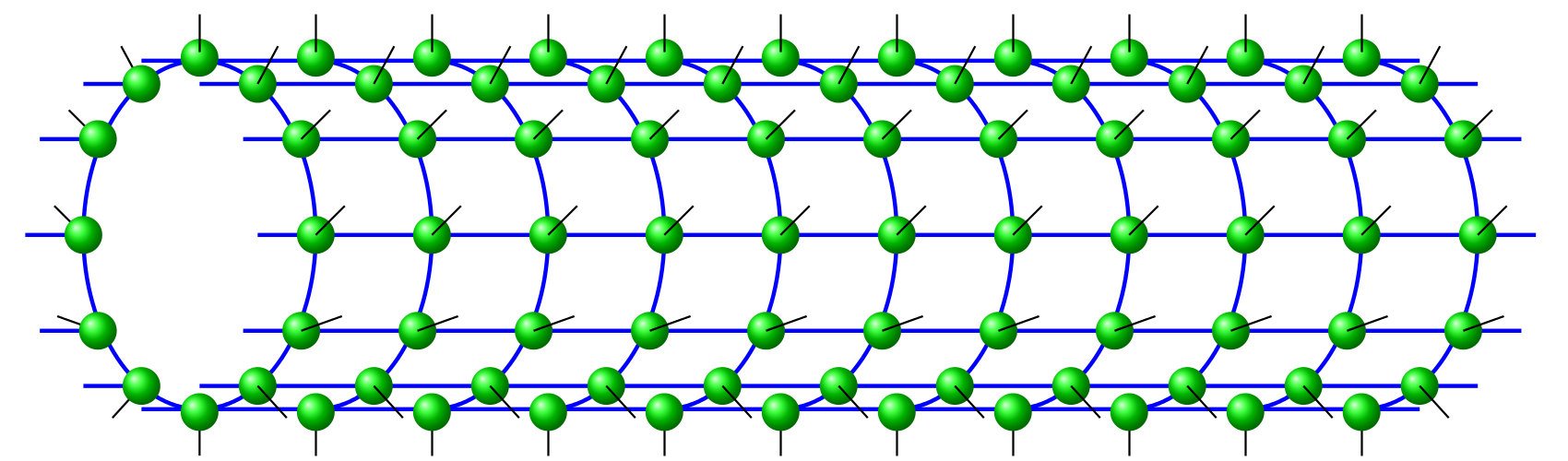}
\caption{Contraction of a translationally invariant PEPS on a cylinder (periodic boundary conditions in vertical direction) with open virtual indices in horizontal direction. The blue thick lines correspond to virtual indices and the thin black lines to the open physical indices. The green balls denote the rank-5 tensor $A_{i,lrud}$ constituting the PEPS.}
\label{fig:cylinder2}
\end{figure}

The set-up we will consider in the remainder of this section is a translationally invariant spin-PEPS with periodic boundary conditions in vertical direction and open virtual indices in horizontal direction. This corresponds to contracting the PEPS on a horizontal cylinder of length, $N$ leaving the leftmost and rightmost virtual indices open (see Fig.~\ref{fig:cylinder2}). This configuration is useful for illustrating the notion of the transfer operator and the bulk-boundary correspondence~\cite{bulk_boundary}, which will be done in the following two subsections.

\subsubsection{Transfer operator of PEPS}\label{sec:PEPS_transfer}

\begin{figure}
\centering\includegraphics[width=0.5\textwidth]{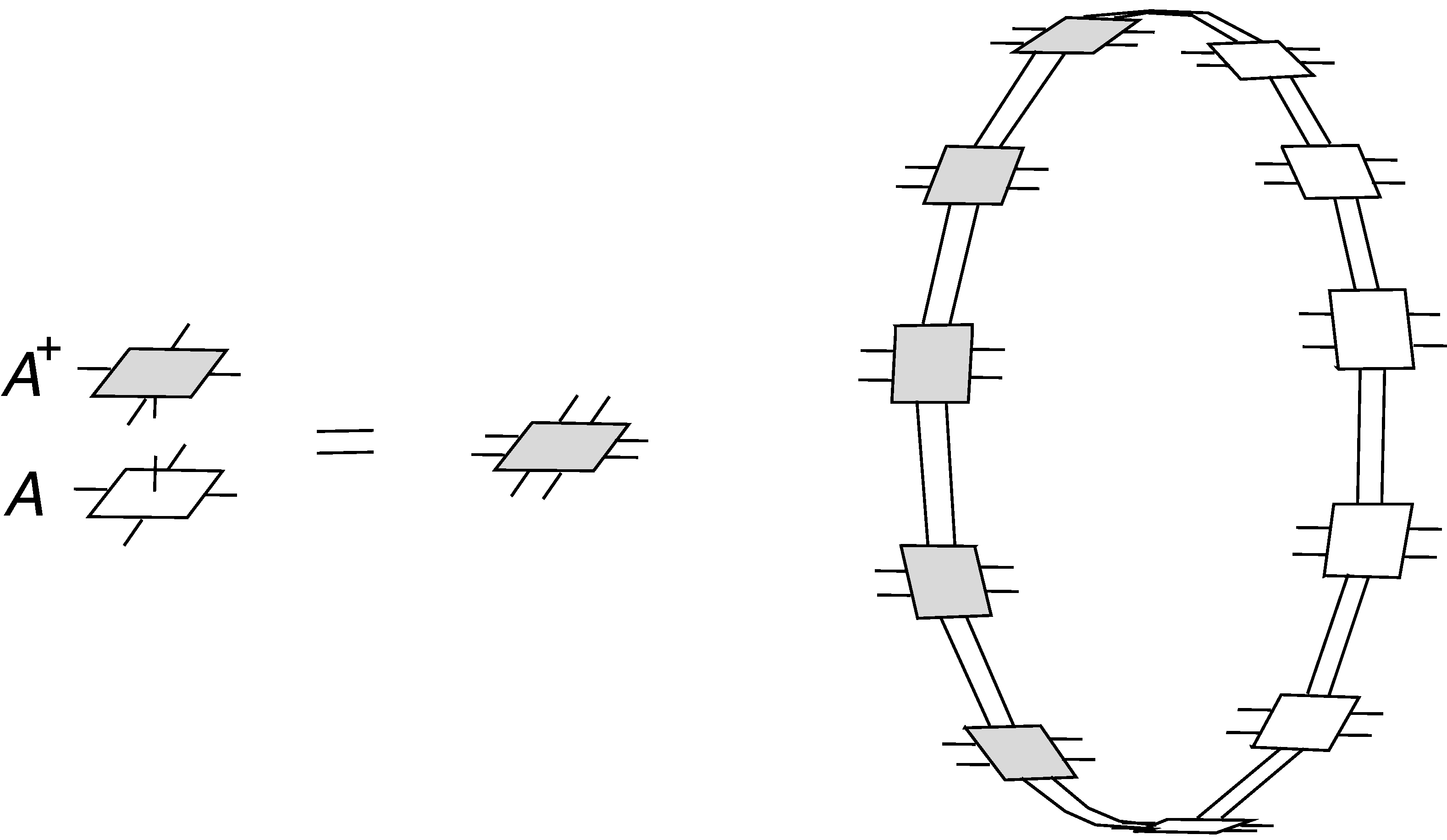}
\caption{Transfer operator of a PEPS on a cylinder. Left: In order to build the transfer operator, the physical index $i$ is contracted, forming the local double tensor $\sum_{i=1}^d A_{i,lrud} A_{i,l'r'u'd'}^\dg$, which has eight virtual legs. Right: After contracting all double tensors of one column, one obtains the transfer operator $\tilde{\mathbb{T}}$.}
\label{fig:transfer}
\end{figure}

The transfer operator can be used to determine whether a PEPS has algebraically or exponentially decaying correlations. First, observe that the configuration displayed in Fig.~\ref{fig:cylinder2} can actually be viewed as an MPS: Blocking all sites in one column to one super-site yields the corresponding rank-3 MPS tensor $\tilde A_{\tilde i,\tilde l \tilde r}$, where $\tilde i = 1, \ldots, d^{N_v}$, $\tilde l = 1, \ldots, D^{N_v}$ and $\tilde r = 1, \ldots, D^{N_v}$ ($N_v$ is the number of sites in vertical direction). As we now are dealing with an MPS, we can bring the tensors $\tilde A$ into the block diagonal form~\eqref{eq:A_blocks}. From that, we might conclude, using the same reasoning as at the end of subsection~\ref{sec:MPS_construction}, that PEPS always have exponentially decaying correlations. However, we need to take into consideration that the bond dimension of $\tilde A$ grows (exponentially) with the circumference $N_v$: The double tensor of one column, usually called the \textit{transfer operator} of the PEPS (see Fig.~\ref{fig:transfer}), is given by $\tilde{\mathbb{T}} = \sum_{\tilde i} \tilde A_{\tilde i} \otimes \tilde A_{\tilde i}^\dg$. If it has a gap between the eigenvalue(s) of magnitude 1 and the next highest absolute eigenvalue $\zeta_1(N_v)$ for $N_v \rightarrow \infty$, we can indeed follow the arguments given for MPS to show that the correlations decay exponentially: Take, e.g., $N = N_v \gg x$ in Eq.~\eqref{eq:exp_corr}; then we obtain the bound
\be
|C(O^n_1, O^{n+x}_2)| < f(N) \zeta^{x-1} + \mc{O}(\zeta^{N-x-1}),
\ee
where $\zeta = |\lim_{N_v \rightarrow \infty} \zeta_1(N_v)|$ and $f(N) \geq 0$. If the operators $O_1$ and $O_2$ act on single sites, $f(N)$ is upper bounded, since then  $\langle L|\mathbb{O}_1 \ldots \mathbb{O}_2 |R\rangle$ (see Eq.~\eqref{eq:exp_corr}) only retains eigenvectors $|R_1\rangle$, $\langle L_1|$ of the transfer operator which are close to $|R\rangle$, $\langle L|$ on all other sites. 

The above arguments can no longer be applied if $\left| \lim_{N_v\rightarrow \infty} \zeta_{1}(N_v) \right| = 1$; in this case, correlations decay algebraically. Thus, determining the spectrum of the transfer operator for different $N_v$ (and from the finite-size scaling its gap) indicates whether a PEPS has exponentially or algebraically decaying correlations.

\subsubsection{Bulk-boundary correspondence}\label{sec:bulk_boundary}

Now consider the PEPS defined on a torus, that is, with periodic boundary conditions in horizontal and vertical direction. If we split the torus into two horizontal cylinders of equal length $N$, say $\mc R$ and $\overline{\mc R}$, and trace out one of them ($\overline{\mc R}$), we obtain a reduced density matrix that is defined only on the physical modes of the other cylinder ($\mc R$). We now use the geometry of Fig.~\ref{fig:transfer}, which defines a physical state $|\Psi_{\alpha_1, \alpha_2, \ldots, \alpha_{2 N_v}}^\mc{R}\rangle$ depending on the choice of the open virtual indices $\alpha_1, \ldots, \alpha_{2 N_v} \in \{1, \ldots, D\}$. From it, the reduced density matrix $\rho_\mc{R}$ of the physical modes in region $\mc R$ can be obtained as in Eq.~\eqref{eq:rho_plaq}. However, this equation shows that the rank of the reduced density matrix $\rho_\mc{R}$ is upper bounded by $(2N_v)^{D}$. Hence, it is isometric to a reduced density matrix $\sigma_\mc{R}$ defined only on the $2 N_v$ virtual boundary modes,
\be 
\rho_\mc{R} = \mc{V}_\mc{R} \sigma_\mc{R} \mc{V}^\dg_\mc{R}
\ee
with an isometry $\mc{V}_\mc{R}$, $\mc{V}^\dg_\mc{R} \mc{V}_\mc{R} = \Id_{(2 N_v)^D \times (2 N_v)^D}$. 
$\sigma_\mc{R}$ has been derived in terms of the reduced density matrices $\tilde \sigma_\mc{R}$ and $\tilde \sigma_{\overline{\mc R}}$ obtained when tracing out the physical indices in Fig.~\ref{fig:cylinder2} and its analogue for region $\overline{\mc R}$, respectively. The result is~\cite{bulk_boundary}
\be
\sigma_\mc{R} \propto \sqrt{\tilde \sigma_{\overline{\mc R}}^\top} \, \tilde \sigma_\mc{R} \sqrt{\tilde \sigma_{\overline{\mc R}}^\top}.
\ee
Note that $\tilde \sigma_\mc{R}$ corresponds to $\tilde{\mathbb{T}}^N$ and $\tilde \sigma_{\overline{\mc R}}$ to $(\tilde{\mathbb{T}}^\top)^N$. If the PEPS changes sign under a horizontal reflection, $\tilde \sigma_{\overline{\mc R}}^\top = \tilde \sigma_\mc{R}$, and thus, $\sigma_\mc{R} \propto \tilde \sigma_\mc{R}^2$. $\sigma_\mc{R}$ is very useful when the entanglement spectrum~\cite{ES_org} of a PEPS is to be calculated, as the computation of the spectrum of the full reduced density matrix $\rho_\mc{R}$ is reduced to a one dimensional problem. The Hamiltonian associated with $\sigma_\mc{R}$ is called the \textit{boundary Hamiltonian} $\mc{H}^\mr{b}$,
\be
\sigma_\mc{R} = \frac{1}{Z} e^{-\mc{H}^\mr{b}}.
\ee
Its energies reproduce all finite energies of the bulk entanglement spectrum.

\paragraph{Boundary theories of topological states.}

Assume that the PEPS is $\mc G$-injective, i.e., it has a symmetry that only exists on the virtual level of the kind shown in Fig.~\ref{fig:PEPS_Ginj}. As in Fig.~\ref{fig:PEPS_strings}, the symmetry of a single site can be concatenated to an operator $U_g$ acting on all open indices of the PEPS defined on the cylinder (Fig.~\ref{fig:cylinder2}) leaving the state (corresponding to a particular combination of the open virtual indices) invariant. This invariance translates to a symmetry of the boundary reduced density matrix $\sigma_\mc{R}$~\cite{transfer},
\be
U_g \sigma_\mc{R} = \sigma_\mc{R} U_g = \sigma_\mc{R}, \label{eq:boundary_Ginj}
\ee
where the symmetry is present on the bra and the ket layer. From Eq.~\eqref{eq:boundary_Ginj} it follows that $\sigma_\mc{R}$ is supported on a proper subspace of the virtual system, that corresponding to the eigenvalue 1 of all $U_g$'s, i.e., 
\be
\sigma_\mc{R} = \frac{1}{Z} P e^{- \mc{H}^\mr{b}} P. \label{eq:project_Ginj}
\ee
%
%
Here, $P$ is a non-local operator which projects onto that subspace. This fact has two consequences: (i) The zero R\'enyi entropy (which is the logarithm of the dimension of that subspace) does not coincide with the logarithm of the dimension of the Hilbert space of the virtual particles on the boundary of ${\cal R}$; (ii) there is a non-local constraint on the boundary Hamiltonian. Those two features are thus related to the topological character of the PEPS. Note that (i) may also imply in some cases that there is a correction to the area law, what is usually called the topological entanglement entropy. That is, the von Neumann entropy of $\sigma_\mc{R}$ scales like the number of virtual particles on the boundary of ${\cal R}$ minus a universal constant, which is directly related to the topological properties of the model under study. The property (ii) acts as a superselection rule in the boundary theories, since any perturbation in the bulk will not change that subspace. Additionally, in the spin lattices studied in Ref.~\cite{transfer}, ${\cal H}^{\rm b}$ is local (contains hoppings that decay exponentially with the distance) whenever the frustration free parent Hamiltonian of the state $\Phi$ is gapped.


%% file: 3.2.tex
\section{Introduction to Chiral Topological Systems}\label{sec:chiral_intro}

\subsection{Background}

The major part of this thesis is devoted to chiral topological PEPS. The purpose of this subsection is to give a brief introduction into the field of topological systems with special emphasis on chiral topological states. Historically, the field of topological systems emerged with the discovery of the Integer Quantum Hall Effect in 1980~\cite{QHE_ex}, where very sharp plateaus in the longitudinal conductance are observed as a function of the magnetic field for sufficiently high magnetic fields, low temperatures and pure samples, as described in the introduction. The sharp quantization of the longitudinal conductance to integer multiples of $h/e^2$ was soon realized to be a topological effect, i.e., the electrons of the system interplay in such a way that the macroscopic observable, the Hall conductance, is only related to a topological feature of the overall wavefunction, which is robust to small perturbations and disorder present in the system. The Integer Quantum Hall Effect can be explained in terms of non-interacting electrons, that is, without taking into account the Coulomb interaction between them.  (Nevertheless, the electrons are not independent, since they occupy a many-body wavefunction, i.e., one first needs to apply a non-local basis transformation between their modes in order to make them independent.) On the other hand, the Fractional Quantum Hall Effect~\cite{FQHE_ex} is a true many-body topological effect and cannot be explained without the Coulomb interaction between the electrons. Interacting topological phases display many intriguing features absent in the non-interacting regime, such as a ground state degeneracy that depends on the topology the system is defined on (such as the fourfold degeneracy of the ground state subspace of the toric code~\cite{toric_code} mentioned in the last section). Moreover, they give rise to Abelian or non-Abelian excitations; in the latter case, the excitations are quasi-particles whose braiding statistics depends on the order in which the individual particles are moved around each other. Such systems have a subleading (negative) correction to the area law, called the \textit{topological entanglement entropy}~\cite{TEE_org,TEE_Levin}, and for this reason are also called \textit{long range entangled}~\cite{SRE_LRE}. On the other hand, non-interacting systems never display such a topological correction to the area law. Such states are denoted as \textit{short range entangled}.

This subsection is divided as follows:  
As is obvious from the above overview, there is an important distinction between non-interacting and interacting topological phases. That is why the next subsection is meant to first define the notion of of non-interacting fermionic systems. As will become clear, their Hamiltonians can be exactly diagonalized by using unitaries whose dimensions grow only linearly with the system size. 
In subsection~\ref{sec:Classification}, advantage will be taken of the possibility to describe free fermionic systems exactly to show that there exists a full classification of free fermionic topological phases. This will also include new examples discovered in the last decade such as \textit{topological insulators}~\cite{top_ins}. Subsection~\ref{sec:CFT} briefly discusses the field of topological phases of interacting fermions. For concreteness, it is restricted to chiral phases, since all results presented in the remainder of this chapter will concern chiral topological systems only.

\subsection{Free fermionic states}\label{sec:free}

Free fermionic states (pure or mixed) are states whose fermionic degrees of freedom can be completely decoupled by applying a linear transformation on the creation and annihilation operators such that the new modes fulfill the fermionic anticommutation relations~\cite{Gaussian_Bravyi}. The transformed modes describe a system of independent fermions, thus the name ``free fermionic state''. All states $\rho$ with such a property turn out to be parameterized according to
\be
\rho = \mc{N} e^{-\tfrac{i}{4} \sum_{l,m} H_{lm} c_l c_m}, \label{eq:def_Gaussian}
\ee
where $\mc{N}$ is a normalization constant and the $c_l$ are Majorana modes defined as 
\begin{align}
c_{2j-1} &= a_j^\dg + a_j \\
c_{2j} &= (-i)(a_j^\dg - a_j).
\end{align}
$a_j$ ($j = 1, \ldots, n$) is the annihilation operator of the fermionic mode with index $j$. Thus, the Majorana operators are Hermitian ($c_l = c_l^\dg$) and fulfill the anticommutation relation
\be
\{c_l,c_m\} = 2 \delta_{lm}. \label{eq:Majorana_anticom}
\ee
Because of that, one restricts the matrix $H$ to be real and antisymmetric, $H = - H^\top$.
This reveals that the strength in the Majorana formalism lies in their equal footing for even and odd indices $l$ (as opposed to the set of creation and annihilation operators). Linear transformations as mentioned above which preserve their anticommutation relations~\eqref{eq:Majorana_anticom} (and thus the anticommutation relations of the corresponding fermionic modes) are simply given by the set of real orthogonal $2n \times 2n$ matrices $O$ via $c_l' = \sum_m O_{lm} c_m$. The state $\rho$ can be decoupled into $n$ independent fermionic modes by setting $O$ such that it block-diagonalizes $H$ into real $2 \times 2$ blocks,
\be
H = O^\top \bigoplus_{j=1}^n \left(\begin{array}{cc}
0&h_j\\
-h_j&0
\end{array}\right) O.
\ee 
The above transformation then gives 
\be 
\rho = \prod_{j=1}^n e^{-\tfrac{i}{2} h_j c_{2j-1}' c_{2j}'}. \label{eq:rho_prod}
\ee

Instead of using the matrix $H$ to characterize free fermionic states, it is more convenient to use the \textit{covariance matrix} (CM) $\Gamma$, defined via the two-point correlations,
\be
\Gamma_{lm} = \frac{i}{2} \tr([c_l,c_m] \rho) = \begin{cases} i \, \tr(c_l c_m \rho) &\mbox{if } l \neq m \\ 
0 & \mbox{if } l = m. \end{cases} \label{eq:CM}
\ee
The description of free fermionic states in terms of CMs is extensively treated in Ref.~\cite{Gaussian_Bravyi}, which shall only very briefly be summarized here. The covariance matrix is real and antisymmetric and fulfills $\Gamma \, \Gamma^\top = - \Gamma^2 \leq \Id$ with equality if and only if $\rho$ is a pure state. Note that a pure Gaussian state (that is not orthogonal to the vacuum) can be created out of the vacuum $|\Omega\rangle$ via an exponential of a sum of second order terms in creation and annihilation operators according to Eq.~\eqref{eq:def_Gaussian}. Up to a normalization factor it is given by
\be
|\psi\rangle \propto e^{-\tfrac{i}{4} \sum H_{lm} c_l c_m} |\Omega\rangle. \label{eq:pure_Gaussian}
\ee
From $\Gamma$ (as from $H$) it is possible to uniquely reconstruct $\rho$. Hence, in the case of free fermions, it is much more convenient to deal with CMs rather than with full states. Higher order correlations can be obtained via Wick's theorem~\cite{Wicks_theorem}. Due to Eq.~\eqref{eq:rho_prod}, the orthogonal transformation $O$ block-diagonalizes $\Gamma$ into real $2 \times 2$ blocks, just as it block-diagonalizes $H$, that is,
\be
\Gamma = O^\top \bigoplus_{j=1}^n \left(\begin{array}{cc}
0&\lambda_j\\
-\lambda_j&0
\end{array}\right) O \label{eq:Gamma_diag}
\ee 
with $-1 \leq \lambda_j \leq 1$ because of $- \Gamma^2 \leq \Id$. For a pure state, we can choose without loss of generality $\lambda_j = 1 \ \forall \ j = 1, \ldots n$. Then, Eq.~\eqref{eq:Gamma_diag} can be used to show that any Hamiltonian $\mc H_0$ of the form
\begin{align}
\mc H_0 &= -\tfrac{i}{4} \sum_{l,m} [H_0]_{lm} \, c_m c_l, \label{eq:Gaussian_parent} \\
H_0 &= O^\top \bigoplus_{j=1}^n \left[ \epsilon_j \left(\begin{array}{cc}
0&1\\
-1&0
\end{array}\right) \right] O \label{eq:Gaussian_parent2}
\end{align}
with $\epsilon_j > 0 \ \forall \ j = 1, \ldots n$ has that pure state as its unique ground state. In particular, $\epsilon_j = 1$, i.e., $H_0 = \Gamma$ yields a flat-band Hamiltonian, that is, a Hamiltonian whose energy eigenvalues are only $+1$ and $-1$. Then, the state described by the CM $\Gamma$ has all negative energy orbitals filled. 

Note that all Hamiltonians which have only second orders of creation and annihilation operators possess eigenstates and thermal states which are free fermionic states.

Free fermionic states are also called \textit{Gaussian states} due to the relation~\eqref{eq:def_Gaussian} (and a similar relation in terms of Grassmann variables). The most general map from Gaussian states to Gaussian states (basically a restriction of the map~\eqref{eq:PEPS_map} to free fermionic systems) is given by
\be
\Gamma_\mr{out} = B(D+\Gamma_\mr{in}^{-1})^{-1} B^\top + A, \label{eq:Gaussian_map}
\ee
where $\Gamma_\mr{in}$ and $\Gamma_\mr{out}$ are the CMs of the input and output state, respectively. They may also describe mixed states, so Gaussian maps may also be used for them. The matrices $A$, $B$ and $D$ are blocks of another CM,
\be
\gamma_1 = \left(\begin{array}{cc}
A&B\\
-B^\top&D
\end{array}\right). \label{eq:M_blocks}
\ee
Correspondingly $- \gamma_1^2 \leq \Id$ with equality indicating that the \textit{Gaussian linear map}~\eqref{eq:Gaussian_map} maps pure states to pure states. Conveniently, maps defined via Eq.~\eqref{eq:Gaussian_map} can be used to define Gaussian fermionic PEPS which are PEPS that are explicitly constructed to be free fermionic.

\subsubsection{Gaussian fermionic PEPS}\label{sec:GFPEPS}

A Gaussian fermionic PEPS (GFPEPS)~\cite{fPEPS_Kraus} is a free fermionic PEPS that can be either pure or mixed and which is obtained by requiring the initial state of maximally entangled pairs before the projection to be Gaussian and by requiring the local projections to be Gaussian maps, Eq.~\eqref{eq:Gaussian_map}:
We start with an $N \times N$ square
lattice with periodic boundaries and $f$ \textit{physical} fermionic orbitals at each site,
with creation (annihilation) operators $a_{\mathbf{r},j }^{\dagger }$
($a_{\mathbf{r},j }$), with $\mathbf{r}=(x,y)$ the site and $j =1, \ldots,
f$ the orbital index;  we will mostly work in the basis of physical
Majorana operators $e_{\mathbf{r},2j-1}=a_{\mathbf{r},j }^{\dagger
}+a_{\mathbf{r},j }$ and $e_{\mathbf{r},2j}=(-i)(a_{\mathbf{r},j
}^{\dagger }-a_{\mathbf{r},j })$.  To obtain a PEPS description of the
system, we start out with maximally entangled \emph{virtual} Majorana
modes $c_{\mathbf{r},w,\kappa}$ (with $\kappa=1,\dots,\chi$ and
$w=L,R,U,D$), which are obtained by acting with $1 + i
c_{\mathbf{r},R,\kappa}^r c_{\mathbf{r}+\hat x,L,\kappa}$ and $1 + i
c_{\mathbf{r},U,\kappa} c_{\mathbf{r}+\hat y,D,\kappa}$ on the
vacuum, yielding a pure state
$\rho_\mathrm{in}$~\cite{fPEPS_Kraus} (thus, a maximally entangled pair of Majorana modes is basically ``half'' of a maximally entangled pair of fermionic modes). Here, the number of Majorana bonds $\chi$
is a parameter which can be used to systematically enlarge the class of accessible physical states. Subsequently, we apply the same Gaussian linear map $\cal{E}$ to
each lattice site $\mathbf r$, which maps the $4\chi$ auxiliary modes
$c_{\mathbf r,w,\kappa}$ to the $2f$ physical Majorana modes $e_{\mathbf
r,s}$ ($s = 1, \ldots, 2f$), as shown in Fig.~\ref{fig:PEPS}. This yields the translationally invariant Gaussian fermionic PEPS
$\rho_{\mathrm{out}}$, which can be described in
terms of its \emph{covariance matrix} (CM) $\Gamma_{\mathrm{out}}$,
defined as $(\Gamma_{\rm out})_{(\mathbf{r},s),(\mathbf{r}',t)} =
\frac{i}{2}{\rm tr}(\rho_{\rm out} [e_{\mathbf{r},s},e_{\mathbf{r}',t}])$;
similarly, for $\rho_\mathrm{in}$ we have $(\Gamma_{\rm
in})_{(\mathbf{r},\kappa),(\mathbf{r}',\kappa')}^{w,w'} = \frac{i}{2}{\rm
tr}(\rho_{\rm in}
[c_{\mathbf{r},w,\kappa},c_{\mathbf{r}',w',\kappa'}])$. The latter can be calculated to be
\begin{align}
\Gamma_{\mathrm{in}} &= \Gamma_h \oplus \Gamma_v,\\
\Gamma_{h/v} &= \bigoplus_{z=1}^N \left( \mathrm{Perm}(N,1,2,...,N-1) \otimes \left(\begin{array}{cc}
0&\Id_{\chi}\\
0&0
\end{array}\right) - \mathrm{Perm}(2,3,...,N,1) \otimes \left(\begin{array}{cc}
0&0\\
\Id_{\chi}&0
\end{array}\right) \right),
\end{align}
where $\mathrm{Perm}(i_1,i_2,...,i_N)$ is the permutation matrix whose $(m,l)$ element is $\sum_{j = 1}^N \delta_{m,i_j} \delta_{j,l}$ and $\Id_n$ denotes the $n \times n$ identity matrix.
The index $z$ denotes the sum over the other direction than the permutation refers to: For $\Gamma_h$ the permutation refers to the $x$ direction and $z$ denotes the direct sum over all rows ($y$ direction), since the horizontal bonds are identical 
for all rows. For $\Gamma_v$ the two directions are simply exchanged. 

Finally, $\mathcal E$ can be expressed using a CM $\gamma_1$ defined on the
$2f+4\chi$ Majorana modes $\{e_{\mathbf r,s},c_{\mathbf r,w,\kappa}\}$
which encodes how $\mathcal E$ correlates the input modes with the output
modes~\cite{Gaussian_Bravyi}. The CM $\gamma_1$ for the Gaussian map $\mathcal E$
is given by Eq.~\eqref{eq:M_blocks}. $A \in \mathbb{R}^{2 f \times 2 f}$, $B \in \mathbb{R}^{2 f \times 4
\chi}$, and $D \in \mathbb{R}^{4 \chi \times 4 \chi}$ are variational parameters corresponding to
physical and virtual modes. Any $\gamma_1$ with $\gamma_1 \gamma_1^\top\leq \Id$ characterizes an admissible $\mathcal E$. 

Thus, the CM of the GFPEPS is
\begin{equation}
\Gamma_{\mathrm{out}} = B' (D' - \Gamma_{\mathrm{in}})^{-1} B'^\top + A',
\label{Gamma_gfpeps}
\end{equation}
where the primed matrices are defined via $\gamma_1' = \bigoplus_{\mb r} \gamma_1$ and we used $\Gamma_\mathrm{in}^2 = - \Id$.

For the calculation of $\Gamma_{\mathrm{out}}$, we employ the formalism introduced in \cite{fPEPS_Kraus}:  $\Gamma_{\mathrm{in,out}}$ are circulant matrices and can thus be block-diagonalized by the unitary matrix representation $\mathcal{F}$ of the Fourier transformation to reciprocal space. The Fourier transformations for the two spacial directions are
\begin{equation}
\mathcal{F}_h = \mc F_v = \frac{1}{\sqrt{N}} \left(\begin{array}{ccccc}
1&1&1&...&1\\
1&e^{-\frac{2 i \pi}{N}}&e^{-\frac{4 i \pi}{N}}&...&e^{-\frac{2 i (N-1) \pi}{N}}\\
1&e^{-\frac{4 i \pi}{N}}&e^{-\frac{8 i \pi}{N}}&...&e^{-\frac{4 i (N-1) \pi}{N}}\\
...\\
1&e^{-\frac{2 i (N - 1) \pi}{N}}&e^{-\frac{4 i (N - 1)\pi}{N}}&...&e^{-\frac{2i (N-1)^2 \pi}{N}}\\
\end{array}\right),
\label{eq:Fourier_matrix}
\end{equation}
such that $\mc F = (\mc F_h \otimes \Id_{2 \chi}) \oplus (\mc F_v \otimes \Id_{2 \chi})$.
If we apply $\mathcal{F}$ from the left and $\mathcal{F}^\dg$ from the right on \eqref{Gamma_gfpeps}, we block-diagonalize $\Gamma_{\mr{in,out}}$ and obtain
\begin{equation}
G_{\mathrm{out}}(\mb k) = B (D - G_{\mathrm{in}}(\mb k))^{-1} B + A, \label{eq:Gout} 
\end{equation}
where $G_{\mathrm{in,out}}(\mb k)$ are the blocks on the diagonals of $\mathcal{F} \Gamma_{\mathrm{in,out}} \mathcal{F}^\dg$, which are labelled by the reciprocal vector $\mb k$. They are the covariance matrices of the Fourier transformed 
virtual Majorana modes  
\begin{equation}
\hat c_{\mb k, w,\kappa} = \frac{1}{N} \sum_\mb{r} e^{i \mb r \cdot \mb k} c_{\mb r,w, \kappa}
\end{equation}
and Fourier transformed physical Majorana modes ($s = 1, \ldots, 2 f$)
\begin{equation}
\hat e_{\mb k,s} = \frac{1}{N} \sum_{\mb r} e^{i \mb r \cdot \mb k} e_{\mb r,s},
\end{equation}
respectively. That is, $(G_\mr{in})_{\kappa,\kappa'}^{w,w'}(\mb k) = \frac{i}{2}\tr(\rho_{\rm in}
[\hat c_{\mb k,w,\kappa},\hat c_{\mb k, w',\kappa'}^\dg])$ and
$(G_\mr{out})_{s,t}(\mb k) =
\frac{i}{2}\tr(\rho_\mr{out} [\hat e_{\mb k,s},\hat e_{\mb k,t}^\dg])$

Hence, we have
\begin{equation}
G_{\mathrm{in}} (\mb k) = \left(\begin{array}{cc}
0&\Id_{\chi} e^{i k_x}\\
-\Id_{\chi} e^{-i k_x}&0\\
\end{array}\right) 
\oplus
\left(\begin{array}{cc}
0&\Id_{\chi} e^{i k_y}\\
-\Id_{\chi} e^{-i k_y}&0\\
\end{array}
\right).
\end{equation}
After applying a Fourier transform on $\Gamma_\mr{out}^\top = - \Gamma_\mr{out}$, one obtains that $G_\mr{out}^\dg (\mb k) = - G_\mr{out}(\mb k)$. Furthermore, if $\Gamma_\mr{out} \Gamma_\mr{out}^\top = \Id$ (a pure state), the Fourier transform yields $G_\mr{out}(\mb k) G_\mr{out}^\dg(\mb k) = \Id$. This is the case whenever the Gaussian map is pure, i.e., $\gamma_1 \gamma_1^\top = \Id$. 

If we write the inverse of $D - G_\mathrm{in}(\mb k)$ in Eq.~\eqref{eq:Gout} in terms of the adjugate matrix $\mathrm{Adj}(\cdot)$
\begin{align}
[D - G_\mathrm{in}(\mb k)]^{-1} = \frac{\mathrm{Adj}(D - G_{\mathrm{in}}(\mb k))}{\det(D - G_{\mathrm{in}}(\mb k))},
\end{align}
it follows that the elements of $G_{\mathrm{out}}(\mb k)$ are fractions of finite-degree trigonometric polynomials,
\begin{equation}
[G_\mathrm{out}(\mb k)]_{s t} = \frac{p_{s t}(\mb k)}{q(\mb k)}.
\end{equation}
$p_{s t}(\mb k)$ and $q(\mb k) \equiv \det(D - G_\mathrm{in}(\mb k))$ are polynomials of $\sin k_{x}$, $\cos k_{x}$, $\sin k_{y}$, and $\cos k_{y}$ of degree $\leq 2 \chi$ each, bounded by the number of virtual Majorana bonds.
Note that $q(\mb k) \in \mathbb{R}$, since $D - G_{\mathrm{in}}(\mb k)$ is anti-Hermitian and has even dimension, implying that its determinant is real. 

According to Eqs.~\eqref{eq:Gaussian_parent} and~\eqref{eq:Gaussian_parent2}, for pure GFPEPS, the class of quadratic Hamiltonians
\begin{equation}
\mc H_0=-\tfrac{i}{4}\sum_{\mathbf{k}} \sum_{s,t} \varepsilon(\mathbf{k}) \lbrack G _{\mathrm{out}}(\mathbf{k})]_{st} \hat e_{\mathbf{k},s} \hat e^\dg_{\mathbf{k},t},  \label{parent_Ham}
\end{equation}%
with spectrum $\varepsilon(\mb k) = {\varepsilon(- \mb k)}\ge0$ has
$\rho_\mathrm{out}$ as its ground state. These Hamiltonians
can have different properties: \emph{(i)} If $q(\mathbf k)>0$, then for
$\varepsilon(\mathbf{k})\equiv 1$, $H_0$ has exponentially decaying
two-body interactions in real space, and by choosing
$\varepsilon(\mathbf{k})=q(\mathbf k)$, one obtains a strictly local
gapped Hamiltonian. \emph{(ii)} If $q(\mathbf{k})=0$ for some $\mathbf k$
and $G _{\mathrm{out}}(\mathbf{k})$ is continuous,
$\varepsilon(\mathbf{k})\equiv 1$ still yields a gapped Hamiltonian. Then,
whether $\mc H_0$ has exponentially decaying terms depends on whether
$G_{\mathrm{out}}(\mathbf{k})$ has any discontinuities in its derivatives
(which give rise to algebraically decaying terms in real space after
Fourier transforming). Such non-analyticities can only appear at $\mb k$-points where the inverse in Eq.~\eqref{eq:Gout} diverges. (Nonetheless, $G_\mr{out}(\mb k)$ is still finite at such points due to $G_\mr{out}(\mb k) G_\mr{out}^\dg(\mb k)= \Id$.) In this case, $\varepsilon(\mb k) = q(\mb k)$ leads to a local, but gapless Hamiltonian. In general, the choice $\varepsilon(\mb k) = q(\mb k)$ will be called the \textit{parent Hamiltonian} in the following.  

Real, physical Majorana modes $\tilde e_\mb{k}$  (Hermitian operators) in reciprocal space can be obtained by taking the Fourier transform of the physical fermionic modes $a_{\mb r,j}$ in the relation
\begin{align}
e_{\mb r, 2 j -1} &= a_{\mb r,j}^\dg + a_{\mb r,j} \\
e_{\mb r, 2j} &= (-i)(a_{\mb r,j}^\dg - a_{\mb r,j}),
\end{align}
and employing the relations $a_{\mb k,j} = \frac{1}{2} (\tilde e_{\mb k,2j-1} - i \tilde e_{\mb k,2j})$ and $a_{\mb k,j}^\dg = \frac{1}{2}(\tilde e_{\mb k,2j-1} + i \tilde e_{\mb k,2j})$. One obtains
\begin{equation}
\left(\begin{array}{c}
\hat e_{\mb k,2 j - 1}\\
\hat e_{\mb k,2j}\\
\hat e_{\mb k,2 j - 1}^\dg\\
\hat e_{\mb k,2j}^\dg
\end{array}\right)
= U 
\left(\begin{array}{c}
\tilde e_{\mb k,2 j - 1}\\
\tilde e_{\mb k,2j}\\
\tilde e_{-\mb k,2 j - 1}\\
\tilde e_{-\mb k,2j}
\end{array}\right),
\ \ 
U = U^\dg = \frac{1}{2}\left(\begin{array}{cccc}
\Id_2 - i W&\Id_2 + i W\\
\Id_2 - i W& \Id_2 + i W
\end{array}\right),
\end{equation}
where  $W = \left(\begin{smallmatrix}
0&1\\
-1&0\end{smallmatrix}\right)$.
The covariance matrix of the real physical Majorana fermions $\tilde e_\mb{k}$ is given by
\begin{align}
\Gamma_{\mathrm{out}}(\mb k > 0) &= \left(\begin{array}{cc}
\Gamma_{++}(\mb k > 0)& \Gamma_{+-}(\mb k > 0)\\
\Gamma_{-+}(\mb k > 0)&\Gamma_{--}(\mb k > 0)
\end{array}\right) \\
&= \frac{1}{2}\left(\begin{array}{cc}
V&\overline V\\
\overline V&V
\end{array}\right)
\left(\begin{array}{cc}
G_{\mathrm{out}}(\mb k)&0\\
0&\overline G_{\mathrm{out}}(\mb k)
\end{array}\right)  \left(\begin{array}{cc}
V&\overline V\\
\overline V & V
\end{array}\right) \\
&= \mathrm{Re} \left[ \left(\begin{array}{cc}
V \, G_{\mathrm{out}}(\mb k) \, V&V \, G_{\mathrm{out}}(\mb k) \, \overline V\\
\overline V \, G_{\mathrm{out}}(\mb k) \, V&\overline V \, G_{\mathrm{out}}(\mb k) \, \overline V
\end{array}\right) \right], \ \ V = \bigoplus_{j = 1}^{f} (\Id_2 - i W). \label{Covariance_real}
\end{align}
Therefore, also 
\begin{align}
(\Gamma_{\mathrm{out}})_{st}(\mb k) = \frac{\tilde p_{st}(\mb k)}{q(\mb k)}
\end{align}
is a fraction of trigonometric polynomials of degree $\leq 2 \chi$.

\subsection{Classification of free fermionic topological phases}\label{sec:Classification}

\subsubsection{Preliminaries}

The fact that free fermionic systems can always be diagonalized efficiently has been used to completely classify all fermionic topological phases (at least under particle-hole symmetry, time-reversal symmetry and chiral symmetry). In order to explain the ideas that led to this classification of topological states, the case of no symmetry (which is called ``symmetry class A'') shall be treated in some detail here. Afterwards, the classification of topological phases including symmetries will be presented, eventually leading to the periodic table of topological phases. 

Before proceeding to the classification of (gapped) fermionic topological phases, let us give a general definition of a topological phase:
A gapped topological phase (including the topologically trivial phase) is denoted as the equivalence class of gapped Hamiltonians which can be connected via an adiabatic path without closing the energy gap in the thermodynamic limit. That is, two Hamiltonians $\mc H_0$ and $\mc H_1$ are in the same topological phase if and only if
there exists a parameterization $\mc H = \mc H(\lambda)$, $\lambda \in [0,1]$ such that
\begin{align}
\mc H(0) = \mc H_0,& \ \mc H(1) = \mc H_1 \label{eq:H_01} \\
E_\mr{gap}(\lambda) \geq c, \label{eq:H_lambda}
\end{align}
where $c > 0$ is a constant and $E_\mr{gap}$ denotes the energy gap between the ground state(s) and the first excited state in the thermodynamic limit. The ground state(s) of those Hamiltonians display characteristic features of the topological phase they are situated in. This definition allows in principle to label topological phases. The phase which is connected to the atomic limit (where the Hamiltonian is a sum of terms that act only on single sites or small disjoint blocks of sites) is called the topologically trivial phase.

If the Hamiltonians involved are required to satisfy a certain symmetry, the corresponding phases are denoted as \textit{symmetry protected topological phases}~\cite{SPT_org}. That is, both $\mc H_0, \mc H_1$ and $\mc H(\lambda)$ are required to satisfy a certain symmetry (such as time reversal symmetry) and if such an $\mc H(\lambda)$ exists that both of them can be connected adiabatically, they are said to lie in the same symmetry protected topological phase. However, such systems are only topologically protected to perturbations and disorder as long as the corresponding symmetry is not broken. Otherwise, an arbitrarily small perturbation can in principle destroy the topological features of the state. An example is the topological insulator~\cite{top_ins}, which is protected by time reversal symmetry. Arbitrarily small time reversal symmetry breaking perturbations, such as magnetic impurities, can destroy its helical edge modes. 

The following is meant to give an idea of the classification of free fermionic topological phases. Details can be found, e.g., in Ref.~\cite{Phases_Schnyder}. In this case, the Hamiltonians $\mc H_0$, $\mc H_1$ in Eq.~\eqref{eq:H_01} and $\mc H(\lambda)$ are required to be free fermionic. However, when interactions are introduced, phases which appear to be topologically distinct in the free fermionic limit might get connected, but also new topological phases will emerge, which only exist in the interacting regime, such as the Fractional Quantum Hall States~\cite{FQHE_th}.

Let us first consider the most general expression for free fermionic Hamiltonians, which reads
\be
\mc H = \sum_{i,j} \left( T_{ij} a_i^\dg a_j + \Delta_{ij} a_i^\dg a_j^\dg + \Delta_{ij}^* a_j a_i \right), \label{eq:Ham_free}
\ee
where $T$ is a Hermitian matrix and $\Delta$ an antisymmetric matrix, $\Delta = -\Delta^\top$, and the asterisk means complex conjugation. The
$a_i$ denote the annihilation operators of the fermionic orbital with index $i$. We shall be concerned only with translationally invariant systems on a $d$-dimensional square lattice, i.e., the Hamiltonian takes the form 
\be
\mc H = \sum_{\mb r, \mb r'} \sum_{i,j=1}^f \left( T_{\mb r - \mb r',ij} a_{\mb r, i}^\dg a_{\mb r', j} + \Delta_{\mb r - \mb r', ij} a_{\mb r,i}^\dg a_{\mb r',j}^\dg + \Delta_{\mb r - \mb r',ij}^* a_{\mb r', j} a_{\mb r, i} \right). \label{eq:Ham_TI}
\ee
In this notation, hermiticity of the Hamiltonian implies $T_{\mb r - \mb r',ij} = T_{\mb r' - \mb r,ji}^*$ and that we can set $\Delta_{\mb r - \mb r',ij} = -\Delta_{\mb r' - \mb r,ji}$. 

The classification turns out to be the same for non-translationally invariant systems, as demonstrated in Ref.~\cite{Phases_Kitaev} using $K$-theory, which is, however, valid only if the number of occupied and empty bands is sufficiently large.

\subsubsection{An example: Characterization of symmetry class A}\label{sec:class_A}

In the absence of all the above symmetries, $\Delta_{\mb r - \mb r',ij} = 0$. Then, the Hamiltonian can be block-diagonalized by Fourier transforming the fermionic modes, $\hat a_{\mb k, j} = \tfrac{1}{N} \sum_\mb{r} e^{i \mb r \cdot \mb k} a_{\mb{r}, j}$, which leads to
\be
\mc H_\mr{A} = \sum_\mb{k} \sum_{i,j=1}^f \hat T_{ij}(\mb k) \hat a_{\mb k,i}^\dg \hat a_{\mb k,j} \label{eq:Ham_k}
\ee
with $\hat T_{ij}(\mb k) = \sum_\mb{r} e^{i \mb r \cdot \mb k} T_{\mb r, ij}$. Hence, translationally invariant free fermionic Hamiltonians in symmetry class A are parameterized by Hermitian matrices $\hat T(\mb k)$, which depend on $\mb k \in \mc T^d$ (the $d$-torus). In order to classify the fermionic phases in class A, let us first consider the diagonalization of $\hat T(\mb k)$, 
\be 
\hat T(\mb k) = U(\mb k) \Lambda(\mb k) U^\dg (\mb k). \label{eq:T_diag}
\ee
The single-particle energies are $\epsilon_j(\mb k) = \Lambda_{jj}(\mb k)$. 

As the system is gapped, we can assume without loss of generality that $\epsilon_{j} (\mb k) < E_F$ (the Fermi energy) for $1 \leq j \leq n$ ($n$ is the number of occupied bands) and $\epsilon_j (\mb k) > E_F$ for $n < j \leq f$ for all $\mb k \in \mc T^d$. In order to classify the topological phases, we set $\epsilon_j' (\mb k) \equiv -1$ for $1 \leq j \leq n$, $\epsilon_{j}' (\mb k) \equiv 1$ for $n < j \leq f$ and $E_F = 0$ by a shift of the energy. This transformation can be done adiabatically and obviously does not close the energy gap (i.e., we stay in the same topological phase). The matrix $\hat T'(\mb k)$ of the new Hamiltonian reads 
\be
\hat T'(\mb k) = U(\mb k) \Lambda'(\mb k) U^\dg(\mb k) = -P(\mb k) + P_\perp(\mb k), \label{eq:flat_trick}
\ee
where $P(\mb k)$ is the projector on the occupied single particle states and $P_\perp(\mb k)$ the projector on the empty single particle states (which obviously has to be the orthogonal projector to $P(\mb k)$). Therefore, the topological phases in class A correspond to the maps $P: \mc T^d \rightarrow Gr(n,f)$ which can be continuously deformed into each other. The \textit{Grassmannian} $Gr(n,f)$ denotes the set of $n$-dimensional complex linear subspaces of $\mathbb{C}^f$ (these are the vector spaces $P(\mb k)$ projects on). The set $Gr(n,f)$ is a topological manifold. Maps from the torus to topological manifolds have been treated by Fox in the 1940's~\cite{torus_homotopy,torus_homotopy2}. 

\begin{figure}
\centering\includegraphics[width=0.3\textwidth]{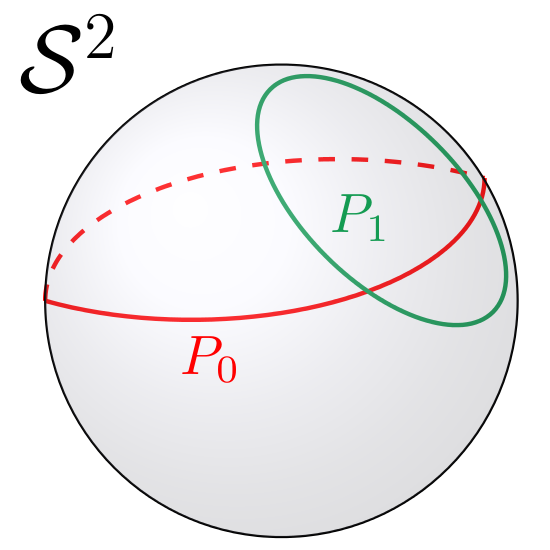}
\caption{Maps from the circle to the sphere appearing in one dimensional systems with one occupied and one empty band. $P_0$ and $P_1$ correspond to the two Hamiltonians $\mc H_0$ and $\mc H_1$ that are to be transformed into each other, respectively. Since the transformation of the two maps into each other can always been done continuously, all such Hamiltonians lie in the same topological phase.}
\label{fig:sphere_circles}
\end{figure}

Before continuing providing abstract  definitions, let us consider the simplest example of two bands, one occupied and one empty band, in $d = 1$ dimension. That is, we have to classify the maps $P: \mc T^1 \rightarrow Gr(1,2)$ which can be continuously deformed into each other. The 1-torus is identical to the 1-sphere $\mc S^1$, so we need to characterize maps from the circle to the set of one dimensional complex subspaces of $\mathbb{C}^2$. The later can be parameterized by $(x, m x) \in \mathbb{C}$, where $m \in \mathbb{C}$ is fixed and $x \in \mathbb{C}$ is variable. Thus, complex one dimensional subspaces of $\mathbb{C}^2$ are parameterized by complex numbers $m$, but we need to identify all $|m| \rightarrow \infty$ with each other. They all correspond to the same one dimensional subspace $(0,y), y \in \mathbb{C}$ of $\mathbb{C}^2$. Hence, instead of considering maps to the set of one dimensional subspaces of $\mathbb{C}^2$ it is also possible to consider maps to the Riemann sphere $\mathbb{C} \cup \{\infty\}$. Thus, the Grassmannian $Gr(1,2)$ is topologically equivalent (isotopic) to the unit sphere $\mc S^2$. However, classifying maps from the circle $\mc S^1$ to the three dimensional unit sphere $\mc S^2$ is simple. They can all be continuously deformed into each other, see Fig.~\ref{fig:sphere_circles}.

In short, we derived the following result: Gapped Hamiltonians with one occupied and one empty band correspond to maps from the circle $\mc S^1$ to set of linear one dimensional subspaces of $\mathbb{C}^2$. Since each subspace of this kind corresponds to one point on the unit sphere, those Hamiltonians correspond to maps $\mc S^1 \rightarrow \mc S^2$. And as the latter can all be continuously deformed into each other, likewise all gapped Hamiltonians with one occupied and one empty band in one dimension can be deformed adiabatically into each other without closing the energy gap. Thus, there is only one topologically trivial phase. 

The same turns out to be true if there are more filled or empty bands. This corresponds to classifying maps $P: \mc S^1 \rightarrow Gr(n,f)$, which can all be continuously deformed into each other. Hence, all systems in symmetry class A are topologically trivial in one dimension. Note that classifications of such maps have been known for a long time in the field of \textit{homotopy groups}. The $d$-th homotopy group $\pi_d$ classifies the maps from the $d-sphere$,  to a topological space $X$, $P: \mc S^d \rightarrow X$. The equivalence classes of maps which can be continuously transformed into each other form the elements of the group. The operation of the group is the concatenation of maps and the inverse is given by performing the map in the counterclockwise sense. The first homotopy group of $Gr(n,f)$ contains only the unit element $0$, which is the equivalence class containing the map to a point, $\pi_1(Gr(n,f)) = \{0\}$.

In $d$ dimensions, we have to classify maps from the torus to the Grassmannian, $P: \mc T^d \rightarrow Gr(n,f)$. The corresponding group of continuously deformable maps is denoted as the $d$-th \textit{torus homotopy group} $\tau_d$~\cite{torus_homotopy,torus_homotopy2}. For all physically relevant target spaces $X$, those groups decompose into a product of powers of the conventional homotopy groups, where the powers are given by the binomial coefficients, $\tau_d (X) = \prod_{\ell = 1}^d [\pi_\ell(X)]^{d \choose \ell}$.

\begin{figure}
\centering\includegraphics[width=0.25\textwidth]{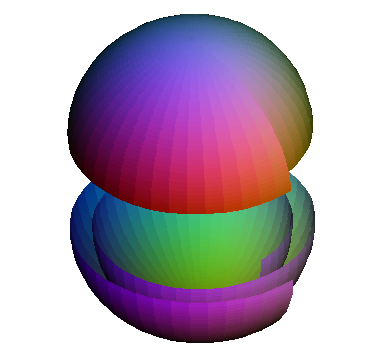}
\caption{Map from the unit sphere unto itself wrapping it twice around itself. (Reproduced from~\cite{wrap_sphere} under the terms of the GNU Free Documentation License.)}
\label{fig:wrap_sphere}
\end{figure}

Let us proceed to the next simplest case of two dimensions and gapped Hamiltonians with one occupied and one empty band. Hence, we are interested in $\tau_2(Gr(1,2)) = [\pi_1(Gr(1,2))]^2 \times \pi_2(Gr(1,2)) = \pi_2(Gr(1,2))$ due to $\pi_1(Gr(1,2)) = \{0\}$. Since $Gr(1,2)$ is topologically equivalent to $\mc S^2$, the goal is to classify maps from the unit sphere unto itself. Now, there are such maps which cannot be continuously deformed into each other: The map to a point on the unit sphere corresponds to the unit element $0$ of $\pi_2(Gr(1,2))$. The identical is denoted by $1$, furthermore it is possible to cut the unit sphere, wrap it twice around itself and to glue the ends back together to obtain the equivalence class $2$ (see Fig.~\ref{fig:wrap_sphere}), etc. Thus, $\pi_2(Gr(1,2)) = \pi_2(\mc S^2) = \mathbb{Z}$. For any number of bands, the second homotopy group of $Gr(n,f)$ turns out to be $\mathbb{Z}$, too, such that there is one integer index labeling the topological phases of systems in class A in two dimensions. This index is the well-known Chern number~\cite{derive_Chern}. 

In three dimensions, we have $\pi_3(Gr(n,f)) = \{0\}$, such that $\tau_3(Gr(n,f)) = [\pi_1(Gr(n,f))]^3 \times [\pi_2(Gr(n,f))]^3 \times \pi_3(Gr(n,f)) = \mathbb{Z}^3$ for $n > 2$ (the case of $n = 1$, $f = 2$ is special, since $\pi_3(Gr(1,2)) = \mathbb{Z}$, where representatives of the corresponding equivalence classes are given by the Hopf map~\cite{Hopf_map}, but this case will not be discussed here.) Although this might indicate that there exist topological systems in class A in three dimensions, this conclusion turns out to be wrong: We have only treated translationally invariant systems, and indeed if we only allow for translationally invariant perturbations, the corresponding systems would be topologically protected. However, if one allows for random disorder terms in the Hamiltonian, topological protection occurs if and only if the index corresponding to the highest homotopy group (in this case $\pi_3(Gr(n,f))$) is nonzero. The latter is also denoted as the \textit{strong topological index}, whereas all other indices are called \textit{weak topological indices}~\cite{strong_weak,strong_weak2}. Therefore, there are no topologically stable systems in three dimensions in symmetry class A. 

\subsubsection{Another example: Characterization of symmetry class D}\label{sec:class_D}

Systems in class D have a non-vanishing $\Delta_{\mb r - \mb r', ij}$ in Eq.~\eqref{eq:Ham_TI} and no particular symmetry requirements (particle-hole symmetry follows automatically, see below). The Hamiltonian reads after using Fourier transformed modes $\hat a_{\mb k, j} = \tfrac{1}{N} \sum_\mb{r} e^{i \mb r \cdot \mb k} a_{\mb r, j}$
\begin{align}
\mc H_\mr{D} &= \sum_{\mb k} \sum_{i,j = 1}^f (\hat{a}_{\mb k,i}^\dg, \hat{a}_{- \mb k, i}) H_{ij}(\mb k) \left(\begin{array}{c}
\hat{a}_{\mb k,j} \\
\hat{a}_{-\mb k,j}^\dg
\end{array}\right), \\
H_{ij}(\mb k) &= \left(\begin{array}{cc}
\tfrac{1}{2} \hat T_{ij}(\mb k) & \hat{\Delta}_{ij}(\mb k) \\
\hat{\Delta}^*_{ji} (\mb k) & -\tfrac{1}{2} \hat T_{ji} (-\mb k)
\end{array}\right) \label{eq:BdG}
\end{align}
with $\hat T_{ij}(\mb k) = \sum_\mb{r} e^{i \mb r \cdot \mb k} T_{\mb r, ij}$ and $\hat \Delta_{ij}(\mb k) = \sum_\mb{r} e^{i \mb r \cdot \mb k} \Delta_{\mb r, ij}$.
$T_{\mb r - \mb r',ij} = T_{\mb r' - \mb r,ji}^*$ and $\Delta_{\mb r - \mb r',ij} = - \Delta_{\mb r' - \mb r,ji}$ imply $\hat{T}_{ij}(\mb k) = \hat{T}_{ji}^*(\mb k)$ and $\hat{\Delta}_{ij}(\mb k) = - \hat{\Delta}_{ji}(-\mb k)$, respectively. 
The Hamiltonian $\mathcal H_\mr{D}$ is diagonalized by solving the eigenvalue equation
\begin{equation}
H(\mb k) |\psi_l (\mb k) \rangle = \varepsilon_l (\mb k) |\psi_l (\mb k) \rangle,
\end{equation}
for $l = 1,2, \ldots, 2 f$ ($H(\mb k)$ denotes the matrix obtained after concatenating the matrices $H_{ij}(\mb k)$). The structure of Eq.~\eqref{eq:BdG} implies for each $l$ 
$\varepsilon_l(\mb k) = - \varepsilon_{l'}(-\mb k)$ for some $l' \in \{1,2,\ldots, 2f\}$, which is the particle-hole symmetry. (Thus, the Fermi energy is at $E = 0$.) Here, $|\psi_l(\mb k)\rangle$ is defined in Nambu space, i.e., in the space within which the matrices $H(\mb k)$ define linear maps. The topological information of the state is encoded in the spectral projector
\begin{equation}
\tilde P(\mb k) = \sum_{\substack{l = 1 \\ \varepsilon_l(\mb k) < 0}}^{2 f} |\psi_l (\mb k) \rangle \langle \psi_l(\mb k)|, \label{eq:Projector_D}
\end{equation}
which is the projector onto the space of occupied states. 

Let us again consider the simplest case, which is $f = 1$. Then
\be
H(\mb k) = \left(\begin{array}{cc}
\tfrac{1}{2} \hat T(\mb k)&\hat \Delta(\mb k)\\
\hat \Delta^*(\mb k)&-\tfrac{1}{2} \hat T(-\mb k)
\end{array}\right)
\ee
with $\hat T(\mb k) = \hat T^*(\mb k)$, i.e., $\hat T(\mb k) \in \mathbb{R}$ and $\hat \Delta(\mb k) = - \hat \Delta(-\mb k)$. Thus, the energy eigenvalues are $\epsilon_{1,2}(\mb k) = \pm \sqrt{\tfrac{1}{4} \hat T^2(\mb k) + |\hat \Delta(\mb k)|^2}$, which is non-zero for a gapped system. We again flatten the energy levels by defining a new Hamiltonian via
\be
H'(\mb k) = \left(\begin{array}{cc}
t(\mb k)&\delta_1(\mb k) + i \delta_2(\mb k)\\
\delta_1(\mb k) - \mr{i} \delta_2(\mb k)&-t(-\mb k)
\end{array}\right), \label{eq:flat_D}
\ee  
$t(\mb k) = \hat T(\mb k) / \sqrt{\hat T^2(\mb k) + 4 |\hat \Delta(\mb k)|^2}$ and $\delta_{1,2}(\mb k) \in \mathbb{R}$ with $\delta_1(\mb k) + i \delta_2(\mb k) = \hat \Delta(\mb k)/ \sqrt{\tfrac{1}{4} \hat T^2(\mb k) + |\hat \Delta(\mb k)|^2}$, which corresponds to a Hamiltonian adiabatically connected to the original one. The constraint $\hat \Delta(\mb k) = -\hat \Delta(- \mb k)$ leads to slight modifications as compared to class A; e.g., there is one topologically non-trivial phase in one dimension, which contains the Kitaev chain~\cite{Kitaev_chain}. In two dimensions, the unit vector $(t(\mb k), \delta_1(\mb k), \delta_2(\mb k))^\top$ parameterizes again a map from the torus $\mc T^2$ to the sphere $\mc S^2$. Despite the restriction $\hat \Delta(\mb k) = - \hat \Delta(-\mb k)$, the topological phases turn out to be labelled by $\mathbb{Z}$ for any number of bands,  and this topological index is also given by the Chern number~\cite{derive_Chern} (see below).

\subsubsection{Classification table}

For other symmetry classes, a similar analysis of maps that can be continuously deformed into each other can be carried out. In each case, all maps are constrained to satisfy the respective symmetry conditions. The corresponding homotopy groups are shown in Table~\ref{tab:classification}.

\begin{table}[ht!]
\centering
\begin{tabular}{| c | c | c | c | c | c | c |} \hline
Symmetry class X& TRS & PHS & CS & $\pi_1(\mr{X})$ & $\pi_2(\mr{X})$ & $\pi_3(\mr{X})$\\ \hline
A&0&0&0&$\left\{0\right\}$&$\mathbb{Z}$&$\left\{0\right\}$\\
AIII&0&0&1&$\mathbb{Z}$&$\left\{0\right\}$&$\mathbb{Z}$\\
\hline
BDI&+1&+1&1&$\mathbb{Z}$&$\left\{0\right\}$&$\left\{0\right\}$\\
D&0&+1&0&$\mathbb{Z}_2$&$\mathbb{Z}$&$\left\{0\right\}$\\
DIII&-1&+1&1&$\mathbb{Z}_2$&$\mathbb{Z}_2$&$\mathbb{Z}$\\
AII&-1&0&0&$\left\{0\right\}$&$\mathbb{Z}_2$&$\mathbb{Z}_2$\\
CII&-1&-1&1&$\mathbb{Z}$&$\left\{0\right\}$&$\mathbb{Z}_2$\\
C&0&-1&0&$\left\{0\right\}$&$\mathbb{Z}$&$\left\{0\right\}$\\
CI&+1&-1&1&$\left\{0\right\}$&$\left\{0\right\}$&$\mathbb{Z}$\\
AI&+1&0&0&$\left\{0\right\}$&$\left\{0\right\}$&$\left\{0\right\}$\\ \hline
\end{tabular}
\caption{Classification of topological insulators and superconductors. The first two symmetry classes have a periodicity of two in the physical dimension, whereas the other eight symmetry classes have a periodicity of 8 in the physical dimension. Moreover, for them, the homotopy group in one physical dimension higher is the same as of the one of the symmetry class in the row below~\cite{Phases_Kitaev}.}
\label{tab:classification}
\end{table}

A few words are in place about the symmetry groups: Particle-hole symmetry (PHS) and time reversal symmetry (TRS) are anti-unitary symmetries, i.e., they can be written as the multiplication by unitaries $U_C$ and $U_T$, respectively, acting on the complex conjugate $\mc H^*$ of the Hamiltonian $\mc H$,
\begin{align}
U_C^\dg {\mc H}^* U_C &= -\mc H, \\
U_T^\dg {\mc H}^* U_T &= \mc H,
\end{align}
where it can be easily checked that $U_C^2 = \pm \Id$ and $U_T^2 = \pm \Id$. If those symmetries are present, there are thus two options for each one of them (denoted by $\pm 1$). The absence of a symmetry is denoted by a $0$, i.e., PHS and THS constitute $3 \times 3 = 9$ options. Chiral symmetry is given by $U_S = U_T^* U_C$ and thus univocally determined unless both PHS and TRS are absent. In the latter case, chiral symmetry can either be present ($1$) or absent ($0$), leading to ten classes in total. These two cases correspond to classes A and AIII. 

It shall be made abundantly clear that for lattice systems Table~\ref{tab:classification} does not provide the topological indices directly, but the full set of topological phases of symmetry class X on the lattice is given by $\tau_2(\mr{X}) = [\pi_1(\mr{X})]^2 \times \pi_2(\mr{X})$ and $\tau_3(\mr{X}) = [\pi_1(\mr{X})]^3 \times [\pi_2(\mr{X})]^3 \times \pi_3(\mr{X})$ in two and three dimensions, respectively. This is very important for class AII corresponding to topological insulators~\cite{Kane_Mele}, which in three dimensions have one strong and three weak topological indices~\cite{strong_weak,strong_weak2}, $\tau_3(\mr{AII}) = \mathbb{Z}^3_2 \times \mathbb{Z}_2$. Note that for continuous systems (without a lattice structure), Table~\ref{tab:classification} can be used directly, since in this case maps from the $d$-sphere to the corresponding topological space have to be classified. The reason is that the periodic boundary conditions of the lattice are absent, such that if the asymptotics of the terms of the Hamiltonian for $| \mb k | \rightarrow \infty$ are fixed (more precisely, the asymptotics of the corresponding projectors $P(\mb k)$), the Brillouin zone is replaced by a sphere.

\subsubsection{A topological invariant: The Chern number}

So far, it has been only said that for certain symmetry classes and spatial dimensions $d$ there exist topological phases, which can be labelled by one or more integer indices. However, nothing has been mentioned so far about how to obtain these labels if one knows the Hamiltonian of a given free fermionic system. Since it is in general very hard to derive such \textit{topological invariants}, only the expressions for the most important one, the \textit{Chern number}, will be given here and we refer the interested reader to Ref.~\cite{derive_Chern} for its derivation (where the Chern number we are interested in is called the Chern number of the first Chern class). The Chern number is the topological index of class A and class D in two dimensions. For class A, we already wrote the the Hamiltonian in such a way that it is explicitly diagonal in reciprocal space, Eq.~\eqref{eq:Ham_k}. After ``flattening'' the energy bands, the matrix $\hat T(\mb k)$ takes the form~\eqref{eq:flat_trick}. The Chern number $C$ is obtained from the projector $P(\mb k)$ on the occupied bands via
\be
C = -\frac{i}{2 \pi} \int \tr\left(\left[P,\frac{\partial P}{\partial k_x}\right]\frac{\partial P}{\partial k_y}\right) \mr{d}^2 \, k, \label{eq:Chern_projector}
\ee
where the range of the integration is the first Brillouin zone. This equation is equivalent to another one that is frequently used,
\be
C = \sum_{j: \mr{occ}} \frac{1}{2 \pi} \int \left(\frac{\partial A^{(j)}_y}{\partial k_x} - \frac{\partial A^{(j)}_x}{\partial k_y}\right) \mr{d}^2 k, \label{eq:Chern_A}
\ee
where $A_{x,y}^{(j)} = \langle u_j (\mb k) | \tfrac{\partial}{\partial k_{x,y}} | u_j(\mb k)\rangle$ is the vector potential associated with band $j$ whose Bloch wave function is $|u_j(\mb k)\rangle$, which is the eigenvector of $\hat T(\mb k)$ with energy $\epsilon_j(\mb k) = \Lambda_{jj}(\mb k)$ (that is, the $j$-th column vector of $U(\mb k)$, cf. Eq.~\eqref{eq:T_diag}). Note that the sum in Eq.~\eqref{eq:Chern_A} is carried out only over occupied bands. 

In the case of two bands, Eq.~\eqref{eq:Chern_projector} and~\eqref{eq:Chern_A}, respectively, can be simplified to~\cite{Chern_ins}
\be
C = \frac{1}{4 \pi} \int \hat{\mb d} \cdot \left(\frac{\partial \hat{\mb d}}{\partial k_x} \times \frac{\partial \hat{\mb d}}{\partial k_y}\right) \mr{d}^2 \, k, \label{eq:Chern_d}
\ee
where $\hat T(\mb k) = \epsilon(\mb k) \Id + \sum_{i = x,y,z} d_i(\mb k) \sigma_i (\mb k)$ with $\sigma_i$ being the Pauli matrices and $\hat d_i(\mb k) = \frac{d_i(\mb k)}{| d(\mb k)|} \in [-1,1]$. The integral counts how many times the unit sphere is wrapped if one covers the first Brillouin zone. The Chern number can be determined in the experiment by measuring the Hall conductivity $\sigma_{xy}$ at finite temperature $\beta= 1/T$~\cite{Chern_ins}
\begin{align}
- \frac{\sigma_{xy}(\beta)}{2 \pi} = \frac{1}{4 \pi} \int \d^2 k \, f(\beta, \mb k) \, \hat {\mb d}(\mb k) \cdot \left(\frac{\partial \hat {\mb d}(\mb k)}{\partial k_x} \times \frac{\partial \hat {\mb d}(\mb k)}{\partial k_y}\right), \label{eq:Hall_thermal}
\end{align}
where $f(\beta, \mb k) = \frac{1}{e^{-\beta d(\mb k)}+1} - \frac{1}{e^{\beta d(\mb k)} + 1}$ and the integration extends over the first Brillouin zone. $-\tfrac{\sigma_{xy}(\beta)}{2 \pi}$ converges to the Chern number $C$ for $T \rightarrow 0$.

In symmetry class D, the topological index in two dimensions is likewise given by the Chern number. It is calculated by replacing $P(\mb k)$ in Eq.~\eqref{eq:Chern_projector} by the projector~\eqref{eq:Projector_D}. In the case of a single fermionic species, $f = 1$, Eq.~\eqref{eq:Chern_d} can also be used if one substitutes $(t(\mb k), \delta_1(\mb k), \delta_2(\mb k))^\top$ (see Eq.~\eqref{eq:flat_D}) for the unit vector $\hat{\mb d}(\mb k)$.

\subsection{Interacting chiral topological systems described by Conformal Field Theory}\label{sec:CFT}

This introduction to Conformal Field Theory (CFT) is by no means meant to be rigorous. Quite to the contrary, derivations will be mostly neglected and only the main results provided. For details, the interested reader is referred to Refs.~\cite{yellow_book} and~\cite{CFT_Ginsparg}.

A CFT  is characterized by a set of fields $\phi_j(x), j = 1, \ldots, n$ that are invariant under \textit{conformal transformations}. Before defining this invariance more precisely, let us first explain the latter term: A conformal transformation is a transformation in $d$-dimensional space-time, $x' = x'(x)$, such that
angles between intersecting curves are preserved. This is the case if and only if the metric tensor $g_{\mu \nu}$ transforms as 
\be
g_{\mu \nu}(x) \rightarrow g_{\mu \nu}'(x') = \frac{\partial x^\alpha}{\partial {x'}^\mu} \frac{\partial x^\beta}{\partial {x'}^\nu} g_{\alpha \beta}(x) = \Omega(x) g_{\mu \nu}(x)
\label{eq:conf_trafo}
\ee
with a scalar positive function $\Omega(x)$ corresponding to the square of the local change of scale. In Eq.~\eqref{eq:conf_trafo} and the remainder of this subsection, repeated indices are summed over. Note also that the infinitesimal length is given by $\d s^2 = g_{\mu \nu} \d x^\mu \d x^\nu$.

The transformation property of scalar conformally invariant fields $\phi_j$ under conformal transformations is
\be
\phi_j(x) \rightarrow \phi_j'(x) = \left| \frac{\partial x'}{\partial x}\right|^{\Delta_j/d} \phi_j(x') \label{eq:trans_phi}
\ee
Such fields are denoted as \textit{primary fields} (see below for details). The constant $\Delta_j \in \mathbb{R}$ is called the \textit{scaling dimension} of the field $\phi_j$.

\subsubsection{Classification of conformal transformations} 
 
We consider flat space-time, i.e., $g_{\mu \nu} = \eta_{\mu \nu}$. In this case it can be shown that for $d > 2$ the only transformations which fulfill Eq.~\eqref{eq:conf_trafo} locally (i.e., for a finite region of space-time) are those, which also fulfill this equation globally, denoted as \textit{global conformal transformations}. There are three types of such transformations,
\begin{align}
1. \ &{x'}^\mu = \Lambda^\mu_\nu x^\nu + a^\mu \label{eq:Poincare} \\
2. \ &{x'}^\mu = \lambda x^\mu \label{eq:dilatation}  \\
3. \ &{x'}^\mu = \frac{x^\mu - b^\mu x_\sigma x^\sigma}{1 - 2 b_\nu x^\nu + b_\nu b^\nu x_\kappa x^\kappa}, \label{eq:special}
\end{align}
where the first set of transformations forms the Poincar\'e group, the second corresponds to dilatations, and the third contains all special conformal transformations parameterized by an arbitrary Lorentz vector $b^\mu$. 

However, in $d = 1 + 1$ dimensions, there is an infinite set of local conformal transformations: Let us consider $g_{\mu \nu} = \delta_{\mu \nu}$, i.e., Euclidean metric (the treatment for Minkowski metric is analogous). Then, conformal transformations are given by holomorphic functions $f(z)$, where $z = x^0 + i x^1$ and $f(z) = {x'}^0 + i {x'}^1$. They locally preserve angles of intersecting lines in Euclidean space-time. If one requires this property to be true for all $z \in \mathbb{C} \cup \{\infty\}$, one recovers Eqs.~\eqref{eq:Poincare} to~\eqref{eq:special}, which in complex notation can be conveniently combined to
\be
f(z) = \frac{a \, z + b}{c \, z + d}
\ee
with $a, b, c, d \in \mathbb{C}$ normalized such that $a d - bc = 1$.

Coming back to local conformal transformations $z \rightarrow f(z)$, we note that $\Omega$ in Eq.~\eqref{eq:conf_trafo} is given by $\Omega(z) = \left| \frac{\partial f}{\partial z}\right|^2$, since
\be 
\d s^2 = (\d x^0)^2 + (\d x^1)^2 = \d z \, \d z^* \rightarrow \left|\frac{\partial f}{\partial z}\right|^2 \d z \, \d z^* \label{eq:metric_trafo}
\ee
(the asterisk denotes complex conjugation).

\subsubsection{Primary fields in $d$ dimensions}

The transformation property of scalar primary fields $\phi_j$ given in Eq.~\eqref{eq:trans_phi} has important physical consequences when considering their correlations
\be
\langle \phi_1(x_1) \ldots \phi_m(x_m) \rangle = \left|\frac{\partial x'}{\partial x}\right|^{\Delta_1/d}_{x = x_1} \ldots \left|\frac{\partial x'}{\partial x}\right|^{\Delta_m/d}_{x = x_m} \langle \phi_1(x_1') \ldots \phi_m(x_m')\rangle,
\ee
where the expectation value refers to the vacuum $|0\rangle$ of the specific CFT under consideration. 
For 2-point correlations, one can use the fact that Eq.~\eqref{eq:Poincare} implies that the correlation function can only depend on $|x_1 - x_2|$. Eq.~\eqref{eq:dilatation} causes it to be of the form
\be
\langle \phi_1(x_1) \phi_2(x_2) \rangle = \frac{c_{12}}{|x_1-x_2|^{\Delta_1 + \Delta_2}}.
\ee
Finally, Eq.~\eqref{eq:special} can be used to show that the correlation vanishes, unless $\Delta_1 = \Delta_2$, i.e.,
\begin{align}
\langle \phi_1(x_1) \phi_2(x_2) \rangle = \begin{cases} \frac{c_{12}}{|x_1-x_2|^{2 \Delta_1}} &\mbox{if } \Delta_1 = \Delta_2 \\ 
0 & \mbox{if } \Delta_1 \neq \Delta_2. \end{cases} \label{eq:conformal_corr}
\end{align}
Similar arguments can be used to derive general conditions on the 3- and 4-point correlation functions.

The dynamics of the specific CFT at hand is determined by the action
\be
S = \int \d^d x \, \mc{L},
\ee 
where $\mc{L}$ is the Lagrangian density. The energy-momentum tensor $T^{\mu \nu}$ quantifies the change of the action induced by an infinitesimal transformation $x \rightarrow x' = x + \epsilon$
\be
\delta S = \int \d^d x \, T^{\mu \nu} \partial_\mu \epsilon_\nu = \tfrac{1}{2} \int \d^d x \, T^{\mu \nu} (\partial_\mu \epsilon_\nu + \partial_\nu \epsilon_\mu) \label{eq:energy_momentum}
\ee
with $\partial_\mu \equiv \frac{\partial}{\partial x^\mu}$. 
The energy-momentum tensor is traceless and can be assumed to be symmetric. Note that for conformal transformations $\delta S = 0$. If we restrict ourselves to such transformations, integrating Eq.~\eqref{eq:energy_momentum} by parts shows that $\partial_\mu T^{\mu \nu} = 0$.

\subsubsection{Primary fields in two dimensions}

In the $1 + 1$ dimensional case it is conventional to express fields in terms of $z$ and $\overline z$ and to treat them as independent variables. The physically relevant sector is accessed by setting $\overline z = z^*$ (the complex conjugate of $z$). If we take a primary field $\phi = \phi(z,\overline z)$ with spin $s$ and scaling dimension $\Delta$, the transformation~\eqref{eq:trans_phi} reads
\be
\phi'(z',\bar z') = \left(\frac{\partial z'}{\partial z}\right)^{-h} \left(\frac{\partial \overline z'}{\partial \overline z}\right)^{-\overline h} \phi(z,\bar z), \label{eq:complex_trafo}
\ee
where $z' = f(z)$, $\bar z' = \bar f(\bar z)$. $h = \tfrac{1}{2}(\Delta + s)$ and $\overline h = \tfrac{1}{2}(\Delta - s)$ ($\bar h$ not being the complex conjugate of $h$) are the \textit{conformal dimensions} of the primary field $\phi$. 

Let us now consider the 2-point correlation function (or \textit{correlator}) of two potentially different primary fields $\phi_1$ and $\phi_2$. Using Eq.~\eqref{eq:complex_trafo} it can be shown to be of the form
\be 
\langle \phi_1(z_1,\bar z_1) \phi_2(z_2, \bar z_2)\rangle = \frac{c_{12}'}{|z_1 - z_2|^{2 h_1} |\bar z_1 - \bar z_2|^{2 \bar h_1}}
\ee 
if $h_1 = h_2$ and $\bar h_1 = \bar h_2$ and zero otherwise, cf. Eq.~\eqref{eq:conformal_corr}.

Due to $\d s^2 = \d z \d \bar z$, $g_{zz} = g_{\bar z \bar z} = 0$ and $g_{z \bar z} = g_{\bar z z} = \tfrac{1}{2}$. $g_{\mu \nu} {g}^{\nu \kappa} = {\delta_\mu}^\kappa$ implies ${g}^{\bar z z} = {g}^{z \bar z} = 2$ and ${g}^{z z} = {g}^{\bar z \bar z} = 0$.
The conservation law $\partial_\mu {T}^{\mu \nu} = 0$, or, equivalently, $\partial^\mu T_{\mu \nu} = {g}^{\mu \kappa} \partial_\kappa T_{\mu \nu} = 0$ therefore yields $\partial_{\bar z} T_{zz} + \partial_z T_{\bar z z} = 0$ and 
$\partial_{\bar z} T_{z \bar z} + \partial_z T_{\bar z \bar z} = 0$. Finally, $0 = {T_\mu}^\mu = {T}_{\mu \nu} {g}^{\nu \mu} = 2 T_{z \bar z} + 2 T_{\bar z z}$ together with the symmetry of the energy-momentum tensor implies $T_{\bar z z} = T_{z \bar z} = 0$ and thus $\partial_{\bar z} T_{zz} = \partial_z T_{\bar z \bar z} = 0$. Hence, we can set
\be
T(z) = T_{zz}(z, \bar z), \ \ \bar T(\bar z) = T_{\bar z \bar z}(z, \bar z).
\ee
Note that $T(z)$ (and $\bar T(\bar z)$) is an operator, as it is derived from the Lagrangian, which contains the fields $\phi_j(z, \bar z)$. In the same way as the energy-momentum tensor, all other relevant quantities decompose into a holomorphic and an antiholomorphic part; in particular we can also assume $\phi(z, \bar z) = \phi(z) + \bar \phi( \bar z)$ and just consider the holomorphic part $\phi(z)$ from now on. (However, one has to keep in mind that the antiholomorphic part comes along with it in all calculations.)

If one calculates the correlation between the energy-momentum tensor $T(z)$ and a field $\phi(w)$, one encounters singular terms as $z \rightarrow w$,
\be 
\langle T(z) \phi(w) \rangle = \frac{h}{(z-w)^2} \langle \phi(w) \rangle + \frac{1}{z-w} \langle \partial_w \phi(w)\rangle + f(z,w),
\ee
where $f(z,w)$ is a function which is regular as $z \rightarrow w$. We write the above expression symbolically as 
\be
T(z) \phi(w)  \sim \frac{h}{(z-w)^2} \phi(w) + \frac{1}{z-w} \partial_w \phi(w), \label{eq:OPE}
\ee
which only accounts for diverging terms and is denoted as the \textit{operator product expansion} (OPE). Hence, the OPE with the energy-momentum tensor can be used to extract the conformal dimension $h$ of a primary field $\phi(z)$. 

If one calculates the OPE of the energy-momentum tensor with itself, the result is different from Eq.~\eqref{eq:OPE},
\be
T(z) T(w) \sim \frac{c}{2 (z - w)^4} + \frac{2 T(w)}{(z-w)^2} + \frac{\partial_w T(w)}{z-w}.
\ee
Due to the anomalous first term, $T(z)$ is not a primary field. The constant $c$ is called the \textit{central charge}. In order to give an idea of its meaning, let us consider two simple cases, the free boson and the free fermion.

\paragraph{The free boson.}

The corresponding action is
\be
S = \frac{1}{8\pi} \int \d^2 x \, \partial_\mu \varphi \partial^\mu \varphi.
\ee
One can check that
\be
\langle \varphi(z) \varphi(w)\rangle = -\log(z-w),
\ee
i.e., $\varphi(z)$ is not itself a primary field. However, $\partial_z \varphi(z)$ turns out to be one with conformal dimension $h = 1$,
\be
T(z) \partial_w \varphi(w) \sim \frac{\partial_w \varphi(w)}{(z-w)^2} + \frac{\partial_w^2 \varphi(w)}{z-w},
\ee
where the energy-momentum tensor can be checked to be 
\be
T(z) = -\tfrac{1}{2} :\partial_z \varphi(z) \partial_z \varphi (z): \equiv -\tfrac{1}{2} \lim_{w \rightarrow z} \left(\partial_z \varphi(z) \partial_w \varphi(w) - \langle \partial_z \varphi(z) \partial_w \varphi(w)\rangle \right)
\ee
(the colons denote normal ordering). The OPE of the energy-momentum tensor with itself yields central charge $c = 1$. If we considered $N$ independent free bosons, the central charge would have been $c = N$, which indicates that it quantifies the number of degrees of freedom. 

If one writes down the Laurent expansion of the primary field $\partial_z \varphi(z)$,
\be
\partial_z \varphi(z) = {-i} \sum_{n=-\infty}^\infty a_n z^{-n-1}, \label{eq:primary_modes}
\ee
the Laurent coefficients $a_n$ fulfill
\be
[a_n,a_m] = n \delta_{n+m,0}, \label{eq:boson_comm}
\ee
which is known as the \textit{Heisenberg algebra}. Note that $a_0$ is a constant of motion (commutes with the Hamiltonian) known as the \textit{momentum}. 

An infinite number of primary fields can be constructed as $\mc{V}_\alpha(z) = : e^{i \alpha \varphi(z)}:$ with $\alpha \in \mathbb{R}$. Those are called \textit{vertex operators}. Their conformal dimensions can be calculated as $h_\alpha = \tfrac{\alpha^2}{2}$.

\paragraph{The free fermion.}

The action is
\be
S = \frac{1}{8 \pi} \int \d^2 x \, \Psi^\dg \gamma^0 \gamma^\mu \partial_\mu \Psi,
\ee
where $\Psi = (\psi,\bar \psi)^\top$ is the two-component spinor and a representation of the Dirac matrices $\gamma^\mu$ is $\gamma^0 = \left( \begin{smallmatrix}0&1\\1&0\end{smallmatrix}\right)$ and $\gamma^1 = i \left( \begin{smallmatrix}0&-1\\1&0\end{smallmatrix}\right)$. We obtain
\begin{align}
\psi(z,\bar z) \psi(w,\bar w) &\sim \frac{1}{2\pi g} \frac{1}{z-w}, \label{eq:OPE_fermion}\\
\bar \psi(z,\bar z) \bar \psi(w,\bar w) &\sim \frac{1}{2\pi g} \frac{1}{\bar z - \bar w}, \\
\psi(z,\bar z) \bar \psi(w,\bar w) &\sim 0,
\end{align}
i.e., holomorphic and antiholomorphic parts decouple. $\psi(z)$ is a primary field for which $T(z) = \tfrac{1}{2} : \psi(z) \partial_z \psi(z):$ can be used to obtain the conformal dimension $h = \tfrac{1}{2}$ and central charge $c = \tfrac{1}{2}$. The latter is consistent with the value of the central charge $c = 1$ for the free boson, as it can be composed of two free fermions.

\paragraph{Radial quantization.}

\begin{figure}
\centering\includegraphics[width=0.8\textwidth]{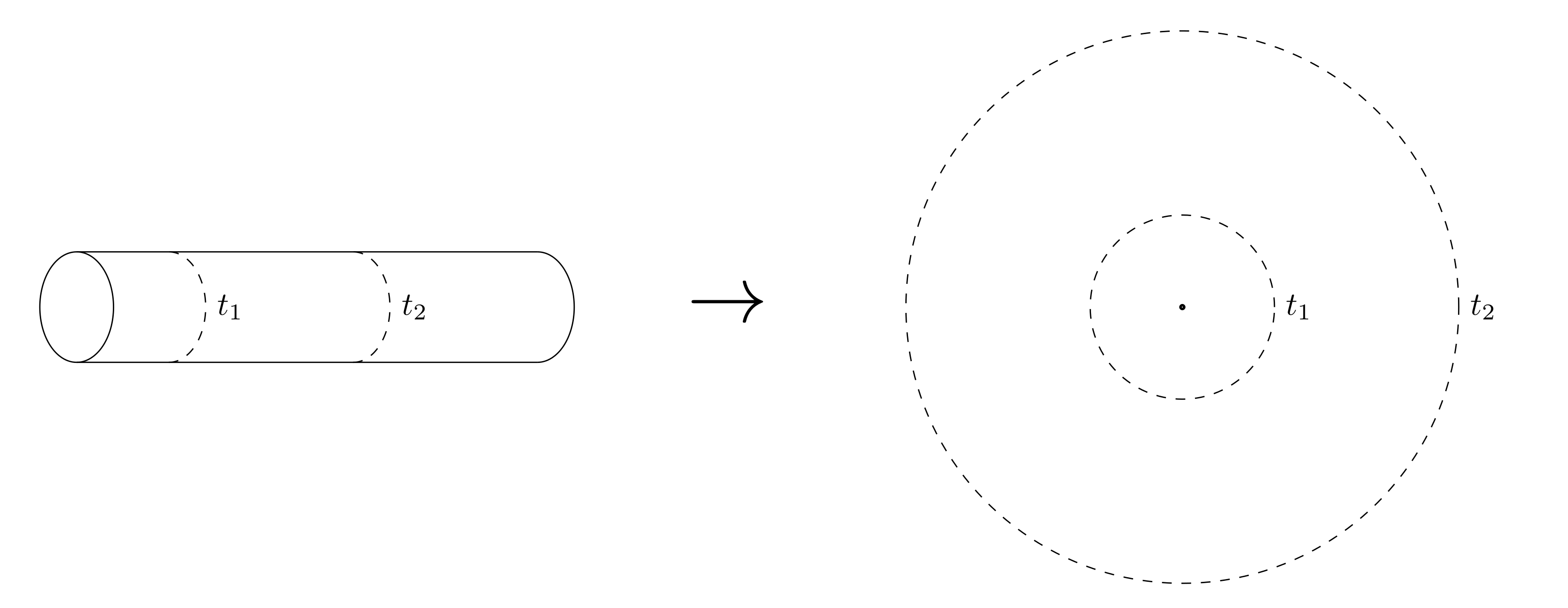}
\caption{The initial system has periodic boundary conditions along the spacial direction and is not confined along the temporal direction. This corresponds to the geometry of a cylinder (left), where the time axis coincides with the cylinder axis. Points with constant times thus correspond to concentric circles. The cylinder is mapped via Eq.~\eqref{eq:cylinder2plane} to the complex plane (right). Points with constant time correspond to circles centered around the origin (which lies at $t = - \infty$), i.e., the temporal direction is now the radial one. The spacial direction is as in the case of the cylinder tangential to those circles.}
\label{fig:radial}
\end{figure}

In order to describe a physical system, it is best to work in a basis given by \textit{radial quantization}: Let us first consider a physical system with periodic boundary conditions along the spacial direction, i.e., 
\be
\phi(x^0,x^1) = \phi(x^0,x^1+L). \label{eq:radial_quant}
\ee
This corresponds to an infinitely long cylinder with circumference $L$. The time axis corresponds to the cylinder axis. Now consider the conformal map
\be
z'(z) = e^{\tfrac{2 \pi z}{L}} = e^{\tfrac{2\pi (x^0 + i x^1)}{L}} \label{eq:cylinder2plane}
\ee
from the cylinder to the complex plane. Note that the infinite past $x^0 \rightarrow -\infty$ corresponds to $z = 0$, whereas the infinite future $x^0 \rightarrow +\infty$ corresponds to $z = \infty$ on the Riemann sphere $\mathbb{C} \cup \{\infty\}$. Points at fixed times correspond to concentric circles around the original cylinder and the origin of the complex plane (see Fig.~\ref{fig:radial}), respectively.

\subsubsection{The Viasoro algebra and descendants of primary fields}

Let us now consider the Laurent expansion of the energy-momentum tensor,
\begin{align}
T(z) &= \sum_{n = -\infty}^\infty z^{-n-2} L_n, \ L_n = \frac{1}{2\pi i} \oint \d z \, z^{n+1} T(z), \\
\bar T(\bar z) &= \sum_{n = -\infty}^\infty \bar z^{-n-2} L_n, \ L_n = \frac{1}{2\pi i} \oint \d \bar z \, \bar z^{n+1} T(\bar z).
\end{align}
Due to the definition of the energy-momentum tensor as the response to changes in the basis, the $L_n$ are \textit{generators} of conformal transformations. Note that as the energy-momentum tensor they are operators. They fulfill the so-called \textit{Viasoro algebra}
\begin{align}
[L_n, L_m] &= (n-m) L_{n+m} + \frac{c}{12} n (n^2-1) \delta_{n+m,0}, \label{eq:Viasoro}\\
[L_n, \bar L_m] &= 0, \\
[\bar L_n, \bar L_m] &= (n-m) \bar L_{n+m} + \frac{c}{12} n (n^2-1) \delta_{n+m,0}. \label{eq:Viasoro_conj}
\end{align}
The $L_n$ are also called the \textit{generators of the Viasoro algebra}. 
$L_{-1}$, $L_0$ and $L_1$ (and analogously $\bar L_{-1}$, $\bar L_0$ and $\bar L_1$) are generators of global conformal transformations and thus leave the vacuum invariant,
\be
L_{-1}|0\rangle = L_0|0\rangle = L_1|0\rangle = 0.
\ee
As $\langle 0| T(z) |0\rangle$ has to be finite as $z \rightarrow 0$ (the expectation value for the infinite past), we even obtain
\be
L_n |0\rangle = 0 \ \mr{for} \ n \geq -1.
\ee
$L_0 + \bar L_0$ creates translations in time and can thus be used to define the Hamiltonian,
\be
\mc H = \frac{2\pi}{L} (L_0 + \bar L_0).
\ee
An excited state of the Hamiltonian is given by
\be
|h\rangle := \phi(z = 0)|0\rangle,
\ee
since one can verify that $L_0 |h\rangle = L_0 \phi(z = 0)|0\rangle = h |h\rangle$ using the Viasoro algebra. Thus, $\mc H|h,\bar h\rangle = h + \bar h$. Similarly, we can verify that $L_n |h\rangle = 0$ for $n \geq 1$ and that
\be
L_{-k_1} L_{-k_2} \ldots L_{-k_m} |h\rangle \label{eq:descendant}
\ee
is an eigenstate of $L_0$ with eigenvalue $h' = h + k_1 + k_2 + \ldots + k_m$. Those states are known as \textit{descendant states} of $|h\rangle$, which is called a \textit{highest weight state}. Let us remark that the Hermitian conjugate of $L_{-n}$ has to be defined as $L_n^\dg$, implying that $L_{-n}$ creates an excitation.
$\sum_{i = 1}^m k_i$ is also denoted as the \textit{level} $\mc N$ of the descendant state \eqref{eq:descendant}. The descendent states can also be created out of the vacuum by acting with \textit{descendent fields}, also called \textit{secondary fields}
\be
L_{-k_1} L_{-k_2} \ldots L_{-k_n} \phi(0).
\ee

The collection of a primary field and its descendants is commonly called a \textit{conformal family}. The highest weight state $|h\rangle$ and its descendants is denoted as a \textit{tower of states} and the subspace spanned by them as a \textit{Verma module}.

It is obvious that at level $\mc N$ the number of descendent states (or fields) is given by the number $P(\mc N)$ of partitions of the integer $\mc N$ into a sum of integers. Using $L_n |h\rangle = 0$ for $n \geq 1$ and the Viasoro algebra, it is a simple exercise to show that descendent states at different levels are orthogonal. 
However, it is not guaranteed that all descendent states at the same level $\mc N$ are linearly independent, so the dimension of the subspace spanned by the descendants of level $\mc N$ might be smaller than $P(\mc N)$. A linear combination $|\chi\rangle$ of descendant fields with zero norm is called a \textit{null state}, $\langle \chi | \chi \rangle = 0$. Descendants of null states are also null states.

\paragraph{Fusion rules.}

The OPE of two primary fields $\phi_i$ and $\phi_j$ is a sum of primary fields $\phi_l$ and their descendants. We write this symbolically as
\be
\phi_i \times \phi_j = \sum_l \mc N_{i,j}^l \phi_l, \label{eq:fusion}
\ee
where we denote by $\phi_{i}$, $\phi_j$, $\phi_l$ the set of the corresponding conformal family, $\times$ represents the OPE of any members of the two conformal families on the left hand side and the sum on the right hand side extends over all conformal families arising in the OPE. For this reason, the elements of $\mc N_{i,j}^l$ are chosen to be only ones and zeros.
Due to Eq.~\eqref{eq:fusion}, the number of conformal families in a CFT is in general infinite. However, physical systems are described by CFTs with a finite number of conformal families.

\subsubsection{Extended chiral algebras}\label{sec:extended_algebras}

While any CFT fulfills the Viasoro algebra, if special symmetries are present, the CFT can give rise to an \textit{extended chiral algebra}. This is very relevant, since a CFT with an infinite number of primary fields in the Viasoro algebra might have a finite number of primary fields with respect to the extended algebra, making it a candidate for the description of physical systems. We will consider two types of such extended algebras in the following, affine Lie algebras and the extended chiral algebra of the compactified boson. 

\paragraph{Affine Lie algebras.}

As shown in the last subsections, the generators $L_n$ of the Viasoro algebra are the generators of conformal transformations of a given CFT. Their commutator (which is the vector product of the algebra) is given by Eqs.~\eqref{eq:Viasoro} to~\eqref{eq:Viasoro_conj}. The $L_n$ are derived from the energy-momentum tensor, which is in turn given by the fields of the CFT. 

Let us now consider the case that the CFT has an additional symmetry (as compared to only conformal symmetry), namely an invariance under a local transformation among the fields given by a Lie group. In this case, apart from the Viasoro algebra, another algebra is fulfilled by a set of generators $\{J_m^a\}_{m \in \mathbb{Z}}^{a = 1, \ldots, d}$, an \textit{affine Lie algebra} or \textit{Kac-Moody algebra},
\be
[J^a_m, J^b_n] = i f_c^{ab} J_{m+n}^c + k m \delta^{a,b} \delta_{m+n,0}, 
\ee
where $f_c^{ab}$ are the structure constants of the underlying Lie algebra $\frak g$ of dimension $d$ (its number of generators)
and $J^a_m$ are the Laurent coefficients of $J^a(z)$. The latter are primary fields called \textit{currents} with conformal dimension $h_a = 1$,
\begin{align}
J^a(z) &= \sum_{m = -\infty}^\infty J^a_n z^{-m-1}\\
T(z) J^a(w) &= \frac{J^a(w)}{(z-w)^2} + \frac{\partial_w J^a(w)}{z-w}.
\end{align}
If the underlying Lie algebra $\frak{g}$ has $SU(2)$ as a subalgebra, one can show that it is always possible to normalize the generators $J^a_m$ such that $k$ is an integer, known as the \textit{level} of the affine Lie algebra denoted as $\frak{g}_k$. In this normalization, the energy-momentum tensor is given by
\be
T(z) = \frac{1}{2(k+g)} \sum_{a=1}^d : J^a(z) J^a(z):
\ee
 and $g$ is its dual Coexter number, which is an integer characterizing the Lie algebra. The central charge is calculated as $c_{\frak{g}_k} = \frac{k d}{k + g}$. Finally, the generators of the Viasoro algebra can be shown to be
\be
L_n = \frac{1}{2(k+g)} \sum_{a=1}^d \sum_{m=-\infty}^\infty :J_m^a J_{n-m}^a:
\ee
fulfilling
\be
[L_m, J^a_n] = -n J^a_{n+m}.
\ee

Let us remark that although the Heisenberg algebra~\eqref{eq:boson_comm} formally has the form of an affine $U(1)$ Lie algebra with level $k=1$, the level is not fixed, as it can be modified via a scale transformation of the generators $a_n \rightarrow \lambda a_n$. Hence, for $U(1)$ the notion of a level has no meaning.

As an example, let us consider the case of $N$ free fermions with OPE (cf., Eq.~\eqref{eq:OPE_fermion})
\be
\psi_i(z) \psi_j(z) = - \frac{\delta_{i,j}}{z-w}.
\ee
If the fermions are real (Majorana fermions), i.e., their fields are self-conjugate, we obtain the Kac-Moody algebra of $SO(N)$ by setting 
\be
J^a(z) = \sum_{i,j} \psi_i(z) t^a_{ij} \psi_j(z),
\ee
where $t^a$ are representation matrices of $SO(N)$. The corresponding level is $k = 1$ and the central charge is calculated as  $c_{SO(N)_1} = \tfrac{N}{2}$ corresponding to $N$ free fermions. 

For $N$ complex fermions, we obtain the $SU(N)_1 \times U(1)$ Kac-Moody algebra by defining
\be
J^a(z) = \sum_{i,j} \psi_i^\dg(z) t^a_{ij} \psi_j(z),
\ee
with central charge $c_{SU(N)_1} + c_{U(1)} = N$ ($c_{U(1)} = 1$ as for the free boson) corresponding to $N$ complex fermions.

\paragraph{The compactified boson.}

An extended chiral algebra, which frequently appears in CFTs describing chiral topological systems, is the \textit{compactified boson}. It is obtained by identifying $\varphi$ with $\varphi + 2\pi R$, i.e., $\varphi$ has the meaning of an angular variable (actually of the length of an arc). In this case, we are allowed to modify Eq.~\eqref{eq:radial_quant} for a bosonic field $\varphi(x_0,x_1)$ in such a way that
\be
\varphi(x_0,x_1+L) = \varphi(x_0,x_1) + 2 \pi \omega R
\ee
with $\omega \in \mathbb{Z}$. $\omega$ is a winding number as $\varphi$ moves around the circle, and $R$ is denoted as the \textit{compactification radius}. 


The extended chiral algebra of the compactified boson is generated by $i\partial_z \varphi$ and some currents. Since the latter are primary fields, the only candidates are the vertex operators $\mc V_\alpha(z) = :e^{i \alpha \varphi(z)}:$. In the case of the Kac-Moody algebra, the currents were required to have conformal dimension one. Here, we require the currents to have integer conformal dimension. Calling the $\alpha$'s of the currents $\alpha_c$, this means $h_{\alpha_c} = \tfrac{\alpha_c^2}{2} \in \mathbb{N}$. The currents need to be well defined as $\varphi$ is replaced by $\varphi + 2 \pi R$, i.e, we have to constrain $\alpha_c$ to $\alpha_c R \in \mathbb{Z}$. Using those conditions, one can derive
\be 
\alpha_c = \pm \sqrt{p p'}, \ \mr{and} \ \ R = \sqrt{\frac{p'}{p}},
\ee
where $p, p' \in \mathbb{N}$. They are fixed for a given CFT by the compactification radius. 
The generators of the extended chiral algebra are
\be
i \partial_z \varphi \ \ \mr{and} \ \ \Gamma^{\pm} = :e^{\pm i \sqrt{p p'}\varphi}:
\ee
For the case $p p' = 2$, they simply generate the $SU(2)_1$ algebra. 

The OPE of two vertex operators (fusion!) $\mc V_\alpha$ and $\mc V_\beta$ is 
\be
\mc V_{\alpha} (z) \mc V_\beta(w) \sim |z-w|^{2 \alpha \beta} \mc V_{\alpha + \beta}(w) + \ldots \label{eq:OPE_vertex}
\ee
For the admissible primary fields, this OPE has to be local, which turns out to constrain them to 
\be
\alpha = \tfrac{\ell}{\sqrt{p p'}} \label{eq:alpha_ell}
\ee 
with $\ell \in \mathbb{Z}$ and conformal dimensions $h_{(\ell)} = \tfrac{\ell^2}{2 p p'}$ . Since according to Eq.~\eqref{eq:OPE_vertex}, shifting $\ell$ by $p p'$ corresponds to the application of a ladder operator $\Gamma^\pm$, leading to a descendent field, the primary fields are restricted by $1 \leq \ell \leq p p'$. The corresponding CFT is denoted by $U(1)_{pp'}$. The central charge is $c = 1$. 


\subsubsection{Relation to chiral topological systems}\label{sec:chiral_physical}

While the whole machinery of CFT is quite intricate, the relation to chiral topological systems is rather simple~\cite{Moore_Read},~\cite{Rev_TopComp}: Given a chiral system described by a CFT, its wave function in the bulk can be expressed as a correlator of primary fields. Moreover, if the system has an edge, its boundary physics is directly described by the CFT. This also applies to a non-physical edge, like in the entanglement spectrum, where the low energy part has the same level counting as the CFT~\cite{ES_org} (with null vectors being removed). 

Let us consider an example of the first aspect, representing the wave function of a chiral state by a correlator of a CFT, the $\nu = \tfrac{1}{q}$ Laughlin state ($q$ odd), which is a model wave function for the fractional Quantum Hall Effect: In first quantization, the wave function can be written as
\begin{align}
\psi_q(z_1, z_2, \ldots, z_N) &= \left\langle \prod_{j=1}^N : e^{i \sqrt{q} \varphi(z_j)} : e^{-i \int \d^2 z \tfrac{\varphi(z)}{2 \pi \sqrt{q}}} \right\rangle \notag \\
&= \prod_{i < j}^N (z_i - z_j)^q \, e^{-\tfrac{1}{4}\sum_{i=1}^N |z_i|^2} \label{eq:Laughlin} 
\end{align}
where $\varphi(z)$ is a free bosonic field and the second exponential function in the correlator serves as a charge neutralizing background. The corresponding CFT is $U(1)_{q}$ with central charge $c = 1$. The anyonic quasi-particle excitations are obtained by inserting primary fields of CFT, i.e., of the extended chiral algebra of the compactified boson (labelled by $l \in \{1,2,\ldots,q\}$), into the correlator in Eq.~\eqref{eq:Laughlin}. 
In the case of the Laughlin wave function, there are thus $q$ quasi-particles, which according to Eq.~\eqref{eq:OPE_vertex} are all Abelian.

The physical relevance of the second aspect lies in the entanglement spectrum: If one traces out the complement $\overline{\mc R}$ of region $\mc R$ of the system, $\rho_{\mc R} = \tr_{\overline{\mc R}}(\rho) := e^{-\mc H_\mr{ent}}$, the low energy spectrum of the \textit{entanglement Hamiltonian} $\mc H_{\mr{ent}}$ is given by the energy levels of the corresponding CFT (up to a non-universal scaling factor). Note that null vectors of the CFT have to be removed and that there might be unifications of the towers of states of different conformal families leading to additional degeneracies in the entanglement spectrum.

In the case of the Laughlin wave function, the level counting is ${1,1,2,3,5,\ldots}$ corresponding to the integer partitions $P(\mc N)$ of the level $\mc N$. There appear several such towers (e.g., at different many-body momenta), which correspond to different conformal families with respect to the Viasoro algebra (emerging from the primary fields for arbitrary $\ell \in \mathbb{Z}$ in Eq.~\eqref{eq:alpha_ell}). However, in the extended algebra these towers of states get unified by the ladder operators $\Gamma^\pm$ to merely $q$ distinct towers.



\paragraph{Topologically ordered chiral systems}

One way to define topologically ordered systems is by the existence of a non-vanishing (negative) subleading correction to the area law, called the \textit{topological entanglement entropy}~\cite{TEE_org,TEE_Levin} $\gamma$. That is,
\be
S(\rho_{\mc R}) = c' \, \partial \mc R - \gamma,
\ee
where $S(\rho_{\mc R})$ is the von Neumann entropy obtained after tracing out the complement of region $\mc R$, $\partial \mc R$ is the length of its boundary and $c'$ is a non-universal positive constant. Topologically ordered systems have anyonic quasi-particle excitations. The number of all quasi-particle species equals the ground state degeneracy if the system is placed on a torus. One can choose the basis of the ground state subspace in such a way that the entanglement entropy between $\mc R$ and its complement is minimized for all ground states $|\psi_\mu\rangle$. The corresponding $|\psi_\mu\rangle$ are denoted as \textit{minimally entangled states}. There is a one to one correspondence to the quasi-particles $\mu$ (which are primary fields of the corresponding CFT), since the topological entanglement entropy is given, respectively, by $\gamma_\mu = \ln(\mc D/ d_\mu)$. $d_\mu$ is the \textit{quantum dimension} of quasi-particle $\mu$. This number characterizes the asymptotic dimension of the Hilbert space per particle given by their braiding statistics. For Abelian anyons, the quantum dimension is one, because many Abelian anyons can be fused together to form a single one (cf., Eq.~\eqref{eq:OPE_vertex}). $\mc D$ is the \textit{total quantum dimension} with $\mc D = \sqrt{\sum_\mu d_\mu^2}$.

\paragraph{Momentum polarization.}

Apart from using the entanglement spectrum, there is another way to verify that a chiral topological system is described by a certain CFT. We assume that the system under consideration is placed on a cylinder and has the ground states $|\Psi_\mu\rangle$, which might have been obtained by a numerical method such as Quantum Monte Carlo simulations. We can calculate the conformal dimensions $h_\mu$ given the central charge $c$ by applying a twist operator $T_L$ on one half of the cylinder (say the left one), which moves the corresponding sites in the wave function it acts on around the cylinder. The overlap with the original ground state, 
\be
\lambda_\mu = \langle \Psi_\mu | T_L | \Psi_\mu\rangle \label{eq:twist_half}
\ee
has been shown in Ref.~\cite{mom-pol} to have the form
\be
\lambda_\mu = \exp\left(\frac{2 \pi i}{N_v} \tau_\mu - \alpha' N_v\right), \label{eq:mom-pol}
\ee
where $N_v$ is the circumference of the cylinder, $\alpha'$ is a complex non-universal parameter and $\tau_\mu = h_\mu - \tfrac{c}{24}$ the so-called \textit{momentum polarization} (see also Refs.~\cite{mom-pol2,mom-pol3,mom-pol4}).

%% file: 3.3.tex
\section[Free Fermionic Chiral Projected Entangled Pair States]{Free Fermionic Chiral Projected Entangled Pair States\footnote{This section is a combination of Refs.~\cite{cPEPS_org,cPEPS_long}, copyright American Physical Society.}}~\label{sec:cGFPEPS}

\subsection{Overview}


Topological states~\cite{TO_org,TO_org2} are quantum states of matter with intriguing properties. They include non-chiral states with topological order, such as the toric code~\cite{toric_code} and string net models~\cite{Levin_Wen}, as well as chiral topological states. The latter have broken time reversal symmetry, and possess non-vanishing topological invariants. They encompass celebrated examples like integer and fractional quantum Hall states as well as Chern insulators~\cite{Haldane_model} and topological superconductors~\cite{Phases_Schnyder,Phases_Kitaev}. They display chiral edge modes which are protected against local perturbations, and cannot be adiabatically connected to states with different values of the topological invariants.

Among others, a remarkable open problem in this field is to classify all topological phases; that is, the equivalence classes of local Hamiltonians that can be connected by a (symmetry preserving) gapped path. For their free fermion versions in arbitrary dimensions, a full classification has been already obtained~\cite{Phases_Schnyder,Phases_Kitaev}. For interacting spins, this goal has only been achieved in one dimension \cite{MPS_classify_sym,MPS_classify_sym2,MPS_classify}, based on the fact that ground states of 1D gapped local Hamiltonians are efficiently represented by MPS~\cite{MPS_faithful,Schollwock}. In dimensions higher than one, this problem  remains open. Still, recent developments reveal that there exist deep intrinsic connections between quantum entanglement and topological states. For instance, topological order is reflected in the universal correction to the entanglement area law, the topological entanglement entropy. A further proposal has been put forward by Li and Haldane~\cite{ES_org}, who suggested that the entanglement spectrum, that is, the eigenvalues of the reduced density operator of a subsystem, contains more valuable information than the topological entanglement entropy.

By construction, PEPS contain the necessary amount of entanglement required by the area law. Furthermore, many known topological models, such as the toric code (and its generalization, the quantum double models~\cite{toric_code}), resonating valence-bond states \cite{RVB_org}, and string nets \cite{Levin_Wen}, possess exact PEPS descriptions~\cite{TNS_Hopf,RVB_PEPS,Levin_PEPS,toric_PEPS}. (For a representation of the toric code ground states in terms of MERA, see Refs.~\cite{toric_MERA,PEPS_MERA_top}). Despite the lack of local order parameters, PEPS nevertheless provide a local description for topological states, with the global topological properties being encoded in a single PEPS tensor. For some of the above examples, the connection of topology and the PEPS tensor has been made precise as originating from a \textit{symmetry} of the PEPS tensor (see subsection~\ref{sec:PEPS_sym}). This symmetry only affects the virtual particles used to build the PEPS, unlike the physical symmetries of PEPS. It can be grown to arbitrary regions, and has several intriguing consequences: (i) it leads to the topological entanglement entropy; (ii) it gives rise to a universal part \cite{transfer} in the boundary Hamiltonian \cite{bulk_boundary} acting on the auxiliary particles at a virtual boundary, whose eigenvalues are related to the entanglement spectrum of the subsystem; (iii) it provides topological protection of the edge modes \cite{edge_theories}; (iv) it gives rise to string operators that provide a mapping between the different topological sectors;  (v) it can also be used to build string operators for anyonic excitations and to determine the braiding statistics; (vi) it determines the ground state degeneracy of the parent Hamiltonian.

In spite of the above indications, there exists a deep reason to believe that
PEPS cannot describe the physics of certain kinds of topological phases,
namely those that have chirality.
Chiral topological states are very different from the above mentioned topological states preserving time-reversal symmetry (known as non-chiral topological states), in that they necessarily have chiral gapless edge modes, which cannot be gapped out by weak perturbations due to the lack of a back-scattering channel.
In fact, despite a significant effort in
the research of Tensor Network States, no PEPS 
corresponding to a 2D chiral topological phase has been found for a long time, not even for the simplest
Chern insulators, see subsection~\ref{sec:class_A}. (Note however that in Ref.~\cite{TNS_NCooper}
it is shown that expectation values of observables in chiral models can be expressed
as a contraction of tensor networks, although the question of whether states can be
expressed as tensor networks was left open). This fact may be
qualitatively understood as follows: Any PEPS is the ground state of a
local parent Hamiltonian $\mc H=\sum_\mb{n} h_\mb{n}$, cf. subsection~\ref{sec:PEPS_parent}.
This Hamiltonian is frustration-free, meaning that the PEPS is
annihilated by each local term $h_\mb{n}$ individually.  The Hamiltonian
terms are of the form $h_\mb{n}=b_\mb{n}^\dagger b_\mb{n}^{\phantom\dagger}$, where the
$b_\mb{n}$ are quasi-particle operators supported in a small region. If $\mc H$ corresponds
to a free-fermion system with a gapped band structure, their
Wannier functions must be localized, which in turn has been proven to
be impossible for systems with a non-trivial Chern number
\cite{exp_Wannier,exp_Wannier2}.  Still, the question whether PEPS can describe chiral
topological insulators was left open: First, even though the parent
Hamiltonians of such PEPS would be gapless, they might still be ground states of other
non-frustration-free gapped Hamiltonians (with non-localized Wannier
functions); second, although they do not provide exact descriptions of all
chiral states, they might still be able to approximate them accurately.

Below, several simple families of GFPEPS with a non-zero Chern number will be provided.
By simple we mean they possess a small number of virtual Majorana bonds ($\chi = 1,2$).
These families of GFPEPS have correlation functions with
a power-law decay, as well as non-localized Wannier functions. In fact,
they are the unique ground states of free-fermion, gapped Hamiltonians
with hopping amplitudes following the same decay. Apart from that, as all
PEPS, they are ground states of local parent Hamiltonians, which however
must be gapless due to the presence of critical correlations.  Indeed, we
prove that there cannot be GFPEPS with a non-trivial Chern number which
have a finite-range \emph{and} gapped parent Hamiltonian, since all such
Hamiltonians are in the trivial phase.

Given the variety of results presented in this section and the different techniques used to derive them, we start with a subsection that gives an overview of all of them, and connects them to known properties of topological PEPS. The specific derivations and explicit statements and proofs are given in the following subsections: In subsection~\ref{sec:Detailed-Analysis}, a general framework for studying the boundary and edge theories of GFPEPS is developed. Their relation to the Chern number is established. In subsection~\ref{sec:examples}, we give different examples of GFPEPS, some of them topological and some of them not, in order to provide comparison between the two cases. Thereafter, we derive necessary conditions for topological chiral GFPEPS, see subsection~\ref{sec:cond_top}.
In subsection~\ref{sec:full-chi1} we completely characterize all PEPS with one Majorana bond with chirality. It turns out that in this case the Chern number can only be $0$ or $\pm 1$; for the latter case we derive necessary and sufficient criteria. In subsection~\ref{sec:sym_chiral} we prove a necessary and sufficient condition on the symmetry that a PEPS tensor has to possess in order to give rise to a chiral edge state for one Majorana bond, and show how those symmetries can be grown to larger regions and how to build string-like operators.
These restrictions on chiral GFPEPS, however, do not rule
out the possibility of using PEPS to {\em approximate} the ground state of
a chiral topological insulator with a local Hamiltonian. In fact, we
investigate this issue in subsection~\ref{sec:numerics} and conclude that this is possible, since the
approximation improves exponentially with the number of fermionic bond modes.  Finally, we show that by using mixed GFPEPS we can approximate
the finite temperature properties of such systems as well.


\subsection{Definitions and Results}\label{sec:Def}

This subsection gives an overview of our main results on free fermionic chiral PEPS, which are detailed in the remainder of this section.
It also reviews in a self-contained way the basic ingredients that are required to derive the results, and to interpret them. It is divided into four parts. The first one contains an alternative definition of GFPEPS, which are the basic objects in our study. It also contains two Hamiltonians for which they are the ground state. The first one is gapped, may have hopping terms decaying as a power-law (which it does for chiral PEPS) and is the one that appears more naturally in the context of topological insulators and superconductors. The second one follows from the PEPS formalism, is local and gapless for chiral PEPS with a degenerate ground state. In the second part we present a simple family of GFPEPS, which we will extensively use to illustrate our findings. The third one contains the construction of boundary and edge theories for GFPEPS, which we explicitly use for the simple family. In the last part, we make a connection between the behavior observed for this family and the one that is known for topological (non-chiral) PEPS (Ref.~\cite{TNS_intersection}). In particular, we show that one can understand it in terms of string operators acting on the virtual particles, which can be moved and deformed without changing the state.

Throughout this subsection we will concentrate on the simplest GFPEPS -- those which have the smallest possible bond dimension (which 
corresponds to one Majorana bond). This will allow us to
simplify the description and formulas. However, all the constructions given here can be easily generalized to larger bond dimensions, and this will be done in the following subsections. Some of the results, however, explicitly apply to one Majorana bond, so that we will specialize to that case in the following subsections too.

\subsubsection{Gaussian Fermionic PEPS and parent Hamiltonians}\label{Sec:Intro-GFPEPS}

We consider a square $N_v \times N_h$ lattice of a single fermionic mode per site, with annihilation operators $a_{\mb r}$, where $\mb r$ is a vector denoting the lattice site. We will consider a state, $\Phi$, of a particular form, and Hamiltonians for which it is the ground state.

\paragraph{Gaussian fermionic PEPS.}

\begin{figure}[!ht]
\begin{center}
\includegraphics[width=0.5\textwidth]{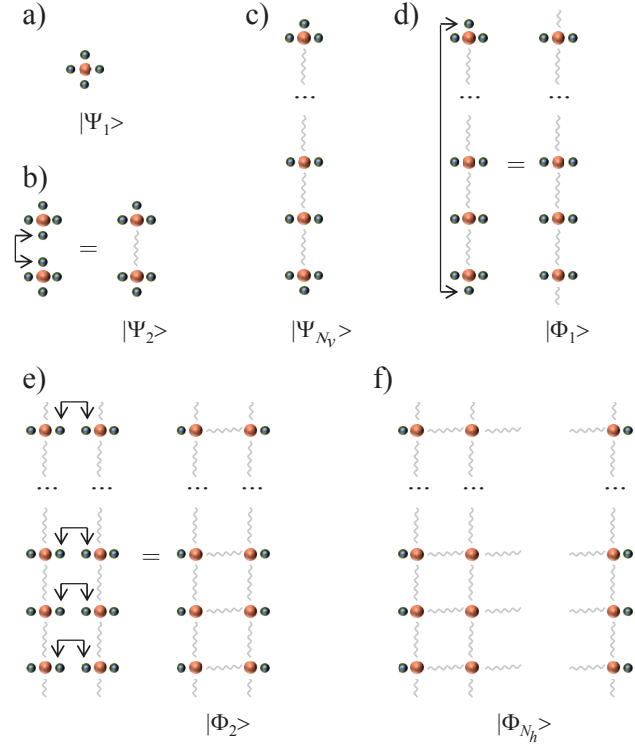}
\end{center}
\caption{Alternative construction of a GFPEPS. (a) We start with a state $\Psi_1$ that is Gaussian and includes one physical fermionic mode (big red ball) and four virtual Majorana fermions (small blue balls) located at site $\mb r$. (b) Two states of this kind in the same column are concatenated by projecting on $\langle \omega_{\mb r, \mb r + \hat y}|$ (see text). (c) Proceeding in the same way one obtains a state $\Psi_{N_v}$ defined on a column with unpaired virtual Majorana modes on the left and right and on the two ends of the column. (d) The remaining up and down modes at the ends are jointly projected out, yielding $\Phi_1$. (e) Afterwards, two columns can be concatenated by pairwise projecting out the left and right virtual Majoranas between them. (f) Continuing in the same way, one obtains a GFPEPS $\Phi_{N_h}$ defined on $N_h$ columns. It can be made completely translationally invariant by pairwise projecting out the remaining left and right virtual Majorana modes, resulting in the final GFPEPS state $\Phi$. This figure was reproduced from Ref.~\cite{cPEPS_long}. \label{Fig:PEPS}}
\end{figure}

We provide an alternative definition of the GFPEPS introduced in subsection~\ref{sec:GFPEPS} corresponding to the same class of physical states. We first show how a GFPEPS, $\Phi$, of the fermionic modes is constructed (see Fig.~\ref{Fig:PEPS}).

The basic object in this construction is a fiducial state, $\Psi_1$, of
one fermionic (physical) mode and four additional (virtual) Majorana
modes, all of them at site $\mb r$
(Fig.\ref{Fig:PEPS}a). The corresponding mode operators,
$c_{\mb r,L},c_{\mb r,R},c_{\mb r,U}$, and $c_{\mb r,D}$, 
fulfill standard anticommutation relations, $\{c_{\mb r,w},c_{\mb r',w'}\}=2
\delta_{\mb r,\mb r'}\delta_{\kappa,\kappa'}$, are Hermitian and anticommute with the
other fermionic operators. The state $\Psi_1$ is arbitrary, except for the
fact that it must be Gaussian and have a well defined parity. This means
that it can be written as
\be
 |\Psi_1\rangle_\mb{r} = e^{\mc H_{\mb r}} |\Omega\rangle, \label{eq:Gaussian-psi1}
 \ee
where $\mc H_\mb{r}$ is a quadratic operator in all the mode operators, and the
$\Omega$ denotes the vacuum of the virtual and physical modes. One can
easily parametrize $\mc H$, and thus $\Psi_1$, but this will not be
necessary here, since we will make use of the fact that the state is
Gaussian, for which a more appropriate parametrization exists.

The state of the physical fermions, $\Phi$, can be obtained by concatenating all the $\Psi_1$ at different sites in the way we explain now and is illustrated in Fig.~\ref{Fig:PEPS}. First, take two consecutive lattice sites in the same column, $\mb r$ and $\mb r + \hat y$ and project the down virtual mode of the first and the up mode of the latter onto a particular state, i.e.,
 \be
 |\Psi_2\rangle_{\mb r, \mb r + \hat y} = \omega_{\mb r, \mb r + \hat y} e^{\mc H_\mb{r}+ \mc H_{\mb r + \hat y}}
 |\Omega\rangle,
 \ee
(see Fig.\ref{Fig:PEPS}b). Here $\omega_{\mb r, \mb r + \hat y}= \tfrac{1}{2}(1+ i c_{\mb r,D} c_{\mb r + \hat y,U})$, which ensures that $c_U$ and $c_D$ are maximally entangled (forming a pure fermionic state)\footnote{The motivation for the name ``maximally entangled Majorana modes'' comes from the following construction (see Ref.~\cite{fPEPS_Kraus}): One can add an extra Majorana operator per bond such that each bond is represented by a full fermionic mode. In that case, the state generated by $\omega_{\mb r,\mb r'}$ out of the vacuum is an entangled state between the two fermionic modes along that bond (for the formal definition of entangled fermionic modes, see Ref.~\cite{ME_fermionic}). The extra Majorana operators do not play any role as long as antiperiodic boundary conditions are imposed on them, and can thus be ignored so that one ends up with the states as they are used throughout this section.}. Since the modes that we project on are in a well defined state after the projection, we can omit them in the following. In order to simplify the notation, we will denote by $\langle \omega_{\mb r, \mb r'}|\Psi\rangle$ the state obtained by applying $\omega_{\mb r,\mb r'}$ and discarding the corresponding modes, and we will say that we have projected onto $\omega_{\mb r, \mb r'}$. We will also omit the indices representing the lattice sites whenever this does not lead to confusion.

We proceed in the same way, concatenating all the sites corresponding to a column by projecting the consecutive up and down virtual modes onto the state defined by $\omega_{\mb r,\mb r+\hat y}$. The resulting state is $\Psi_{N_v}$, since we have $N_v$ sites in a column (see Fig.\ref{Fig:PEPS}c). This state contains $N_v$ physical fermionic modes as well as $2N_v+2$ virtual Majorana modes, $N_v$ on the left, $N_v$ on the right, one up and one down mode. As we will consider here periodic boundary conditions along the vertical direction, we also project out the up and down virtual modes, obtaining $\Phi_1$, a state that corresponds to one column (and thus the subindex). Such a state contains $N_v$ physical fermionic modes and $2N_v$ virtual Majorana modes (see Fig.\ref{Fig:PEPS}d). By construction, the state is translationally invariant along the vertical direction.

In order to obtain the state on the lattice, we have to follow a similar procedure in the horizontal direction (see Fig.\ref{Fig:PEPS}e). For that, we take the states of two consecutive columns, and project each of the right virtual modes (at site $\mb r$) of one and the corresponding left virtual mode (at site $\mb r + \hat x$) of the other onto  $\omega_{\mb r, \mb r + \hat x}'= \tfrac{1}{2}(1+ i c_{\mb r,R} c_{\mb r+\hat x,L})$. The resulting state, $\Phi_2$, contains $2N_v$ physical fermionic modes, as well as $2N_v$ virtual Majorana modes. We continue adding columns in the same way, until we obtain $\Phi_{N_h}$, containing $N_v\times N_h$ physical fermionic modes and $2N_v$ virtual Majorana ones (see Fig.\ref{Fig:PEPS}f).

In order to obtain a translationally invariant state in the horizontal direction too, we have to project each remaining virtual pair of modes on the left and the right onto the state defined by $\omega'_{\mb r + N_h \hat x,\mb r}$. In this case, we will say that we have a state, $\Phi$, on the torus. Otherwise, we can project the virtual modes on the left and the right onto some other state. If we took a product state (of left and right virtual modes) that is translationally invariant in vertical direction itself, we would have kept that property in the vertical direction and the state $\Phi$ would have been defined on a cylinder. A subtle point is that when we perform this last projection in order to generate the physical state $\Phi$, the result may vanish. This happens, for instance, in some of the examples considered in this section in the torus case. There, we will have to introduce a string operator in the virtual modes to render the final state non-vanishing.

The state $\Phi$ on the torus is fully characterized by the fiducial state $\Psi_1$ (and therefore by $\mc H_\mb{r}$), since the construction is carried out by concatenating them with a specific procedure. For the cylinder, $\Phi$ also depends on the states we choose to close the virtual boundaries. From now on we will work on the torus, unless explicitly stated otherwise.

Since the fiducial state $\Psi_1$ is Gaussian and our construction keeps the Gaussian nature, all the states defined above will be Gaussian. For that reason, instead of expressing $\Psi_j$ and $\Phi$ in the Hilbert space on which the mode operators act, we characterize them in terms of their covariance matrices (CMs).  Thus, the original state $\Psi_1$ has a CM with four blocks,
 \be
 \label{eq:gamma1}
 \gamma_1 = \left(\begin{array}{cc} A & B \\ - B^\top & D \end{array} \right)
 \ee
where $A,D$ are $2\times 2$ and $4\times 4$ antisymmetric matrices, respectively, $B$ is a $2\times 4$ matrix, and they are constrained by $\gamma_1^2=-\Id$ (since the state $\Psi_1$ is pure). Hence, the state $\Phi$ is completely characterized by those matrices. Concatenating states as explained above can be easily done in terms of the CMs (see Ref.~\cite{fPEPS_Kraus} and subsection~\ref{sec:construct-GFPEPS} below).

As outlined in subsection~\ref{sec:GFPEPS}, the CM $\gamma$ of the translationally invariant final state $\Phi$ can be block-diagonalized via a discrete Fourier transform. The blocks on the diagonal are given by
\begin{equation}
G_\mr{out}(\mb k) = \left(\begin{smallmatrix}
i \hat d_x(\mb k)&\hat d_z(\mb k) + i \hat d_y(\mb k)\\
-\hat d_z(\mb k) + i \hat d_y(\mb k)&- i \hat d_x(\mb k)
\end{smallmatrix}\right) \label{eq:G-param}
\end{equation}
with 
$\hat d_i(\mb k) \in \mathbb{R}$ and $|\hat d(\mb k) | = 1$ due to $\gamma \gamma^\top = \Id$, i.e., $G_\mr{out}(\mb k) G^\dg_\mr{out}(\mb k) = \Id$.

The above construction can be trivially extended to more general GFPEPS, where there are $4\chi$ virtual Majorana modes and $f$ fermions per site. In subsection~\ref{sec:Detailed-Analysis} we will show how to carry out such a construction for that general case. The case considered in this subsection, $\chi=1$, is much simpler to describe and already possesses all the ingredients to give rise to topological chiral states.

\paragraph{Parent and flat band Hamiltonians}

One can easily construct Hamiltonians for which $\Phi$ is the ground state. For that, we can follow two different approaches. The first one takes advantage of the fact that $\Phi$ is a Gaussian state, whereas the second uses that it is a PEPS.

Our first Hamiltonian is the ``flat band'' Hamiltonian
 \be
 \label{eq:flat}
 {\cal H}_{\rm fb} = -\frac{i}{4} \sum_{l,m} \gamma_{lm} e_l e_m
 \ee
where $\gamma$ is the CM of the state $\Phi$, and $e$ are the Majorana modes
built out of the physical fermionic modes. Since $\Phi$ is pure,
$\gamma^2=-\Id$, and thus, it has eigenvalues $\pm i$. Hence, ${\cal
H}_\mr{fb}$ contains two bands separated by a bandgap of magnitude
$2$, which are flat. As $\gamma$ is antisymmetric,
there exists an orthogonal matrix $O$ such that $O^\top \gamma O$ is block
diagonal. Using this, one can easily convince oneself that $\Phi$ is the
unique ground state of $\mc H_\mr{fb}$. Note also that the Hamiltonian
${\cal H}_\mr{fb}$ will not be local in general, since $\gamma_{lm}\ne 0$
for all $l,m$. We also remark that for general $\gamma$ the single
particle spectrum of a Hamiltonian of the form~\eqref{eq:flat} is given by
the eigenvalues of $-i \gamma$.

We transform Eq.~\eqref{eq:flat} to reciprocal space
and write it 
in terms of the Fourier transformed Majorana modes 
\be
\hat e_{\mb k,s} = \frac{1}{\sqrt{N_h N_v}} \sum_{\mb r} e_{\mb r,s} e^{i k \cdot r}
\ee
(with $s = 1,2$ and $(\mb r,s)$ corresponding to the joint index $l$ above),
so that it takes the form
\be
\mc{H}_\mr{fb} = -\tfrac{i}{4} \sum_{\mb k} \sum_{r,s=1}^2 [G_\mr{out}]_{rs}(\mb k) \hat e_{\mb k,r} \hat e_{\mb k,s} \label{eq:H-Gout},
\ee
where $G_{\mr{out}}(\mb k)$ is given in Eq.~\eqref{eq:G-param}.

The second Hamiltonian can be constructed by invoking the general theory of PEPS (see subsection~\ref{sec:PEPS_parent}). We can always find a local, positive operator, $h\ge 0$, acting on a sufficiently large plaquette, that annihilates our state, i.e. $h_\mb{r}|\Phi\rangle=0$. Here $\mb r$ denotes the position of the plaquette. In the case of a GFPEPS, $h_\mb{r}$ can be chosen to be local. Furthermore, since the state is translationally invariant, we can take
 \be
 \label{eq:HPEPS}
 {\cal H}_{\rm ff} = \sum_{\mb r} h_\mb{r}.
 \ee
Now, this Hamiltonian is local (i.e., a sum of terms acting on finite regions, the plaquettes), frustration free (thus the subscript), and it is clear that $\Phi$ is a ground state. However, there may still be other ground states, and, additionally, ${\cal H}_{\rm ff}$ may have a gapless continuous spectrum in the thermodynamic limit.

For the topological states considered later on, we will see that ${\cal H}_{\rm fb}$ is intimately connected to the chiral properties at the edges, as it is well known for topological insulators and superconductors \cite{Topo_rev, Topo_rev2}. The other one, ${\cal H}_{\rm ff}$ will share other topological properties that makes it akin to Kitaev's toric code~\cite{toric_code} and its generalizations.

\subsubsection{A family of topological superconductors}\label{sec:Intro-Examples}

\paragraph{Parameterization of the GFPEPS}

Now, we review a family of chiral topological GFPEPS which is characterized by a parameter, $\lambda\in[0,1]$. The fiducial state $\Psi_1$ is given by
 \be
 \label{eq:Psi1ex}
 |\Psi_1\rangle=\left(\sqrt{1-\lambda}\Id+\sqrt{\lambda}a^\dagger b^\dagger\right)|\Omega\rangle.
 \ee
Here, $b$ is an annihilation operator acting on the virtual modes as
follows
 \be
 b= \frac{1}{\sqrt{2}}(h+v)
 \ee
where
\be
h= \frac{c_L - i c_R}{2} e^{\frac{i \pi}{4}} \ \mr{and} \ v= \frac{c_U - i c_D}{2}. \label{eq:def-hv}
\ee
The corresponding CM $\gamma_1$ [Eq.~(\ref{eq:gamma1})] is given by
\begin{align}
A &= \left(\begin{array}{cc}
0&1-2\lambda\\
-1+2\lambda&0
\end{array}\right), \notag \\
B &=  \sqrt{\lambda - \lambda^2}\left(\begin{array}{cccc}
1&-1 &0&-\sqrt{2}\\
-1&-1&-\sqrt{2}&0
\end{array}\right), \notag \\
D &= \left(\begin{array}{cccc}
0&     1-\lambda     &-\frac{\lambda}{\sqrt 2}&-\frac{\lambda}{\sqrt{2}}\\
-1+\lambda&0&\frac{\lambda}{\sqrt 2}&-\frac{\lambda}{\sqrt{2}}\\
\frac{\lambda}{\sqrt 2}&-\frac{\lambda}{\sqrt 2}&0&1-\lambda\\
\frac{\lambda}{\sqrt 2}&\frac{\lambda}{\sqrt 2}&-1+\lambda&0
\end{array}\right) \label{eq:Chern-1}
\end{align}
We have sorted the Majorana mode operators as $e_1,e_2,c_{L},c_{R},c_U,c_D$.

Later on we will consider other states, topological or not, to illustrate the properties of the boundary theories. However, the family of states given here will be a central object of our analysis, since it already possesses all the basic ingredients. As is evident from the definition, the fiducial state $\Psi_1$ in Eq.~\eqref{eq:Psi1ex} is an entangled state between the physical and one virtual mode, except for $\lambda=0,1$. For $\lambda=1/2$ it is maximally entangled. It has certain symmetries, which will be of utmost importance to understand the topological features of the state $\Phi$ it generates. Explicitly,
 \begin{subequations}
 \label{eq:symmetries}
 \bea
 \left(\sqrt{\lambda} a^\dagger - \sqrt{1-\lambda}b \right) |\Psi_1\rangle &=&0,\\
 \left(\sqrt{1-\lambda} a + \sqrt{\lambda}b^\dagger \right) |\Psi_1\rangle &=&0,\\
 d_1 |\Psi_1\rangle &=& 0 \label{eq:dPsi0}
 \eea
 \end{subequations}
with
 \be
 \label{eq:d}
 d_1 =  \frac{1}{\sqrt{2}}(-h+v).
 \ee
The operators $a,b,$ and $d_1$ define three fermionic modes (one physical and two virtual ones). Eqs. (\ref{eq:symmetries}) just reflect the fact that for a Gaussian state the physical mode can be entangled at most to one virtual mode, since we can always find a basis in which one virtual mode is disentangled. The latter is precisely the one annihilated by $d_1$. In fact, Eqs. (\ref{eq:symmetries}) completely define the state $\Psi_1$.

\paragraph{Algebraic decay of correlations} 

\begin{figure}[!ht]
\begin{center}
\includegraphics[width=0.5\textwidth]{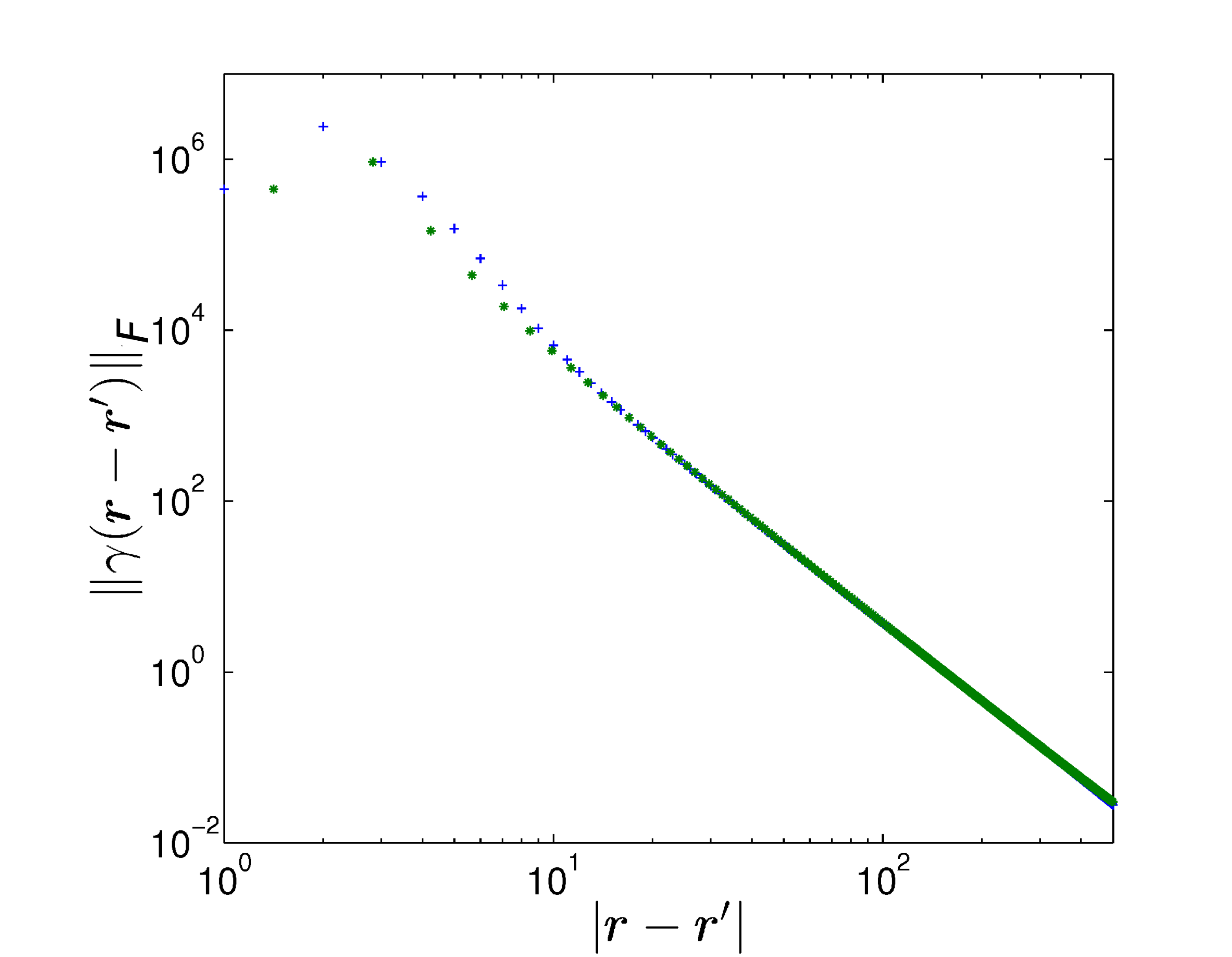}
\end{center}
\caption{Frobenius norm $\|\gamma(\mb r- \mb r')\|_\mr{F}$ of the block of the covariance matrix $\gamma_{lm}$ for $\lambda = 1/2$ in Eq.~\eqref{eq:Psi1ex} corresponding to sites $l$ and $m$ at positions $\mb r$ and $\mb r'$, respectively, as a function of distance $|\mb r - \mb r'|$. Blue crosses correspond to $\mb r- \mb r'$ aligned along the $x$- or $y$-axis (both lie on top of each other), and green stars indicate the case of $\mb r- \mb r'$ aligned along the diagonal of both axes. In this plot, $\gamma$ has been calculated for a $2000 \times 2000$ lattice, and it decays as $\tfrac{1}{|\mb r- \mb r'|^{3.05}}$. The exponent converges to $3$ with increasing lattice size. This figure was reproduced from Ref.~\cite{cPEPS_long}.
\label{fig:corr}}
\end{figure}

The correlation functions of the PEPS defined via Eq.~\eqref{eq:Psi1ex} decay algebraically in real space, see Fig.~\ref{fig:corr}. This is most easily understood by considering the Fourier transform~\eqref{eq:G-param}. 
All $\hat d_i(\mb k)$ are continuous for all $\mb k$. 
However, the $\hat d_i(\mb k)$ have a non-analyticity at $\mb k = (0,0)$, where
the first derivatives of $\hat d_x$ and $\hat d_y$ are discontinuous.  For
instance, for $\lambda = 1/2$ in the example of Eq.~\eqref{eq:Psi1ex}, one
obtains
\begin{align}
\hat d_x(\mb k) &= -\frac{2 \sin(k_x) (1 - \cos(k_y))}{3 - 2 \cos(k_x) - 2\cos(k_y) + \cos(k_x) \cos(k_y)}, \label{eq:dx}\\
\hat d_y(\mb k) &= \frac{2 \sin(k_y) (1 - \cos(k_x))}{3 - 2 \cos(k_x) - 2\cos(k_y) + \cos(k_x) \cos(k_y)}, \label{eq:dy} \\
\hat d_z(\mb k) &= \frac{1 - 2 \cos(k_x) - 2 \cos(k_y) + 3 \cos(k_x) \cos(k_y)}{3 - 2 \cos(k_x) - 2\cos(k_y) + \cos(k_x) \cos(k_y)}. \label{eq:dz}
\end{align}
At $\mb k = (0,0)$ both the numerators and the common denominator are zero.
In Appendix~\ref{app:corr}, we show that due to this non-analyticity, correlations in real space decay like the inverse of the distance cubed (up to possible logarithmic corrections).

\paragraph{Frustration free Hamiltonian: fragility}

\begin{figure}[!ht]
\begin{center}
\includegraphics[width=0.5\textwidth]{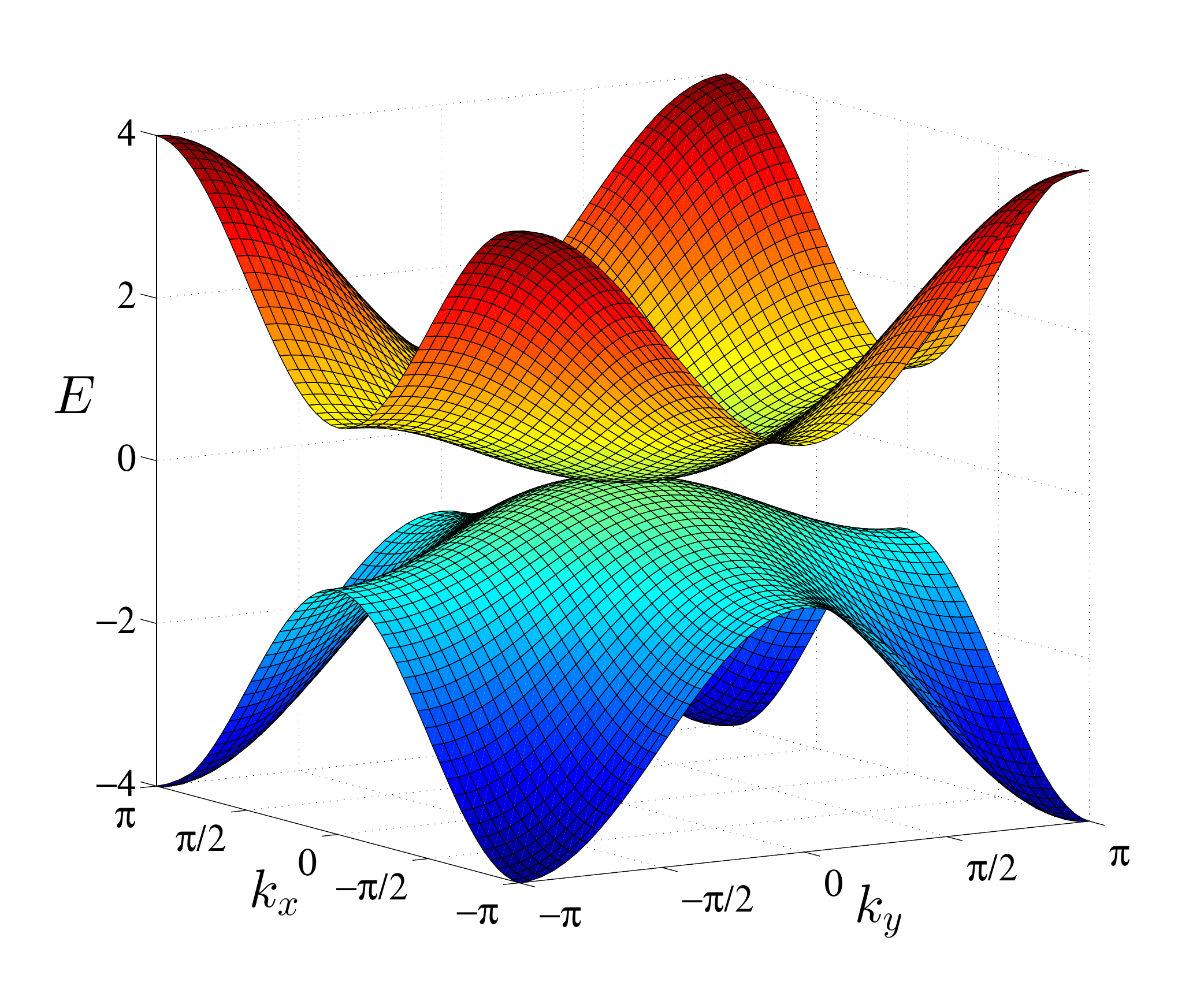}
\end{center}
\caption{Single-particle energy spectrum of the frustration free parent
Hamiltonian $\mc H_\mr{ff}$ of the GFPEPS defined via
Eq.~\eqref{eq:Psi1ex} for $\lambda = 1/2$. The band-touching point is at
$\mb k = (0,0)$. This figure was reproduced from Ref.~\cite{cPEPS_long}.
\label{fig:parent-Ham}}
\end{figure}

The frustration free parent Hamiltonian for this model is obtained by
explicitly calculating the state $\Psi_{2,2}$ obtained when four $\Psi_1$
on a $2 \times 2$ plaquette are concatenated without closing the
boundaries in horizontal or vertical direction.  Thereafter, one
calculates the fermionic operator $a_{\Box}$, acting only on the physical level, which annihilates $\Psi_{2,2}$, $a_{\Box} |\Psi_{2,2}\rangle = 0$
(it turns out that exactly one such operator exists for any $\lambda \in (0,1)$). This can be done conveniently in the CM formalism. The parent Hamiltonian, $\mc H_\mr{ff}$, can then be obtained by setting 
\be
h_\mb{r}(\lambda) \propto a_{\Box,\mb r}^\dg(\lambda) \, a_{\Box,\mb r}(\lambda) \label{eq:Ham_ex}
\ee 
in Eq.~\eqref{eq:HPEPS}.
For $\lambda=1/2$, for instance, we have 
\begin{align} 
a_{\Box} &= e_{a,1,1} (2 + i) + e_{b,1,1} - e_{a,1,2} (1 + 2i) + i e_{b,1,2} - e_{a,2,1} \notag \\
&+ e_{b,2,1} (-2 + i) + i e_{a,2,2} + e_{b,2,2} (1 - 2i), \label{eq:annihilation_ex}
\end{align}
where $e_{a,x,y}$ denotes the first physical Majorana mode located at the site with coordinates $(x,y)$ and $e_{b,x,y}$ the second one. The single-particle spectrum for that case is displayed in Fig.~\ref{fig:parent-Ham}. Note that there is a band-touching point at $\mb k = (0,0)$, and thus, this Hamiltonian is gapless, i.e., it has a continuous spectrum above the ground state energy. Furthermore, the ground state energy is exactly two-fold degenerate for finite systems, and in the thermodynamic limit it possesses a continuous spectrum right on top.

\begin{figure}[!ht]
\begin{center}
\includegraphics[width=0.4\textwidth]{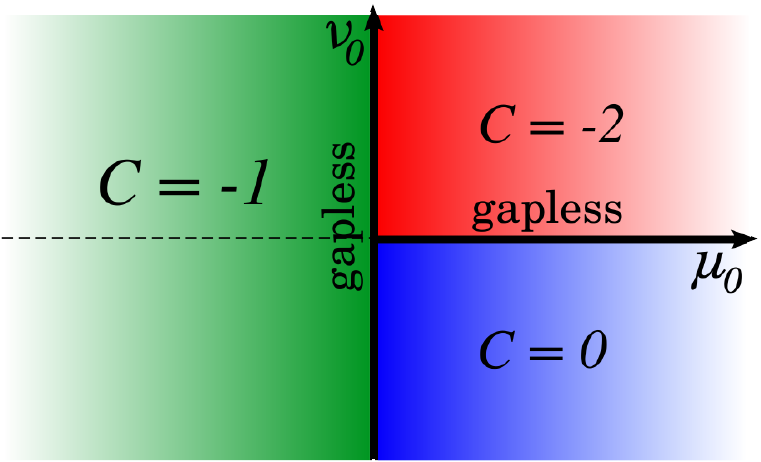}
\end{center}
\caption{Phase diagram of the perturbed Hamiltonian $\tilde{\mc H}_\mr{ff}(\lambda, \mu_0,\nu_0)$ (see text) for $\mu_0$, $\nu_0$ close to zero and $\lambda \in (0,1)$ arbitrary. The vertical gapless line corresponds to a quadratic band touching, whereas the horizontal gapless line ($\mu_0 > 0$) corresponds to four Dirac points. 
All other points in the phase diagram are gapped with the shown Chern numbers. This figure was reproduced from Ref.~\cite{cPEPS_long}.
\label{fig:phases}}
\end{figure}

The frustration free Hamiltonian $\mc H_\mr{ff}$ does not have a protected chiral edge mode, as it is gapless in the bulk: Let us add a translationally invariant perturbation [with variable GFPEPS parameter $\lambda \in (0,1)$],
\begin{align}
&\tilde{\mc H}_\mr{ff}(\lambda,\mu_0,\nu_0) = H_\mr{ff}(\lambda) - \tfrac{i}{4}\sum_{x,y} [\mu_0 e_{a,x,y} e_{b,x,y} + \nu_0 (e_{a,x+1,y} e_{b,x,y} \notag \\ 
&- e_{a,x,y} e_{b,x+1,y} + e_{a,x,y+1} e_{b,x,y} - e_{a,x,y} e_{2,x,y+1})]
\end{align}
where $\mu_0, \nu_0 \in \mathbb{R}$. Note that only $\mu_0 = \nu_0 = 0$ corresponds to a GFPEPS ground state. After carrying out a Fourier transform, the Hamiltonian can be brought into the form 
\be 
\tilde{\mc{H}}_\mr{ff}(\lambda,\mu_0,\nu_0) = \sum_{i=x,y,z} \sum_k d'_i(\mb k) (a_{\mb k}^\dg, a_{-\mb k}) \sigma_i \left(\begin{smallmatrix}
a_{\mb k} \\
a_{-\mb k}^\dg
\end{smallmatrix}\right)
\ee 
with $\sigma_i$ the Pauli matrices, the Chern number can be calculated by inserting $\hat d_i'(\mb k) = \tfrac{d_i'(\mb k)}{| d'(\mb k)|}$ into~\eqref{eq:Chern_d}.
The Hamiltonian can be driven by infinitesimally small $\mu_0$ and $\nu_0$ to gapped phases with Chern number $C = 0$ (trivial), $C = -1$ or $C = -2$ as shown in Fig.~\ref{fig:phases}. This phase diagram does not depend on the parameter $\lambda$ as long as $|\mu_0|$ and $|\nu_0|$ are sufficiently small. 
Hence, with respect to the frustration free Hamiltonian, the states defined by Eq.~\eqref{eq:Psi1ex} describe
critical points at the transition between different topological phases with Chern number $C = -2$ and $C = -1$ and a topologically trivial phase ($C = 0$). 

We conclude that the frustration free Hamiltonian is gapless and thus not topologically protected. Instead, it is at the critical point between free fermionic topological phases with different Chern numbers. 

\paragraph{Flat band Hamiltonian: robustness}

Let us now consider the stability of the flat band Hamiltonian $\mc
H_{\mr{fb}}$ against perturbations.  First, we will show analytically that
the Hamiltonian is robust even against long ranged translationally
invariant perturbations; and second, we will demonstrate numerically the
stability against local disorder.  This shows that the Hamiltonian is
topologically protected and its Chern number is therefore a meaningful
quantity.

Let us first consider translational invariant perturbations, where we
assume that the perturbation decays faster than $1/r^3$ in real space (with
$r = | \mb r|$). Then, it can be shown (see, e.g., Ref.~\cite{Grafakos},
Proposition 3.2.12) that the perturbation $\mathcal H$ is differentiable
in Fourier space, and thus, the perturbed flat band Hamiltonian 
$\tilde{\mc{H}}_\mr{fb} = \mc{H}_\mr{fb} + \epsilon \mc H$ 
is differentiable as well.  This makes the Chern number a well-defined quantity for small $\epsilon$, see Eq.~\eqref{eq:Chern_d}.
Moreover, since the Fourier components of
$\mathcal H$ are uniformly bounded,
the gap of $\tilde{\mathcal H}_\mathrm{fb}$ stays open for sufficiently small
$\epsilon$. Hence, the bands of $\tilde{\mathcal H}_\mathrm{fb}$ are a
smooth function of $\epsilon$, and thus, the Chern number cannot change
under sufficiently small perturbations.

\begin{figure}[!ht]
\begin{center}
\includegraphics[width=0.55\textwidth]{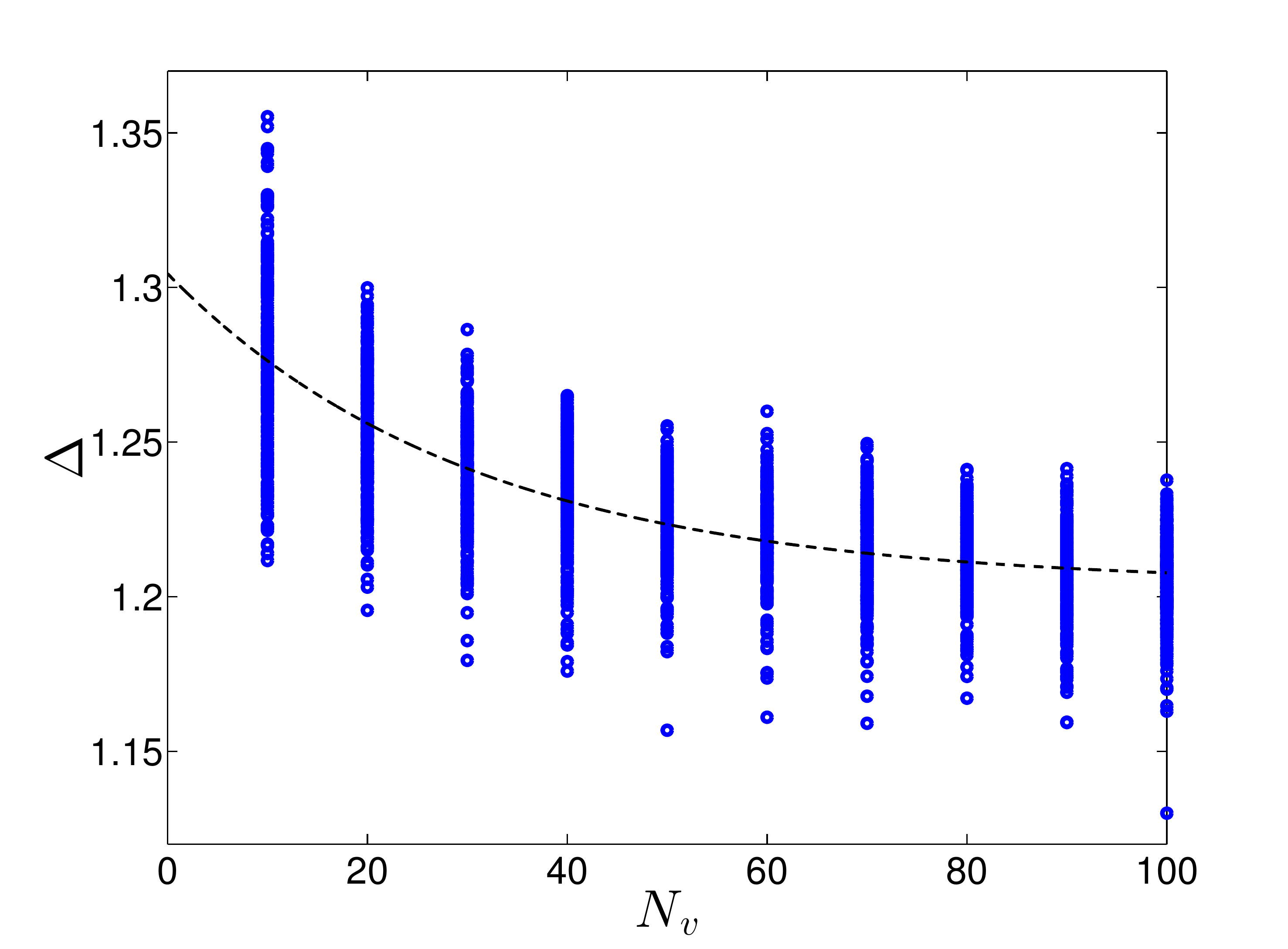}
\end{center}
\caption{Energy gap $\Delta$ of the flat band Hamiltonian $\mc H_\mr{fb}$ after the addition of disorder terms (see text). The Hamiltonian is defined on a torus of size $N_v \times N_v$ ($N_h = N_v$) and for each system size 225 random samples have been considered. The dashed line represents a fit with the function $f(N_v) = a \exp(-b N_v) + c$, which gives $a = 0.101\pm 0.005$, $b = 0.033\pm 0.004$ and $c = 1.204 \pm 0.003$ ($95 \%$ confidence intervals), i.e., the gap saturates at a value that is roughly $60 \%$ of the unperturbed gap. This figure was reproduced from Ref.~\cite{cPEPS_long}. \label{fig:disorder}}
\end{figure}

Let us now turn towards the stability of $\mathcal H_\mathrm{fb}$ against
random disorder, which we have verified numerically. To this end, we randomly
added local disorder terms $\sum_{\mb r} \mu_\mb{r} a_\mb{r}^\dg a_\mb{r}$ ($\mu_\mb{r} \in [-1,1]$) to the flat band Hamiltonian for $\lambda = {1}/{2}$ defined on
an $N_v \times N_v$ torus ($N_h = N_v$) as a function of its length $N_v$.
In Fig.~\ref{fig:disorder} we plot the energy gap obtained for 225 random realizations
 for each system size $N_v$. As can be
gathered from the figure, the gap stays non-vanishing in the thermodynamic
limit, indicating that it is topologically protected against disorder.

To summarize, the gap of the flat band Hamiltonian $\mc H_\mr{fb}$ is topologically protected against the addition of on-site disorder and (small) translationally invariant perturbations whose hoppings decay faster than the inverse of the distance cubed. Its Chern number is $-1$.

\subsubsection{Boundary and Edge Theories}\label{sec:boundary-edge}

In subsection~\ref{sec:bulk_boundary}, a formalism was introduced for spin PEPS to map the state in some region ${\cal R}$ to its boundary. This bulk-boundary correspondence associates to each PEPS a boundary Hamiltonian, ${\cal H}^{\rm b}$, that acts on the virtual particles. The Hamiltonian faithfully reflects the properties of the original PEPS. In particular, for the toric code \cite{toric_code}, or the resonating valence-bond states \cite{RVB_org}, that boundary Hamiltonian features their topological character \cite{transfer}. In this subsection, we review that theory for GFPEPS and show how one can determine ${\cal H}^{\rm b}$ for GFPEPS.

Chiral topological insulators and superconductors, on the other hand, are characterized by the presence of chiral edge modes, featuring  robustness against certain bulk perturbations. Here, we also analyze how those features are reflected in ${\cal H}^{\rm b}$, as well as the relation of that Hamiltonian with that found for the toric code.

\paragraph{Boundary Theories.}

Given the GFPEPS $\Phi$, let us take a region ${\mc R}$ of the lattice, trace all the degrees of freedom of the complementary region, $\bar{\mc R}$, and denote by $\rho_{\mc R}$ the resulting mixed state. As explained in subsection~\ref{sec:bulk_boundary}, $\rho_\mc{R}$ can be isometrically mapped onto a state of the virtual particles (or modes) that are at the boundary of the region ${\mc R}$. That is, there exists an isometry $\mc V_{\mc R}$, such that $\rho_\mc{R} = \mc V_{\cal R} \sigma_\mc{R} \mc V_{\cal R}^\dagger$, where $\sigma_\mc{R}$ is a mixed state defined on those virtual modes.

\begin{figure}[!ht]
\begin{center}
\includegraphics[width=0.4\textwidth]{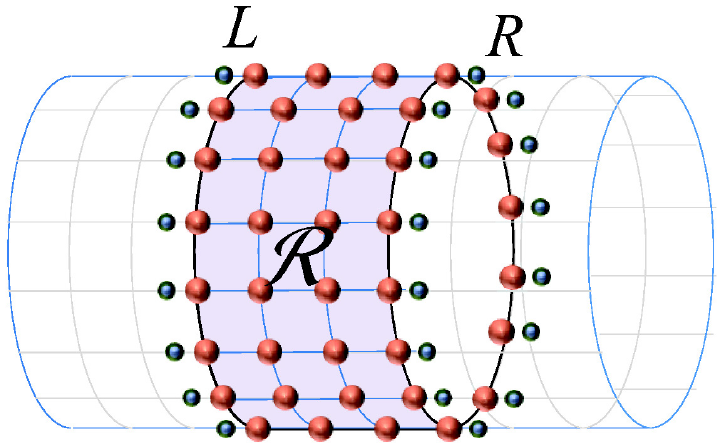}
\includegraphics[width=0.4\textwidth]{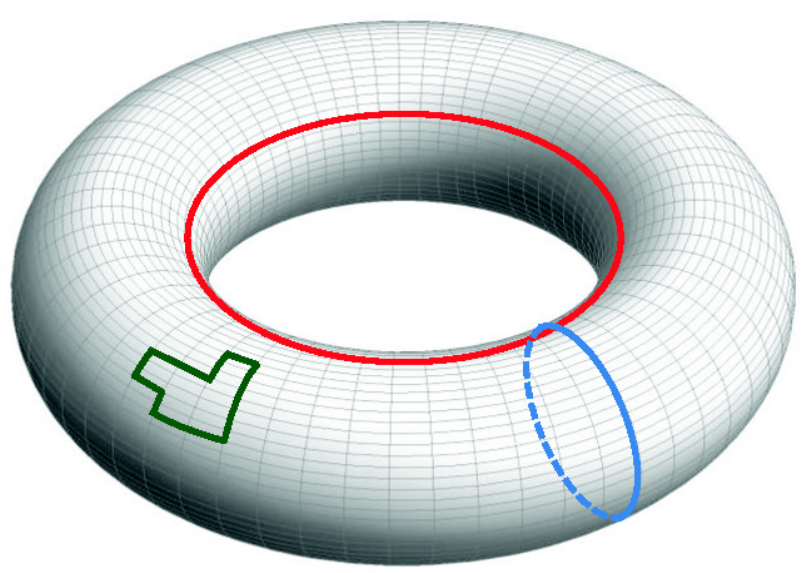}
\end{center}
\caption{Left: The state obtained after cutting out $N$ columns (region $\mc R$) from a translationally invariant GFPEPS is still translationally invariant in the vertical direction. Hence, it can be understood as being defined on a cylinder. The Majorana modes on the left and the right boundary (small blue balls) remain unpaired.
Right: Illustration of string operators. Those are defined as operators acting on the virtual Majorana modes lying on a closed string (such as the blue, red and green examples). The projection onto the final physical state $\Phi$ is carried out after applying one or more of those string operators. This figure was reproduced from Ref.~\cite{cPEPS_long}.}
\label{fig:cylinder}
\end{figure}

We take as region ${\mc R}$ a cylinder with $N$ columns, see Fig.~\ref{fig:cylinder}. There are shown the physical fermions (red) as well as the virtual Majorana modes (blue), as they appear in the construction explained above (Fig. \ref{Fig:PEPS}).

The state $\sigma_\mc{R}$ is Gaussian and is thus also characterized by a CM, which we will denote by $\Sigma_N$. In subsection~\ref{sec:Detailed-Analysis} we will show how to determine it in terms of $\gamma_1$. Here, we just quote the results. We can write
 \be
 \sigma_\mc{R} = \frac{1}{Z_N} e^{-{\cal H}^{\rm b}_{N}},
 \ee
where
 \be
 {\cal H}_N^{\rm b} = -\frac{i}{4} \sum_{j,k} (H_N^{\rm b})_{jk} c_j c_k, \label{eq:boundary-Ham}
 \ee
is the so-called \textit{boundary Hamiltonian}, with $c_j$ the Majorana operators acting on the left and right boundaries, and $H_N^{\rm b}$ a $2N_v\times 2N_v$ antisymmetric matrix, given by
 \be
 \label{eq:HN}
 H_N^{\rm b}= 2 \arctan ( \Sigma_N).
 \ee
The spectrum of ${\cal H}^{\rm b}_N$ coincides with the entanglement spectrum~\cite{ES_org}. Here, we are interested in the corresponding single-particle spectrum, i.e. that of $H^\mr{b}_N$.

Since ${\cal H}^{\rm b}_N$ is translationally invariant in the vertical direction, we can easily diagonalize it by using Fourier transformed Majorana modes. It is convenient to define
 \be
 \hat c_{k_y}= \frac{1}{\sqrt{N_v}} \sum_{y=1}^{N_v} e^{i k_y y} c_y,
 \ee
separately for the left and right virtual modes, so that ${\cal H}_N^{\rm b}$ displays a simple form in their terms. Here, the quasi-momentum is $k_y=2\pi n_y/N_v$, with $n_y=-{N_v}/{2}+1,\ldots,{N_v}/{2}$. Up to a factor of two, the operators $\hat c_{k_y}^\dagger = \hat c_{-k_y}$ fulfill canonical commutation relations for fermionic operators, $\{\hat c_{k_y},\hat c_{k_y'}^\dagger\}=2\delta_{k_y,k_y'}$, for $k_y\ne 0,\pi$. For $k_y=0,\pi$, they are Majorana operators (i.e., $\hat c_0^\dagger=\hat c_0$, and $\hat c_\pi^\dagger=\hat c_\pi$). This latter fact is crucial to understand the topological properties of the original state $\Phi$, as we will discuss in subsection~\ref{sec:full-chi1}.

The single-particle spectrum (dispersion relation, since we have translational invariance) will be labelled by $k_y$. For the GFPEPS determined by Eq.~\eqref{eq:Psi1ex}, for $\lambda \in (0,1)$ we will show that in the limit $N\to\infty$ one can write
 \be
 H_\infty^{\rm b} = \bigoplus_{k_y\ne 0,\pi} \left(\hat H_\infty^L(k_y) \oplus \hat H_\infty^R(k_y)\right) \oplus \hat H^{LR}_\infty(0) \oplus \hat H^{LR}_\infty(\pi), \label{eq:decouple-H}
 \ee
where $\hat H_\infty^{L}(k_y)$ and $\hat H_\infty^{R}(k_y)$ correspond to virtual fermionic modes on the left and right, respectively, which are decorrelated from each other. For $k_y=0$ and $k_y=\pi$, however, there is a single unpaired Majorana mode in each boundary. For the above family of chiral GFPEPS, the $k_y = 0$ Majorana modes
pair up, giving rise to an entangled state between the left and the right boundaries.

The Chern number $C$ is (up to a sign) given by the number of right-movers minus the number of left-movers on one of the boundaries. For the simple case considered in this subsection, with one Majorana bond, $|C|=0,1$. For GFPEPS with more Majorana bonds, one can build the boundary Hamiltonian in the same fashion, as we will show in the next subsection. In that case, the Chern number is determined ditto, but it may be larger than one.

\begin{figure}[!ht]
\centering
\includegraphics[width=0.49\textwidth]{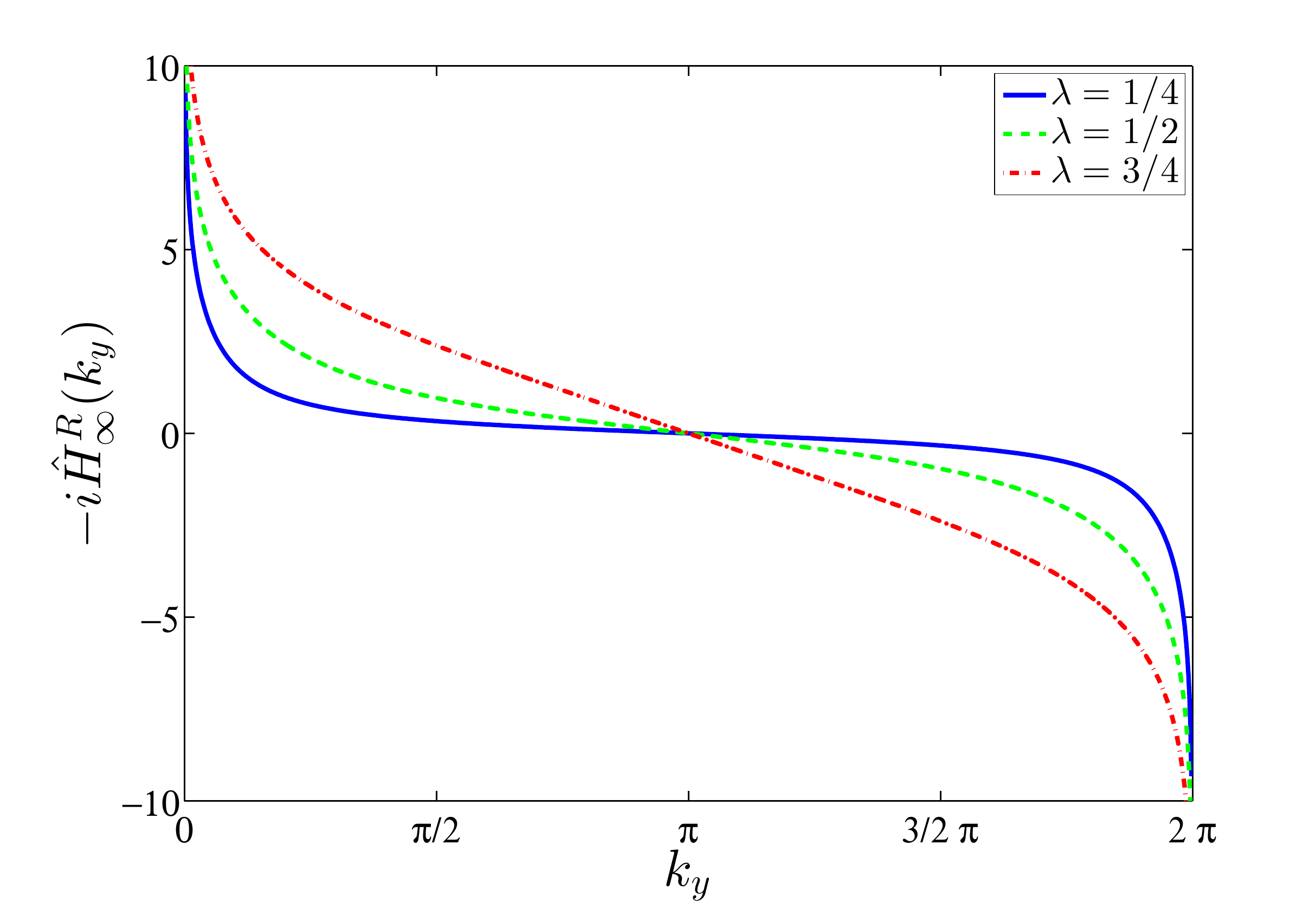}
\includegraphics[width=0.49\textwidth]{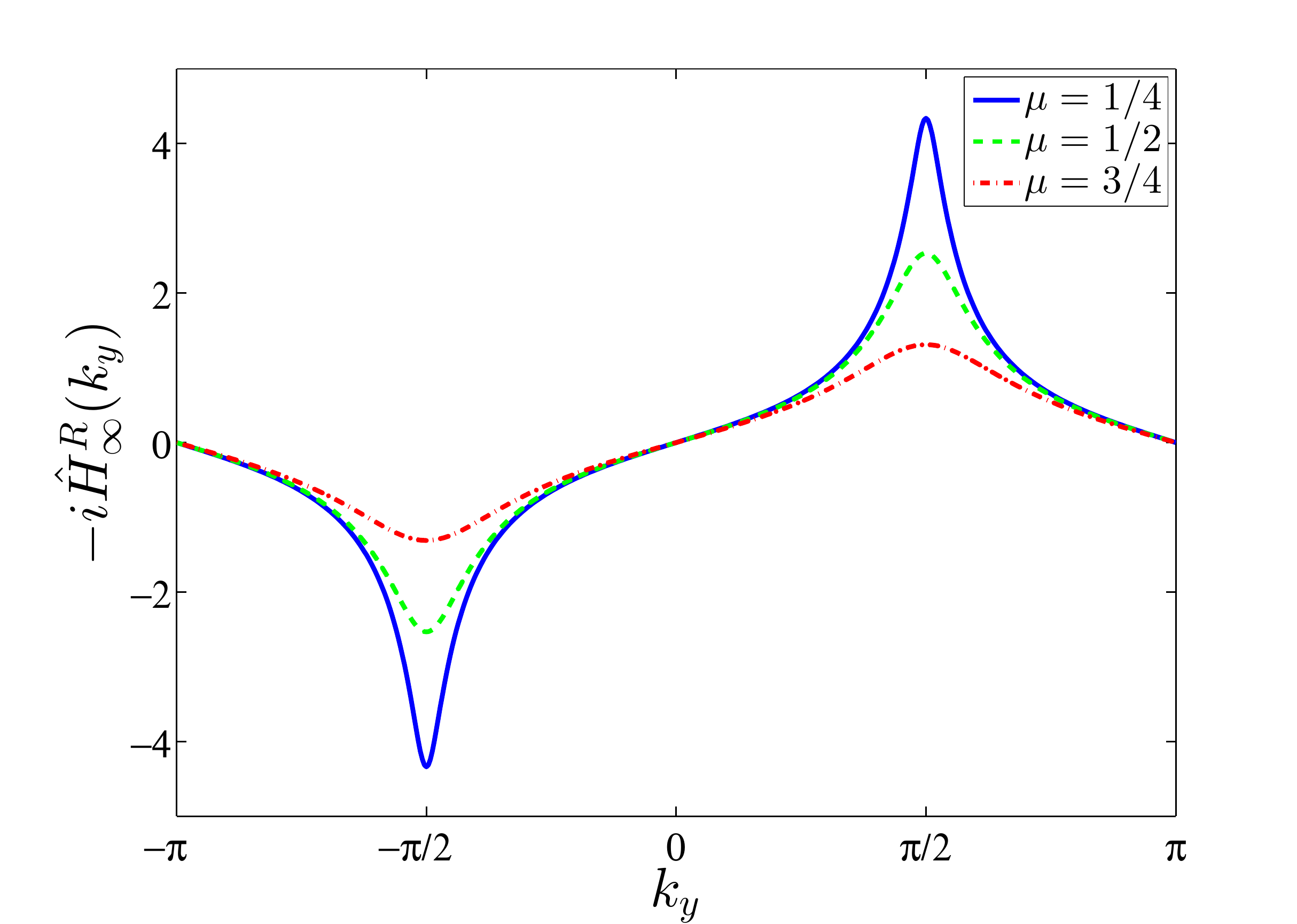}
\caption{
Left: Dispersion relation corresponding to the right boundary Hamiltonian of the chiral state defined via Eq.~\eqref{eq:Psi1ex}. 
We plot $-i \hat H^R_\infty(k_y)$ (which is a $1 \times 1$ matrix), for $\lambda = {1}/{4}$ (blue solid line), $\lambda = {1}/{2}$ (green dashed line) and $\lambda = {3}/{4}$ (red dash-dotted line) and $N \rightarrow \infty$. For convenience, we have plotted it for $k_y\in[0,2\pi)$. Note the divergence at $k_y = 0$, where there is a maximally entangled virtual Majorana pair between the left and the right boundary. The lines cross
the Fermi level from above at $k_y = \pm \pi$ corresponding to Chern number $C = -1$. 
Right: 
Dispersion relation at the right boundary for the non-chiral
state defined via Eq.~\eqref{eq:D-chi0}. We plot $-i \hat H^R_\infty(k_y)$ for $\mu = {1}/{4}$ (blue solid line),
$\mu =
{1}/{2}$ (green dashed line) and $\mu = {3}/{4}$ (red dash-dotted line)
and $N \rightarrow \infty$. It crosses the Fermi level twice with slopes
of different signs, hence $C = 0$. This figure was reproduced from Ref.~\cite{cPEPS_long}.
\label{Gamma_R1}}
\end{figure}

In Fig.~\ref{Gamma_R1} we plot the single-particle dispersion relation of
the right boundary as a function of $k_y$, for the state generated by
(\ref{eq:Psi1ex}) for different values of $\lambda$ and $N\to\infty$ (we
will provide an analytic formula for that limit in
subsection~\ref{sec:full-chi1}). It displays chirality, and the Chern number is
$-1$. The mode at $k_y=\pi$ has zero ``energy", indicating that the state
of the left and right Majorana modes with such a momentum is
completely mixed. If we construct a fermionic operator using those
two modes, the boundary state $\sigma_\mc{R}$ at momentum $\pi$ has
infinite temperature, and thus is an equal mixture of zero and one
occupation. If we do the same with the modes at $k_y=0$, the opposite is
true, namely they are in a pure state (the vacuum mode of the fermionic mode
built out of the two Majorana modes from the left and the right). Thus, as
anticipated, the left and right boundaries are in an entangled state,
which reflects the topological properties of the state. In
subsection~\ref{sec:Detailed-Analysis} we will show that all the features
displayed by this example are intimately related.

As a second example, we take a state that does not display any topological features. Its explicit form is given in subsection~\ref{sec:C0-nontrivial}. The dispersion relation for the right boundary is shown in Fig.~\ref{Gamma_R1}. Since the energy band of the boundary Hamiltonian does not connect the valence and conduction band for any value of its parameter $\mu$, the Chern number is zero. Furthermore, both at $k_y=0$ and at $k_y =\pi$ the ``energy'' vanishes, showing that the right and left boundaries are unentangled.

In subsection~\ref{sec:examples}, we present further examples: We give an example of a GFPEPS displaying $C=2$. We also investigate an example of a Chern insulator as well as the non-chiral state introduced in Ref.~\cite{fPEPS_Kraus}.

\paragraph{Edge theories.}

The definition of the boundary theory used above may look a bit artificial; the Hamiltonian ${\cal H}_N^{\rm b}$ does not generate any dynamics, but is just the logarithm of the density operator, and thus comes from the interpretation of the boundary operator as a Gibbs state. However, it is well known~\cite{edge_entanglement} that for free fermionic (i.e., Gaussian) states, its spectrum is intimately related to the one of another Hamiltonian that indeed generates the dynamics at the physical edges of the system in question. In the PEPS representation, there is a way of constructing such an \textit{edge Hamiltonian}~\cite{edge_theories}, which we review here and we explicitly illustrate such a relation.

Let us consider the flat band Hamiltonian (\ref{eq:flat}), but in the case of a cylinder with open boundary conditions. For that, we restrict the sum in Eq.~\eqref{eq:flat} to the modes that correspond to region ${\cal R}$ (the cylinder in Fig.~\ref{fig:cylinder}), and denote by ${\cal H}_{\cal R}$ the corresponding Hamiltonian. The state $\Phi_N$ (see Fig. \ref{Fig:PEPS}f) has extra (virtual) modes, which we can project onto an arbitrary state, say $\phi_v$. 
There is a subspace spanned by all the states resulting from this construction with an energy lying within the gap. By choosing a set of linearly independent vectors $\phi_v$, and orthonormalizing the resulting state, we can project ${\cal H}_{\cal R}$ onto that subspace. This is precisely the procedure given in Ref.~\cite{edge_theories}, and the resulting Hamiltonian, which has as many degrees of freedom as there are virtual Majorana modes, is the edge Hamiltonian, ${\cal H}_N^{\rm e}$. We now write
 \be
 \label{eq:Hedge}
 {\cal H}_N^{\rm e} = -\frac{i}{4} \sum_{l,k} (H_N^{\rm e})_{lk} c_l c_k,
 \ee
and in subsection~\ref{sec:edge} we show that one obtains that $H_N^{\rm e}=\Sigma_N$. Thus, up to a scale transformation (cf. Eq.~\eqref{eq:HN}), we see that the edge Hamiltonian is nothing but the boundary Hamiltonian, whenever we take the flat band Hamiltonian as the Hamiltonian of our GFPEPS. This agrees with the statement of Ref.~\cite{edge_entanglement}, and indicates that our results on the boundary Hamiltonian can be translated to the edge Hamiltonian constructed in the outlined way.

\subsubsection{Symmetries, degeneracy, and topological entanglement entropy}\label{sec:symm-top}


Here, we show how the topological properties of GFPEPS are reflected in the symmetries of the fiducial state $\Psi_1$ and of the intermediate state contracted on a cylinder with open legs, cf. subsections~\ref{sec:PEPS_sym} and~\ref{sec:bulk_boundary}, respectively. It turns out that our chiral topological models possess a symmetry in $\Psi_1$, which is inherited for larger regions and that gives rise to the properties (i) the zero R\'enyi entropy is not the logarithm of the dimension of the Hilbert space of the virtual particles and (ii) there is a non-local constraint on the boundary and edge Hamiltonian, see subsection~\ref{sec:bulk_boundary}. 
Besides that, the parent Hamiltonian ${\cal H}_{\rm ff}$ is degenerate on the torus, and the different ground states can be obtained by attaching to the virtual modes string operators around the torus. The strings can be deformed, without changing the state. However, there are some differences as compared to $\mc G$-injective spin-PEPS, too. First of all, the von Neumann entropy of $\sigma_\mc{R}$ does not display a universal correction, which we attribute to the long range properties of the flat band Hamiltonian $\mc H_\mr{fb}$ of the state $\Phi$ (cf. Ref.~\cite{cPEPS_Read}). For the same reason, the hoppings in ${\cal H}^\mr{b}_N$ decay according to a power law. Furthermore, the ground state subspace of the parent Hamiltonian ${\cal H}_{\rm ff}$ is doubly degenerate on the torus, and some topologically inequivalent string configurations give rise to the same state.

Let us consider any region ${\cal R}$, and denote by $\Psi_{\cal R}$  the state obtained by projecting all the virtual modes within region ${\cal R}$ onto the states generated by $\omega$ and $\omega'$, as they appear in the PEPS construction. We arrive at a state of the physical modes in ${\cal R}$ and the virtual ones sitting at the boundary of $\mc R$. For instance, if we take as ${\cal R}$ a cylinder with $N$ columns, the state is $\Psi_{\cal R} =\Phi_N$ (see Fig.~\ref{fig:cylinder}). In general, we can write
 \be
 |\Phi\rangle = \langle \omega_{\partial \mc R, \partial \bar{\mc R}} |\Psi_{\cal R},\Psi_{\bar{\cal R}}\rangle,
 \ee
where $\omega_{\partial \mc R, \partial \bar{\mc R}}$ projects out all the virtual modes at the boundaries of ${\cal R}$ and its complement $\bar{\cal R}$.

If a contour ${\cal C}$ encloses a connected region ${\cal R}$, for chiral GFPEPS with one Majorana bond, there is a fermionic operator $d_\mc{C}$ such that
 \be
 \label{eq:dC}
 d_{\cal C} |\Psi_{\cal R}\rangle =0.
 \ee
For any contour, we will say that the state
 \be
 |\Phi_{\cal C}\rangle = \langle \omega_{\partial \mc R, \partial \bar{\mc R}} | d_{\cal C}| \Psi_{\cal R},\Psi_{\bar{\cal R}}\rangle,
 \ee
is a GFPEPS with a string along the contour ${\cal C}$. In subsection~\ref{sec:symmetry-GS} we will show how this string operator can be deformed continuously for a chiral GFPEPS without changing the state we are building. However, if a contour wraps up around one of the sections of the torus, we cannot get rid of it by continuous deformations.

Let us denote by ${\cal C}_{h,v}$ contours wrapping the torus horizontally and vertically, respectively.
We show in subsection~\ref{sec:symmetry-GS}
that if we build the family of chiral GFPEPS starting out from $\Psi_1$ according to Eq.~\eqref{eq:Psi1ex}, we obtain $\Phi=0$ after the last projection. However, the states obtained if we add a certain string along any of those contours coincide, $\Phi_{{\cal C}_h} \propto \Phi_{{\cal C}_v}$, and in the following that is the state that we will consider. We also show that if we insert string operators along the two contours $\mc{C}_h$ and $\mc{C}_v$,
the state $\Phi_{{\cal C}_h,{\cal C}_v}$ we obtain is orthogonal to the previous one, but it is also a ground state of ${\mc H}_{\rm ff}$.

The frustration free Hamiltonian has certainly very interesting properties, although we cannot determine them unambiguously given our results. It is
not only at a quantum phase transition point between free fermionic (gapped) phases with Chern numbers $C = 0, -1$ and $-2$, but it furthermore carries features of states described by PEPS with long range topological order: Its ground state manifold is obtained by inserting strings along the non-trivial loops of the torus. Hence, our results also allow to interpret the local parent Hamiltonian as being at the edge of a topologically ordered interacting phase.

The existence of the operators $d_{\cal C}$ in Eq.~\eqref{eq:dC} for any
simply connected region $\mc R$ has another important consequence. It
follows that we can build a unitary operator $U= \Id-2 d_{\mc C}^\dagger
d_{\mc C}$ such that Eq.~\eqref{eq:boundary_Ginj} is fulfilled for the boundary operator.
As a consequence, we also have Eq.~\eqref{eq:project_Ginj} with $P = \Id - d_{\mc
C}^\dagger d_{\mc C}$. Note that in our case $\mc G = \mathbb{Z}_2$ is
represented by $\{\Id, U\}$. Thus, we conclude that the properties of subsection~\ref{sec:bulk_boundary} (i) (topological correction to zero R\'e{}nyi entropy)
and (ii) (non-local constraint on the boundary and edge Hamiltonian) are
fulfilled as in the standard PEPS case. Note that if $\mc R$ lies on a
cylinder as in
Fig.~\ref{fig:cylinder}, we can also give the interpretation that, as in
the case of a Majorana chain~\cite{Kitaev_chain}, there are two Majorana modes at the
boundaries building a fermionic mode in the (pure) vacuum state. As a
consequence, we can write $\sigma_\mc{R}$ for the cylinder as in Eq.~\eqref{eq:project_Ginj},
where $P$ projects onto the subspace where that mode
is in the vacuum.

\begin{figure}
\centering
\includegraphics[width=0.58\textwidth]{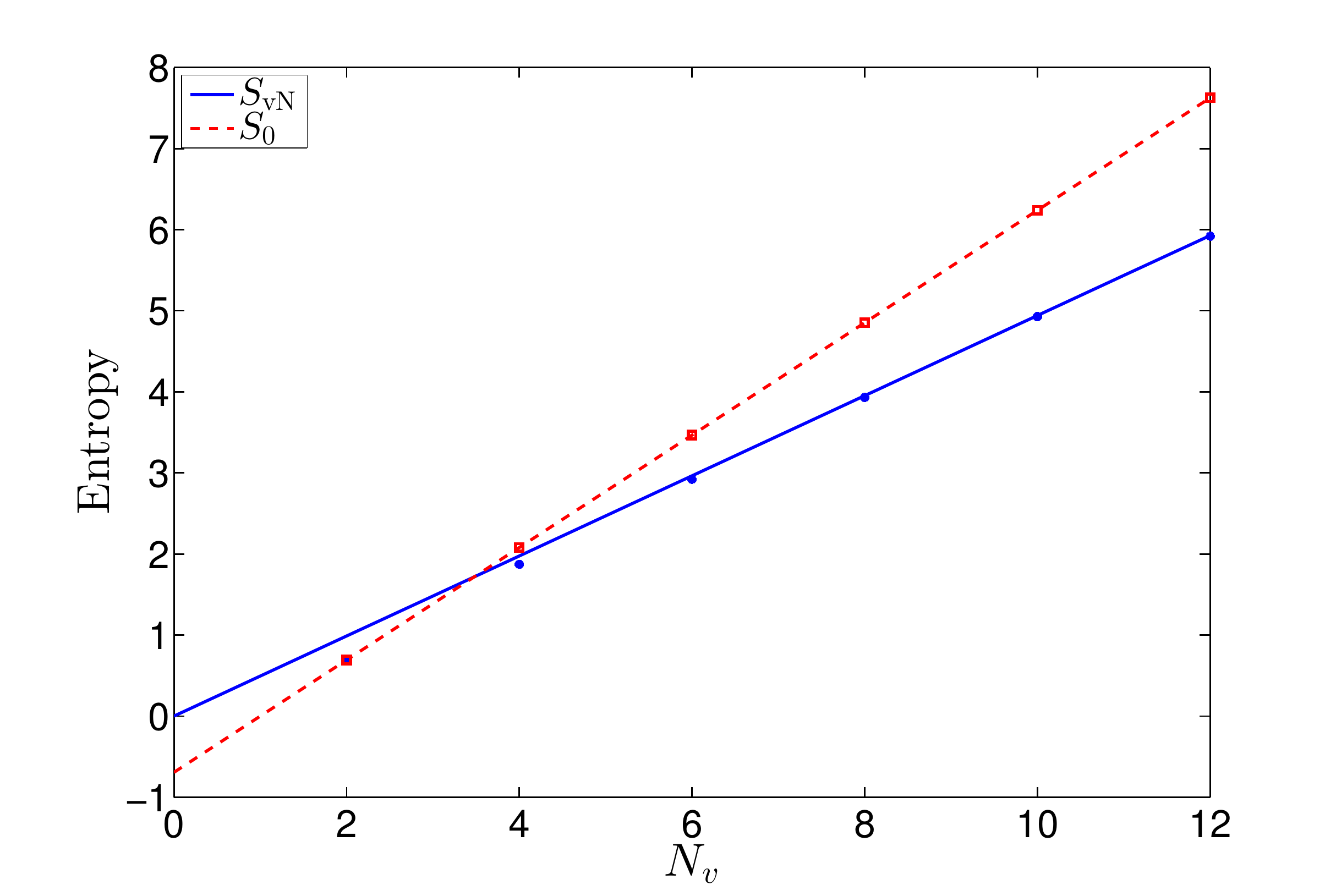}
\caption{Von Neumann entropy $S_\mr{vN}$ (blue circles) and zero R\'enyi entropy $S_0$ (red squares) versus the length of the cylinder in vertical direction, $N_v$, for the example given in Eq.~\eqref{eq:Psi1ex} for $\lambda = 1/2$. The lines indicate linear fits, which have been done for 31 data points distributed uniformly between $N_v = 4000$ and $N_v = 4600$. They yield $S_\mr{vN} = 0.49401 N_v - 2.0 \cdot 10^{-7}$ (the constant converges to zero for intervals containing increasing $N_v$'s)  and $S_0 = \mathrm{ln}(2) N_v - \mathrm{ln}(2)$. This figure was reproduced from Ref.~\cite{cPEPS_long}.}
\label{fig:entropy}
\end{figure}

In addition to the zero R\'e{}nyi entropy $S_0(N_v)$, we have also
numerically computed the von Neumann entropy $S_{\mathrm{vN}}(N_v)$ for
the example given in Eq.~\eqref{eq:Psi1ex} for $\lambda=1/2$. Both are shown in Fig.~\ref{fig:entropy} as
a function of $N_v$: While the zero R\'e{}nyi entropy clearly shows a
topological correction of $\ln(2)$, similar to the toric code model, the
von Neumann entropy does not exhibit such a correction. As we prove in
Appendix~\ref{app:euler-maclaurin}, this follows from the fact
that $S_{\mathrm{vN}}(N_v)$ forms a discrete approximation to the integral
over the modewise entropy, which is sufficiently smooth in $k_y$ to ensure
fast convergence. The same happens for all R\'enyi entropies $S_\alpha$ except for $\alpha = 0$. This is 
consistent with the result of Ref.~\cite{topological_Renyi} (where, however, only non-chiral topological
states have been considered).

In order to further investigate the topological properties of our model,
we have also computed the momentum polarization (see subsection~\ref{sec:chiral_physical}), 
which is given by the conformal dimension and
chiral central charge, cf., Eq.~\eqref{eq:mom-pol}. Due to its definition via the overlap $\lambda(N_v)$ of the original ground state with itself twisted on half of the cylinder (see Eq.~\eqref{eq:twist_half}), it can be rephrased in terms of the (many-body) entanglement spectrum $\varepsilon_\ell$ of the left half, which implies that
in the framework of PEPS, it can be naturally evaluated on the virtual
boundary between the two parts of the system (note that there is only one topological sector in our case, since it is free fermionic). In particular, for GFPEPS it
can be expressed as a function of the (single-particle) spectrum of the
boundary Hamiltonian $\mc H^{\mr b}_N$, as shown in Fig.~\ref{Gamma_R1}.
In Ref.~\cite{mom-pol}, it has been shown that (for systems with CFT edges)
$\lambda(N_v)=\exp(-\alpha' N_v-2\pi i \tau/N_v+\dots)$, with a non-universal
$\alpha'$, and a universal $\tau$, which carries information about the
topological properties of the system.  In
appendix~\ref{app:euler-maclaurin}, we prove that for GFPEPS, $\lambda(N_v)$
exactly follows the above behavior, and $\tau$ is indeed universal:
Remarkably, it only depends on whether the boundary Hamiltonian exhibits a
divergence, but not at all on its exact form. In particular, for our
example, we analytically obtain a $\tau$ which corresponds to a chiral
central charge of $c=1/2$, independent of $\lambda$, corresponding to the CFT of one real fermion.

Finally, an interesting behavior is also observed for the boundary Hamiltonian,
Eq.~\eqref{eq:boundary-Ham}, for $N \rightarrow \infty$. On the right
boundary we perform the Fourier transformation to position space
$[H_\infty^R]_{n,m}$. Then, for $y \gg 1$, $|[H_\infty^{R}]_{n,n+y}|
\propto \log(y)/y + \mathcal{O}(1/y)$, which is demonstrated in subsection~\ref{sec:bnd-theories-on-torus} .

\begin{figure}
\centering
\includegraphics[width=0.5\textwidth]{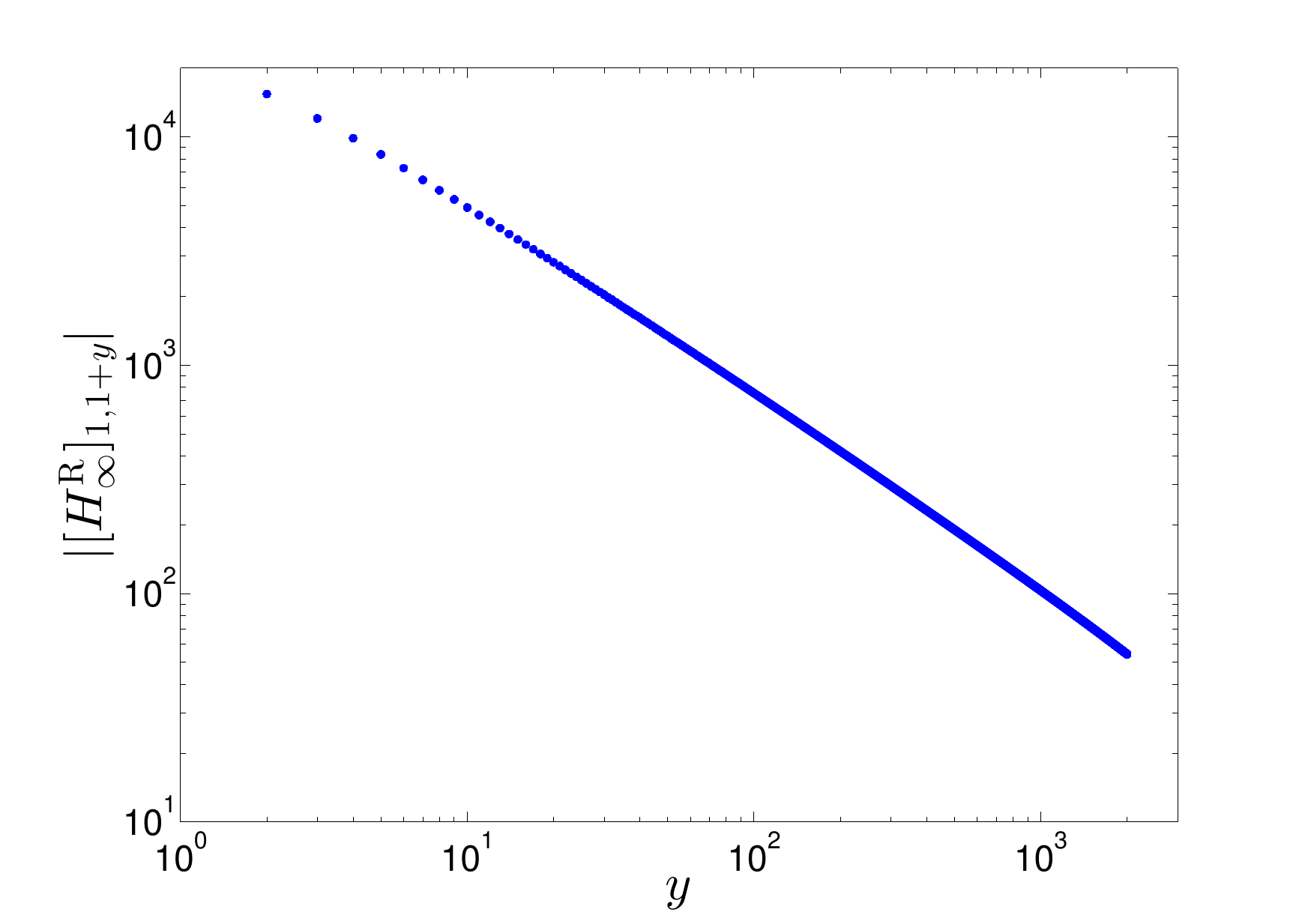}
\caption{Hopping amplitudes $|[H_\infty^{R}]_{1,1+y}|$ of the boundary Hamiltonian of the example given in Eq.~\eqref{eq:Psi1ex} for $\lambda = 1/2$ versus $y$. For large $y$ the curve has an inclination of $-1$ (on the log-log scale) indicating a decay as $1/y$, consistent with the fact that the logarithmic correction gets less important. The plot has been generated for $N \rightarrow \infty$ and $N_v = 2 \cdot 10^4$ sites in vertical direction. This figure was reproduced from Ref.~\cite{cPEPS_long}.}
\label{fig:Ham}
\end{figure}

Thus, the decay of the matrix elements of the boundary Hamiltonian with distance is not exponential, as is the case for gapped spin-phases, but follows a power law. We
plot the hopping amplitudes $|[H_\infty^{R}]_{1,1+y}|$ of the above chiral
family for $\lambda = 1/2$ in Fig.~\ref{fig:Ham}.

\subsection{Detailed analysis}\label{sec:Detailed-Analysis}

In the following, we provide a detailed derivation of the boundary and edge
theories for GFPEPS. We start in subsection~\ref{sec:construct-GFPEPS} by
formally introducing GFPEPS and then provide the derivation of boundary
theories (\ref{sec:boundary}) and edge theories (\ref{sec:edge}) for GFPEPS.

\subsubsection{GFPEPS}\label{sec:construct-GFPEPS}

The alternative construction of GFPEPS given in subsection~\ref{Sec:Intro-GFPEPS} can be
defined more generally for $f$ physical fermionic modes per site and
$\chi$ Majorana bonds between them. We start again with an $N_v \times
N_h$ lattice, now with $\chi$ left, right, up and down Majorana modes per
site, $c_{\mb r,L,\kappa}, c_{\mb r,R,\kappa}, c_{\mb r,U,\kappa}$ and
$c_{\mb r,D,\kappa}$, respectively, where $\kappa = 1, \ldots, \chi$ is the
index of the Majorana bonds. At each site $\mb r$, they are jointly with the
physical modes in a Gaussian state as in Eq.~\eqref{eq:Gaussian-psi1}. The
procedure to construct the GFPEPS is the same, except that there are
now $\chi$ virtual bonds between any two neighboring sites, i.e., here we have
to set 
\begin{subequations}
 \label{eq:omegas}
 \bea
 \omega_{\mb r, \mb r + \hat y} &=& \frac{1}{2^\chi} \prod_{\kappa=1}^\chi (1 + i c_{\mb r,D,\kappa} c_{\mb r + \hat y,U,\kappa}),\\
 \omega'_{\mb r, \mb r + \hat x} &=& \frac{1}{2^\chi} \prod_{\kappa=1}^\chi (1 + i c_{\mb r,R,\kappa} c_{\mb r + \hat x,L,\kappa}),
 \eea
 \end{subequations}
for the vertical and horizontal bonds, respectively. We will again denote
by $\langle \omega_{\mb r, \mb r + \hat y}|$ ($\langle \omega_{\mb r, \mb r + \hat x}'|$) the map which applies $\omega_{\mb r, \mb r + \hat y}$ ($\omega_{\mb r, \mb r + \hat x}'$) and discards the corresponding virtual
modes.  For simplicity, in the following, we call the states generated
by the operators~\eqref{eq:omegas} out of the vacuum {\em maximally
entangled states}.  The remaining procedure of how to concatenate them is
the same as in subsection \ref{Sec:Intro-GFPEPS}, cf.~also Fig.~\ref{Fig:PEPS}.

In this scenario, the CM is likewise given by Eq.~\eqref{eq:gamma1}, just that
the blocks $A$, $B$, and $D$ now have sizes $2 f \times 2 f$, $2f
\times 4 \chi$, and $4 \chi \times 4 \chi$, respectively. We are
interested in how to determine the CMs of the different states $\Psi_{N_v}$,
$\Phi_N$, and $\Phi$ involved in the construction of the GFPEPS.
It is based on two operations (see Fig.~\ref{Fig:PEPS}): \emph{(i)} building
the state of $l+m$ modes out of two states of $l$ and $m$ modes, respectively, i.e.,
taking tensor products; \emph{(ii)} projecting some of the modes onto some
state (given by $\omega$ and/or $\omega'$). Apart from that, we
will also extensively use: \emph{(iii)} tracing out some of the modes.

In terms of the CM, those operations are performed as follows
\cite{Gaussian_Bravyi}. \emph{(i)---joining two systems}: the resulting CM is
a $2\times2$ block diagonal matrix, where the two diagonal  
blocks are given by the CM of the state of the $l$ and $m$ modes,
respectively. 
The operation
\emph{(ii)---projecting out some of the modes}, is slightly more elaborate.
Let us consider an arbitrary state (pure or mixed) with CM $\gamma_1$ with
blocks $A,B,D$ [as in Eq.~\eqref{eq:gamma1}], and we want to project
the last modes (corresponding to matrix $D$) onto some other state
of CM $\omega$. The resulting CM is given by~\cite{Gaussian_Bravyi,fPEPS_Kraus}
\be
\gamma_1' = A + B(D+\omega^{-1})^{-1}B^\top. \label{eq:proj-ME}
\ee
Typically, we will have to project onto the states generated by (\ref{eq:omegas}). Their CM is very simple,
 \begin{equation}
    \label{eq:max-ent-def}
    \omega = \left(\begin{matrix}0&-\Id\\
    \Id&0\end{matrix}\right).
 \end{equation}
Finally, in the case of operation \emph{(iii)---tracing out some of the
modes}, one simply has to take the corresponding subblock of the CM. This
block is the CM of the reduced state. For instance, if one traces out the
physical degrees of freedom of the state described by the CM
\eqref{eq:gamma1}, one obtains a (generally mixed) state defined on the
virtual degrees of freedom with CM $D$.  Conversely, one can also build
the CM of a purification of a mixed state $D$, as
\be
 \left(\begin{array}{cc}
 -D&\sqrt{\Id + D^2}\\
  -\sqrt{\Id + D^2}&D
 \end{array}\right) \ .
 \ee
Operations \textit{(i)} and \textit{(ii)} can be used to build the CM of
the state $\Phi$ out of that of $\Psi_1$. In this subsection we will
extensively use all presented operations to construct the boundary and
edge states and Hamiltonians.

\subsubsection{Boundary Theories}\label{sec:boundary}

\paragraph{Boundary Theories in GFPEPS}

\begin{figure}
\centering\includegraphics[width=0.85\textwidth]{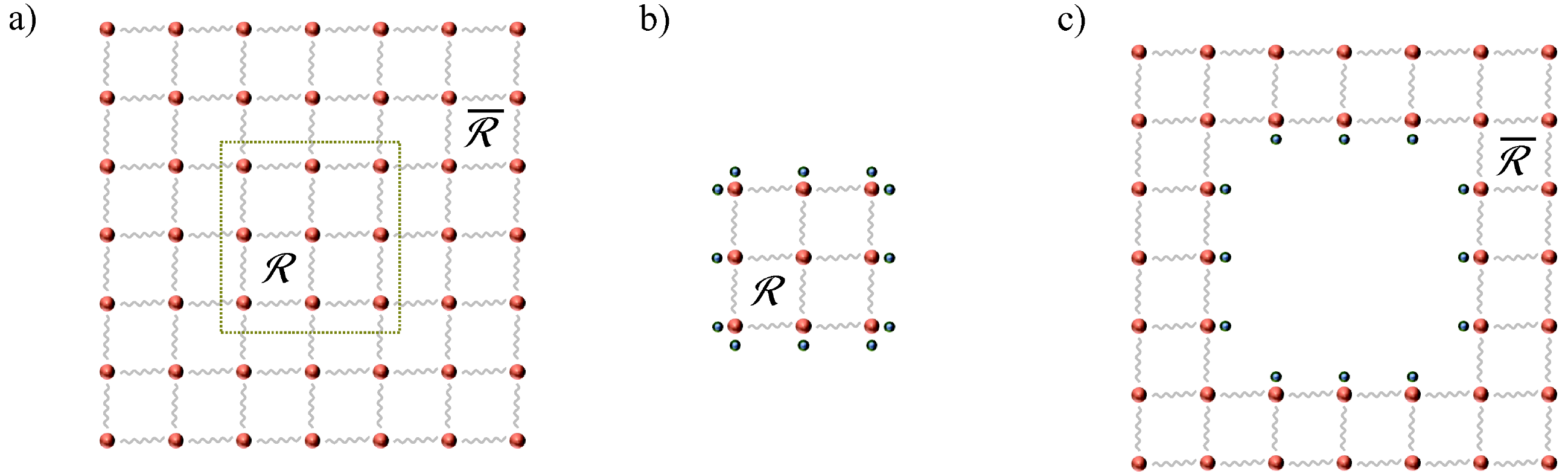}
\caption{(a) Partition of a lattice with a GFPEPS defined on it into a
region $\mc R$ and its complement $\bar{\mc R}$. 
The blue balls represent 
the virtual Majorana modes and the big red balls (connected
by wavy lines indicating the prior projection on maximally entangled pairs of virtual Majorana modes) the physical fermions.
 (b) After cutting the bonds as indicated in (a), the
region $\mc R$ has unpaired virtual indices at its boundary which are
collectively denoted by $\partial \mc R$. (c) The same is true for the
(inner) boundary of region $\bar{\mc R}$, whose virtual degrees of freedom
are denoted by $\partial \overline {\mc R}$. This figure was reproduced from Ref.~\cite{cPEPS_long}.\label{fig:peps-bipartition}}
\end{figure}

We will now show how to derive boundary theories in the
framework of fermionic Gaussian states by only using their description in
terms of CMs rather than the full state.  We construct a bipartition of the
PEPS $\Phi$ into two regions $\mc R$ and $\bar{\mc R}$
(Fig.~\ref{fig:peps-bipartition}) and are interested in the reduced state
$\rho_{\cal R}={\rm tr}_{\bar{\cal R}} (|\Phi\rangle\langle\Phi|)$. We
proceed as follows. First, we consider the states where all virtual
bonds \textit{within} those regions have been projected out, leaving only
virtual particles at the boundaries of those regions (which are denoted by $\partial
\mc R$ and $\partial \bar{\mc R}$, respectively) unpaired. Hence, we are
left with two states, which are defined on the physical degrees of freedom
of these regions plus the virtual degrees of freedom of the respective
boundaries (see Fig.~\ref{fig:peps-bipartition}b,c). We define their CMs as
\begin{equation} \label{eq:Gamma-R-Rbar}
\Gamma=\left(\begin{matrix}
L&F\\-F^\top & G\end{matrix}\right) \mbox{\ \ and\ \ }
\bar\Gamma=\left(\begin{matrix} \bar L&\bar F\\-\bar F^\top & \bar
G\end{matrix}\right)\ ,
\end{equation}
respectively, where the first (second) block corresponds to the physical
(virtual) degrees of freedom. The whole \mbox{GFPEPS} $\Phi$ could be obtained by
pairwise projecting their virtual degrees of freedom on maximally
entangled states, and thus, according to Eq.~\eqref{eq:proj-ME}, its CM is
 \be
 \gamma =
 \left(\begin{array}{cc}
 L&0\\
 0&\bar L
 \end{array}\right) +
 \left(\begin{array}{cc}
 F&0\\
 0&\bar F
 \end{array}\right)
 \left(\begin{array}{cc}
 G&\Id\\
 -\Id&\bar G
 \end{array}\right)^{-1}
 \left(\begin{array}{cc}
 F&0\\
 0&\bar F
 \end{array}\right)^\top. \label{eq:CM-GFPEPS}
 \ee
The CM of $\rho_\mc{R}$ is given by the (1,1) block of
Eq.~\eqref{eq:CM-GFPEPS}, that is
\be
\gamma_\mc{R} = L + F(G + \bar G^{-1})^{-1} F^\top. \label{eq:CM-R}
\ee

As explained in subsection~\ref{sec:boundary-edge}, we are interested in a state
$\sigma_{\partial \mc{R}}$ defined on the virtual degrees of freedom located on
$\partial {\cal R}$, which is isometric to $\rho_\mc{R}$. Naively, one could think
that its CM is given by the (2,2) block of $\Gamma$, i.e., $G$, which
corresponds to a reduced state acting on that boundary. However, this is
not the case in general, since the state described by the CM $G$ is usually not isometric
to $\rho_\mc{R}$. As outlined in Ref.~\cite{bulk_boundary},
$\sigma_{\partial \mc{R}}$ is given by a symmetrized version which takes into account
$\partial {\cal R}$ and $\partial \bar{\cal R}$. In fact, we can construct $\sigma_{\partial \mc{R}}$
by first finding
the appropriate purification of $\rho_{\mc{R}}$ and then tracing the
physical modes. We carry out this task in two steps. First, we 
conveniently rotate the basis of the physical Majorana modes in
region $\mc R$ and afterwards truncate the redundant degrees of freedom
(projection). Both taken together correspond to the application of an isometry on $\rho_\mc{R}$.

We start with an orthogonal basis change in the basis of physical Majorana
modes $\{e_l\}$ in region $\mc R$. The new ones are given by an orthogonal
matrix $M$,
\begin{equation}
\label{eq:Q-trafo}
e_m' = \sum_{l} M_{ml} e_l\ .
\end{equation}
This obviously does not change
the spectrum of $\rho_\mc{R}$.  By performing this basis change, the CM
$\Gamma$ gets modified to
 \[
 \Gamma' =
 \left(\begin{array}{cc}
 M&0\\
 0&\Id
 \end{array}\right)
 \left(\begin{array}{cc}
 L&F\\
 -F^\top&G
 \end{array}\right)
 \left(\begin{array}{cc}
 M^\top&0\\
 0&\Id
 \end{array}\right).
 \]
Note that this CM corresponds to a pure state, as $\Gamma$ does. We choose
$M$ in such a way that $\Gamma'$ decouples into a purification of the
virtual state and a trivial part on the remaining physical level. This is
always possible if the region $\mc R$ contains more degrees of freedom
than $\partial \mc R$ and can be done in practice by using a singular value
decomposition of $F$. Then,
\be
 \Gamma' = \left(\begin{array}{ccc}
 Z&0&0\\
 0&-G&\sqrt{\Id + G^2}\\
 0&-\sqrt{\Id + G^2}&G
 \end{array}\right), \label{eq:P-CM}
 \ee
where $Z$ is the CM of a pure state defined on the physical level and the
remaining non-trivial part of $\Gamma'$ corresponds to a purification of
$G$ (note that the first and second block correspond to the physical
degrees of freedom and only the third block to the virtual ones). We
discard the decoupled physical part and project the virtual degrees of
freedom (together with those of region $\bar{\mathcal R}$, given by $\bar
G$) on the maximally entangled state. This yields the relevant part of
Eq.~\eqref{eq:CM-R}, which is the CM of $\sigma_\mc{R}$,
\begin{equation}
    \label{eq:symmetrized-es}
 \Sigma_N = -G + \sqrt{\Id + G^2} (G+\bar G^{-1})^{-1}
    \sqrt{\Id + G^2}
 \ ,
\end{equation}
which is defined on the modes at the boundary. (We denote it by
$\Sigma_N$, since $\mc R$ will be typically taken to lie on a cylinder,
cf.~Fig.~\ref{fig:cylinder}, with $N$ columns. However,
Eq.~\eqref{eq:symmetrized-es} is true for any bipartition $\mathcal R$,
$\bar{\mathcal R}$.)

In order to obtain the boundary Hamiltonian $\mc H_N^{\rm b} =
-\tfrac{i}{4}\sum_{l,m} [H_N^{\rm b}]_{l,m} c_l c_m$, which reproduces the
entanglement spectrum, we can then use the relation
$H_N^\mathrm{b} = 2\arctan(\Sigma_N)$, Eq.~(\ref{eq:HN}). Note that for
$G=\bar G$, Eq.~(\ref{eq:symmetrized-es}) yields a trivial entanglement
spectrum, $\Sigma_N=0$, while for $G=-\bar G$, one finds $\Sigma_N = - 2 G
(\Id-G^2)^{-1}$, which gives a factor
of $\tfrac12$ in the entanglement temperature (i.e., the effective
strength of $H_N^{\rm b}$) with respect to $G$, $H_N^{\rm b} =
4\arctan(G)$, corresponding to the case $\sigma_L=\sigma_R^\top$
in Ref.~\cite{bulk_boundary}.

A crucial point to observe in the result for the boundary theory is that
$\Sigma_N$ only depends on the CMs $G$ and $\bar G$, which characterize
the reduced state of the virtual degrees of freedom at the boundaries of
$\mathcal R$ and $\bar{\mathcal R}$. We can therefore trace the physical
degrees of freedom from the beginning and only ever need to consider $G$ and
$\bar G$. While this observation is also true for general PEPS, it is
particularly useful when working with GFPEPS in terms of CMs, as it allows
us to completely neglect the physical part of the CM right from the
beginning. 

Let us finally briefly comment on the relation of the boundary theory as
given by $\Sigma_N$ to the construction of the boundary theory for general
PEPS derived in Ref.~\cite{bulk_boundary}. There, the part of the
PEPS which describes $\mathcal R$ (corresponding to the CM $\Gamma$) is
interpreted as a linear map $\mathcal X_\mc{R}$ from the boundary to the bulk
degrees of freedom, which is then decomposed as $\mathcal X_\mc{R}=\mathcal
V_\mc{R}\,\mathcal P_\mc{R}$, with $\mathcal V_\mc{R}$ an isometry and $\mathcal
P_\mc{R}=\sqrt{\tau_{\mc{R}}^\top}$, where $\tau_{\mathcal R}$ is the reduced
density matrix of $\mathcal R$ on the virtual system (corresponding to
$G$). This is exactly identical to the decomposition (\ref{eq:P-CM}); in
particular, $M$ describes the isometry $\mathcal V_\mc{R}$, and the (2+3,2+3)
block of $\Gamma'$ describes the map $\nu\rightarrow \sqrt{\tau_{\mathcal
R}^\top} \nu \sqrt{\tau_{\mathcal{R}}^\top}$ (realized by projecting the (3,3)
part onto $\nu$).  Finally, $\bar G$ describes the analogous state
$\tau_{\bar{\mathcal R}}$ obtained from the part $\bar{\mathcal R}$, and
thus, $\Sigma_N$ exactly corresponds to the boundary theory
$\sqrt{\tau_{\mathcal R}^\top}\tau_{\bar{\mathcal{R}}}
\sqrt{\tau_{\mathcal R}^\top}$ derived in
Ref.~\cite{bulk_boundary}.

\paragraph{Boundary Theories on the torus.\label{sec:bnd-theories-on-torus}} 

We focus now on the situation where the GFPEPS is placed on a long torus, where we take the length of the torus to
infinity. The two regions $\mc R$ and $\bar{\mc R}$ are then obtained
by cutting the torus into two halves, and are thus given by (identical)
long cylinders with diameter $N_v$ and length $N \rightarrow \infty$,
cf.~Fig.~\ref{fig:cylinder}. As we have seen, the central object in the
description is the CM $G$ at the boundary of region $\mc R$, $\partial \mc{R}$, obtained after
tracing out the physical system (and correspondingly for $\bar{\mc R}$).
In the case of a cylinder, $\mc R$ is given by the left
and right boundary of the cylinder together.  In the following, we will
show how to determine $G$ given the CM $\gamma_1$ defining the GFPEPS
without having to construct the CM of the whole state $\Phi_N$.

As we have seen in the preceding paragraph, the boundary theory is
entirely determined by the CM of the virtual part of the initial state
$\Psi_1$.  We thus start by decomposing the CM of the virtual system of
$\Psi_1$ into
 \begin{equation}
    \label{eq:CM-singlesite-virtonly-def}
 D=\left(\begin{matrix} H & K \\ -K^\top & V\end{matrix}\right)\ .
 \end{equation}
Here, $V$ corresponds to the vertical and $H$ to the horizontal Majorana
modes, respectively.  We now concatenate one column of tensors, closing
its vertical boundary, leaving us with a CM which describes the left and
right virtual indices of the column (cf.~Fig.~\ref{Fig:PEPS}b-d). This is done by employing
Eq.~\eqref{eq:proj-ME} for the corresponding subblocks of $V$ of each pair
of (cyclically)
consecutive states $\Psi_{1,\mb r}$ and $\Psi_{1,\mb r + \hat y}$. Due to translational
invariance, this is conveniently expressed in the Fourier basis (with
$k_y$ the quasi-momentum in $y$-direction): In this basis the $D$'s of one
column form a block-diagonal matrix, while
 \[
 \hat \omega(k_y) = \left(\begin{matrix}
    0 & e^{ik_y} \Id_\chi \\ -e^{-ik_y} \Id_\chi & 0
 \end{matrix}\right)
 \]
($\Id_\chi$ denoting the $\chi \times \chi$ identity matrix) since the
$\omega$'s of one column form a circulant matrix with the two blocks
coupling the ``up'' and ``down'' indices of adjacent $V$'s.  In
Fourier space, the CM describing the left and right virtual modes of one
column is thus
 \begin{equation}
 \label{eq:hat-D1-fromHVK}
 \hat D_1 = H + K \,(V+\hat \omega^{-1})^{-1}\, K^\top \ .
 \end{equation}
(We use the hat to denote dependence on $k_y$ in the following; the
subscript $N$ of $\hat D_N$ indicates the number of columns.)
Taking advantage of the fact that the matrix inverse can be written in terms of determinants, one immediately finds that each entry of $\hat D_1$ is a complex ratio of trigonometric polynomials (i.e., polynomials in $e^{\pm i k_y})$ with a degree bounded by the dimension of $\hat\omega$, i.e., $2\chi$.

The matrix $\hat D_1$ consists itself of four blocks,
 \be
 \hat D_1=\left(\begin{matrix} \hat R_1 & \hat S_1
 \\ -\hat S_1^\dagger & \hat T_1 \end{matrix}\right)\ , \label{eq:defD1}
 \ee
corresponding to the left and right indices, respectively.  Let us now see
what happens if we contract two columns. We will consider the general case
where the two columns can be different -- for instance, each of them could
have been derived by contracting some number of single columns $\Phi_1$;
this will allow us to easily derive recursion relations. We thus have two
columns described by
\[
 \hat D=\left(\begin{matrix} \hat R & \hat S \\
 -\hat S^\dagger & \hat T \end{matrix}\right)
 \mbox{\ \ and \ \ }
 \hat D'=\left(\begin{matrix} \hat R' & \hat S'
 \\ -(\hat S')^\dagger & \hat T' \end{matrix}\right)\ ,
 \]
with a column of maximally entangled states connecting them: The CM of both blocks concatenated is then according to Eq.~\eqref{eq:proj-ME}
 \begin{equation}
 \label{eq:general-D-recursion}
 \hat D'' =
 \left(\begin{matrix} \hat R & 0 \\ 0 & \hat T' \end{matrix}\right)
 +
 \left(\begin{matrix} -\hat S & 0 \\ 0 & (\hat S')^\dagger \end{matrix}\right)
 \left(\begin{matrix} \hat T & \Id \\
 -\Id & \hat R' \end{matrix}\right)^{-1}
 \left(\begin{matrix} -\hat S^\dagger & 0 \\ 0 & \hat S' \end{matrix}\right)\ .
 \end{equation}
Using the Schur complement formula for the matrix inverse in the middle,
this gives a recursion relation for the blocks $\hat R$, $\hat S$, and
$\hat T$, which serves several purposes. In particular, by choosing $\hat
D=\hat D'$, we can obtain an iteration formula for $\hat D_{2^\ell}$
describing $2^\ell$ columns, which quickly converges towards the infinite
cylinder limit $\hat D_\infty$, thus being very useful for numerical study.
Moreover, as we will see in subsection~\ref{sec:full-chi1}, in certain cases, Eq.~\eqref{eq:general-D-recursion}
can also be used to analyze the convergence of the transfer operator, or, by
choosing $\hat D'=\hat D''$
and $\hat D=\hat D_1$, to determine the explicit form of the fixed point
$\hat D_\infty$.

Finally, given the fixed point $\hat D_\infty$, as well as $\hat{\bar
D}_\infty$ corresponding to the boundary  $\partial \bar{\mc R}$, it is
now straightforward to determine the boundary Hamiltonian using
Eqs.~(\ref{eq:symmetrized-es}) and \eqref{eq:HN} for $N \rightarrow
\infty$. Note that in the particular case of a torus, which we will consider,
$\hat{\bar D}_\infty$ can be obtained from $\hat D_\infty$ by exchanging
the blocks corresponding to the left and right boundary.

\paragraph{Proof of the polynomial decay of the hopping amplitude of the boundary Hamiltonian}

Here, we prove that the amplitudes of the boundary Hamiltonian decay as $|H^R_\infty| \propto \log(y)/y + \mc{O}(1/y)$ for $y \gg 1$.

We start by calculating the single-particle entanglement spectrum on the right boundary:
For that we employ Eq.~\eqref{eq:hat-D1-fromHVK} to calculate $\hat D_1(k_y)$ for the topological superconductor defined by Eq.~\eqref{eq:Psi1ex} and from it $\hat \Sigma^R_\infty(k_y)$ (the part entirely localized on the right edge) via Eqs.~\ref{eq:general-D-recursion} and~\eqref{eq:symmetrized-es} as a function of $\lambda$. The result is
\be
\hat \Sigma^R_\infty(k_y) = i \frac{2 \lambda^2 \sin(k_y)}{\sqrt{\frac{g_\lambda^2(k_y)}{|1-\lambda - e^{i k_y}|^4} + 4 \lambda^4 \sin^2(k_y)}}
\ee
with $g_\lambda(k_y)$ some second order polynomial in $\cos(k_y)$. For $\lambda
\neq 0$, $\hat \Sigma_\infty^R(k_y)$ is obviously analytic as long as $k_y$ is not an integer
multiple of $\pi$. One can check that $g_\lambda(\pi) \neq 0$ for any $\lambda \in
(0,1)$, so $\hat \Sigma^R_\infty(\pi) = 0$, i.e., the only possible
non-analytical point is $k_y = 0$.  One can check from the explicit function
$g_\lambda(k_y)$ that $g_\lambda(\delta k_y) = g_\lambda(-\delta k_y) = g_{\lambda,0} \, \delta k_y^{2} (1 +
\mathcal{O}(\delta k_y^2))$. Therefore,
\begin{align}
\hat \Sigma^R_\infty(\delta k_y) &= i \frac{2 \lambda^2 \delta k_y}{\sqrt{\frac{g_{\lambda,0}^2 \delta k_y^{4}(1 + \mathcal{O}(\delta k_y^2))}{(2-\lambda)^4} + 4 \lambda^4 \delta k_y^2}} \notag \\
&= i \left(1 - \frac{g_{\lambda,0}^2}{8 \lambda^4 (2-\lambda)^4} \delta k_y^{2}\left(1 + \mathcal{O}(\delta k_y^2)\right)\right) \mr{sgn}(\delta k_y)
\end{align}
Owing to Eq.~\eqref{eq:HN} for $N \rightarrow \infty$, the single-particle spectrum is given by
\be
- i \hat H_\infty^R(k_y) = \ln \left( \frac{1 - i \hat \Sigma^R_\infty (k_y)}{1 + i \hat \Sigma^R_\infty(k_y)}\right).
\ee
Henceforth, we can expand
\begin{align}
- i \hat H_\infty^R(\delta k_y) = \big[\ln\left(\frac{16 \lambda^4 (2-\lambda)^4}{g_{\lambda,0}^2}\right) - 2\ln(\delta k_y) - \ln(1 + \mathcal{O}(\delta k_y^2)) \big] \mr{sgn}(\delta k_y),
\end{align}
and we see that the non-analyticity is only due to the term $2\ln(\delta k_y)$, the other ones being analytical around $k_y = 0$. The Fourier coefficients of an analytical function defined on $(-\pi,\pi]$ decay exponentially. Thus, the algebraic decay of $|[H_\infty^{R}]_{1,1+y}|$ is due to the diverging term we singled out,
\begin{align}
|[H_\infty^{R}]_{1,1+y}| \xrightarrow[y \rightarrow \infty]{} \frac{4}{\sqrt{N_v}} \int_0^\pi \sin(k_y y) \ln(k_y) \d k_y \xrightarrow[y \rightarrow \infty]{} \frac{4}{\sqrt{N_v}} \left[\frac{\ln(y)}{y} + \mc{O}\left(\frac{1}{y}\right)\right]
\end{align}
with the prefactor of the ${1}/{y}$ contribution being dependent on whether $y$ is even or odd but constant otherwise. \qed

\subsubsection{Edge theories}\label{sec:edge}

\paragraph{Derivation of edge theory}

We will now turn our attention towards the edge Hamiltonian which
describes the effective low-energy physics obtained at an edge of the
system.

As explained in subsection~\ref{Sec:Intro-GFPEPS}, the GFPEPS $\Phi$ is the
ground state of the flat band Hamiltonian $\mc H_\mr{fb}=
-\tfrac{i}{4}\sum_{l,m} \gamma_{lm} e_l e_m$, Eq.~(\ref{eq:flat}), where
$\gamma$ is the CM of the whole state $\Phi$, Eq.~(\ref{eq:CM-GFPEPS}).
The restriction of $\mathcal H_\mathrm{fb}$ to a region $\mathcal R$ of
the system is then given by
\be
\mc H_\mc{R}= -\tfrac{i}{4} \sum_{l,m} [\gamma_\mathcal{R}]_{lm} e_l e_m,
\label{eq:Ham-truncated}
\ee
where the sum now only runs over modes in $\mathcal R$, and
$\gamma_{\mathcal R}$ is determined by Eq.~(\ref{eq:CM-R}).

Let us now perform the basis transformation $M$, Eq.~(\ref{eq:Q-trafo}):
Following Eq.~(\ref{eq:CM-R}), the CM of $\mathcal R$,
$\gamma_\mathcal{R}$, is transformed to
\be
\gamma_\mathcal{R}' = \left(\begin{matrix}
    Z & 0 \\ 0 & \Sigma_N\end{matrix}\right)\ , \label{eq:gamma_trans}
\ee
with $\Sigma_N$ given by Eq.~(\ref{eq:symmetrized-es}), and, at the same time,
$\mathcal H_\mathcal{R}$ is transformed into an isomorphic Hamiltonian
$\mc H_\mc{R}'= -\tfrac{i}{4} \sum_{l,m} [\gamma'_\mathcal{R}]_{lm} e_l
e_m$.  We see that the spectrum of $\mathcal H'_\mathcal R$ (and thus
of $\mathcal H_\mathcal R$) consists of two parts: First, the (1,1)-block of
$\gamma_\mathcal{R}'$ corresponds to bulk modes at energy $\pm 1$.
Second, the (2,2) bock $\Sigma_N$ corresponds to modes at generally
smaller energy, which are hence related to restricting $\mathcal
H_{\mathrm{fb}}$ to region $\mathcal R$; those modes correspond to the
boundary degrees of freedom via the purification in the (3,3) block of
$\Gamma'$, Eq.~\eqref{eq:P-CM}.  We thus find that the edge Hamiltonian,
i.e., the low-energy part of the truncated flat band Hamiltonian, is given
by
\begin{equation}
\label{eq:Hedge-eq-Sigma}
 H_N^\mr{e} = \Sigma_N \ ,
\end{equation}
with $\mathcal H_N^\mathrm{e}=-\tfrac{i}{4}\sum_{j,k} [H_N^\mathrm{e}]_{jk} c_j
c_k$, Eq.~\eqref{eq:Hedge}.  Except for additional bulk modes with energy
$\pm 1$, $H_N^\mr{e}$ in fact exactly reproduces the spectrum of
the truncated flat band Hamiltonian.  The relation
(\ref{eq:Hedge-eq-Sigma}) allows us to transfer the results on the
boundary theory $\Sigma_N$ of GFPEPS one-to-one to their edge Hamiltonians
$H_N^\mr{e}$.  Note that the resulting relation between entanglement
spectrum and edge Hamiltonian, $H_N^\mr{b} = 2 \arctan(H_N^\mr{e})$,
corresponds to the one derived by Fidkowski~\cite{edge_entanglement}.

The derivation of the edge Hamiltonian in this subsection is again identical
to the edge Hamiltonian introduced for general PEPS in Ref.~\cite{edge_theories}.
Using the same notation as in the last paragraph, the edge Hamiltonian for general
PEPS is obtained by projecting the physical Hamiltonian onto the boundary
using the isometry $\mathcal V_\mc{R}$. This projection is exactly accomplished
by rotating with $M$ and subsequently considering only the (2,2) block of
$\gamma_{\mathcal R}'$ [see Eq.~\eqref{eq:gamma_trans}], and thus, the edge Hamiltonian obtained here is
identical to the one of Ref.~\cite{edge_theories}, with the bulk Hamiltonian
taken to be the flat band Hamiltonian.

\subsubsection{Localization of edge modes}

In the case of a cylinder, on which we focus, the edge Hamiltonian
$H^{\mathrm{e}}_N$ is supported on the auxiliary modes both on the left
and the right edge (cf.~Fig.~\ref{Fig:PEPS}f). However, as we will show in
the following, the edge Hamiltonian (as well as the boundary theory) on
the two edges decouples for almost all $k_y$, and moreover, the
corresponding physical edge modes are localized at the same edge as the
virtual modes.  An important consequence of that is that we can use the
\emph{virtual} edge Hamiltonian to compute the Chern number of the system,
as it is known that the Chern number corresponds to the winding number of
the edge modes localized at one of the edges of the system~\cite{Chern_edge,honeycomb}.

In order to answer both of these questions, we first need to
demonstrate some properties of the CM $\Gamma\equiv \Gamma_{N}$,
Eq.~\eqref{eq:Gamma-R-Rbar}, which describes the GFPEPS $\Phi_{N}$
(Fig.~\ref{Fig:PEPS}f) on a cylinder of length $N\gg1$. Since the system
is translational invariant in vertical direction, we can equally well
carry out our analysis in Fourier space, and we will do so in the
following. By combining Eqs.~\eqref{eq:gamma1},
(\ref{eq:CM-singlesite-virtonly-def}) and \eqref{eq:hat-D1-fromHVK}, we
immediately find that $\Phi_1$ is described by a CM of the form
\[
\hat\Gamma_1 = \left(\begin{array}{c|cc}
    \hat A_1 & \hat B_{1,R} & \hat B_{1,T} \\\hline
    -\hat B_{1,R}^\dagger & \hat R_1 & \hat S_1 \\
    -\hat B_{1,T}^\dagger & -\hat S_1^\dagger & \hat T_1
\end{array}\right)\ ,
\]
with $\hat R_1$, $\hat S_1$, and $\hat T_1$ defined in Eq.~\eqref{eq:defD1}. The concatenation of $N$ columns is then given by the Schur complement
\begin{equation}
\label{eq:PhiN-bnd-coupling}
\hat \Gamma_{N} =  \hat P_{N} +
    \hat Q_{N} \hat V_{N}^{-1} \hat Q_{N}^\dagger\ ,
\end{equation}
with
\begingroup
\allowdisplaybreaks
\begin{align*}
\hat P_{N} & = \left(\begin{array}{c|cccc|c}
    \hat R_1 &  -\hat B_{1,R}^\dagger & 0 & \cdots &  \\\hline
    \hat B_{1,R} & \hat A_1  & 0 & & \\
      0		 & 0 & \hat A_1 & & \\
      \vdots & & \ddots & \ddots& \\
           &  &   & 0 & \hat A_1 & -\hat B_{1,T}^\dg \\\hline
		 & &  &   &  \hat B_{1,T}  & \hat T_1
\end{array}\right)\ ,\\[1ex]
\hat Q_{N} & = \left(\begin{matrix}
    \hat S_1 \\\hline
    \hat B_{1,T} \\
    & \hat B_{1,R} & \hat B_{1,T} \\
    & & & \cdots \\
    & & &  & \hat B_{1,R} & \hat B_{1,T} \\
    & & & & & & \hat B_{1,R} \\\hline
    & & & & & & -\hat S_1^\dagger
\end{matrix}\right)\ ,\\[1ex]
\hat V_{N} &= \left(\begin{matrix}
    \hat T_1 & \Id \\
    -\Id & \hat R_1 & \hat S_1 \\
    0 & -\hat S_1^\dagger & \hat T_1 & \Id \\
    & & -\Id & \ddots & \ddots  \\
    & & & \ddots & \ddots & \Id \\
    & & & & -\Id & \hat R_1 & \hat S_1 \\
    & & & & & -\hat S_1^\dagger & \hat T_1 & \Id \\
    & & & & & & -\Id & \hat R_1
\end{matrix}\right)\ ,
\end{align*}
\endgroup
where we have moved the virtual modes on the left (right) boundary to the
left (right) corner of the CM, as indicated by the lines above.

Let us now first show that the two virtual edges are decoupled.  To this
end, we consider the reduced state of the virtual system of
$\Phi_{N}$, which is given by the CM $G$ in Eq.~\eqref{eq:Gamma-R-Rbar};
evidently, vanishing off-diagonal blocks in $G$ (and $\bar G$) imply that
any coupling between the two boundaries in $\Sigma_{N}$,
Eq.~\eqref{eq:symmetrized-es}, vanishes as well.
$G$ is given by the two outer blocks of $\hat\Gamma_{N}$. Obviously, the
only way in which these two blocks can couple is via $\hat V_{N}^{-1}$.

We now invoke a result on the inverse of banded
matrices~\cite{banded_matrices}: Given a banded matrix $A_b$, it holds that
$|(A_b^{-1})_{ij}|\le \mr{const.} \times \beta^{|i-j|}$, where $\beta <1$ depends on the
ratio of the largest and smallest eigenvalue of $A_b A_b^\dagger$ (and
$\beta\rightarrow1$ if the ratio diverges). Using this result, we find
that the coupling between the two edges in $G$ is exponentially suppressed
in the
length $N$ of the cylinder, as desired, as long as the ratio of the
eigenvalues of $\hat V_N \hat V_N^\dagger$ does not diverge.  Its largest
eigenvalue is clearly bounded by $4$,
since $\hat V_{N}$ is the sum of two CMs.  To lower bound the smallest eigenvalue,
observe that $\hat V_{N}\hat V_{N}^\dagger$ is again a banded Toeplitz
matrix, which we can regard as a subblock of a larger circulant matrix.
This
circulant matrix can in turn be block-diagonalized using a Fourier transform,
and we find that its blocks are of the form $(\hat D_1 +
\hat\omega^{-1}(\mb k))(\hat D_1+\hat\omega^{-1}(\mb k))^\dagger$.  On
the other hand, $\mathrm{det}(\hat D_1+\hat \omega^{-1}(\mb k))$
is exactly the energy spectrum of the local parent Hamiltonian as
constructed in Ref.~\cite{fPEPS_Kraus}, and thus, $\hat
V_{N}^{-1}\equiv \hat V_{N}^{-1}(k_y)$ has exponentially decaying entries
if and only if the energy spectrum has a gap for all $k_x$ and for the given value of
$k_y$ (this turns out to be the case for almost all $k_y$).

As we have seen, (almost) all virtual edge modes on the left and the right of the
cylinder decouple. In the following, we will show that also the physical
modes corresponding to these edge modes are exponentially localized around
the respective boundary.  To this end, we fix $N\gg1$ and consider the CM
$\Gamma'$, Eq.~(\ref{eq:P-CM}), which is obtained by an orthogonal
transformation from the original CM $\Gamma\equiv \Gamma_{N}$.
In $\Gamma'$, the edge modes are supported on the $(2,2)$ block, and we
need to figure out how the inverse of the orthogonal transformation $M$,
Eq.~(\ref{eq:Q-trafo}), maps these back to the physical
modes.

To this end, note that in order to prepare an arbitrary state in
the $(2,2)$ block of $\Gamma'$, Eq.~\eqref{eq:P-CM}, we just need to project the $(3,3)$
block on an (unphysical) CM $X$ via the Schur complement formula
Eq.~\eqref{eq:proj-ME}. In particular, we can use this to occupy or
deplete a specific mode. (We assume $\chi$ to be even; otherwise, one can simply
group pairs of modes.) Consequently, by projecting the original CM
$\Gamma_{N}$ onto the very same $X$, we will exactly occupy or deplete the
corresponding physical mode. Now, we can make use of
Eq.~(\ref{eq:PhiN-bnd-coupling}), together with the aforementioned result
on inverses of banded matrices: Take $X$ and $X'$ such that projecting
onto $X$ ($X'$) occupies (depletes) a certain mode, where $X$ and $X'$ are non-trivial
only on one virtual boundary. If we denote by  $\hat \gamma(X)$ [$\hat \gamma(X')$] the respective CMs after the
projection, we have that
\[
\gamma(X) - \gamma(X') = \hat Y
    \left[
	(\hat Z+X^{-1})^{-1}
	-(\hat Z+X'^{-1})^{-1}
    \right]
\hat Y^\dagger\ ,
\]
where we have set $\hat \Gamma_N = \left(\begin{smallmatrix} \hat W&\hat Y\\ - \hat Y^\dg &\hat Z\end{smallmatrix}\right)$ 
with the first block corresponding to the physical and the second block corresponding to the virtual modes. 
The coupling between the two boundaries mediated by $\hat Y$ and the off-diagonal block of $\hat Z$ (correlating the left and 
right virtual boundaries) are exponentially  small in $N$.
Since we also have that
\[
[\hat \gamma(X)-\hat \gamma(X')]_{lm} = 2i( v_l v^*_m - w_l w^*_m)
\]
with $\sum_l v_l \hat c_l$ and $\sum_l w_l \hat c_l$ the creation/annihilation operator
for the corresponding physical mode, it follows that $v_l$ and $w_l$ decay
exponentially with the distance from the corresponding boundary, i.e., the
physical edge mode corresponding to a given virtual edge mode is localized
around that edge (for all $k_x$ for which the energy spectrum of $\mc H_\mr{ff}$ has a gap).

\subsection{Further examples}\label{sec:examples}

In the following, we will present further examples for both chiral and
non-chiral GFPEPS and discuss their respective boundary theories.
In the next subsection, we discuss a Chern insulator with $C=-1$; in subsection
\ref{sec:Chern2}, we introduce a model with $C=2$, which has entangled edge modes at
incommensurate values of $k_y$. Finally, two non-chiral models will be considered in 
 subsections \ref{sec:C0-nontrivial} and \ref{sec:ex:nochern-flatspec}.

\subsubsection{GFPEPS describing a Chern insulator with $C = -1$}\label{sec:Chern}

In the following, we study a family of chiral GFPEPS which are particle number conserving and describe a
Chern insulator with $C = -1$.  They can be decoupled into two
copies of a topological superconductor which is closely related to the family
given by Eq.~\eqref{eq:Psi1ex}.
The family of Chern insulators has $f = 2$ physical fermionic modes per site, $\chi
= 2$ Majorana bonds, and $\gamma_1$ [Eq.~\eqref{eq:gamma1}] is defined via
\begin{align}
\begin{split}
A &= (-1 + 2 \eta) \left(\begin{array}{cc}
W&0\\
0&-W
\end{array}\right), \\[0.5ex]
B &= \sqrt{\frac{\eta - \eta^2}{2}} \left(\begin{matrix}
\Id - W&\Id + W&-\sqrt{2}\,W&\sqrt{2}\,\Id\\
\Id - W&-\Id - W&\sqrt{2}\,\Id&-\sqrt{2}\,W
\end{matrix}\right), \\[0.5ex]
D &= \left(\begin{matrix}
0&(-1+\eta)\,\Id&-\frac{\eta}{\sqrt{2}}\,\Id&\frac{\eta}{\sqrt{2}}\,\Id\\\
(1-\eta)\,\Id&0&-\frac{\eta}{\sqrt{2}}\,\Id&-\frac{\eta}{\sqrt{2}}\,\Id\\
\frac{\eta}{\sqrt{2}}\,\Id&\frac{\eta}{\sqrt{2}}\, \Id&0&(-1+\eta)\,\Id\\
-\frac{\eta}{\sqrt{2}}\,\Id&\frac{\eta}{\sqrt{2}}\,\Id&(1-\eta)\,\Id&0
\end{matrix}\right),
\end{split}
\label{eq:Chern-ins}
\end{align}
where $\Id = \left(\begin{smallmatrix} 1&0\\  0&1
\end{smallmatrix}\right)$,
$W = \left(\begin{smallmatrix} 0&1\\ -1&0
\end{smallmatrix}\right)$ and $\eta \in (0,1)$. The ordering of the physical Majorana modes is
$(c_{1\uparrow},c_{2\uparrow},c_{1\downarrow},c_{2\downarrow})$, and
the blocks of $D$ are ordered according to left, right, up, and down virtual modes.

\begin{figure}[!ht]
\centering
\includegraphics[width=0.32\columnwidth]{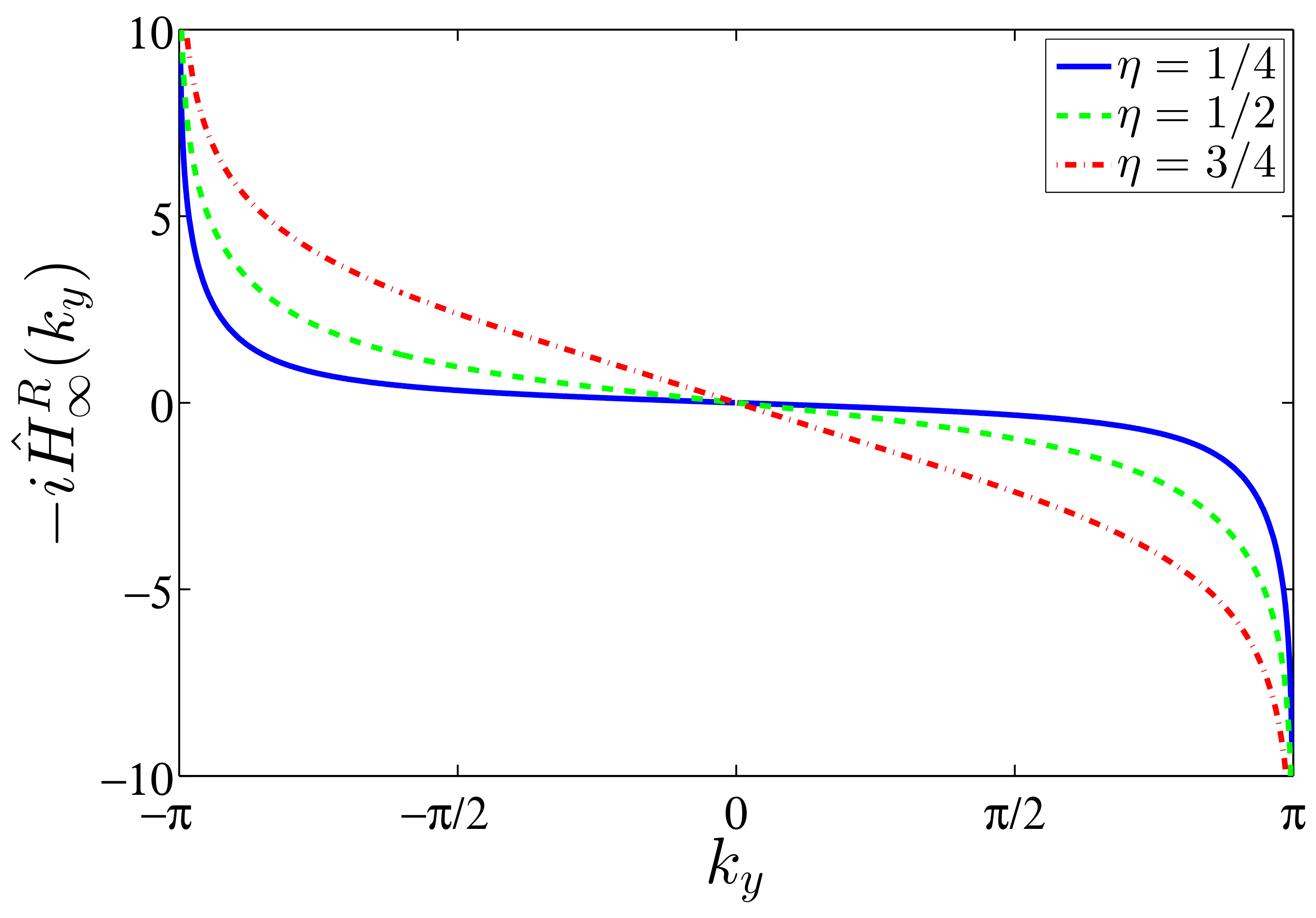}
\includegraphics[width=0.33\columnwidth]{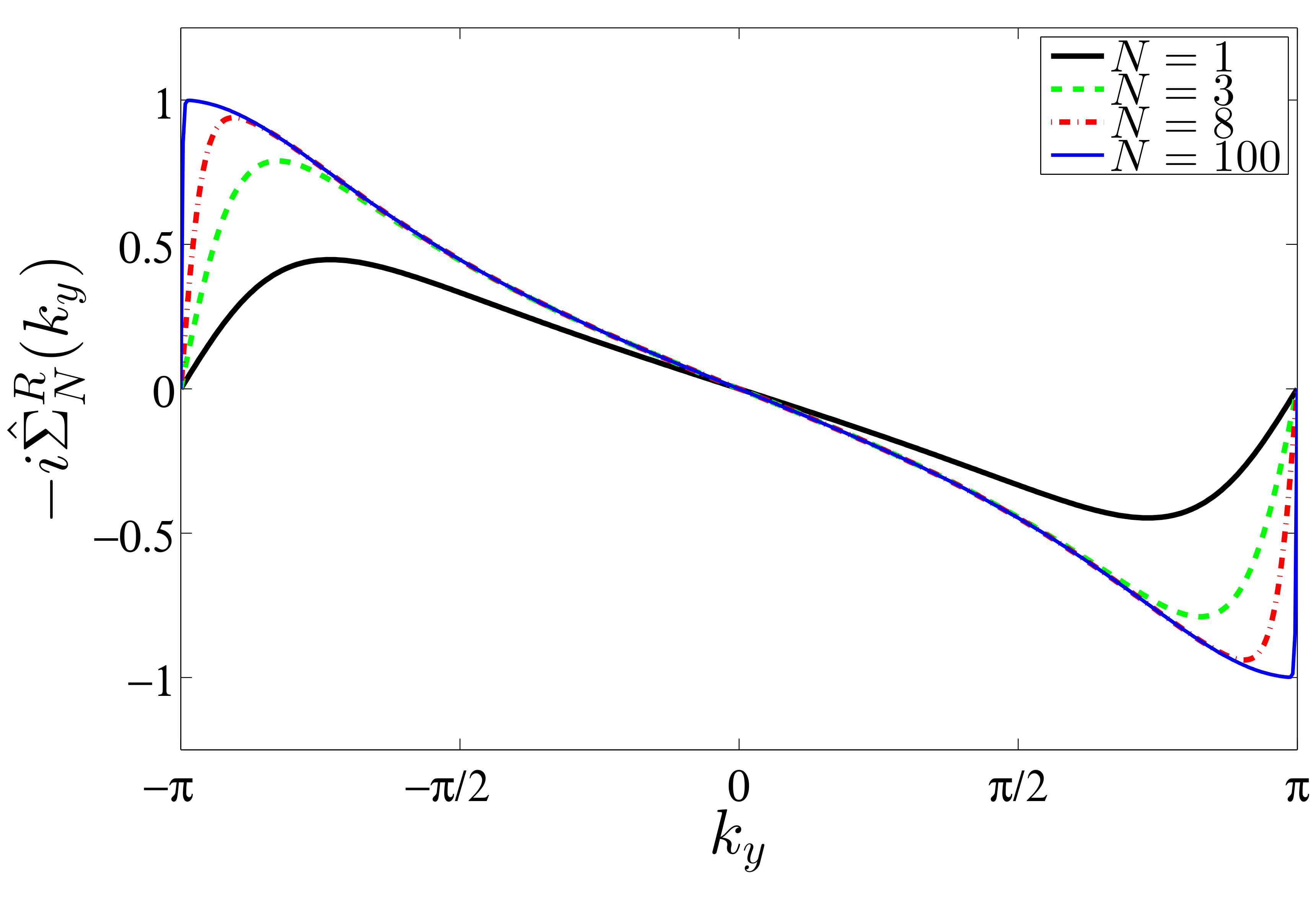}
\includegraphics[width=0.33\columnwidth]{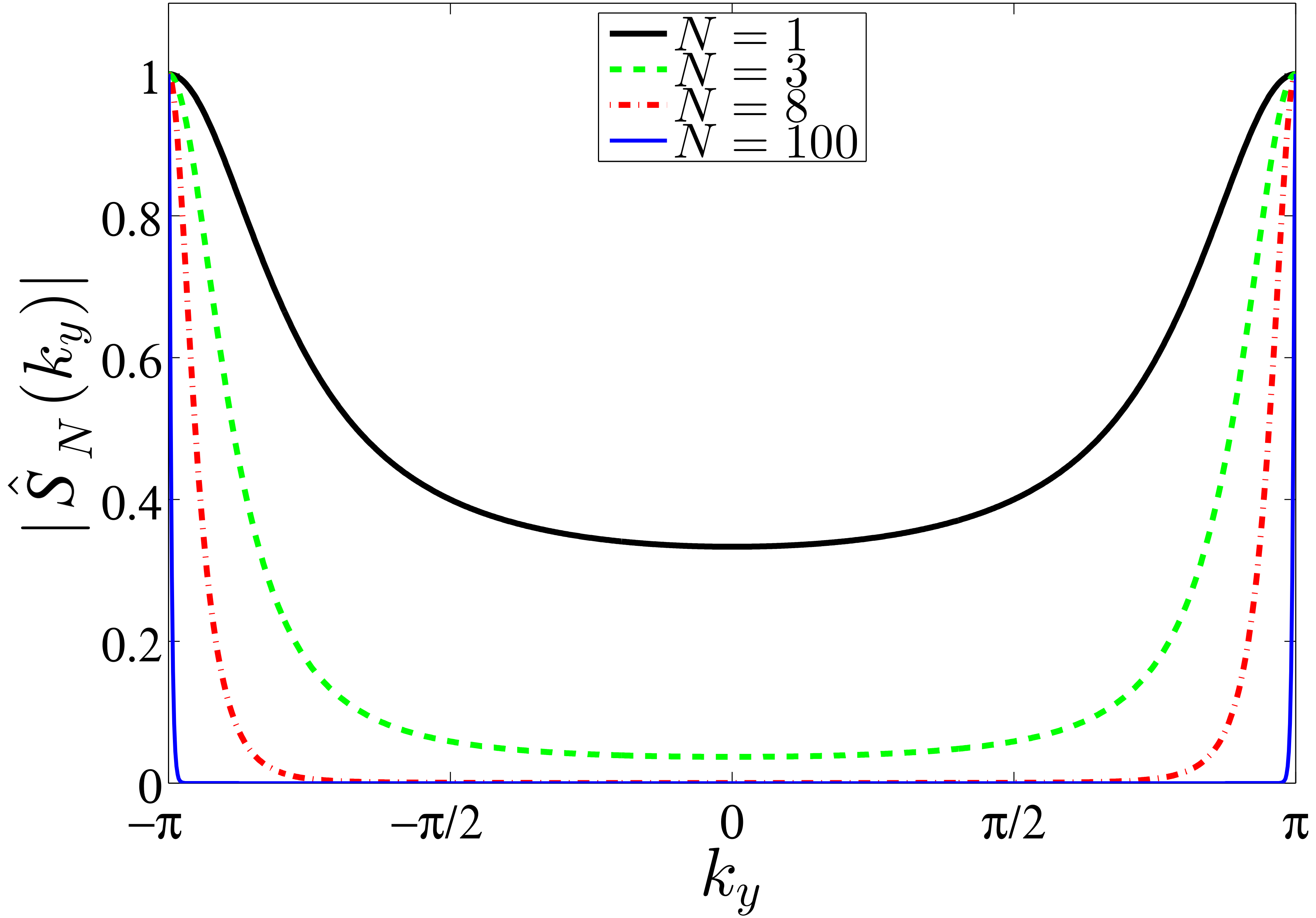}
\caption{
\label{fig:Convergence_R1}
Analysis of the boundary and edge Hamiltonian of the model of
subsection~\ref{sec:Chern}, using the decomposition of Eq.~\eqref{eq:sigma-chernins}.
Left: Boundary Hamiltonian $-i\hat H_\infty^{R}(k_y)$ for different values
of $\eta$. Middle: Convergence of the right edge spectrum $\hat
\Sigma^{R}_N(k_y)$ (i.e., the block of $\hat\Sigma_{N}(k_y)$ corresponding
to the right edge) for $\eta = 1/2$ with increasing cylinder length
$N=1,3,8,100$.  For $N = 100$, the spectrum is already well 
converged.  Right: Magnitude of the corresponding off-diagonal element $\hat S_N(k_y)$ of
$\hat D_{N} (k_y)$ which describes the coupling of the two boundaries
(cf.~subsection~\ref{sec:bnd-theories-on-torus}), for cylinder lengths
$N=1,3,8,100$, illustrating the exponential decoupling with increasing $N$. The dispersionless bulk bands of the truncated flat band Hamiltonian $\mc H_\mc{R}$ correspond to energies of $\pm \infty$.
This figure was reproduced from Ref.~\cite{cPEPS_long}.}
\end{figure}

The boundary CM $\hat \Sigma_\infty(k_y)$ can be computed using the
results of subsection~\ref{sec:full-chi1}, and we find it to be of the form
\begin{equation}
\label{eq:sigma-chernins}
\Sigma_\infty = \left(\bigoplus_{k_y \neq \pi} \left(\hat
\Sigma^L_\infty(k_y) \oplus \hat \Sigma^R_\infty(k_y) \right)\oplus \hat
\Sigma^{LR}_\infty(\pi)\right) \otimes \Id
\end{equation}
with $\hat \Sigma^{LR}_N(\pi) = \left(\begin{smallmatrix} 0&\pm 1 \\ \mp
1& 0 \end{smallmatrix}\right)$, the sign depending on whether the
horizontal length $N$ of the cylinder is even or odd.
In Fig.~\ref{fig:Convergence_R1}, we show the spectrum of the boundary
Hamiltonian of the above model (left panel). Moreover, we illustrate how
for $N\rightarrow\infty$ the spectrum of the edge Hamiltonian for a single edge converges
(middle panel) and how the coupling between the two edges vanishes (right
panel).

The Chern number can now be determined by counting the number of times the
bands of $-i \hat \Sigma_\infty^R(k_y) \otimes \Id$ (or, alternatively, of
$-i \hat H_\infty^R(k_y) \otimes \Id$) cross the Fermi level.  Obviously,
the spectrum of the boundary and edge Hamiltonian consists of two bands
lying on top of each other. In the language of topological
superconductors, this would give rise to a Chern number of $-2$. However,
since we assume particle number conservation (as we deal with a Chern
insulator), the Chern number is given by the number of \textit{fermionic}
chiral modes of the edge or boundary Hamiltonian, respectively. There is
only one such fermionic chiral mode (annihilation operator $\hat
a_{k_y}$), which is obtained by combining the two chiral Majorana modes on
the right edge, $\hat c_{k_y,1}$ and $\hat c_{k_y,2}$, with equal dispersion to
$\hat a_{k_y} = \frac{1}{2} (\hat c_{k_y,1} - i \hat c_{k_y,2})$. Therefore, the
(particle number conserving) Chern number is $C = -1$.

\subsubsection{GFPEPS with Chern number $C = 2$}\label{sec:Chern2}

In the following, we provide an example of a topological superconductor
with $\chi=2$ and Chern number $C=2$. The model has been constructed
numerically such that it exhibits discontinuities in $\hat\Sigma^R_\infty(k_y)$,
and thus maximally entangled modes between the edges at $k_y=\pm1$;
it thus demonstrates that for $\chi>1$, there is no constraint (in terms
of simple fractions of $\pi$) on the possible values of $k_y$.  The CM $D$
of the example is given by
\[
D \!\approx\!\! \left(\begin{smallmatrix}
    0     &  -0.326 &  -0.250 &   0.510  &  0.295  &  0.071 &  -0.434 &  -0.030  \\
    0.326 &        0&   0.044 & -0.074   & -0.513  &  0.032 &  -0.051 &   0.577  \\
    0.250 &  -0.044 &   0     &  -0.467  &  0.036  &  0.603 &  -0.423 &  -0.125 \\
   -0.510 &   0.074 &   0.467 &       0  &  0.148  &  0.156 &   0.169 &   0.216 \\
   -0.295 &   0.513 &  -0.036 &  -0.148  &  0      & -0.161 &  -0.296 &   0.237 \\
   -0.071 &  -0.032 &  -0.603 &  -0.156  &  0.161  &  0     &   0.042 &   0.521 \\
    0.434 &   0.051 &   0.423 &  -0.169  &  0.296  & -0.042 &   0     &   0.047 \\
    0.030 &  -0.577 &   0.125 &  -0.216  & -0.237  & -0.521 &  -0.047 &   0
\end{smallmatrix}\right).
\]
It has been obtained by numerically optimizing $D$ such that one of the
eigenvalues of $\hat\Sigma_N^R(k_y)$ (where $N=2^{29}$) jumps from $\pm i$
to $\mp i$ for some $k_y\in[0.999,1.001]$, while restricting half of the
eigenvalues of $D$ to be between $-0.6i$ and $+0.6i$ such as to prevent $D$ from
converging to a pure state. As
$\hat\Sigma_N^R(-k_y)=[\hat\Sigma_N^R(k_y)]^*$ (with $^*$ indicating the complex
conjugate), this automatically yields another identical discontinuity at
$k_y=-1$.
Note that $D$ can be purified to a state with $f = 2$ physical fermions.

\begin{figure}[!ht]
\centering
\includegraphics[width=0.6\textwidth]{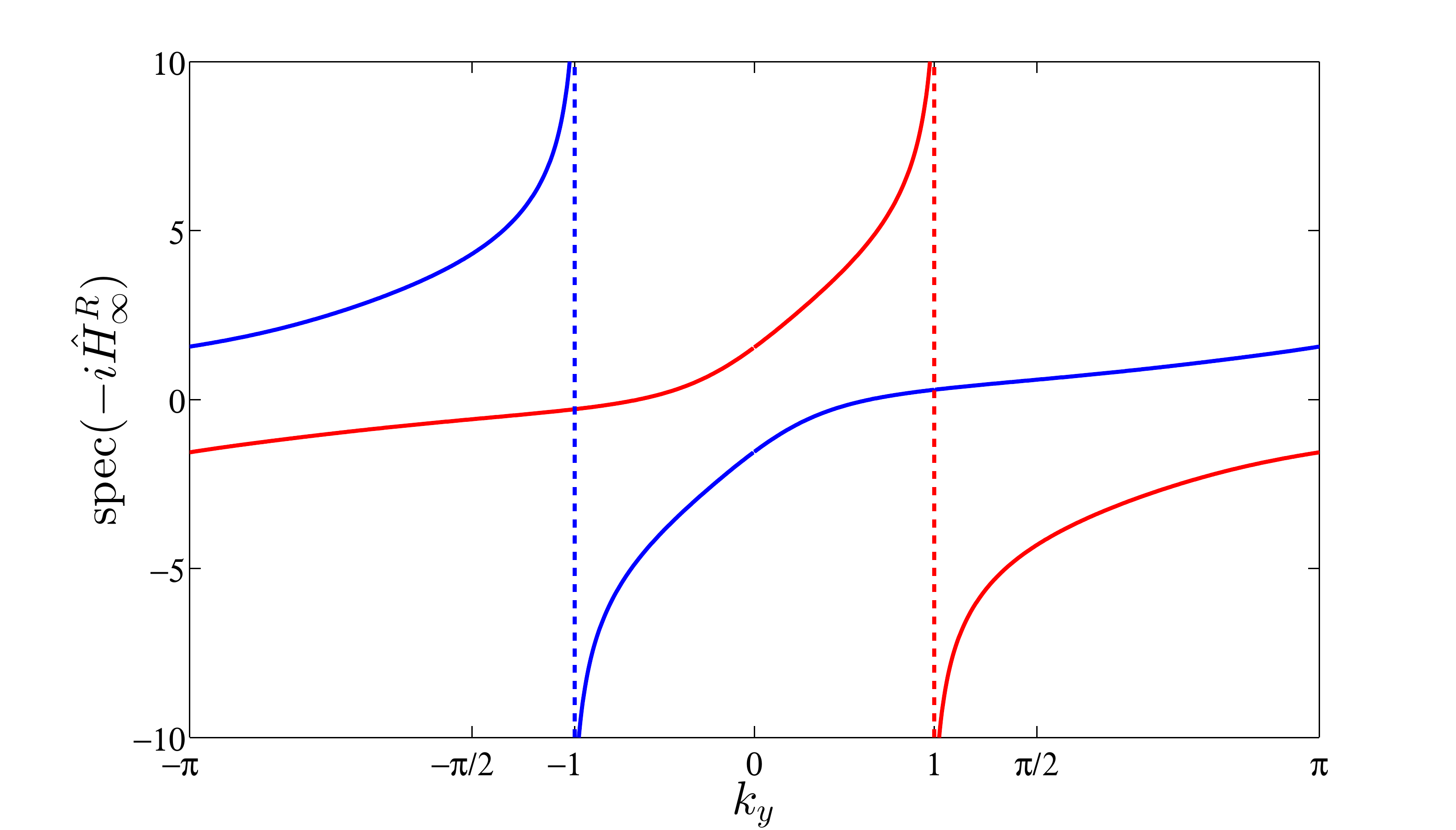}
\caption{Eigenvalue spectrum of $- i \hat H^{R}_\infty(k_y)$ for the
example of subsection~\ref{sec:Chern2}. Since $\chi = 2$, there are two bands.
They diverge at $k_y = \pm 1$, respectively, where only one mode of $\hat
H^{R}_\infty(k_y)$ is defined. Since the Fermi level at $E = 0$ is crossed
two times from below, $C = 2$. This figure was reproduced from Ref.~\cite{cPEPS_long}.} \label{Gamma_L2}
\end{figure}

The spectrum of $-i \hat H^R_\infty(k_y)$ is plotted in
Fig.~\ref{Gamma_L2}. Due to the discontinuities at $k_y\pm1$, it crosses
the Fermi energy twice from below, thus describing a topological
superconductor with Chern number $C = 2$.  At $k_y = \pm 1$, one of
the eigenvalues of $-i \hat H_\infty^R(k_y)$ diverges, and hence $\hat
\Sigma^{LR}_\infty(\pm 1)$ is non-trivial, coupling one of the two virtual
Majorana modes between the left and the right end of the cylinder.

\subsubsection{GFPEPS with Chern number $C = 0$}\label{sec:C0-nontrivial}

The following example provides a family of non-topological GFPEPS with Chern
number $C=0$.  It has one parameter $\mu$, and its matrix $D$ is given by
\be
D = \left(\begin{array}{cccc}
0&0&-\frac{\mu}{2}&f(\mu)\\
0&0&f(\mu)&-\mu\\
\frac{\mu}{2}&-f(\mu)&0&0\\
-f(\mu)&\mu&0&0
\end{array}\right) \label{eq:D-chi0}
\ee
with $f(\mu) = \sqrt{1-\frac{3 \mu}{2}+\frac{\mu^2}{2}}$ and $\mu \in
(0,1)$. ($A$ and $B$ are irrelevant of the Chern number and can be 
generated by choosing an arbitrary
purification.)

We find that the left and right boundary, Eq.~\eqref{eq:decouple-H},
decouple for all $k_y$. The dispersion relation for the right boundary is
shown in Fig.~\ref{Gamma_R1}(right). Since the energy band of the boundary
Hamiltonian crosses the Fermi energy once with positive and once with
negative slope for all $\mu \in (0,1)$, the Chern number is always zero.

\subsubsection{GFPEPS with a flat entanglement spectrum and $C = 0$}\label{sec:ex:nochern-flatspec}

The last example we consider is taken from Ref.~\cite{fPEPS_Kraus}. It
does not display any topological features and is given by
\be
\label{eq:ex:nochern-flatspec}
\gamma_1 = \left(\begin{array}{cccccc}
0&0&\frac{1}{\sqrt 2}&-\frac{1}{\sqrt 2}&0&0\\
0&0&0&0&\frac{1}{\sqrt 2}&-\frac{1}{\sqrt 2}\\
-\frac{1}{\sqrt 2}&0&0&0&\frac{1}{2}&\frac{1}{2}\\
\frac{1}{\sqrt 2}&0&0&0&\frac{1}{2}&\frac{1}{2}\\
0&-\frac{1}{\sqrt{2}}&-\frac{1}{2}&-\frac{1}{2}&0&0\\
0&\frac{1}{\sqrt 2}&-\frac{1}{2}&-\frac{1}{2}&0&0
\end{array}\right)\ .
 \ee
Since $\hat D_N=\hat{\bar D}_N$ due to the reflection symmetry, and thus $G=\bar G$, the entanglement
spectrum and edge Hamiltonian of this model are totally flat, i.e.,
$\Sigma_N=0$ according to Eq.~\eqref{eq:symmetrized-es}, and the Chern
number is zero.


\subsection{Conditions for topological GFPEPS}\label{sec:cond_top}

Let us now show that
topological GFPEPS are very special. In particular, we will prove that any
GFPEPS which is injective (which is generically the case for PEPS, cf. subsection~\ref{sec:PEPS_parent}), or more generally, for which $q(\mathbf{k}) = \det(D - G_\mr{in}(\mb k))$ introduced in subsection~\ref{sec:GFPEPS} is
non-zero everywhere, has a gapped local parent Hamiltonian which is connected to
a trivial state via a gapped path and therefore cannot be topological;
this implies that the parent Hamiltonians defined via $\epsilon(\mb k) = q(\mb k)$ has to be gapless. (This shows also that injectivity in GFPEPS is
much stronger than for general PEPS, where it does not have implications
about the spectrum or the phase except for 1D systems.)

Let us first define injectivity for GFPEPS: By blocking $n_v \times n_h$
sites to a new super-site (by tracing over the virtual particles), we can
reach a point where the number of physical Majorana modes $n_{\mathrm{ph}}
= 2f n_h n_v$ is larger than the number of virtual modes $n_{\mathrm{vir}}
= 2 \chi (n_h + n_v)$. Then, $G_{\mathrm{out} \Box}(\mathbf k) = B_{\Box}
(D_\Box - G_\mathrm{in \Box}(\mathbf k))^{-1} B_\Box^\top + A_\Box, $
where $\Box$ denotes the corresponding matrices after blocking. We say
that a GFPEPS is \emph{injective} if there is a finite blocking size such
that $\mathrm{rank}(B_\Box) = n_{\mathrm{vir}}$, i.e., the virtual system
$G_{\mathrm{in} \Box}(\mathbf k)$ is fully mapped onto the physical space.
In this case we can use an SVD of $B_\Box = \mathbb{V}^\top \Sigma U$,
where $\mathbb{V}$ is an isometry, $\mathbb{V} \mathbb{V}^\top =
\Id_{n_{\mathrm{vir}}}$, and $\Sigma$ is a diagonal strictly
positive matrix, to obtain from Eq.~\eqref{eq:Gout} $\mathbb{V}
\left(G_{\mathrm{out}\Box}(\mathbf k) - A_\Box \right) \mathbb{V}^\top =
\Sigma U (D_\Box - G_{\mathrm{in} \Box}(\mathbf k))^{-1} U^\top \Sigma$,
which implies
\begin{align}
\det\left(\mathbb{V} \left(G_{\mathrm{out}\Box}(\mathbf k) - A_\Box \right) \mathbb{V}^\top\right)
= \frac{\det^2(\Sigma)}{\det(D_\Box - G_{\mathrm{in} \Box}(\mathbf
k))}\ .
\label{determinants}
\end{align}
Since all terms on the left hand side arise from entries of CMs and are thus bounded, it
follows that {$q_\Box(\mathbf k):= \det(D_\Box - G_{\mathrm{in}\Box}(\mathbf k))
\ge \delta>0$} [in particular, the parent Hamiltonian of the blocked GFPEPS
with {$\varepsilon(\mathbf k)= q_\Box(\mathbf k)$} in eq. \eqref{parent_Ham} is gapped and local].

It is now exactly this property which allows us to construct a gapped
interpolation from $G_{\mathrm{out}\Box} (\mathbf k)$ to the
topologically trivial state by adiabatically disentangling pairs of Majorana bonds (we can take $n_v$, $n_h$ to be even, since injectivity is stable under blocking) via
\begin{align}
\Gamma_{\mathrm{in}}^\varphi = \left(\begin{array}{cc}
W \sin\varphi & \Id_2\cos\varphi\\
-\Id_2\cos\varphi & W \sin\varphi
\end{array}\right)\ .
\end{align}
Here, $\Gamma_{\mathrm{in}}^\varphi$ is the CM of pairs of Majorana bonds
connecting horizontally or vertically adjacent sites, which for $\varphi = 0$
describes a maximally entangled state corresponding to the initial
GFPEPS, while $\Gamma_\mathrm{in}^{\pi/2}$ corresponds to a product state
and thus $G_{\mathrm{out}\Box}^{\pi/2}$ describes a topologically trivial
state (note that this construction does not work in one dimension if the number of Majorana bonds is odd, such as for the Kitaev chain~\cite{Kitaev_chain}). Since from Eq.~\eqref{determinants} for $G_\mr{out}(\mb k) \rightarrow G_\mr{out}^\varphi(\mb k)$, $\det(D_\Box - G^\varphi_{\mathrm{in}\Box}(\mathbf k)) > 0$ for all
$\varphi \in [0, \tfrac{\pi}{2}]$, this interpolation corresponds
to a smooth gapped local Hamiltonian, showing that any injective GFPEPS is
in the trivial phase.


This argument can be generalized to the case where
$\mathrm{rank}(B_\Box)<\mathrm{n_\mathrm{vir}}$ (i.e., the state is
non-injective), as long as $n_{\mathrm{vir}} \leq n_{\mathrm{ph}}$ and
{$q_\Box(\mb k) = \det(D_\Box - G_{\mathrm{in} \Box}(\mathbf k)) > 0$}.  In this case,
define $\Delta := \mathrm{min}_{\mathbf k} \det(D_\Box - G_{\mathrm{in} \Box
}(\mathbf k))$. It is always possible to rotate $\gamma_{1\Box}$ into ${\gamma'_{1\Box}}
= e^{- \epsilon Z} \gamma_{1\Box} e^{\epsilon Z}$ with some appropriate real $Z = - Z^\top$ to
obtain $\mathrm{rank}(B_\Box) = n_{\mathrm{vir}}$, while keeping
{$\det(D'_\Box - G_{\mathrm{in} \Box}(\mathbf k)) > 0$} if $\epsilon$
is sufficiently small compared to $\Delta$. From there, it is again
possible to perform an adiabatic evolution to the trivial state as before.
If the initial GFPEPS was particle number conserving, this symmetry can be
kept along the path by using a particle number conserving interpolation
$\Gamma_{\mathrm{in}}^\varphi$, cf., subsection~\ref{sec:GFPEPS_sym}. 
Thus, our proof applies both to topological insulators and topological superconductors.


\subsection{Full solution for $\chi=1$}\label{sec:full-chi1}

In this subsection, we use the recursion relation
(\ref{eq:general-D-recursion}) to explicitly
derive the boundary and edge theories for GFPEPS with one Majorana mode
per bond, $\chi=1$. We will then use this result to show that the presence
of chiral edge modes is related to the occurrence of Majorana modes maximally
correlated between the two edges, i.e., a fermionic mode in a pure state
shared between the two edges.

We start by deriving a closed expression for the boundary and edge
Hamiltonian for $\chi=1$.  In this case,
\be
\hat D = \left(\begin{matrix} i \hat r & i\hat s \\
    i\hat s^* & i\hat t\end{matrix}\right), \label{eq:D-column}
\ee
with scalar functions $\hat r\equiv \hat r(k_y)\in \mathbb R$, $\hat
t\equiv \hat t(k_y)\in \mathbb R$, and $\hat s\equiv \hat s(k_y)$.  Note
that for given $k_y$, the eigenvalues need not to come in complex conjugate
pairs. However, they are still bounded by one, which implies that for
$\hat r\,\hat t\ge0$,
\begin{equation}
\label{eq:sqrtacb-eq}
1-\sqrt{\hat r\,\hat t} \ge |\hat s|
\mbox{\ with equality iff $|\hat s|=1$,}
\end{equation}
which in turn implies that for all $\hat r$ and $\hat t$,
\begin{equation}
\label{eq:acb-eq}
1-\hat r\,\hat t \ge |\hat s|
\mbox{\ with equality iff $|\hat s|=1$.}
\end{equation}
(For $\hat r\ge0$ and $\hat t\ge0$, Eq.~\eqref{eq:sqrtacb-eq} follows from
$2\ge -i\,(1,\tfrac{\hat s}{|\hat s|})\,\hat D_1 \,(1,\tfrac{\hat s}{|\hat
s|})^\dagger = \hat r + \hat t +2 |\hat s| \ge
2\sqrt{\hat r\hat t} +2|\hat s|$,  and similarly for
$\hat r\le0$ or $\hat t\le0$.)

Let us now study what happens if we concatenate cylinders; for the CM of
$N$ columns, we will write $\hat D_N$ and $\hat r_N$, $\hat
s_N$, and $\hat t_N$.  The iteration relation
(\ref{eq:general-D-recursion})
yields the following iteration relations for the matrix elements:
\begin{subequations}
\label{eq:onemode-recursion}
\begin{align}
\hat r''  & = \hat r + \hat r'\, \frac{|\hat s|^2}{1-\hat r'\hat t},
    \label{eq:onemode-recursion-a}\\
\hat t''  & = \hat t' + \hat t\, \frac{|\hat s'|^2}{1-\hat r'\hat t},
    \label{eq:onemode-recursion-c}\\
\hat s''  & = \frac{\hat s\hat s'}{1-\hat r'\hat t}.
    \label{eq:onemode-recursion-b}
\end{align}
\end{subequations}
For $\hat z= \hat z'= \hat z_N$ ($\hat z= \hat r, \hat s, \hat t$), and $\hat z''= \hat z_{2N}$, with $N$ a power of
$2$ (i.e., doubling the number of columns in each step), we obtain
\begin{subequations}
\begin{align}
\hat r_{2N}  & = \hat r_N\, (1+\hat \xi_N),\\
\hat t_{2N}  & = \hat t_N\, (1+\hat \xi_N),\\
\hat s_{2N}  & = \frac{\hat s_N}{\hat s_N^*} \,\hat \xi_N,
\end{align}
\end{subequations}
with
\begin{equation}
\hat \xi_N = \frac{|\hat s_N|^2}{1-\hat r_N \hat t_N}.
\end{equation}
Assume for now $|\hat s_N|<1$: Then, \eqref{eq:acb-eq}
$\Rightarrow\;\hat \xi_N<1\;\Rightarrow \;  |\hat s_{2N}|<1$, and thus
$|\hat s_1|<1$ implies  $|\hat s_N|<1$. Moreover, Eq.~\eqref{eq:acb-eq}
implies $|\hat s_{2N}|<|\hat s_N|$, which in turn implies that $|\hat
s_N|$ converges; similarly, since $\hat \xi_N\ge0$, $\hat
r_N$ and $\hat t_N$ monotonously move away from zero and thus
converge. We therefore find that for $|\hat s_N|<1$, all matrix elements
converge.

On the other hand, $|\hat s_1|=1$ implies that $\hat r_1=\hat t_1=0$ (as
$\hat D_1$ must have spectral radius $\le 1$), and thus $\hat r_\infty=\hat
t_\infty=0$, while $|\hat s_\infty|=1$.  An explicit analysis of the
possible $\hat D_1$ for $\chi=1$, using Eq.~\eqref{eq:hat-D1-fromHVK},
shows that $\hat r_1(k_y)=\hat t_1(k_y)=0$ can only be the case for
$k_y=0$ or $k_y=\pi$, unless both vanish identically (in which case the
fixed point and the GFPEPS are trivial).

In order to determine the fixed point for $|\hat s_1| < 1$ ($N\rightarrow\infty$),
 we now consider the scenario where
$\hat z=\hat z''= \hat z_\infty$ and $\hat z'= \hat z_1$, i.e., where we append a single column to an
infinite cylinder. From (\ref{eq:onemode-recursion-a}), we find that
\[
\hat r_\infty = \hat r_\infty + \
    \hat r_1 \frac{|\hat s_\infty|^2}{1-\hat r_1\hat t_\infty}
\]
and thus $\hat s_\infty=0$ for $\hat r_1\ne0$ (and similarly if $\hat
t_1\ne 0$); if $\hat r_1 = \hat t_1 =0$,
(\ref{eq:onemode-recursion-b}) yields $\hat s_\infty=\hat s_\infty \hat
s_1$ which as well implies $\hat s_\infty=0$ as long as $|\hat s_1|<1$, otherwise $|\hat s_\infty| = 1$. Hence, let us assume $\hat s_\infty = 0$. On the
other hand, Eq. (\ref{eq:onemode-recursion-c})
yields a quadratic equation for $\hat t_\infty$,
\be
    \label{eq:cinfty-quad-eq}
\hat r_1\hat t_\infty^2 -(1+\hat r_1\hat t_1-|\hat s_1|^2)\hat t_\infty
+\hat t_1=0,
\ee
and similarly for $\hat r_\infty$ by exchanging $\hat r$ and $\hat t$.
 Of the two solutions
\[
\hat t_\infty^{\pm} = \frac{\hat\Delta_1 \pm
\sqrt{\hat\Delta_1^2-4\hat r_1\hat t_1}}{2\hat r_1}
\]
[where $\hat\Delta_1 =
1+\hat r_1\hat t_1 -|\hat s_1|^2 = \det(i\hat D_1)+1\ge 0$], the fixed
point is always given by $\hat t_\infty^-$.  This is seen by noting that
$-1\le\hat t_\infty^+\le1$ implies that $\pm 2\hat r_1-\hat
\Delta_1\ge\sqrt{\hat\Delta_1^2-4\hat r_1\hat t_1}$ (with $\pm$ the sign
of $\hat r_1$). Squaring yields $0\ge(\hat r_1\mp1)(\hat
t_1\mp1)-|\hat s_1|^2 = \det(i\hat D_1\mp\Id)$, and thus $t_\infty^+$
can only be physical if $i\hat D_1$ has an eigenvalue $\pm1$; and these
remaining cases can be easily analyzed by hand.
We thus find that the fixed point CM is of the form
\[
\hat D_\infty = \left(\begin{matrix} i \hat r^-_\infty
& 0 \\ 0 & i\hat t^-_\infty\end{matrix}\right)\ ,
\]
except when $|\hat s_1|=1$, which we found can only happen at $k_y=0,\pi$
(in which case $\hat s_1(k_y)$ is real).

In order to obtain the boundary theory, we need to combine the expression
for $\Sigma_N$, Eq.~(\ref{eq:symmetrized-es}), with the fact
that $\hat G$ and $\hat{\bar G}$ are given by $\hat G=i\hat r_\infty\oplus i \hat t_\infty$
and $\hat{\bar G} = i \hat t_\infty\oplus i \hat r_\infty$, with the exception of the
singular points in $k_y$-space where $|\hat s_\infty|=1$. In particular, the
two boundaries can be described independently almost everywhere, and we
obtain for the edge theory of the right edge ($\hat t_\infty = \hat t_\infty^-$, $\hat r_\infty = \hat r_\infty^-$)
\begin{align}
\hat \Sigma^{R}_\infty(k_y)  = -i\hat t_\infty-i(1-\hat t_\infty^2)
    (\hat t_\infty - \hat r_\infty^{-1})^{-1} = i\frac{\hat r_1-\hat t_1}{\sqrt{ {\hat\Delta}_1^2-4\hat r_1 \hat t_1}}, \label{eq:sigma_es_chi1}
\end{align}
with the boundary Hamiltonian given by $\hat H^R_\infty(k_y)=
2\arctan(\hat \Sigma_\infty^R(k_y))$;
for the
opposite edge, $\hat r$ and $\hat t$ need to be interchanged.
For the points with $|\hat s_1|=|\hat s_\infty|=1$, on the other hand,
the two boundaries are in a maximally entangled state of the
Majorana modes with the corresponding $k_y$.

Clearly, $\hat \Sigma_{\infty}^R(k_y)$ [Eq.~(\ref{eq:sigma_es_chi1})] is
continuous unless the denominator becomes zero.  For the latter to happen,
one first needs that $\hat r_1\hat t_1\ge 0$, and with this, $\hat
\Delta_1^2-4\hat r_1\hat t_1=0$ is equivalent to $1-\sqrt{\hat r_1\hat
t_1}=|\hat s_1|$, which using Eq.~(\ref{eq:sqrtacb-eq}) implies that
$|\hat s_1|=1$, which can only be the case for $k_y=k_y^0=0,\pi$. In order
to analyze how $\hat \Sigma^{R}_\infty(k_y)$
behaves around such a point, we expand to first order in $\delta
k_y=k_y-k_y^0$:
Then, $\hat r_1=\hat r_1'\,\delta k_y+O(\delta k_y^2)$, $\hat t_1=\hat
t_1'\,\delta k_y+O(\delta k_y^2)$, and $|\hat s_1|=1+O(\delta k_y^2)$
(since $|\hat s_1|\le 1$). One immediately finds that
\begin{align*}
\hat \Sigma^{R}_\infty(k_y) = i\frac{(\hat r_1'-\hat t_1')\delta k_y + O(\delta k_y^2)}{
    \sqrt{-4\hat r_1'\hat t_1'\,\delta k_y^2 + \mc O(\delta k_y^4)}}
 =i\,\mathrm{sgn}(\delta k_y)\,
    \frac{\hat r_1'-\hat t_1'}{\sqrt{-4\hat r_1'\hat t_1'}}+ \mc O(\delta k_y),
\end{align*}
this is, $\hat \Sigma^{R}_\infty(k_y)$ exhibits a discontinuity unless $\hat
r_1'=\hat t_1'$. In order to relate $\hat r'_1$ and $\hat t'_1$, we
observe that the eigenvalues of $\hat D_1$ around $k_y^0$ are $i(\pm
1+\tfrac12(\hat r_1'+\hat t_1') \delta k_y +O(\delta k_y^2))$, and thus $\hat r_1'+\hat
t_1'=0$, which implies that
\[
\hat \Sigma^{R}_\infty(k_y) = i\,\mathrm{sgn}(\delta k_y) \,
    \mathrm{sgn}(\hat r_1'(k_y^0)).
\]
Therefore, the edge Hamiltonian exhibits a jump between $\pm1$, and the
boundary Hamiltonian derived from the entanglement spectrum diverges, as
we have seen in the examples.  The case of vanishing
first order terms, $\hat r_1'= \hat t_1'=0$, can be dealt with using the
explicit form of $\hat D_1$ for $\chi=1$, which yields that $\hat r_1=\hat
t_1=0$ vanish identically for all $k_y$, making the fixed point trivial; 
if $\hat r'_1$ changes its sign, this corresponds to a transition point
between $C=+1$ and $C=-1$.  Note that according to
Eq.~\eqref{eq:hat-D1-fromHVK},  $\hat r_1=\hat t_1=0 \ \forall \ k_y$ happens if and only
if $K$ is either diagonal or off-diagonal (as the other terms are
antihermitian $2\times 2$ matrices). This means that the virtual CM $D$
does not couple the left with the down Majorana mode and the right 
with the up Majorana mode (or the other way around), i.e., the GFPEPS is topologically trivial.

We thus find that $|\hat s_1(k_y^0)|=1$ at $k_y^0=0$ or $k_y^0=\pi$ is
equivalent to having a discontinuity in the edge Hamiltonian, which jumps
between $\pm 1$. Since $H^\mathrm{e}_N$ is otherwise continuous, and we
will see that for $\chi=1$, $|\hat s_1(k_y^0)|=1$ can occur for at most one
$k_y$ (see subsection~\ref{sec:necessity}),  it follows that $|\hat
s_1(k_{y}^0)|=1$, i.e., the existence of a maximally entangled mode
between the left and right edge of the cylinder 
at $k_y = k_y^0$ is equivalent to having a chiral mode at the edge.


\subsection{Symmetry and chirality}\label{sec:sym_chiral}

As we have seen in the preceding subsection, for $\chi = 1$, the existence of a
chiral edge mode is equivalent to the existence of a maximally entangled
Majorana mode
between the left and right edge of the cylinder at $k_y^0=0$
or $k_y^0=\pi$. In the following, we will show that this mode can be
understood as arising from a \emph{local} symmetry of the state $\Psi_1$
which defines the GFPEPS [Eq.~\eqref{eq:Gaussian-psi1}].

Concretely, in subsection~\ref{sec:sufficiency-symmetry} we will demonstrate that a certain symmetry of
$\Psi_1$ leads to a maximally entangled Majorana pair between the left and
right edge and thus to a chiral edge state. In subsection~\ref{sec:necessity} we will show the
opposite -- that a maximally entangled Majorana pair between the left and
the right implies $\Psi_1$ having a certain symmetry.  In the subsequent subsection, we
uncover these kinds of symmetries in the examples presented in subsection~\ref{sec:examples}. In subsection~\ref{sec:symmetry-GS}, we consider again the example given by
Eq.~\eqref{eq:Psi1ex} and outline how strings of symmetry operators can be
used to construct all ground states of its frustration free parent
Hamiltonian $\mc H_\mr{ff}$.

We will generally restrict the discussion in this subsection to the case of
$\chi=1$ Majorana mode per bond, though some of the results (in particular
those of the following subsection) readily generalize to larger $\chi$.

\subsubsection{Sufficiency of local symmetry}\label{sec:sufficiency-symmetry}

\begin{figure}[!hb]
\begin{center}
\includegraphics[width=0.55\textwidth]{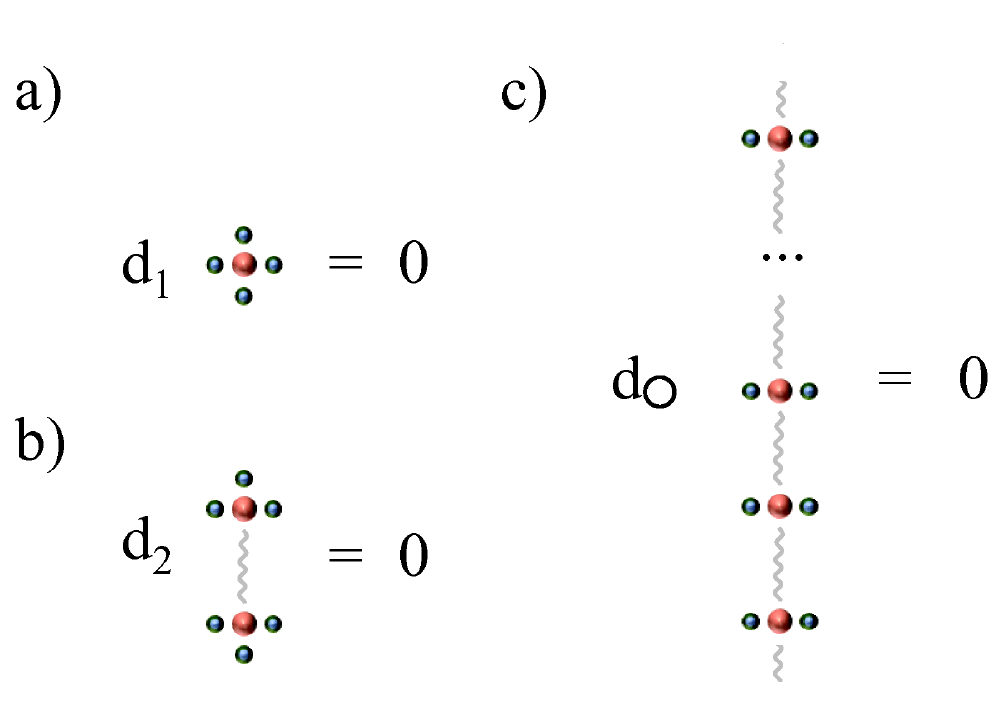}
\caption{Concatenation of a symmetry. (a) Symmetry $d_1$ annihilating the state $\Psi_1$ of virtual and physical Majorana modes on one site. For a chiral state with $\chi = 1$ it can be concatenated as described in the text to a symmetry $d_{2}$ annihilating the state $\Psi_2$ defined on two sites (b). Proceeding in the same manner and closing the vertical boundary, one obtains a symmetry $d_\bigcirc$ annihilating one column $\Phi_1$ (c). This figure was reproduced from Ref.~\cite{cPEPS_long}. \label{fig:symmetry}}
\end{center}
\end{figure}

We start by showing how a symmetry in $\Psi_1$ induces a symmetry on a
whole column, $\Phi_1$, and how this subsequently gives rise to a maximally
correlated mode between the two edges of a cylinder.
Since $\Psi_1$ is a pure Gaussian state
in which four virtual Majorana modes are entangled with one physical
fermionic mode, there must be a virtual fermionic mode which is in the vacuum, i.e., some
\begin{equation}
\label{eq:d-sym-def}
d_1=\alpha_L c_L +\alpha_R c_R + \alpha_U c_U + \alpha_D c_D
\end{equation}
acting on the virtual system which annihilates $\Psi_1$,
\begin{equation}
\label{eq:P-annih-d}
d_1\,\ket{\Psi_1} = 0,
\end{equation}
as already indicated in subsection~\ref{sec:symm-top}.
[$d_1$ corresponds to the eigenvector of $D$, Eq.~\eqref{eq:gamma1},
with eigenvalue $-i$, and describes a fermionic mode].  We will refer
to $d_1$ as a \textit{symmetry}, since it corresponds to a $\mathbb{Z}_2$
symmetry of $\Psi_1$ with $U_1 = \Id - 2 d_1^\dg d_1$.
On the other hand, for the virtual fermionic bonds
$\omega_{12}$ (the indices denoting the vertical positions),
Eq.~\eqref{eq:omegas}, it
holds that $\bra{\omega_{12}}(1-i c_{1,D} c_{2,U})=0$, and thus,
\begin{equation}
\label{eq:annih-omega}
\bra{\omega_{12}}(c_{1,D}+i c_{2,U})=0.
\end{equation}
By combining Eqs.~\eqref{eq:P-annih-d} and Eq.~\eqref{eq:annih-omega}, we can now
study how the symmetry (\ref{eq:d-sym-def}) behaves if we concatenate
two or more sites by projecting onto $\bra{\omega_{12}}$
(we assume $\alpha_U\ne0$ for now and define
$\theta:=i\alpha_D/\alpha_U$):
\begin{align*}
0   &=   \bra{\omega_{12}}\Big[
	(\alpha_L c_{1,L} + \alpha_R c_{1,R} + \alpha_U c_{1,U} + \alpha_D c_{1,D})|\Psi_1,\Psi_1\rangle_{1,2} \\
    &\qquad+ \theta
    (\alpha_L c_{2,L} + \alpha_R c_{2,R} + \alpha_U c_{2,U} + \alpha_D c_{2,D})|\Psi_1,\Psi_1\rangle_{1,2} \Big] 
\\
&=  d_2 \langle\omega_{12} \ket{\Psi_1, \Psi_1}_{1,2} \ \equiv \
d_2\ket{\Psi_2}
\end{align*}
with
\[
d_2=\alpha_L (c_{1,L} + \theta c_{2,L})
         + \alpha_R (c_{1,R} + \theta c_{2,R})
	 + \alpha_U c_{1,U}
	 +\theta \alpha_D c_{2,D}
\]
the symmetry of the concatenated state $\Psi_2$, see
Figs.~\ref{Fig:PEPS}b and~\ref{fig:symmetry}b.  The argument can be
easily iterated, and we find that
\[
d_{N_v} \ket{\Psi_{N_v}} = 0
\]
with
\begin{align*}
d_{N_v} =
    \alpha_L \sum_{y=1}^{N_v}\theta^{y-1} c_{y,L}
    +\alpha_R \sum_{y=1}^{N_v}\theta^{y-1} c_{y,R}
	 + \alpha_U c_{1,U} +\theta^{N_v-1} \alpha_D c_{N_v,D}.
\end{align*}
Let us now see what happens if we close the boundary between sites $N_v$
and $1$, which yields $\ket{\Phi_1}\equiv \langle\omega_{N_v,1}\ket{\Psi_{N_v}}$, Fig.~\ref{Fig:PEPS}d: Since
$\bra{\omega_{N_v,1}}(c_{N_v,D}+i c_{1,U})=0$, we find that
\be
\d_{\bigcirc}\ket{\Phi_1} = 0 \label{eq:d-column}
\ee
with
\begin{equation}
\label{eq:d-column-2}
d_{\bigcirc}=
    \alpha_L \sum_{y=1}^{N_v}\theta^{y-1} c_{y,L}
    +\alpha_R \sum_{y=1}^{N_v}\theta^{y-1} c_{y,R}
\end{equation}
(Fig.~\ref{fig:symmetry}c) whenever $\theta^{N_v}=1$.  This leads to two
requirements for the existence of $d_\bigcirc$ fulfilling
Eq.~\eqref{eq:d-column}: First, $|\alpha_U|=|\alpha_D|$, and
second, the momentum $k_y^0$ of $d_\bigcirc$ (defined via $e^{ik_y^0} =
\theta$) must be commensurate with the lattice size.  Whenever these
requirements are fulfilled, we thus find that the local symmetry $d_1$,
Eq.~\eqref{eq:d-sym-def}, gives rise to a symmetry $d_{\bigcirc} \propto
\alpha_L \hat c_{L,k_y^0} + \alpha_R  \hat c_{R,k_y^0}$,
Eq.~\eqref{eq:d-column}, on the whole column (i.e., on $\hat D_1$), at
momentum $e^{ik_y^0} = i \alpha_D/\alpha_U$.  Note that we only need to
assume that either $\alpha_U$ or $\alpha_D$ is non-zero; if both are zero,
the condition \eqref{eq:P-annih-d} implies that the horizontal virtual modes
entirely decouple from the physical system, and the GFPEPS describes a
product of one-dimensional vertical chains.

We have thus demonstrated that a certain local symmetry induces a symmetry on one
column $\Phi_1$, which forces the Majorana modes with a specific momentum on
both ends of the column to be correlated.  This is equivalent to demanding
that for this $k_y = k_y^0$, $\hat D_1(k_y^0)$ has an eigenvalue $-i$. For
$k_y^0=0,\pi$, this implies that $|\hat s_1(k_y^0)|=1$, as the diagonal
elements of $\hat D_1(k_y^0)$ are zero due to $\hat D_1(-k_y) = \hat
D_1^*(k_y)$.

The symmetry of a single column is passed on when concatenating columns, this
is, when going from $\Phi_1$ to $\Phi_N$, Fig.~\ref{Fig:PEPS}e,f, in
analogy to the arguments given before. In order for this to lead to a
coupling between the two edge modes in the limit of an infinite cylinder,
as observed in the examples with chiral edge modes, it is additionally
required that $|\alpha_L|=|\alpha_R|$. Otherwise, the symmetry becomes
localized at a single boundary.  This can be understood by exchanging
horizontal and vertical directions, leading to $e^{i k_x^0} = i \alpha_R /
\alpha_L$ (and $k_x^0 = 0, \pi$, too).  As we have seen in the last
subsection, a coupling between the left and the right edge for $\chi = 1$ can
only emerge, if $k_y^0 = 0, \pi$ (and analogously $k_x^0 = 0, \pi$). Thus,
we have to require $\alpha_D / \alpha_U = \pm i$, $\alpha_R / \alpha_L =
\pm i$ for a symmetry leading to a chiral edge state. Since there can only
be one such symmetry for $\chi = 1$ (otherwise the virtual and physical
system decouple), we conclude that there can be a maximally entangled
 Majorana mode only for $k_y^0 = 0$ or $k_y^0 = \pi$, 
but not for both of them (and similarly for $k_x$). 
We thus find that $d_1$ must be of the form
\begin{equation}
d_1 = \alpha_L(c_L \pm i c_R) + \alpha_U (c_U \pm i c_D) \label{d_max_ent}
\end{equation}
($\alpha_L, \alpha_U \neq 0$)
in order to be stable under concatenation.

Let us finally show that in order to have a non-trivial Chern number,
there has to be an additional constraint on $\alpha_L$ and $\alpha_U$, namely
that
\be
\label{eq:pm1-pmi}
\mathrm{arg}(\tfrac{\alpha_L}{\alpha_U}) \not\in \{0,\pi, \pm \tfrac\pi2\}
\ee
This can be directly verified by explicitly constructing $D$ (given $\hat
d_1$, the only remaining freedom is the eigenvalue of the non-pure mode),
where one finds that the diagonal (off-diagonal) elements of $K$
[cf.~Eq.~\eqref{eq:CM-singlesite-virtonly-def}] vanish exactly if
$\mathrm{arg}(\tfrac{\alpha_L}{\alpha_U}) = 0,\pi$ 
[$\mathrm{arg}(\tfrac{\alpha_L}{\alpha_U}) = \pm\tfrac\pi2$]. As we have
seen in the last subsection, this in turn is equivalent to a trivial
(completely flat) edge spectrum, and thus to a vanishing Chern
number. 

In summary, we find that the Chern number is non-trivial whenever there is 
exactly one symmetry $d_1$ which satisfies Eqs.~\eqref{d_max_ent} and
\eqref{eq:pm1-pmi}.

\subsubsection{Necessity of on-site symmetry}\label{sec:necessity}

Let us now show the converse statement of the previous subsection: We
will prove that for $\chi=1$ a maximally entangled Majorana pair between
the left and the right boundary of a cylinder at
$k_y^0=0,\pi$, which is equivalent to the presence of a chiral edge mode (assuming a non-trivial boundary spectrum),
implies the existence of a local symmetry of the form Eq.~\eqref{d_max_ent}.

Following the results of subsection~\ref{sec:full-chi1}, the presence of a
maximally entangled Majorana pair on the boundary of 
a cylinder of arbitrary length is
equivalent to the presence of the symmetry on a single column (i.e., a
cylinder of length $N=1$), that is,
\begin{equation}
\hat D_1 (k_y^0) = \left(\begin{array}{cc}
0&\pm 1 \\
\mp 1&0
\end{array}\right)\ .
\label{eq:D1_fixpoint}
\end{equation}
According to Eq. \eqref{eq:hat-D1-fromHVK}, we also have
\begin{equation}
\hat D_1 (k_y^0) = H + K\left(V-\left(\begin{array}{cc}
0&\pm 1\\
\mp 1&0
\end{array}\right) \right)^{-1} K^\top,
\label{eq:D1_realmap}
\end{equation}
where the upper sign is for $k_y^0 = 0$ and the lower one for $k_y^0
= \pi$ (and is unrelated to the sign in Eq.~\eqref{eq:D1_fixpoint}). We
choose in both cases the upper sign; the other cases can be treated
analogously. Then, Eq.~\eqref{eq:D1_realmap} tells us that $\langle
\omega_v| \Psi_1 \rangle$ (with $\langle \omega_v|$ corresponding to the
projection on $\tfrac12(1 + i c_D c_U)$) is in a maximally entangled state of
the two horizontal Majorana modes. This maximally entangled state fulfills
\be
\langle \omega_h, \omega_v | \Psi_1 \rangle = 0 \label{eq:proj-zero}
\ee
with $\langle \omega_h |$ corresponding to the projection on $\tfrac12(1 +
i c_R c_L)$.

We now parameterize the reduced density matrix of the virtual system $\rho_\mr{vir}$ in the basis \\
$\{|\Omega_\mr{vir} \rangle, |\omega_h\rangle, |\omega_v\rangle, |\omega_h, \omega_v\rangle\}$ ($|\Omega_\mr{vir} \rangle$ denoting the vacuum of the virtual particles). According to Eq.~\eqref{eq:proj-zero} its matrix representation is
\begin{equation}
\rho_\mr{vir} = \frac{1}{4} \left(\begin{array}{cccc}
\rho_{00}&0&0&0\\
0&\rho_{hh}&\rho_{hv}&0\\
0&\rho_{hv}^*&\rho_{vv}&0\\
0&0&0&0
\end{array}\right). \label{eq:Rho-proj}
\end{equation}
From it, we can calculate the elements of $D$ via $D_{pq} =
\frac{i}{2} \tr\left(\rho_\mr{vir} [c_p, c_q]\right)$ with
$p, q = L, R, U, D$, cf.~Eq.~\eqref{eq:CM}, and obtain
\be
D = \left(\begin{smallmatrix}
0&\rho_{00}-\rho_{hh}+\rho_{vv}&2\, \mr{Im}(\rho_{hv})&-2\, \mr{Re}(\rho_{hv})\\
-\rho_{00}+\rho_{hh}-\rho_{vv}&0&2\, \mr{Re}(\rho_{hv})&2\, \mr{Im}(\rho_{hv})\\
-2\, \mr{Im}(\rho_{hv})&-2\, \mr{Re}(\rho_{hv})&0&\rho_{00}+\rho_{hh} - \rho_{vv}\\
2\, \mr{Re}(\rho_{hv})&-2\, \mr{Im}(\rho_{hv})&-\rho_{00}-\rho_{hh}+\rho_{vv}&0
\end{smallmatrix}\right).  \label{eq:D-projected}
\ee
The fact that $\rho_\mr{vir}$ describes a Gaussian state is used by
inserting this into Eq.~\eqref{eq:D1_realmap}, which results in
\be
\rho_{00} = 1 - \sqrt{(\rho_{hh} - \rho_{vv})^2 + 4 |\rho_{hv}|^2}.
\ee
Given this restriction, one can check that $D$ in
Eq.~\eqref{eq:D-projected} has an eigenvalue $-i$ with the corresponding
symmetry $d_1 = \alpha_L(c_L - i c_R) + \alpha_U (c_U - i c_D)$ fulfilling
Eq.~\eqref{eq:P-annih-d}, where $\alpha_L = 2 \rho_{hv}^*$, $\alpha_U =
-\rho_{hh} + \rho_{vv} - \sqrt{(\rho_{hh} - \rho_{vv})^2 + 4
|\rho_{hv}|^2}$. After considering all possible sign cases in
Eqs.~\eqref{eq:D1_fixpoint} and \eqref{eq:D1_realmap}, one arrives at Eq.~\eqref{d_max_ent}.
We thus find that for a GFPEPS with $\chi=1$, a (unique) symmetry of this
form with $\mathrm{arg}(\tfrac{\alpha_L}{\alpha_U})\notin \{0,\pi,\pm \tfrac\pi2\}$ and $\alpha_L, \alpha_U \neq 0$
is both necessary and sufficient to give rise to a divergence in the boundary
spectrum, and thus a Chern number $C = \pm 1$.  The states simultaneously
fulfilling Eq.~\eqref{d_max_ent} and
$\mathrm{arg}(\tfrac{\alpha_L}{\alpha_U})\in \{0,\pi,\pm \tfrac\pi2\}$, on
the other hand,  are the transition points between
GFPEPS with Chern number $C = -1$ and $C = +1$.

\subsubsection{Symmetries in the considered examples}\label{sec:symm-examples}

We will now study the symmetries in the examples given in
subsection~\ref{sec:examples} and relate them to chiral edge modes in the light
of the results of the previous subsections.

\paragraph{Chern insulator with Chern number $C = -1$.}

For the Chern insulator introduced in subsection~\ref{sec:Chern}, we consider
only one copy of the two superconductors constituting the Chern insulator.
Doing so is trivial on the virtual level, since the matrix $D$ is
block-diagonal with two identical blocks. Each of those blocks has an
eigenvalue $-i$ corresponding to the symmetry
\be
d_1 = -e^{-i\frac{\pi}{4}} c_L - e^{i\frac{\pi}{4}} c_R - i c_U + c_D \label{eq:symm-Chern}
\ee
for any $\eta \in (0,1)$.
Thus, the state $\Psi_1$ possesses two symmetries, $d_1^{(1,2)}$, with
$d_1^{(1)}\ket{\Psi_1}=d_1^{(2)}\ket{\Psi_1}=0$.  Both of them are the form
\eqref{eq:symm-Chern} with one containing only the first type of Majorana modes and the other only the second type.

\paragraph{GFPEPS with Chern number $C = 2$.} 

Let us now consider the topological superconductor with Chern number $C =
2$ introduced in subsection~\ref{sec:Chern2}; in the following, all equalities
are to be understood up to numerical accuracy.  By diagonalizing the
CM $D$ of the virtual system, one obtains two
linearly independent eigenvectors with eigenvalue $-i$,
 i.e., there exist two operators
$d_1^{(1,2)}$ such that $(x_1 d_1^{(1)}+x_2 d_1^{(2)})\ket{\Psi_1}=0$ for
all $x_i$. In order to find a basis $d_1^{(\pm)}$ of operators which
reveals the symmetries of the model, we start from the state
$\Phi_1$ of one column, which has zero modes at momenta
$k_y=\pm1$. We first focus on the symmetry at $k_y = k_y^0=1$, where
we find that the horizontal virtual modes of $\Phi_1$ at momentum $k_y^0$ are
annihilated by an operator $\sum_{\kappa=1}^2 \alpha^{(+)}_{L,\kappa}
[\hat c_{k_y^0,L,\kappa} - i e^{i k_x^0} \hat c_{k_y^0,R,\kappa}]$
with $k_x^0=-2.58$.
This suggests to try to construct a $d^{(+)}_1$ which contains the above
operator: it turns out that $x_1 d_1^{(1)}+x_2 d_1^{(2)}$ indeed contains
an operator of this form, which at the same time acts on the vertical
modes as $\sum_\kappa\alpha^{(+)}_{U,\kappa}(c_{U,\kappa} - i e^{i k_y^0}
c_{D,\kappa})$. We proceed identically for $k_y = -k_y^0=-1$, and obtain a
pair of (non-orthogonal) symmetries\footnote{The values of the coefficients are
\begin{align*}
\alpha_{L,1}^{(+)} &= -0.267 + 0.422i, & \alpha_{L,2}^{(+)} &= 0.488 + 0.105i, \\
\alpha_{U,1}^{(+)} &= 0.076 - 0.251i, & \alpha_{U,2}^{(+)} &= 0.147 + 0.259i, \\
\alpha_{L,1}^{(-)} &= 0.607, & \alpha_{L,2}^{(-)} &= -0.358 - 0.052i, \\
\alpha_{U,1}^{(-)} &= -0.058 + 0.412i, & \alpha_{U,2}^{(-)} &= 0.196 + 0.406i.
\end{align*}}
\begin{align}
d_1^{(\pm)} = \sum_{\kappa=1}^2 [ \alpha^{(\pm)}_{L,\kappa} (c_{L,\kappa}
- i e^{\pm i k_x^0} c_{R,\kappa}) + \alpha^{(\pm)}_{U,\kappa} ( c_{U,\kappa} - i e^{\pm i k_y^0} c_{D,\kappa})]. \label{annihilate_pm}
\end{align}
We thus find that also for this model, the existence of divergences in the
entanglement spectrum and thus of chiral edge modes is closely related to
local symmetries of $\Psi_1$ with the corresponding momenta. Note that
since all coefficients $\alpha$ are different, the only way to grow those
symmetries following the procedure of subsection~\ref{sec:sufficiency-symmetry}
is to concatenate either exclusively $d_1^{(+)}$ or exclusively
$d_1^{(-)}$, which therefore gives rise to maximally entangled Majorana
pairs between the two boundaries with definite momenta $\pm k_y^0$ and $\pm k_x^0$,
respectively.

\paragraph{Generic GFPEPS with Chern number $C = 0$.}

Let us now consider the non-chiral family of states discussed in
subsection~\ref{sec:C0-nontrivial}. As it has only one physical mode, there must
be a symmetry $d_1$ such that $d_1\ket{\Psi_1}=0$.  It can be calculated
to be
\be d_1 = - \frac{2 i \sqrt{f(\mu)}}{2-\mu} c_L + i c_R
-\frac{2 \sqrt{f(\mu)}}{2-\mu} c_U + c_D.  
\ee
Since it is not of the form Eq.~\eqref{d_max_ent} required for chiral edge
states, the Chern number of the family is zero.

\paragraph{GFPEPS with flat entanglement spectrum and $C = 0$.}

Let us finally consider the example of subsection~\ref{sec:ex:nochern-flatspec},
which has a flat entanglement spectrum. It has a symmetry
$d_1\ket{\Psi_1}=0$ with $d_1 = c_L + c_R - i c_U - i c_D$, i.e., with
momentum $k_x^0=k_y^0=\tfrac\pi2$.  Since it is not at momentum $0$ or $\pi$,
there cannot be entangled Majorana modes between the left and the right
edge of a long cylinder.  However, since the amplitudes $\alpha_R/\alpha_L$ and $\alpha_D/\alpha_U$ have magnitude 1, the
symmetry is stable under concatenation, and must therefore still be
present in an infinite cylinder. Nonetheless, the state is topologically trivial, because in the limit
$N\rightarrow\infty$, a second symmetry at $k_y^0=\tfrac\pi2$ arises, such that on 
each edge the two modes at $k_y = \pm \tfrac\pi2$ can pair up locally.

\subsubsection{Symmetry and ground space}\label{sec:symmetry-GS}

The GFPEPS models discussed in this section appear as ground states of two
types of Hamiltonians: On the one hand, there is is the flat band
Hamiltonian $\mathcal H_\mr{fb}$, Eq.~\eqref{eq:flat}, which by
construction has the GFPEPS $\Phi$ as its unique ground state. On the
other hand, we can construct the local parent Hamiltonian $\mc H_\mr{ff}$,
Eq.~\eqref{eq:HPEPS}, which is gapless for the chiral examples considered,
i.e., for any finite 
system size, it is exactly doubly degenerate with energy splittings to higher 
energies that are inverses of polynomials of the system size.
In the following, we will
show how this ground space can be parametrized by using the virtual
symmetry $d_1$ of the local state $\Psi_1$. This is in close analogy to
the case of conventional PEPS with topological order, where the ground
space can be parametrized by putting loops of symmetry operators on the
virtual bonds in horizontal and vertical direction around the torus on
which the GFPEPS is defined.

For concreteness, we consider the example of
subsection~\ref{sec:Intro-Examples} and show how to parametrize its doubly 
degenerate ground space
in terms of strings of symmetry operators. For simplicity, we will set
$\lambda = 1/2$. Let us start by recalling Eq.~(\ref{eq:symmetries}),
which defines operators $u$, $w$, and $d_1$ such that $u\ket{\Psi_1} =
w\ket{\Psi_1} = d_1\ket{\Psi_1}=0$, where
$u=\tfrac{1}{\sqrt2}(a^\dagger-b)$ and $w=\tfrac{1}{\sqrt2}(a+b^\dagger)$,
with $a$ the physical mode, and $b=\tfrac{1}{\sqrt{2}}(h+v)$,
$d_1=\tfrac{1}{\sqrt2}(-h+v)$, with $h=\exp(i\tfrac\pi4)(c_L-i c_R)/2$ and
$v=(c_U-ic_D)/2$, cf.~Eqs.~\eqref{eq:def-hv} and \eqref{eq:d}.

Let us now consider a lattice of size $N_v \times N_h$, and concatenate all
the $\Psi_1$ in this region by projecting onto $\omega_{\mb r, \mb r + \hat y}$ and
$\omega_{\mb r, \mb r + \hat x}'$
on all the horizontal and vertical links,
respectively, but without closing either of the boundaries, resulting in a
state $\Psi_{N_v \times N_h}$. Following the arguments given in
subsection~\ref{sec:sufficiency-symmetry}, projecting onto the maximally
entangled states concatenates the symmetry operators $u$, $w$ and $d_1$,
which gives rise to three symmetries of the obtained state,
\[
\tilde u \ket{\Psi_{N_v \times N_h}} = \tilde w \ket{\Psi_{N_v\times
N_h}} = \tilde d_1 \ket{\Psi_{N_v \times N_h}}=0,
\]
where
\begin{subequations}
\label{eq:k0-syms}
\begin{align}
\tilde u & =
\tfrac{1}{\sqrt{2}}(\tilde a^\dg -\tilde b), \\
\label{eq:k0-syms-2}
\tilde w & =
\tfrac{1}{\sqrt{2}}(\tilde a + \tilde b^\dagger) , \\
\tilde d_1 & =\tfrac{1}{\sqrt{2}}(-\tilde h+\tilde v) ,
\label{eq:k0-syms-3}
\end{align}
\end{subequations}
where again
\begin{align}
\label{eq:k0-b-def}
\tilde b &= \tfrac{1}{\sqrt{2}} (\tilde h + \tilde v), \\
\nonumber
\tilde h= \tfrac{1}{2} e^{i \tfrac{\pi}{4}}(\tilde c_L-i \tilde c_R) &\mbox{\ and\ }
\tilde v= \tfrac{1}{2}(\tilde c_U-i \tilde c_D)\ ,
\end{align}
Here, $\tilde c_p = \sum_y c_{y,p}$ and $\tilde c_q = \sum_x
c_{x,q}$ ($p = L, R$, $q = U, D$) 
are the zero-momentum
(center-of-mass) modes of the virtual Majorana modes on the corresponding
boundary, and $\tilde a = \sum_\mb{r} a_\mb{r}$ is the center-of-mass mode of the
physical fermion.

In order to close the boundary, we first transform the entangled states
across the boundary to the Fourier basis (since they are translational
invariant, they are of the same form in $k$-space), and project onto all
entangled states at the boundary except those with momentum $k_x=0$ and
$k_y=0$.  This leaves us with the zero-momentum part of the state
$\Psi_{N_v \times N_h}$, which we denote by $\tilde \Psi_{N_v \times N_h}$,
where we disregard additional physical modes which are unentangled to the
boundary degrees of freedom. This state is exactly characterized by the
three symmetries of Eq.~(\ref{eq:k0-syms}), and thus,
\begin{align*}
\tilde \Pi_{N_v \times N_h} & =
    \ket{\tilde\Psi_{N_v \times N_h}} \bra{\tilde \Psi_{N_v \times N_h}} =
    \tilde d_1 \tilde d_1^\dagger \tilde u \tilde u^\dg \tilde w \tilde w^\dagger\\
    & = \tfrac{1}{4} \tilde d_1 \tilde d_1^\dagger (\tilde a^\dg -\tilde b)(\tilde a-\tilde
b^\dagger)(\tilde a+\tilde b^\dagger)(\tilde a^\dg +\tilde b)\\
    & = \tfrac{1}{2} \tilde d_1 \tilde d_1^\dagger (-\tilde a^\dg \tilde b^\dagger +
	 \tilde a^\dg \tilde a \tilde b^\dagger \tilde b +
	 \tilde b \tilde b^\dagger \tilde a \tilde a^\dg +
	 \tilde a \tilde b).
\end{align*}
Following Eq.~(\ref{eq:omegas}), the projection onto the remaining zero
momentum bonds is $\tilde \omega_h=\tfrac12(1 + i \tilde c_R \tilde c_L)$
and $\tilde \omega_v=\tfrac12(1 + i \tilde c_D \tilde c_U)$. Hence,
$\tilde h^\dagger \ket{\tilde\omega_h} = \tilde v^\dagger
\ket{\tilde\omega_v} = 0$,  and thus, using Eqs.~(\ref{eq:k0-syms-3}) and
(\ref{eq:k0-b-def}), also $\tilde d_1^\dagger
\ket{\tilde\omega_h,\tilde\omega_v} = \tilde b^\dagger\, \ket{\tilde
\omega_h, \tilde \omega_v}  = 0$, i.e.,
\begin{equation}
\label{eq:k0-bond-fockrep}
\ket{\tilde\omega_h,\tilde \omega_v} = \tilde d_1^\dagger \tilde b^\dagger
\ket{\Omega_\mathrm{vir}}
\end{equation}
(with $\ket{\Omega_\mathrm{vir}}$ the vacuum of $\tilde d_1$ and $\tilde
b$, or equivalently of $\tilde h$ and $\tilde v$; the phase can be
absorbed into $\ket{\Omega_\mathrm{vir}}$).

Let us now see what happens if we close the remaining $(k_x, k_y) = (0,0)$ boundary. 
Since $\tilde \Pi_{N_v \times N_h}$ is proportional to $\tilde d_1\tilde
d_1^\dagger$, we find
that
\[
\langle \tilde \omega_h, \tilde \omega_v| \, \tilde \Pi_{N_v \times N_h}
\, | \tilde \omega_h, \tilde \omega_v\rangle = 0
\]
-- the success probability for constructing the GFPEPS by projecting onto
entangled states is zero!  Indeed, this comes as no surprise, since the
success probability of any such projection is related to
$\sqrt{\det(D+\omega^{-1})}$ in Eq.~\eqref{eq:proj-ME}~\cite{Gaussian_Bravyi}, which
in turn is the square root of the spectral function of the parent
Hamiltonian constructed via Eqs.~\eqref{eq:HPEPS}, \eqref{eq:Ham_ex} and \eqref{eq:annihilation_ex}:
Having a gapless parent Hamiltonian requires the GFPEPS to vanish when
performing the projections. This raises the question of how to obtain a
proper PEPS description of the ground state subspace.

Fortunately, this problem can be overcome exactly by using the virtual
symmetry of $\Psi_1$. To this end, let us place a string of symmetry
operators
\[
\tilde c_L = \tfrac{1}{\sqrt{2}} e^{-i \tfrac{\pi}{4}} (-\tilde d_1+\tilde b)+\tfrac{1}{\sqrt{2}} e^{i \frac{\pi}{4}} (-\tilde d_1^\dagger+\tilde b^\dagger)
\]
at the left edge before closing the boundary, i.e., we replace the state
$| \tilde \omega_h, \tilde \omega_v\rangle=\tilde d_1^\dagger \tilde
b^\dagger \ket{\Omega_\mathrm{vir}}$ by
\[
\ket{\tilde \omega_L}=\tilde c_L | \tilde \omega_h, \tilde \omega_v\rangle =
\tfrac{-1}{\sqrt{2}} e^{-i \tfrac{\pi}{4}} (\tilde b^\dagger+\tilde
d_1^\dagger) \ket{\Omega_\mathrm{vir}}.
\]
Using that $\tilde \Pi_{N_v \times N_h}$ is proportional to $\tilde
d_1\tilde d_1^\dagger$, this immediately yields
\begin{equation}
\bra{\tilde \omega_L} \tilde \Pi_{N_v \times N_h}\ket{\tilde \omega_L} =
\tfrac12\bra{\Omega_\mathrm{vir}} \,\tilde b \,
 \tilde \Pi_{N_v \times N_h} \tilde b^\dagger \ket{\Omega_\mathrm{vir}} =
\tfrac{1}{4} \tilde a^\dg \tilde a,
\label{eq:sym-gs:gs1}
\end{equation}
i.e., this way we obtain a GFPEPS for one of the ground states of the
parent Hamiltonian $\mc H_\mr{ff}$, namely the one with the gapless
center-of-mass mode occupied.  It is easy to see that we obtain the same
result if we insert a horizontal string instead, e.g., $\tilde c_U$. In
terms of the notation introduced in subsection~\ref{sec:symm-top}, $\ket{\Phi} =
0$ and $\ket{\Phi_{{\mc C}_h}} \propto \ket{\Phi_{\mc C_v}}$. Note that
the string operators $\tilde c_L$ and $\tilde c_U$ can be deformed
without changing the state: This can be seen by consecutively using
Eq.~\eqref{eq:dPsi0} to deform the string as
\begin{align}
 \ket{\Phi_{{\mc C}_h}} &= \langle \omega_{\partial \mc R, \partial \bar{\mc R}} | \tilde c_L | \Psi_{\mc R}, \Psi_{\bar{\mc R}}\rangle = \langle \bar \omega | \tilde c_L | \bar \Psi\rangle \notag \\
 &= \langle \bar \omega | \left(\sum_{y =2}^{N_v} c_{y,L} + i c_{1,R} + e^{-i\tfrac{\pi}{4}} c_{1,U} - e^{i \tfrac{\pi}{4}} c_{1,D}\right)|\bar \Psi\rangle  \notag\\
 &= \ket{\Phi_{\mc C'_h}}
\end{align}
etc., where we defined $|\bar \Psi\rangle$ as the state of all virtual and
physical particles before any projection is applied, and $\langle \bar
\omega |$ denotes the projection on all virtual modes.

Let us now finally see what happens if we insert both a horizontal and a
vertical string: Then, we must replace
$\ket{\tilde \omega_h, \tilde \omega_v}$ by
\[
\ket{\tilde \omega_{UL}}=
\tilde c_U \tilde c_L | \tilde \omega_h, \tilde \omega_v \rangle =
-e^{-i\pi/4} |\Omega_\mr{vir}\rangle,
\]
and we find
\begin{equation}
\label{eq:sym-gs:gs2}
\bra{\tilde\omega_{UL}} \tilde\Pi_{N_v \times N_h}\ket{\tilde\omega_{UL}} =
    \bra{\Omega_\mr{vir}} \,
\tilde \Pi_{N_v \times N_h}\ket{\Omega_\mr{vir}} = \tfrac12 \tilde a \tilde a^\dg,
\end{equation}
which is the second ground state, where the gapless center-of-mass mode is
in the vacuum.

Note that the second ground state can equivalently be obtained using that
$\tilde a \ket{\tilde \Psi_{N_v \times N_h}} = -\tilde b^\dg \ket{\tilde
\Psi_{N_h \times N_v}}$ [Eq.~\eqref{eq:k0-syms-2}], i.e, applying $a \ldots a^\dg$
on the left hand side of
 Eq.~(\ref{eq:sym-gs:gs1}) exactly cancels the $b \ldots b^\dg$, and thus yields
Eq.~\eqref{eq:sym-gs:gs2}.

In summary, we found that it is possible to parametrize the two dimensional ground
state subspace of the model using the string operators given by the virtual symmetry
of $\Psi_1$: One of the ground states is obtained by inserting a single
string (either horizontally or vertically), while the other ground state is
obtained by inserting \emph{both} a horizontal and a vertical string.

\subsection{Numerical approximation of a Chern insulator with short range hoppings}\label{sec:numerics}

In this subsection, our results on the approximation of a given local Chern insulator by GFPEPS are presented. In subsection~\ref{sec:GFPEPS_sym}, it will be elaborated how to impose the particle number conservation present in the Chern insulator onto the class of our ansatz states. Subsection~\ref{sec:num_results} contains our numerical results, which include the minimization of the error of the CM as a function of the number of Majorana bonds, the minimization of the free energy as a function of temperature and the detection of a quantum phase transition by minimizing the variational energy. 

We performed numerical calculations on a $10 \times 10$ lattice for the model
\be
\mc H=\sum_{\mathbf{k}} (a_{\mathbf{k},\uparrow}^\dagger,
a_{\mathbf{k},\downarrow}^\dagger)
\,
(\mathbf{\bm\sigma}\cdot
\mathbf{d}(\mathbf{k}))
\,
(a_{\mathbf{k},\uparrow},
a_{\mathbf{k},\downarrow})^\top, \label{eq:Ham_Chern-ins}
\ee
with $\bm\sigma=(\sigma_x,\sigma_y,\sigma_z)$ the Pauli matrices, and
\begin{align}
\mathbf d (\mathbf k) = \left(\sin k_y,-\sin k_x,2 - \cos k_x - \cos k_y - e_S\right).
\end{align}
This model has Chern number $C = -1$ for $0 < e_S < 2$, $C = 1$ for $2 <
e_S < 4$ and $C = 0$ otherwise~\cite{Chern_ins}. The symmetries of the Hamiltonian~\eqref{eq:Ham_Chern-ins} are best taken advantage of by parameterizing the CM $\gamma_1$ of the Gaussian map [Eq.~\eqref{eq:gamma1}] in terms of quaternions, as explained in the following subsection.

\subsubsection{Particle number conserving Gaussian fermionic PEPS}\label{sec:GFPEPS_sym}

Symmetries are imposed on the variational states used in the numerical approximation as follows:
For an arbitrary number $f$ of physical modes per site, particle number conservation is equivalent to
\begin{align}
\frac{i}{2} \tr( \rho [a_{\mb r,s}^\dg, a_{\mb r',t}^\dg]) &= 0, \\
\frac{i}{2} \tr( \rho [a_{\mb r,s}, a_{\mb r',t}]) & = 0,
\end{align}
with $s,t = 1, \ldots, f$.
If we express this in terms of Majorana operators and perform a Fourier transformation to reciprocal space, we arrive at
\begin{equation}
[G_{\mathrm{out}}]_{2s-1,2t-1}(\mb k) \mp i \left([G_{\mathrm{out}}]_{2s,2t-1}(\mb k) + [G_{\mathrm{out}}]_{2s-1,2t}(\mb k)\right) - [G_{\mathrm{out}}]_{2s,2t}(\mb k) = 0
\end{equation}
with $G_\mr{out}(\mb k)$ given by Eq.~\eqref{eq:Gout}.
This is fulfilled, if and only if $G_{\mathrm{out}}(\mb k)$ can be written as 
$G_{\mathrm{out}}(\mb k) = \Id_2 \otimes G_1(\mb k) + W \otimes G_2(\mb k)$, where $G_{1,2}(\mb k)$ are complex $f \times f$ matrices. Inserting this into \eqref{Covariance_real} yields
\begin{align}
\Gamma_{+-}(\mb k > 0) &= \Gamma_{-+}(\mb k > 0) = 0 \label{pm} \\
\Gamma_{++}(\mb k > 0) &= \Gamma_{--}(- \mb k) = \Id_2 \otimes \mathrm{Re}(G_1(\mb k)) + W \otimes \mathrm{Re}(G_2(\mb k)) + W \otimes \mathrm{Im} (G_1(\mb k)) - \Id_2 \otimes \mathrm{Im}(G_2(\mb k)) \label{pp}
\end{align}
with $W = \left(\begin{smallmatrix} 0&1\\-1&0\end{smallmatrix}\right)$. 
Comparing this to
$G_1(\mb k) + i G_2(\mb k) = \mathrm{Re}(G_1(\mb k)) + i \, \mathrm{Re}(G_2(\mb k)) + i \, \mathrm{Im}(G_1(\mb k)) - \mathrm{Im}(G_2(\mb k))$, we note that the complex 
\begin{equation}
\Gamma_{\mathrm{out}}^{\mathbb{C}} (\mb k) := G_1(\mb k) + i G_2(\mb k)
\end{equation}
contains all the information about the CM in the case of particle number conservation. This property is due to the fact that $\Id_2$ and $W$ are a real representation of complex numbers.

In order ensure the GFPEPS has particle number conservation, it is best to impose the above symmetry already on the virtual level. This is done by grouping pairs of Majorana modes together,
which in complex representation can be compactly written as 
\begin{equation}
\Gamma^{\mathbb{C}}_{\mathrm{in}}(\mb k) = \left(\begin{array}{cc}
0&\Id_{\chi/2}e^{i k_x}\\
-\Id_{\chi/2}e^{- i k_x}&0
\end{array}\right) \oplus 
\left(\begin{array}{cc}
0&\Id_{\chi/2} e^{i k_y}\\
-\Id_{\chi/2}e^{- i k_y}&0
\end{array}\right).
\end{equation}
The Gaussian map parameterized by $\gamma_1$ (see subsection~\ref{sec:free}) needs to preserve that symmetry, which is achieved by  $\gamma_1 = \Id_2 \otimes \gamma_{1,1} + W \otimes \gamma_{1,2}$, or in complex representation $\gamma_1^{\mathbb{C}} = \gamma_{1,1} + i \gamma_{1,2}$. Then,
\begin{equation}
\Gamma^{\mathbb{C}}_{\mathrm{out}}(\mb k) = B^{\mathbb{C}} (D^{\mathbb{C}} - \Gamma_{\mathrm{in}}^{\mathbb{C}}(\mb k))^{-1} B^{\mathbb{C} \dg} + A^{\mathbb{C}}.
\end{equation}
Again $\gamma_1^{\mathbb{C}} = -\gamma_1^{\mathbb{C} \, \dg}$ and $\gamma_1^{\mathbb{C}} \gamma_1^{\mathbb{C} \, \dg} \leq \Id_{f + 2 \chi}$ with equality for pure states. We note that particle number conservation of the output state can also be achieved by using normal virtual fermions for the bonds and applying a Gaussian map keeping the fermion number conservation symmetry.


Let us consider a Chern insulator with Hamiltonian \eqref{eq:Ham_Chern-ins} and arbitrary $\mb d(\mb k)$
corresponding to two bands with energies $\pm d(\mb k)$, where $d(\mathbf k) = |\mathbf d(\mathbf k)|$. We assume that it is gapped and
has one occupied band. In this two-orbital case, the Chern number can be calculated via Eq.~\eqref{eq:Chern_d}. Moreover, we would like to characterize its properties at finite temperature.
For a given temperature $T=\frac{1}{\beta}$, the thermal density matrix of the two-band Hamiltonian $\mc H = \oplus_{\mb k} \mc H_\mb{k}$ is written as
\begin{align}
\rho(\beta) = \frac{e^{- \beta \mc H}}{\tr(e^{-\beta \mc H})} = \bigotimes_{\mb k} \frac{e^{- \beta \mc H_{\mb k}}}{\tr(e^{-\beta \mc H_{\mb k}})} := \bigotimes_{\mb k} \rho_{\mb k}(\beta) \label{rho_k},
\end{align}
The von Neumann entropy of this thermal state is given by
\begin{align}
S_\mr{vN}(\rho(\beta)) = \sum_{\mb k}S(\rho_{\mb k}(\beta)) = \sum_{\mb k} - \tr(\rho_{\mb k}(\beta) \ln \rho_{\mb k}(\beta))
\end{align}
and the covariance matrix of $\rho(\beta)$ calculated via Eq.~\eqref{eq:CM} is 
\begin{align}
\Gamma_{++}(\mb k) = \frac{\sinh(\beta d(\mb k))}{1 + \cosh(\beta d(\mb k))} \left(\begin{array}{cccc}
0&\hat d_z(\mb k)& \hat d_y(\mb k) & \hat d_x(\mb k)\\
-\hat d_z(\mb k)&0&-\hat d_x(\mb k)&\hat d_y(\mb k)\\
-\hat d_y(\mb k)& \hat d_x(\mb k)& 0 & -\hat d_z(\mb k)\\
-\hat d_x(\mb k)& -\hat d_y(\mb k)& \hat d_z(\mb k)&0
\end{array}\right)\ ,
\label{Gamma_thermal}
\end{align}
where due to particle number conservation the remaining parts of the covariance matrix are given by Eqs. \eqref{pm} and \eqref{pp}. $\Gamma_{++}(\mb k)$ could be written in complex representation; however it has one more symmetry arising from the fact that there are only two bands.
Both symmetries can be incorporated simultaneously be representing $\Gamma_{++}(\mb k)$ by quaternions,
\begin{align}
\Gamma^{\mathbb{H}}(\mb k) = \frac{\sinh(\beta d(\mb k))}{1 + \cosh(\beta d(\mb k))} (i \hat d_z(\mb k) + j \hat d_y(\mb k) + k \hat d_x(\mb k)).
\end{align}
In order to see that, observe that a matrix representation of the quaternionic units is
\begin{equation}
\underline r = \left(\begin{array}{cccc}
1&0&0&0\\
0&1&0&0\\
0&0&1&0\\
0&0&0&1
\end{array}\right), \
\underline i = \left(\begin{array}{cccc}
0&1&0&0\\
-1&0&0&0\\
0&0&0&-1\\
0&0&1&0
\end{array}\right), \
\underline j = \left(\begin{array}{cccc}
0&0&1&0\\
0&0&0&1\\
-1&0&0&0\\
0&-1&0&0
\end{array}\right), \
\underline k = \left(\begin{array}{cccc}
0&0&0&1\\
0&0&-1&0\\
0&1&0&0\\
-1&0&0&0
\end{array}\right)
\end{equation}
where $\underline r$ denotes the real unit. 

Again, the desired symmetry of the output state is obtained by imposing it on the virtual level and ensuring that the Gaussian map keeps it. We realize that the virtual modes already have the desired symmetry, since their covariance matrix $\Gamma_{\mathrm{in}}^{\mathbb{C}}$ after regrouping the virtual indices can be written directly in quaternionic representation as
\begin{align}
\Gamma^{\mathbb{H}}_{\mathrm{in}}(\mb k) = \left(\begin{array}{cc}
\Id_{\chi/2} (j \cos(k_x) + k \sin(k_x))&0 \\
0& \Id_{\chi/2} (j \cos(k_y) + k \sin(k_y))
\end{array}\right),
\end{align}
because the representation of a quaternion $q = a + i b + j c + k d$ by complex matrices is
$\underline q =  \left(\begin{array}{cc}
a + i b& c + i d\\
-c + i d& a - i b
\end{array}\right)$. The Gaussian map keeps the symmetry by requiring that $\gamma_1$ can also be represented by quaternions, that is
\begin{equation}
\gamma_1^{\mathbb{H}} = \gamma_{1,r} + i \gamma_{1,i} + j \gamma_{1,j} + k \gamma_{1,k} \in \mathbb{H}^{(1 + \chi) \times (1 + \chi)} \label{eq:M_quaternions}
\end{equation}
(note that $\chi$ has to be even). Then, the output covariance matrix is simply
\begin{align}
\Gamma_{\mathrm{out}}^\mathbb{H}(\mb k) = B^\mathbb{H} (D^\mathbb{H} - \Gamma_{\mathrm{in}}^\mathbb{H}(\mb k))^{-1} B^{\mathbb{H}\, \dg} + A^\mathbb{H}\ ,
\end{align}
where $\gamma_1^\mathbb{H} = -\gamma_1^{\mathbb{H}\,\dg}$ and $\gamma_1^\mathbb{H} \gamma_1^{\mathbb{H}\,\dg} \leq \Id_{1 + \chi}$ with equality for pure states. The quaternionic representation has been used in the numerical calculations in order to exploit the symmetries of the Hamiltonian.

Given a mixed or pure GFPEPS (obtained, e.g., after performing a minimization of the free energy), its Hall conductivity can be calculated by setting
\begin{align}
\Gamma_{\mathrm{out}}^{\mathbb{H}}(\mb k) = \frac{\sinh(\beta d(\mb k))}{1 + \cosh(\beta d(\mb k))} (i \hat d_z(\mb k) + j \hat d_y(\mb k) + k \hat d_x(\mb k))
\end{align}
to extract $\hat {\mb d}(\mb k)$ and $\beta d(\mb k)$. The $\mb d$-vector from the GFPEPS gives rise to the Hall conductivity $\sigma_{xy}(\beta)$ of Eq.~\eqref{eq:Hall_thermal}, which only depends on $\hat {\mb d}(\mb k)$ and the product $\beta d(\mb k)$ . 

\subsubsection{Results}\label{sec:num_results}

\paragraph{Optimization of the covariance matrix.}

\begin{figure}[!ht]
\centering\includegraphics[width=0.8\textwidth]{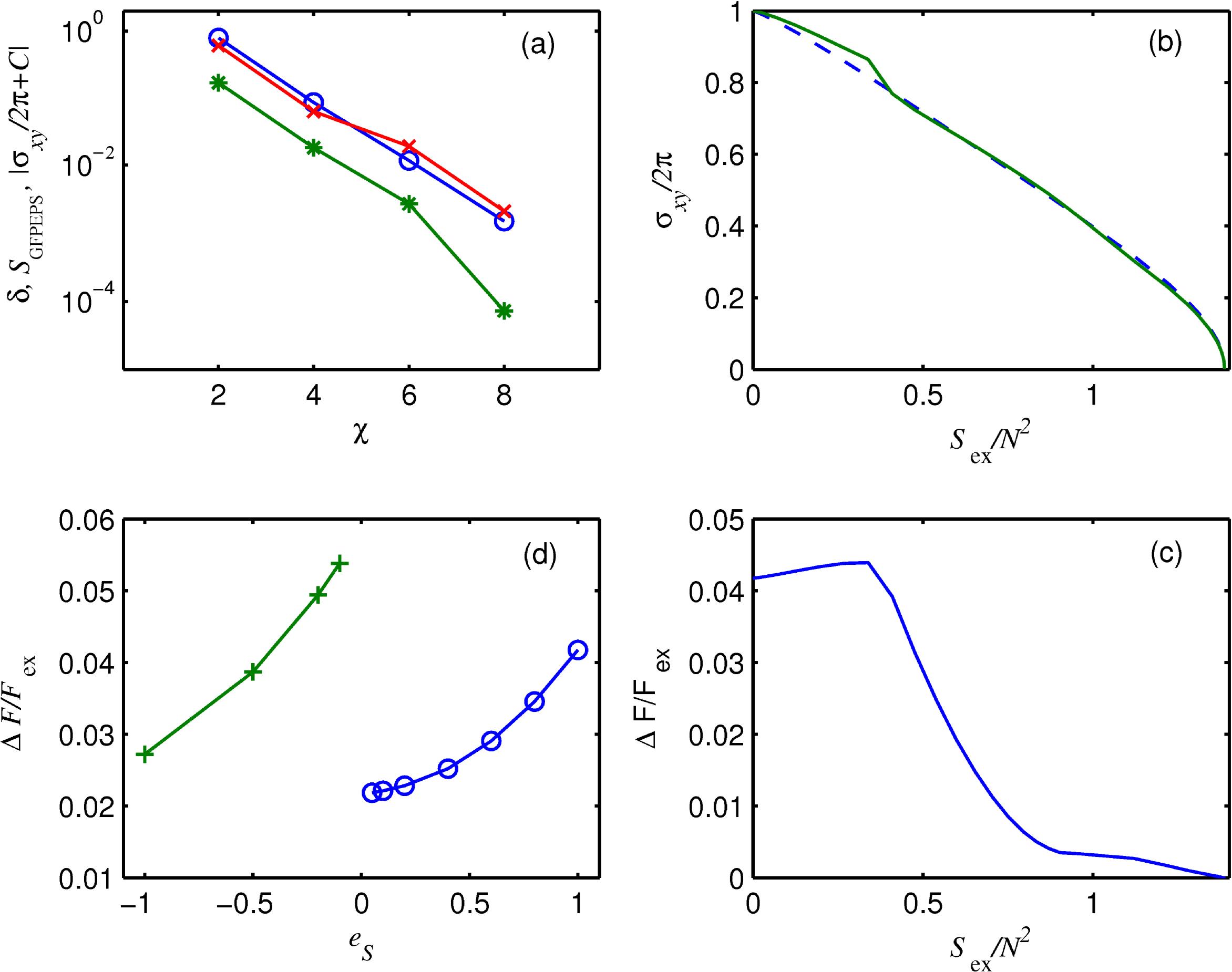}
\caption{(a) Error $\delta$ (see text) of the covariance matrix for $T = 0$, $e_S = 1$ (blue circles), entropy $S_{\mathrm{GFPEPS}}$ of the optimized GFPEPS (red crosses), difference between the Hall conductivity $-\frac{\sigma_{xy}}{2 \pi}$ and the Chern number $C = -1$ (green stars) for the optimized GFPEPS as a function of the number of Majorana modes $\chi$. 
(b) The blue dashed curve and solid green curve denote the Hall conductivity of the exact thermal state and
GFPEPS, respectively, as a function of the von Neumann entropy per site of the exact state, $S_{\mathrm{ex}}/N^2$. (c) Relative error of the free energy per site of the optimized
GFPEPS as a function of $S_{\mathrm{ex}}/N^2$. The entropies of the optimized GFPEPS were roughly proportional to $S_{\mathrm{ex}}$. The Chern numbers of the Hamiltonians of which they are thermal states were always $C = -1$. (d) Relative error of the free energy per site of
the optimized GFPEPS as a function of the parameter $e_S$ compared to the exact
state at $T = 0$. The optimized GFPEPS with Chern number $C = 0$ are represented by green crosses and those with $C = -1$ by open blue circles. This figure was reproduced from Ref.~\cite{cPEPS_long}.
} \label{numerics}
\end{figure}

The following numerical results were obtained using a parameterization of the CM $\gamma_1$ in terms of quaternions according to Eq.~\eqref{eq:M_quaternions}. 
First, we minimized the distance $\delta := \mathrm{max}_{\mathbf
k} \| G_{\mathrm{ex}}(\beta, \mathbf k) -
G_{\mathrm{GFPEPS}}(\mathbf k)\|_{\mathrm{F}}$ ($\| \cdot \|_\mr{F}$ denoting the Frobenius norm) between the CM of $e^{-\beta H}/\tr(e^{-\beta H})$ denoted by $G_\mr{ex}(\beta, \mb k)$ at
$\beta = \infty$ and the one of the GFPEPS
$G_{\mathrm{GFPEPS}}(\mathbf{k})$ for a
given $\chi$ and $e_S = 1$. The results are shown in Fig.~\ref{numerics}a: We
find that the error, $\delta$, in the CM decreases exponentially
with the number Majorana bonds $\chi$. Since all physical quantities
depend solely on the CM, our results indicate that if
$\chi$ is increased, all relevant observables can be approximated by a
GFPEPS with exponentially decreasing error. Most importantly, the Hall
conductivity $- \frac{\sigma_{xy}}{2 \pi}$ reaches $C = -1$ with
exponentially decreasing difference, and the entropy of the
optimal GFPEPS approximation decreases exponentially with $\chi$.

\paragraph{Approximation at finite temperatures and detection of quantum phase transitions.}

We also investigated the power of GFPEPS to describe Chern
insulators at finite temperature. This was done by minimizing the free energy
functional $F(\rho_{\mathrm{GFPEPS}}) = \tr(H \rho_{\mathrm{GFPEPS}}) - T
S_\mr{vN}(\rho_{\mathrm{GFPEPS}})$ of the above model at $e_S = 1$ with a GFPEPS
with $\chi=2$, i.e., one bond fermion per physical fermion, cf. Fig.~\ref{numerics}b,c:  For $T \rightarrow 0$, the entropy of the GFPEPS does indeed converge to
zero, that is, it approaches a pure state; its analytical form is very similar to the one
given in Eq.~\eqref{eq:Chern-ins}. This
shows that by minimizing the free energy as a function of $T$, one can
converge to pure states which are chiral even for small $\chi$. Thus,
pure GFPEPS are well suited to describe chiral insulators in numerical
simulations. We have substantiated this claim further by minimizing the free
energy $F(\rho_{\mathrm{GFPEPS}})$ for $\chi = 2$ as a function of $-1
\leq e_S \leq 1$ for $T = 0$, see Fig.~\ref{numerics}d. The entropies of the optimal GFPEPS were of the order of
$\leq 10^{-10}$ which is why their Chern numbers coincide with their Hall
conductivities, which jump from $0$ to $-1$ at $e_S \approx 0$. These results indicate that quantum phase transitions
can be detected by minimizing the ground state energy with GFPEPS.

Thus, GFPEPS can efficiently approximate both ground and thermal states of chiral
Hamiltonians with a small bond dimension, making them a well-suited tool
for the numerical study of chiral free fermionic systems.

%
%


\subsection{Conclusions and outlook}

In this section, we provided several families of chiral GFPEPS, i.e., PEPS with a non-zero Chern number. This defies previous arguments which seemed to rule out their very existence. We showed that such PEPS necessarily have a singularity in the Gaussian map at some point in reciprocal space, which generically gives rise to long range decaying correlations. Due to that, the frustration free parent Hamiltonians of chiral GFPEPS are gapless; they also turn out to have two ground states that can be obtained by inserting string operators along non-contractible loops of the torus. On the other hand, chiral GFPEPS are also unique ground states of long range gapped Hamiltonians. We verified for one family of chiral PEPS that the frustration free Hamiltonians are located at quantum phase transition points
between local gapped Hamiltonians with different Chern numbers, while the long range Hamiltonians are robust to the addition of local disorder and translationally invariant perturbations decaying faster than $1/r^3$.

Furthermore, we established a framework for boundary and edge
theories of GFPES and applied it to the study of chiral GFPEPS, in particular
of their underlying symmetry structure.
We introduced two different kinds of corresponding Hamiltonians, the boundary
Hamiltonian $\mc H_N^\mr{b}$ and the edge Hamiltonian $\mc H_N^\mr{e}$.
The former reproduces the entanglement spectrum of the reduced density
matrix of a region as a thermal state $\exp(-\mc H_N^\mr{b})$, while the
latter contains the low energy physics of the truncated flat band
Hamiltonian $\mc H_\mr{fb}$. We have shown that in the context of GFPEPS,
both of these Hamiltonians act on the auxiliary degrees of freedom at the
boundary, which naturally imposes a one dimensional structure, and that
they are related in a simple way.  As the physical edge modes
corresponding to $\mathcal H_N^\mr{e}$ are localized at the same edge of a
cylinder, the number of chiral edge modes and thus the Chern number of a
GFPEPS can be read off the virtual boundary and edge Hamiltonian.  We have
also provided constructive methods for analytically and numerically
determining $\mc H_N^\mr{b}$ and $\mc H_N^\mr{e}$ for general GFPEPS, 
in particular, on infinite cylinders and tori.

We also provided a full analysis of the edge and boundary
Hamiltonian for the case of GFPEPS with one Majorana mode per bond,
$\chi=1$. We have put particular emphasis on the case of GFPEPS with
chiral edge modes, where we have shown that the presence of chiral edge
modes is equivalent to a maximally entangled state 
between the virtual Majorana modes at 
the two boundaries of a cylinder, which leads to a
divergence in the entanglement spectrum at the corresponding momentum.
Subsequently, we related this global virtual symmetry in the GFPEPS
to a local virtual symmetry in the PEPS tensor $\Psi_1$. Identifying such
symmetries has proven extremely powerful in the case of non-chiral
topological models, where it has allowed for a comprehensive understanding
of ground state degeneracy, topological entanglement entropy, excitations, and more,
from a simple local symmetry and the strings formed by it. We have shown
that the virtual symmetry of chiral GFPEPS is similarly powerful, as it
explains the origin of chiral edge modes, the topological correction to
the zero R\'{e}nyi entropy, and it allows to parametrize the ground state space of
the gapless parent Hamiltonian using strings formed by the symmetry. It is
an interesting endeavor to try to understand further implications of the
symmetry, such as the excitations obtained from open strings, or the role
played by symmetries in GFPEPS with a higher number of Majorana bonds $\chi$.
Our numerical results indeed suggest that the same type of symmetries
underlie chiral edge modes for $\chi>1$.

Understanding the local symmetries giving rise to chiral topological order is
of particular interest when proceeding to interacting models, since these local
symmetries will still give rise to maximally entangled Majorana 
modes between
distant edges even for interacting models; keeping the symmetry structure
of the local PEPS tensor untouched thus seems to be a crucial ingredient
when adding interactions. This can in particular be achieved by taking
several copies of a chiral GFPEPS and coupling the copies on the physical
level without changing the auxiliary modes, for instance by a Gutzwiller
projection (cf. Ref.~\cite{cPEPS_Read}), similar to the way in which fractional
Chern insulators are constructed. This is the subject of the following section.

%% file: 3.4.tex
\section[Interacting Chiral Projected Entangled Pair States]{Interacting Chiral Projected Entangled Pair States\footnote{This section is a slight modification of Ref.~\cite{interacting_cPEPS}, copyright American Physical Society.}}~\label{sec:cPEPS}

\subsection{Overview}

States corresponding to topologically ordered phases in lattices, like the Kitaev toric code \cite{toric_code}, resonating valence-bond states \cite{RVB_org}, Levin-Wen string nets \cite{Levin_Wen}, or their generalizations \cite{TNS_string_nets} have very simple descriptions in terms of PEPS \cite{toric_PEPS,RVB_PEPS,Levin_PEPS}. Despite being a global property, the topological character of such states can be identified in terms of a symmetry group of the auxiliary indices of such a tensor (see subsection~\ref{sec:top_PEPS}). The symmetry allows one to construct strings of operators that can be moved without changing the state, in terms of which one can identify the degeneracy of the parent Hamiltonian, the topological entanglement entropy \cite{TEE_org,TEE_Levin}, or even the braiding properties of the excitations \cite{TNS_intersection}. This, together with the bulk-boundary correspondence for PEPS (subsection~\ref{sec:bulk_boundary}) as well as the existing numerical algorithms to determine correlation functions, the entanglement spectrum \cite{ES_org} and boundary Hamiltonians \cite{bulk_boundary,TE_RVB,transfer} provides us with a very powerful technique to analyze a variety of topologically ordered states (TOS).

The examples above and the cases of free fermionic chiral PEPS presented in the last section do not contain any chiral TOS. Those states are utterly important, as they naturally appear in the Fractional Quantum Hall Effect~\cite{FQHE_ex}, one of the most intriguing phenomena of modern physics. Furthermore, they can be associated to a topological quantum field theory, which dictates their coarse-grained properties and relates them to other areas of mathematical and high energy physics. Whether PEPS (or other tensor network states) can describe examples of such states or not is a relevant question. An affirmative answer would open up the possibility of using the PEPS techniques to describe them, offering a new perspective to such important states and establishing a connection with other non-chiral TOS, like the ones mentioned above. A negative one would indicate that the family of PEPS should be extended in order to describe some of the most relevant strong-correlation phenomena.


The mere existence of chiral GFPEPS gives a strong expectation of the existence of PEPS describing chiral TOS. On the one hand, it is well known that some TOS can be constructed by applying the Gutzwiller projector technique on several copies of some particular Chern insulators or superconductors \cite{chiral_SC} (see also \cite{Pfaffian_bosons,non_abelian_antiferromag,BCS_WZW}). On the other, it is rather obvious that this technique does not change the PEPS character of the states. Consequently, the states obtained by taking two or more copies of the state described by Eqs. \eqref{eq:Psi1ex} and \eqref{eq:Chern-1} and applying the Gutzwiller projector offer us a very promising candidate for a PEPS with TOS.

In this section, we analyze the state created by applying the Gutzwiller projector on two copies of chiral GFPEPS. The resulting state is a spin-1/2 PEPS with fermionic bonds; that is, the auxiliary particles involved in the PEPS projectors are fermions \cite{fPEPS_Kraus,sim_fPEPS,RVB_fPEPS}, with one fermion per bond. We develop methods to determine the boundary theory (including the boundary Hamiltonian and the corresponding entanglement spectrum) for such a PEPS defined on a cylinder and to extract the corresponding conformal dimensions of the associated conformal field theory (CFT), which characterizes the topological order \cite{TO1,TO2}. The symmetries of the tensor and the boundary theory indicate that there are four primary fields with conformal dimensions coinciding with those of the SO$(2)_1$ CFT [which is equivalent to the $U(1)_4$ CFT, see, e.g., Ref.~\cite{SO2_U1}]. This, together with the value of the topological entanglement entropy, confirms that our state is indeed a TOS, thus providing the first example of a chiral PEPS with such a property.

The nature of the chiral GFPEPS used in the construction is akin to the $p+ip$ superconductor \cite{p_ip}, as it belongs to the same class according to the standard classification \cite{Phases_Schnyder,Phases_Kitaev}. This also explains why the state we obtain is associated to the SO$(2)_1$ CFT, since using the techniques of \cite{BCS_WZW}, it can be shown that this also happens for the state obtained by Gutzwiller projecting two copies of the $p+ip$ state. However, as opposed to that state, the chiral GFPEPS possesses correlations decaying as $1/r^3$ with the distance, $r$, see subsection~\ref{sec:Intro-Examples}. This indicates that it corresponds to a state at a topological quantum phase transition for any local parent Hamiltonian. At the same time, there exists another Hamiltonian with long range hoppings (which have the same algebraic decay), which is gapped and which protects the chiral edges against perturbations. By analyzing the gap in the transfer matrix \cite{FCS_org}, we conclude that the PEPS we construct has an infinite correlation length, and thus, one can also view it as either at a topological quantum phase transition or as the ground state of a gapped long-range Hamiltonian, which gives the topological robustness. While it is possible to determine a local parent Hamiltonian, we have not been able to identify the one with long range interactions.

\subsection{Spin-1/2 PEPS} 

\begin{figure}[!ht]
\centering\includegraphics[width=0.8\textwidth]{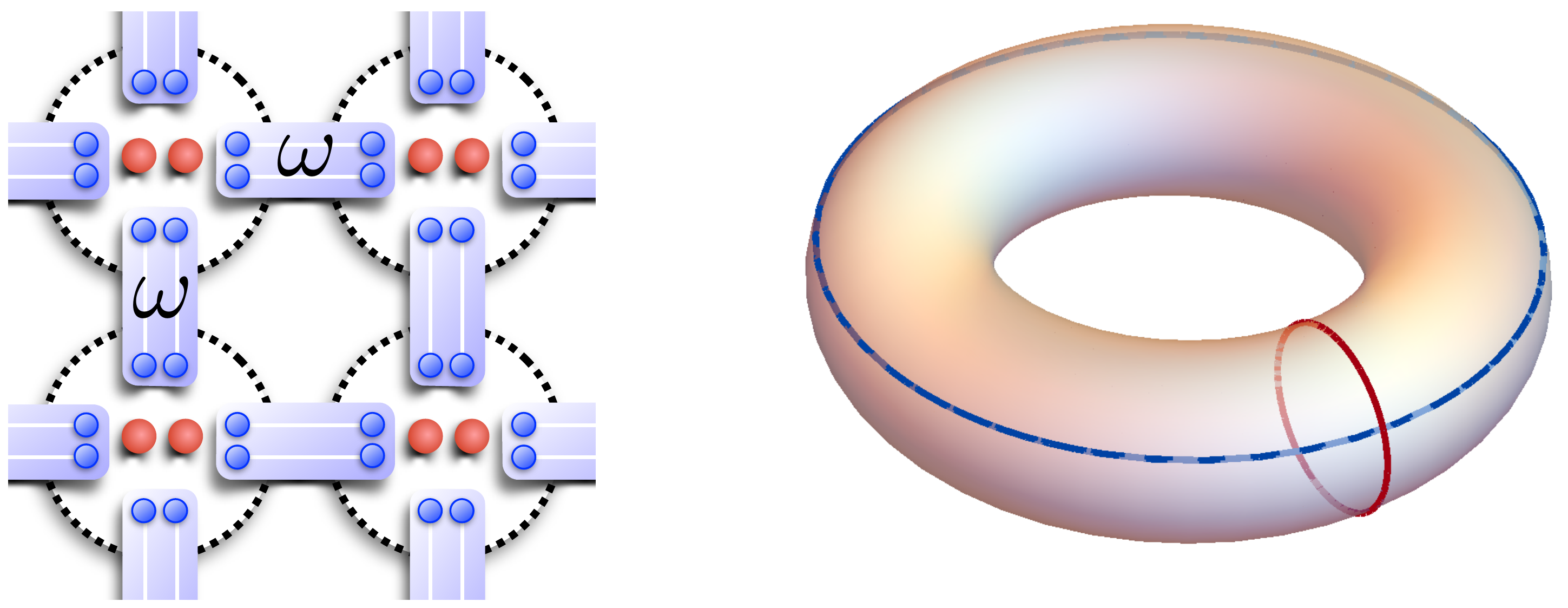}
\caption{Left: Construction of the PEPS defined via
Eqs.~\eqref{eq:projector} and~\eqref{eq:PEPS}.  $|\Psi_1\rangle$ is defined on
two physical fermionic modes (red) and eight virtual Majorana modes (blue)
for each site. Sites are marked by dashed circles.
Afterwards, the projectors $\omega_{\mb r, \mb r + \hat x}'$ and
$\omega_{\mb r, \mb r + \hat y}$ are applied on the four Majorana modes
located between neighboring sites, as indicated by light blue boxes,
yielding the PEPS $|\Phi\rangle$.
Right: Non-contractible loops on the torus. This figure has been reproduced from Ref.~\cite{interacting_cPEPS}.}
\label{fig:PEPS_torus}
\end{figure}

Let us start with an $N_v \times N_h$ square lattice with periodic boundary conditions and spin-$\tfrac12$ \textit{physical} fermions at each site with annihilation operator $a_{\mathbf{r},\kappa}$, where $\mathbf{r}$ denotes the lattice site and $\kappa=1,2$ the spin index. To construct the PEPS, we allocate eight additional \emph{virtual} Majorana modes at each site, denoted by $c_{\mathbf{r},w,\kappa}$ (with $w=L,R,U,D$
and $\kappa=1,2$). Below, we suppress the site index $\mathbf{r}$ when considering a single site, for ease of reading. To each site, we associate a fiducial state $\ket{\Psi_{1}}_\mb{r} \equiv\ket{\Psi_1}$ (corresponding to the ``PEPS projector''), which we choose site-independent in order to ensure translational invariance. It can be written as an entangled state out of physical and virtual fermionic modes
\begin{equation}
|\Psi_1\rangle=P |\Omega \rangle =\left(a_1^{\dag}b_1^{\dag}+a_2^{\dag}b_2^{\dag}\right)|\Omega
\rangle,  \label{eq:projector}
\end{equation}
where $b_{\kappa}=\frac{1}{2\sqrt{2}}[(c_{L,\kappa}+ic_{R,\kappa})e^{i \tfrac{\pi}{4}} -(c_{U,\kappa}-ic_{D,\kappa})]$ is an annihilation operator acting on the virtual modes, and the vacuum $|\Omega\rangle$ is annihilated by $d_{\kappa}=\frac{1}{2\sqrt{2}} [(c_{L,\kappa}+ic_{R,\kappa})e^{i \tfrac{\pi}{4}}+(c_{U,\kappa}-ic_{D,\kappa})]$ in addition to $a_{\kappa}$ and $b_{\kappa}$. To complete the PEPS construction, we project the virtual Majorana modes between neighboring sites onto maximally entangled Majorana bonds $\omega_{\mathbf{r},\mathbf{r}+\hat x}'= \prod_{\kappa=1}^{2}\tfrac{1}{2}(1+ ic_{\mathbf{r},R,\kappa} c_{\mathbf{r}+\hat x,L,\kappa})$ and
$\omega_{\mathbf{r},\mathbf{r}+\hat y}= \prod_{\kappa=1}^{2} \tfrac{1}{2}(1+ ic_{\mathbf{r},D,\alpha} c_{\mathbf{r}+\hat y,U,\kappa})$, see Fig.~\ref{fig:PEPS_torus}~\cite{fPEPS_Kraus}. After discarding the virtual modes, we arrive at the PEPS wave function
\begin{align}
|\Phi\rangle = \langle \Omega^{\prime}| \left( \prod_{\mathbf{r}} \omega_{\mathbf{r},\mathbf{r}+\hat x}'  \omega_{\mathbf{r},\mathbf{r}+\hat y}\right)\left(\prod_\mathbf{r} P_\mathbf{r}\right) |\Omega \rangle, 
\label{eq:PEPS}
\end{align}
where $|\Omega^{\prime} \rangle$ indicates the vacuum of the auxiliary fermions.

The state (\ref{eq:PEPS}) describes spin-1/2 particles in a lattice, as there is always a single fermion per site. Furthermore, it is not Gaussian, as $|\Psi_1\rangle$ is itself not Gaussian, and thus it must correspond to a theory with interactions. In fact, $|\Psi_1 \rangle$ can be viewed as a Gaussian state $|\tilde \Psi_{1} \rangle =\prod_{\kappa =1}^{2}(\sqrt{1-\lambda }+\sqrt{\lambda }a_{\kappa }^{\dag}b_{\kappa }^{\dag })|\Omega \rangle $ (with $0<\lambda <1$), followed by a Gutzwiller projection $P_{\mathrm{G}}=a_{1}^{\dagger}a_{1}a_{2}a_{2}^{\dagger }+a_{1}a_{1}^{\dagger }a_{2}^{\dagger }a_{2}$ enforcing \emph{single} occupancy of physical fermions, $|\Psi_1\rangle =P_{\mathrm{G}}|\tilde \Psi_{1} \rangle $. Without $P_{\mathrm{G}}$, $|\Phi_1\rangle $ is a product of two identical GFPEPS, which individually describe the topological superconductor of spinless fermions given by Eq.~\eqref{eq:Psi1ex}. The long range Hamiltonian of this GFPEPS has Chern number $C=-1$ and supports a single chiral Majorana edge mode at each boundary, implying that it belongs to the same class as the $p+ip$ topological superconductor \cite{p_ip}.

The close relation between the GFPEPS used in our construction and the $p+ip$ state suggests that $|\Phi\rangle$ could have the same associated CFT as the wave function constructed by Gutzwiller projecting two copies of the $p+ip$ state. As argued in \cite{BCS_WZW}, the latter is a chiral topological state with Abelian anyons, whose edge theory is an SO(2)$_{1}$ [or equivalently, U(1)$_{4}$, see, e.g., Ref.~\cite{SO2_U1}] CFT. Such a CFT describes a compactified boson (cf. subsection~\ref{sec:extended_algebras}) with central charge $c = 1$ and four primary fields, which are denoted as $\boldsymbol{I}$, $\boldsymbol{s}$, $\bar{\boldsymbol{s}}$, and $\boldsymbol{v}$. They correspond to the domain  $\ell \in \{-1,0,1,2\}$ in Eq.~\eqref{eq:alpha_ell} with conformal dimensions $h_{(-1)} = h_{\bar{\boldsymbol{s}}}=1/8$, $h_{(0)} = h_{\boldsymbol{I}}=0$, $h_{(1)} = h_{\boldsymbol{s}} = 1/8$ and $h_{(2)} = h_{\boldsymbol{v}}=1/2$, respectively.
This serves as a guide for our interacting PEPS construction and provides sharp predictions which we can compare with numerics. Of course, one has to keep in mind that, unlike the $p+ip$ state, the GFPEPS has power-law decaying correlations, so we do not have \emph{a priori} knowledge whether topological order is still present in the PEPS (\ref{eq:PEPS}).

\subsection{Symmetries} 

The symmetries of the local PEPS description are known to be important for topological states. For $|\Psi_1\rangle$ in (\ref{eq:projector}), we find three $\mathbb{Z}_2$ gauge symmetries
\begin{align}
U_0 |\Psi_1\rangle := X_{L}X_{R}X_{U}X_{D}|\Psi_1\rangle &= -|\Psi_1\rangle ,  \label{eq:flux} \\
U_{\kappa} |\Psi_1\rangle &= |\Psi_1\rangle , \label{eq:symmetry}
\end{align}%
where $\kappa = 1,2$, $X_{w}=ic_{w,1}c_{w,2}$ ($w=L,R,U,D$) and $U_{\kappa} = {\Id} - 2d_\kappa^\dg d_\kappa$. $X_{w}$ and $U_\kappa$ are unitaries and $X_{w}^{2}=U_\kappa^2 = {\Id}$. In fact, $U_0$ measures the fermion parity of the virtual modes, arises due to the single-occupancy constraint of physical fermions and is absent without the Gutzwiller projection $P_{\mathrm{G}}$. The symmetry $U_{\kappa}$ is inherited from the free fermionic chiral PEPS, see subsection~\ref{sec:Intro-Examples}.

A bosonic version of $U_0$ appears in the toric code as a $\mathbb{Z}_2$-injectivity~\cite{TNS_intersection}, whose generalizations allow to explain the properties of all known non-chiral topological phases~\cite{TNS_intersection,twisted_inj,Top_PEPS}. A similar formalism allows us to also uncover the topological character from the symmetry in the present scenario. For instance, a \textit{closed} loop of operators $\prod_{w\in C}X_{w}$, where the loop $C$ crosses the virtual bonds, can be inserted between the virtual bonds and the PEPS projectors in (\ref{eq:PEPS}). If $C$ is contractible, the symmetry condition $U_0$ [Eq.~\eqref{eq:flux}] allows us to remove the loop operator. In contrast, if the PEPS is defined on a manifold with nontrivial topology, \textit{non-contractible} loops $C$ exist and nontrivial loop operators $W(C)=\prod_{u\in C}X_{u}$ can be defined (see Fig.~\ref{fig:PEPS}). The symmetry condition $U_0$ implies that such loop operators can be moved around in the PEPS without affecting the physical state.  In particular, states both with and without loops are locally described by the same ``bare'' PEPS. Thus, they are all ground states of local \emph{parent Hamiltonians} $\mc H=\sum_{\mb r}h_{\mb r}$, where the $h_\mb{r}$ enforce the PEPS structure locally (see subsection~\ref{sec:PEPS_parent}).

Instead of inserting loop operators defined by the symmetry $U_0$ (called ``fluxes''), it is also possible to insert loop operators constructed from $U_\kappa$ (called ``strings''), which can likewise be deformed without changing the state (subsection~\ref{sec:symmetry-GS}) as long as they do not cross $U_0$-type loops. 

\subsection{Five degenerate ground states on the torus}

By combining all non-equivalent possibilities of inserting fluxes and strings, five different ground states of the local parent Hamiltonian can be constructed, as will be demonstrated in the following.  First, we show how they are constructed, and, thereafter, we provide numerical evidence that they characterize the full ground state subspace of the local parent Hamiltonian in the thermodynamic limit. Later on, in subsection~\ref{sec:four_topological}, we will argue that only four ground states correspond to the topological sectors, while the fifth one arises from the fact that for the local parent Hamiltonian we have a gapless continuum of excitations.

\subsubsection{Construction of five non-vanishing PEPS on the torus}

We will consider states which are generated out of the original PEPS by
inserting non-contractible loops of operators constructed from the
symmetries \eqref{eq:flux} and \eqref{eq:symmetry}.
Eq.~\eqref{eq:flux} gives rise to loops $C$ of the form $\prod_{w \in C} X_w$, which we call \emph{fluxes}, because they correspond to threading a $\pi$ flux through the torus, which gives rise to a phase of $\pi$ when moving an electron around the torus.
Similarly, Eq.~\eqref{eq:symmetry} gives rise to loops of string operators of the form $\sum_{w \in C}
c_{w,\kappa}$, which can be moved freely as well. Due to the construction of parent Hamiltonians as operators which enforce that the
state is locally described by the corresponding PEPS, any PEPS with such movable loops will
still be a ground state. By combining all possible loops in horizontal and vertical
direction, we can in principle construct $2^{2 \cdot 3} = 64$ states on the
torus; we will denote these states by $|s_{1h}, s_{1v}, s_{2h}, s_{2v}, f_h,
f_v\rangle$ (string in first layer along horizontal/vertical direction,
string in second layer along horizontal/vertical direction, flux along
horizontal/vertical direction; e.g., $s_{1h} = 1 (0)$ denotes the presence
(absence) of a string along the horizontal direction in the first layer).

Yet, as we will see in the following, this only gives rise to five linearly
independent ground states. This stems from three facts: First, certain
states vanish (i.e., have norm zero); second, states can be linearly
dependent; and third, different loops might not commute with each other,
which makes it impossible to move their crossing point. 

Let us first consider the state defined on a single layer $\kappa = 1, 2$
prior to the Gutzwiller projection: As shown in subsection~\ref{sec:symmetry-GS}, if one
contracts the PEPS on a horizontal cylinder with open virtual indices on its
ends, the state of physical and virtual modes has a virtual fermionic mode
in the vacuum state delocalized between the two edges of the cylinder at $%
k_y = 0$. We are allowed to consider single-particle momenta in
vertical direction, since the state is a Gaussian fermionic state. However, the
projection on maximally entangled Majorana modes when closing the horizontal
boundary corresponds to a projection on occupied fermionic modes delocalized
between the two edges of the cylinder. Thus, the state on the torus
vanishes. In contrast, the insertion of one or two strings occupies the
fermionic mode at $k_y = 0$ (for a horizontal string it occupies the one at $%
k_x = 0$ of a vertical cylinder), rendering the final state non-vanishing.
Note that the state is the same for an insertion of a horizontal or a
vertical string and has a different parity as the state obtained for two
strings. Those arguments are also valid after the application of the
Gutzwiller projector, as it only acts on the physical modes.

Let us now turn to the state after the Gutzwiller projection and consider
the case without any fluxes. We are left with four different possibilities,
as on each layer either one or two strings can be inserted. Since the
Gutzwiller projector keeps only states with total parity even, that is, with
the same parity on both layers, there are only two different states without
fluxes, which are $|101000\rangle \propto |100100\rangle \propto
|011000\rangle \propto |010100\rangle$ and $|111100\rangle$.

Next, let us consider a state with one flux along the vertical direction.
One can easily verify that a flux crossing a string cannot be moved freely
(it induces a phase jump of $\pi$ in the string), raising its energy and
thus ruling out horizontal strings. We are thus left with the possibility of
inserting vertical strings in either layer. But if one does insert any of
these, the final state is vanishing: The insertion of a string occupies the
fermionic mode at $k_y = 0$ on the corresponding layer, as pointed out
above, while the flux inverts the final projection on occupied modes to a
projection on empty modes. The same argument of course applies to a
horizontal flux, leaving us with two non-vanishing states $|000010\rangle$
and $|000001\rangle$.

Finally, a state with two fluxes does no longer allow for the insertion of
strings, since the fluxes could no longer be moved freely, and thus yields $%
|000011\rangle$ as the last ground state.
We have verified numerically that those five remaining states are indeed
linearly independent on the $4 \times 4$, $5 \times 4$, and $6 \times 4$ torus. 

\subsubsection{Parent Hamiltonian}\label{sec:Ham_5GS}

The parent Hamiltonian can be constructed using standard PEPS techniques (see subsection~\ref{sec:PEPS_parent}). For that, we contracted a $4 \times 4$ plaquette with open boundaries and found a maximal operator, $h=h^2 \ge 0$, that annihilates the PEPS on that plaquette for all values of the virtual indices (we left them unprojected). The parent Hamiltonian is obtained by adding all horizontal and vertical translations of $h$.

Starting from the operator $h$ acting on a $4 \times 4$ plaquette, we constructed a Hamiltonian for the $4 \times 4$, $5 \times 4$, and $6 \times 4$ lattice. We verified that the ground space of that Hamiltonian is five-fold degenerate with the ground states exactly corresponding to the ones constructed in the previous subsection.

\subsection{Topological sectors}\label{sec:four_topological}

Let us argue that four ground states correspond to distinct
topological sectors, while the fifth one is the lowest state in a gapless
continuum of excitations. To this end, consider the construction of an $%
\mathrm{SO}(2)_{1}$ theory by Gutzwiller projecting two $p+ip$ states \cite{BCS_WZW}. These $p+ip$ states are ground states of the pairing Hamiltonian%
\begin{equation}
\mc H_{p+ip}=\sum_{\mathbf{k},\kappa }[2(\cos k_{x}+\cos k_{y})-\mu' ] a_{\mathbf{k}%
\kappa }^{\dagger }a_{\mathbf{k}\kappa }+\Delta' \sum_{\mathbf{k}\kappa
}(\sin k_{x}+i\sin k_{y})(a_{\mathbf{k}\kappa }^{\dagger }a_{-\mathbf{k}%
\kappa }^{\dagger }+a_{-\mathbf{k}\kappa }a_{\mathbf{k}\kappa }),
\label{eq:p+ip}
\end{equation}%
where $\kappa $ denotes the two layers ($\kappa =1,2$) and we choose $0<\mu'
<4$ and $\Delta' \neq 0$ so that the model is in the topological phase. The $%
p+ip$ Hamiltonian (\ref{eq:p+ip}) can have periodic or anti-periodic
boundary conditions in either direction, but the boundary conditions must be
the same in each layer to ensure translational invariance. All four states
have even fermion parity and thus survive after Gutzwiller projection. These
Gutzwiller projected states are four topologically distinct states
corresponding to the quasi-particle types of the $\mathrm{SO}(2)_{1}$ theory.
On the other hand, in the PEPS, the choice of anti-periodic boundary
conditions corresponds to the insertion of a flux loop, leading us to the
conclusion that in principle, four topological ground states can be obtained by inserting fluxes. However, as explained above, the state without any fluxes or strings vanishes.

It remains to understand why there are two states in the flux free sector.
To this end, consider a single layer of the gapless topological
superconductor used to construct the PEPS. Its parent Hamiltonian has a
gapless band of excitations with minimum at $\mathbf{k}=(0,0)$. On the other
hand, anti-periodic boundary conditions shift the momentum by $\pi /N$,
i.e., only in the flux free sector the gapless band will give rise to a
second ground state. This carries over through the Gutzwiller projection,
since it enforces equal parity for both copies of the superconductor.
Following this reasoning and subsection~\ref{sec:Intro-Examples}, we find the state with an
empty $\mathbf{k}=(0,0)$ mode, $|101000\rangle $, to be the ground state,
while $|111100\rangle $ arises from the gapless continuum above it. Note
that this is also compatible with the construction of a Gutzwiller projected
state from the fully gapped $p+ip$ Hamiltonian (\ref{eq:p+ip}) with periodic
boundary conditions in both directions (corresponding to the case without fluxes).

\subsection{Boundary theory} 

In order to characterize the topological nature of the state, we compute the entanglement spectrum of the minimally entangled states (MES)~\cite{mom-pol2}.  We consider a bipartition of a torus into cylinders: There, the four MES are characterized by the presence or absence of a $U_0$ flux \emph{along} the cylinder, and are eigenstates of a $U_0$ loop \emph{around} the cylinder~\cite{transfer}.  In the context of PEPS, the entanglement spectrum (this is, the spectrum of the reduced density operator of the cylinder) can be determined from the reduced state of the \emph{virtual} fermions at the boundaries of the cylinder~\cite{bulk_boundary}. For the MES, the two boundaries of the cylinder decouple in the limit of a long cylinder, and the entanglement spectrum for, e.g., the left boundary is given by $\sqrt{\sigma_L^\top} \sigma_R \sqrt{\sigma_L^\top} \propto \rho_L =:e^{-\mc H_L}$ (see subsection~\ref{sec:bulk_boundary}), where $\sigma_{L}$ ($\sigma_R$) is the reduced density matrix for the virtual modes on the left (right) boundary, $\rho_L$ is normalized to $\mathrm{tr} (\rho_L)=1$, and we have implicitly defined the entanglement Hamiltonian $\mc H_L$. In the following subsection, we show how $\sigma_L$ and $\sigma_R$ can be determined numerically for each topological sector.

\subsubsection{Construction of minimally entangled states}\label{sec:construct_MES}

In this subsection, we show that the fixed point density matrices $\sigma_L$ and $\sigma_R$ of each topological sector are determined numerically by (i) inserting or not inserting the $\mathbb{Z}_2$ flux operator parallel to the cylinder axis, and (ii) imposing even- or odd-parity boundary conditions at the ends of the cylinder.

The insertion of a flux is equivalent to changing one row of maximally
entangled operators $\omega_{\mb r, \mb r + \hat y}$ to
\begin{align}
\omega^\pi_{\mb r,\mb r + \hat y} = \frac{1}{4} \prod_{\kappa=1}^2 (1 - i c_{\mb r,D,\kappa} c_{\mb r + \hat y,U,\kappa}),
\end{align}
while all the other $\omega$ and all $\omega'$ remain the same. 
By imposing even (odd) boundary conditions at the ends of the cylinder, we mean that the eigenvalues of the virtual fermionic particle number $\sum_{y=1}^{N_v} (2-2 i c_{y,w,1} c_{y,w,2})$ ($w=L,R$) are even (odd) for both bra and ket layers. 

We focus on the case of an even number of sites $N_v$ along the circumference direction hereafter.
In the presence of a flux, we obtain $\sigma_{\boldsymbol{I}}$ when imposing even boundary conditions for $N_v=4m$ or imposing odd boundary conditions for $N_v=4m+2$, where $m$ is an integer. Meanwhile, we obtain $\sigma_{\boldsymbol{v}}$ when imposing odd boundary conditions for $N_v=4m$ or imposing even boundary conditions for $N_v=4m+2$. 
In the absence of a flux, the fixed point reduced density matrix $\sigma$ is in general a linear superposition of two MES corresponding to the topological sectors $\boldsymbol{s}$ and $\overline{\boldsymbol{s}}$. We may recover the true MES by minimizing the rank of the reduced density matrix numerically. When imposing even parity boundary conditions ($+$) on both the bra and the ket layers at both ends, we obtain the MES $\sigma_{\boldsymbol{s}+}$ and $\sigma_{\overline{\boldsymbol{s}}+}$. When imposing odd parity boundary conditions ($-$), we obtain MES $\sigma_{\boldsymbol{s}-}$ and $\sigma_{\overline{\boldsymbol{s}}-}$ in the same topological sectors $\boldsymbol{s}$ and $\overline{\boldsymbol{s}}$, respectively.

Once we obtain the fixed point reduced density matrices $\sigma_L$ and $\sigma_R$, the boundary reduced density matrix $\rho_L$ which reproduces the entanglement spectrum is given by $\rho_L \propto \sqrt{\sigma_L^{\top}} \sigma_R \sqrt{\sigma_L^{\top}}$ with $\mathrm{tr}(\rho_L)=1$. For our system we find $\sigma_L^{\top}=\sigma_R$, so that $\rho_L \propto \sigma_R^{2}$.

These numerical calculations do not account for the sector corresponding to the $|111100\rangle$ state on the torus, since this would require the insertion of strings along the cylinder axis. This state, however, turns out to be non-topological and stems from the gapless continuum above the ground state energy of the parent Hamiltonian, see subsection~\ref{sec:four_topological}. 
On the other hand, the states we are able to access by inserting or not inserting a flux parallel to the axis of a long cylinder belong to the topological sectors; they correspond to the topological ground states $|001000\rangle \propto |000100\rangle, |000010\rangle, |000001\rangle$ and $|000011\rangle$ on the torus. 

\subsubsection{Extraction of conformal dimensions}

From the spectrum of the entanglement Hamiltonian, we are able to extract the conformal dimensions of the CFT primary fields, based on the theory developed in Ref.~\cite{ES_extract}: For each MES $|\Psi _{\mu}\rangle $ ($\mu$ labels the primary field giving rise to the corresponding tower of states), the reduced density matrix of a cylinder cut from a torus is a thermal state of two \emph{chiral} CFTs restricted to the sector $\mu$, $\rho _{\mu}\propto e^{-\mc H_{L}-\mc H_{R}}|_{\mu}$, where $\mc H_{L}$ and $\mc H_{R}$ live at the left and right boundaries of the cylinder, respectively. When constraining to the left boundary, the spectrum of the PEPS boundary Hamiltonian for the MES $|\Psi _{\mu}\rangle $, at least its low-energy part, should correspond to the chiral CFT spectrum of $\mc H_{L}$ restricted to the sector $\mu$. Once this is settled, the procedure of extracting the conformal dimensions of the primary fields is the following: (i) The sector with the lowest entanglement energy $\xi _{0}^{\boldsymbol{I}}$ corresponds to the CFT identity field $\boldsymbol{I}$ with $h_{\boldsymbol{I}}=0$. (ii) The differences of the two lowest entanglement energies $\xi _{0}^{\mu}$ and $\xi _{1}^{\mu}$ in the same sector $\mu$, $\Delta _{\mu}=\xi_{1}^{\mu}-\xi _{0}^{\mu}$, set the energy scale of the CFT. This energy scale is the same for all sectors, $\Delta _{\mu}=\Delta $. (iii) The difference of the lowest entanglement energies in sectors $\mu$ and $\boldsymbol{I}$ divided by the energy scale $\Delta$ gives the conformal dimension of the primary field $\mu$, $h_{\mu}=(\xi _{0}^{\mu}-\xi _{0}^{\boldsymbol{I}})/\Delta$.

\begin{figure*}[!ht]
\centering
\includegraphics[width=\linewidth]{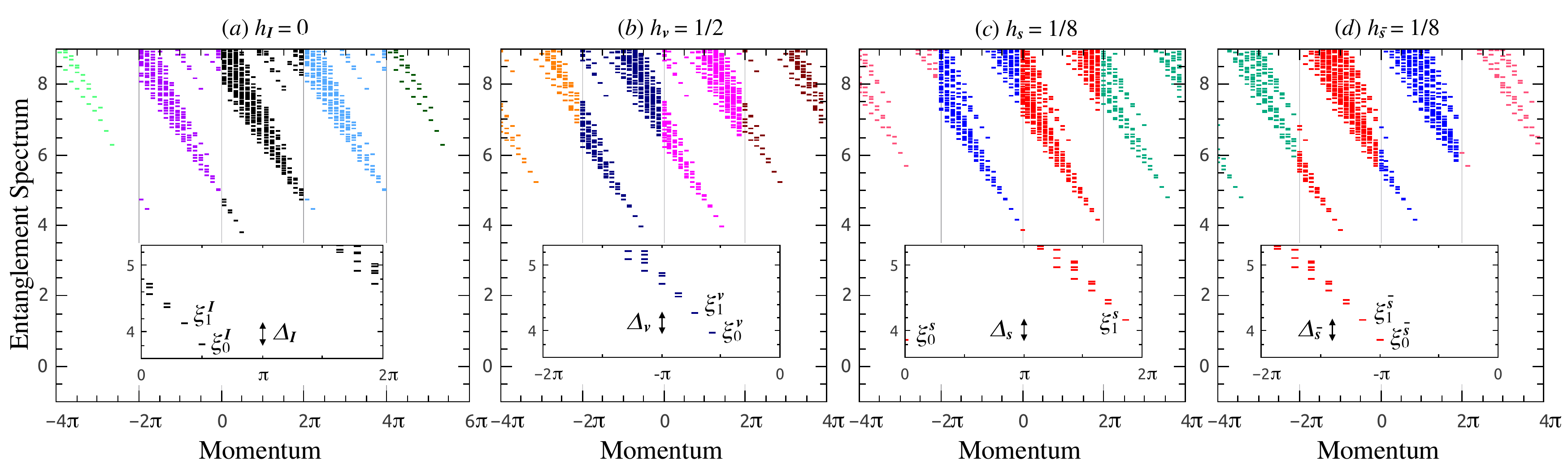}
\caption{The entanglement spectra for an $N_v=14$ infinite cylinder split into its four topological sectors. The momentum in horizontal direction is extended beyond its $2\protect\pi$ periodicity to distinguish the spectra corresponding to different particle number subspaces. The entanglement spectra with the same particle number obey the degeneracy pattern $\left\{1,1,2,3,5,\cdots \right\}$, a fingerprint for $c=1 $ chiral Luttinger liquid theory, which is equivalent to a compactified boson. Insets: Zoom-in of the low-lying part of the spectra. This figure has been reproduced from Ref.~\cite{interacting_cPEPS}.}
\label{fig:ES}
\end{figure*} 

\begin{figure}[!ht]
\centering
\includegraphics[width=0.75\columnwidth]{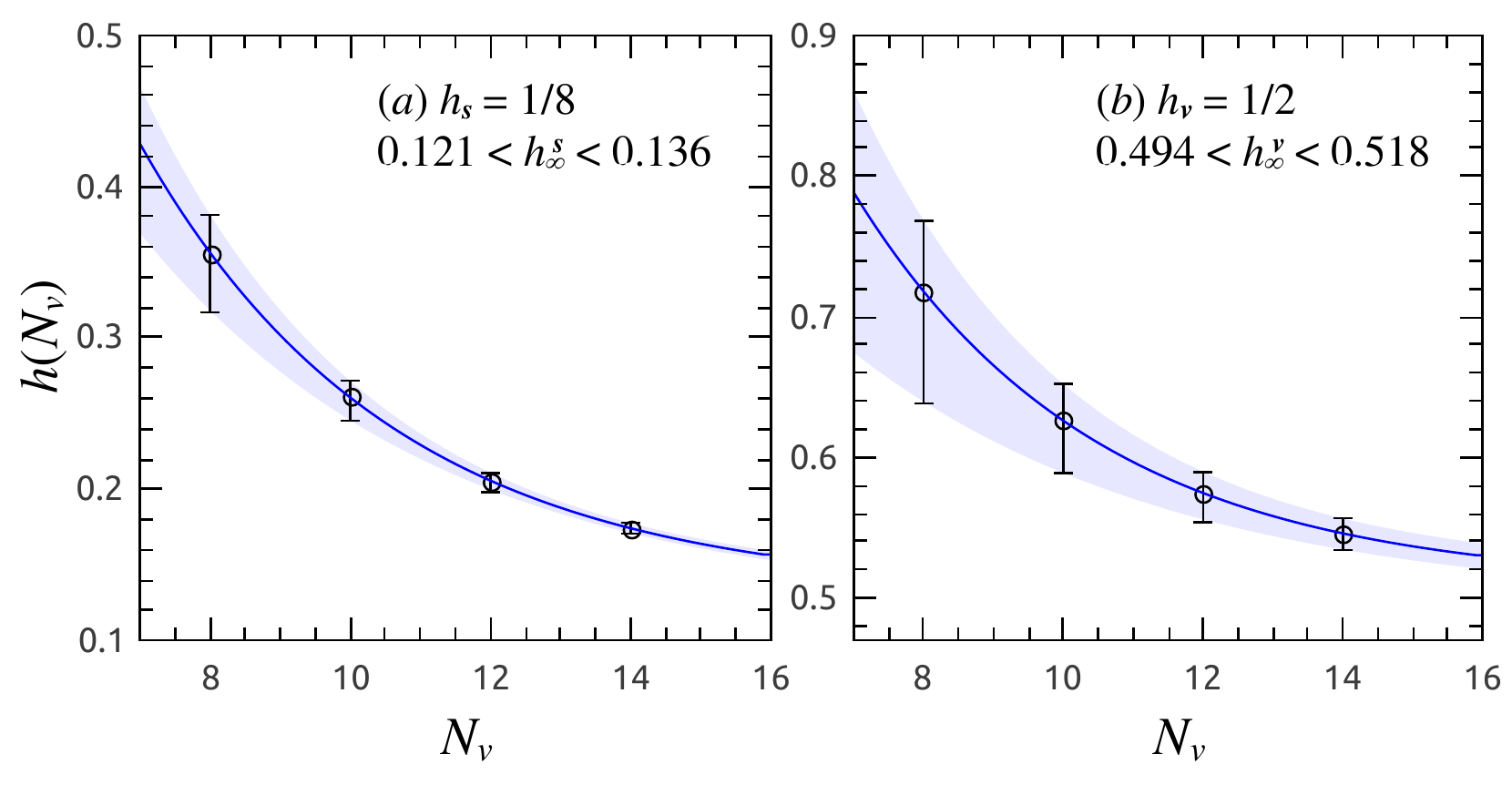}
\caption{
\label{fig:spin}
Conformal dimensions of the primary fields (a) $\boldsymbol{s}$
and (b) $\boldsymbol{v}$ through exponential decay fitting. This figure has been reproduced from Ref.~\cite{interacting_cPEPS}.}
\end{figure}

The entanglement spectra of the four MES $|\Psi _{\mu}\rangle $ with $N_{v}=14$ and $N_{h}=\infty$ are shown in Fig.~\ref{fig:ES}. In all four sectors, the state counting of the spectra shows clear chiral Luttinger liquid behavior (which corresponds to a compactified boson) with the characteristic degeneracy pattern $\{1,1,2,3,5,\ldots\}$. Following the described procedure, we extracted the conformal dimension $h_{\mu}$ for each sector as shown in Fig.~\ref{fig:spin}. Since the $\Delta _{\mu}$ are slightly different due to finite-size effects, we use their average and indicate the minimal and maximal values by error bars. To obtain an estimate of the conformal dimensions in the thermodynamic limit,  we use a fit $h(N_{v})=h_{\infty}+A\exp (-N_{v}/t)$ and find that $h_{\text{avg},\infty }^{\boldsymbol{s}}=0.131$, $0.121<h_{\infty }^{\boldsymbol{s}}<0.136$ (same for $\bar{\boldsymbol{s}}$), $h_{\text{avg},\infty }^{\boldsymbol{v}}=0.510$ and $0.494<h_{\infty }^{\boldsymbol{v}}<0.518$, in good agreement with the SO(2)$_{1}$ CFT prediction.

\subsubsection{Topological entanglement entropy}

\begin{figure}[!ht]
\centering
\includegraphics[width=0.65\columnwidth]{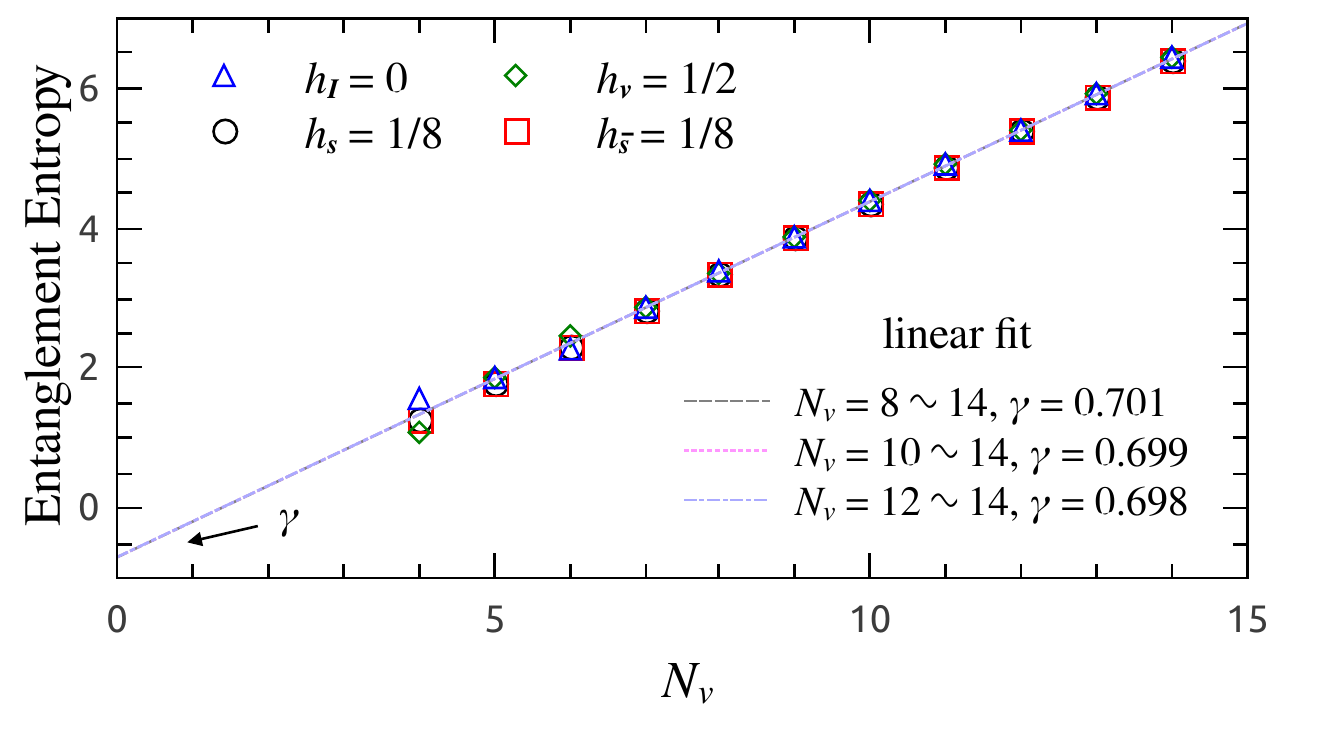}
\caption{Entanglement entropy versus perimeter $N_{v}$ for
different topological sectors. The dashed lines are linear fits based on the
average of the data points for $N_{v}=8\sim 14$, $N_{v}=10\sim 14$, and
$N_{v}=12\sim 14$, respectively. This figure has been reproduced from Ref.~\cite{interacting_cPEPS}.}
\label{fig:TEE}
\end{figure}

The spectra of the boundary Hamiltonians for the MES also give direct access to the von Neumann entropy $S_{\mathrm{vN}}$ of the reduced density matrix on a half-infinite cylinder. For sectors with anyonic excitations, the von Neumann entropy of the MES $|\Psi _{\mu}\rangle $ contains the usual area law contribution and a universal subleading constant $\gamma_{\mu}=\ln(\mathcal{D}/d_{\mu})$~\cite{mom-pol2}, where $d_{\mu}$ is the quantum dimension of the anyonic quasi-particle $\mu$ and $\mathcal{D}$ the total quantum dimension, $\mathcal{D}=\sqrt{\sum_{\mu}(d_{\mu})^{2}}$. For the $\mathrm{SO(2)}_{1}$ CFT, there exist only four Abelian anyons with $d_{\mu}=1$, and thus $\mathcal D=2$.  Fig.~\ref{fig:TEE} shows the von Neumann entropies of the MES of the PEPS given by Eq.~(\ref{eq:PEPS}) as a function of $N_{v}$. We find that the difference between the sectors vanishes as $N_v$ increases, and fitting $S_{\mathrm{vN}}(N_{v})=c N_{v}-\gamma_\mu$ \cite{TO_TEE} gives $\gamma_\mu\approx\ln 2$, in agreement with the prediction of $\mathrm{SO(2)}_{1}$ theory.

\subsection{Correlations in the bulk and boundary}

\subsubsection{Transfer operator}\label{sec:transfer_cPEPS}

\begin{figure}[!ht]
\centering
\includegraphics[width=0.65\columnwidth]{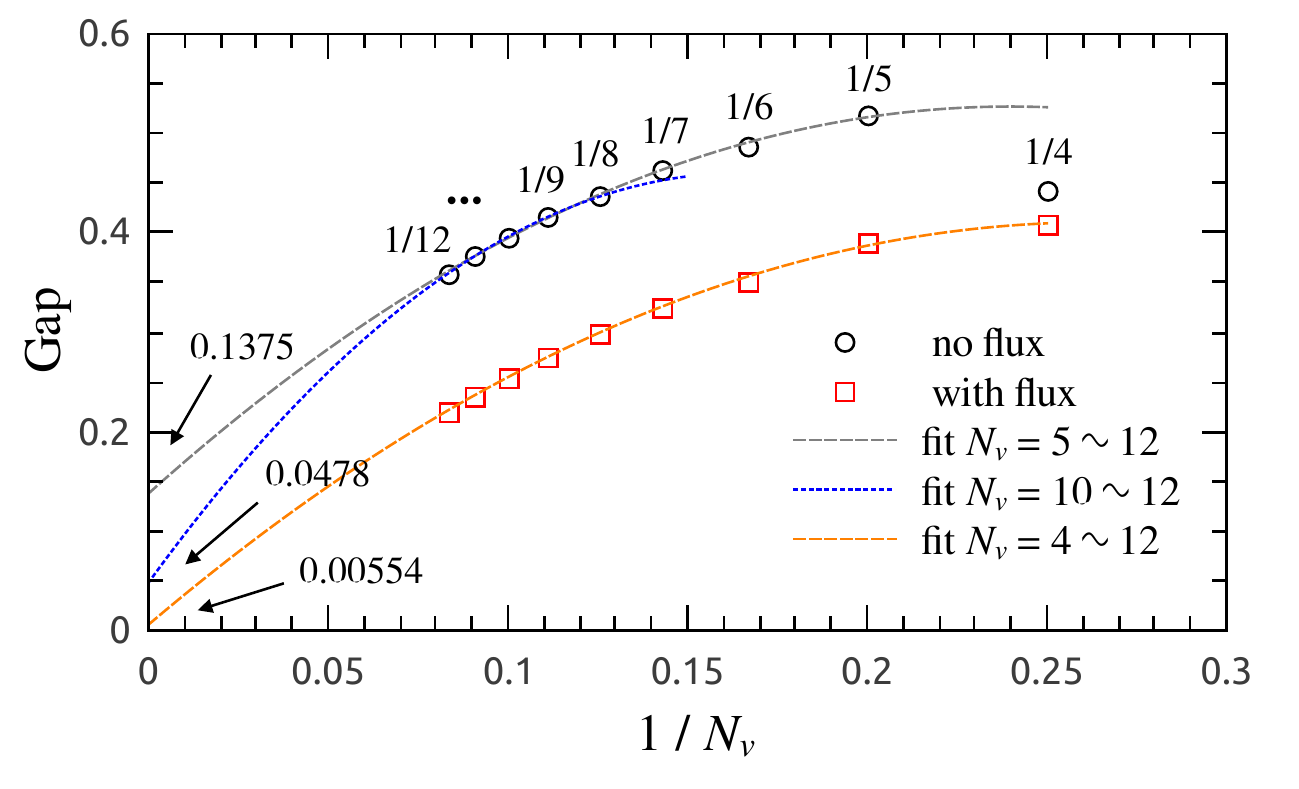}
\caption{ \label{fig:Gap}
Finite-size scaling of the gap of the transfer operator with and without flux. The fitting function is $a/N_{v}^{2}+b/N_{v}+c$, with $c$ given in the plot (for the ``no flux'' sectors, fits for two different ranges of $N_v$ are given) and suggests that the gap vanishes as $1/N_v$ in the thermodynamic limit. This figure has been reproduced from Ref.~\cite{interacting_cPEPS}.}
\end{figure}

Let us now address whether our interacting chiral PEPS has infinite correlation length, as it has been found for the GFPEPS describing topological superconductors and Chern insulators (see subsection~\ref{sec:Intro-Examples} and Ref.~\cite{cPEPS_Read}). In the PEPS formalism, this is related to the absence of a gap of the transfer operator in the limit $N_v \rightarrow \infty$ (see subsection~\ref{sec:PEPS_transfer}). Using the numerical technique sketched in appendix~\ref{app:numerical}, we have determined such a gap for different values of $N_v$ (both with and without a horizontal flux) for the PEPS (\ref{eq:PEPS}). The results are shown in Fig.~\ref{fig:Gap}, and suggest that the gap of the transfer operator vanishes polynomially in the thermodynamic limit, indicating a divergent correlation length. The same conclusion can be drawn by studying the interaction range of the boundary Hamiltonian.

\subsubsection{Interaction range of the boundary Hamiltonian}

In the following, we explore the locality of the boundary Hamiltonian for the interacting system and compare the result with its non-interacting counterpart. 

From subsection~\ref{sec:construct_MES}, we know that for even $N_v$ in the absence of a flux, the boundary reduced density matrix is a mixture of $\rho_{\boldsymbol{I}}$ and $\rho_{\boldsymbol{v}}$, which can be extracted by imposing even and odd boundary conditions at the boundary, respectively. In the presence of a flux, by minimizing the rank of the reduced density matrix, one obtains $\rho_{\boldsymbol{s}+}$, $\rho_{\overline{\boldsymbol{s}}+}$, $\rho_{\boldsymbol{s}-}$ and $\rho_{\overline{\boldsymbol{s}}-}$, where $+$ ($-$) corresponds to an even (odd) fermion parity of both bra and ket layers. (We assume the $\rho$'s to be normalized to $\tr(\rho) = 1$.)

\begin{figure}[!ht]
\centering
\includegraphics[width=0.85\columnwidth]{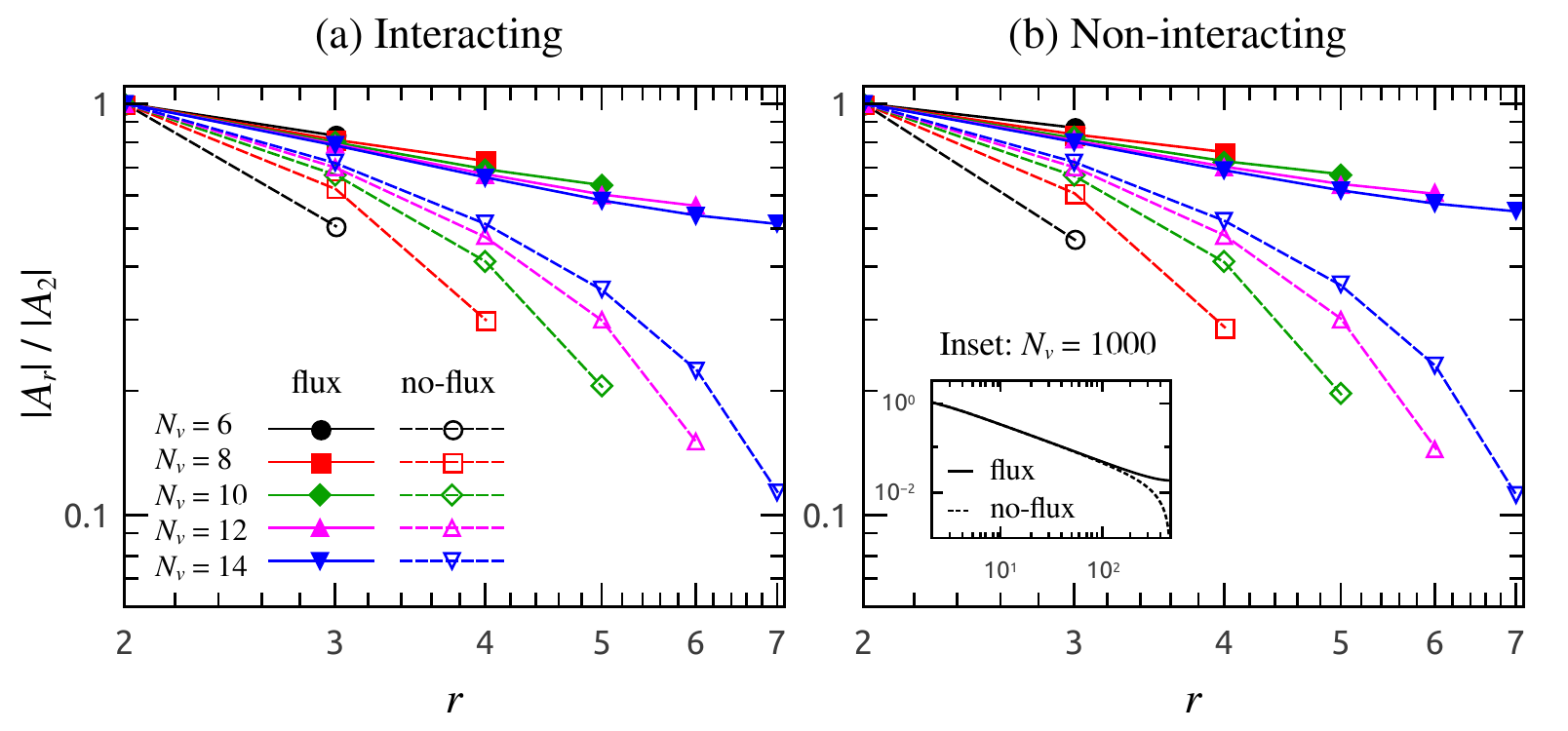}
\caption{Interaction range of the boundary Hamiltonian with
and without flux plotted using a log-log scale. (a) Interacting case. (b)
Non-interacting case. Inset: Non-interacting case with $N_v=1000$.}
\label{fig:IntRange}
\end{figure}

To make the boundary Hamiltonians as local as possible, we choose the equal-weight combinations of the reduced density matrices of the flux sector and of the no-flux sector~\cite{transfer}. We define two Hamiltonians $\mc H_{\textrm{flux}}$ and $\mc H_{\textrm{no-flux}}$,
\begin{align}
\mc H_{\textrm{flux}}&=-\log \left[ (\rho_{\boldsymbol{I}}+\rho_{\boldsymbol{v}})/2 \right], \notag \\
\mc H_{\textrm{no-flux}}&=-\log \left[ (\rho_{\boldsymbol{s}+}+\rho_{\overline{\boldsymbol{s}}+}+\rho_{\boldsymbol{s}-}+\rho_{\overline{\boldsymbol{s}}-})/4 \right].
\end{align}
Then we decompose the boundary Hamiltonian in terms of Majorana modes,
\begin{align}
\mc H=\sum_{i_{1},\cdots,i_{N_v}=0}^{3}f_{i_{1},\cdots,i_{N}} c^{i_{1}} c^{i_{2}} \cdots c^{i_{N_v}}.
\end{align} 
Here $c^{0}=\Id$, $c^{1}=c_{1}$, $c^{2}=c_{2}$, $c^{3}=c_{1}c_{2}$, where $c_{1}$ and $c_{2}$ are the virtual Majorana modes of the two copies. 
We say the locality of a term is $r$ if it spans $r$ nearest neighbor sites (taking the periodic boundary conditions into consideration) \cite{bulk_boundary}. Thus $r=1$ denotes the on-site term, $r=2$ denotes the nearest neighbor terms, $r=3$ includes both next-nearest neighbor terms and the terms acting on three contiguous sites, \textit{etc}. The interaction strength $|A_{r}|$ is defined as the $2$-norm of all the weights of terms with locality $r$ \cite{bulk_boundary,transfer,TE_RVB,edge_theories}.

In Fig.~\ref{fig:IntRange}(a) we show the relative interaction strength $|A_{r}|/|A_{2}|$ as a function of the locality $r$. We find that the interactions in $\mc H_{\textrm{flux}}$ and $\mc H_{\textrm{no-flux}}$ indeed decay with distance. Moreover, the two curves of $\mc H_{\textrm{flux}}$ and $\mc H_{\textrm{no-flux}}$ will converge with increasing $N_v$. Although the finite size effect is strong, we may get some hints by comparing it with the non-interacting case, subsection~\ref{sec:Intro-Examples}. In Fig.~\ref{fig:IntRange}(b), the non-interacting case shows almost the same behavior for $N_v=6,8,\ldots,14$. We expect a power law decay for large system sizes in the interacting case, since such a decay is present in the non-interacting case (see the inset of Figs.~\ref{fig:IntRange}b and~\ref{fig:Ham}). This is consistent with the infinite correlation length indicated in subsection~\ref{sec:transfer_cPEPS}.

\subsection{Conclusions}

In this section, we investigated to which extent PEPS can describe chiral topological order by constructing and studying an explicit example. We have identified the local symmetries in the PEPS known to be responsible for the topological character of all non-chiral topological states.  Based on these symmetries, we further showed that our example exhibits the characteristic properties of topologically ordered chiral states, such as ground state degeneracy, an entanglement spectrum described by a chiral CFT, and a non-vanishing topological entanglement entropy. Finally, we have provided numerical evidence suggesting that the state has a diverging correlation length and therefore a gapless local parent Hamiltonian. This raises several intriguing questions: Can one construct an interacting chiral topological PEPS which has exponentially decaying correlations and is the ground state of a gapped local Hamiltonian? Can our PEPS describe  a topological quantum phase transition? And, is it possible to determine a Hamiltonian for the PEPS which has long range interactions and stabilizes the topological phase?

%% file: 4.tex
\chapter{Conclusion and Perspectives}

This thesis contains our results on two independent subjects, Matrix Product States (MPS) with long range localizable entanglement (LRLE) and on chiral fermionic Projected Entangled Pair States (PEPS). Regarding the first, we provided an expression for the localizable entanglement of a translationally invariant MPS. We showed that in the thermodynamic limit, the problem can be reduced to qubits at the ends of the spin chain. We used this result to derive a necessary and sufficient criterion on the rank-3 tensors of the MPS to give rise to LRLE. This criterion has the form of a set of polynomial equations, which can analyzed for possible solutions with a computational cost that is not dependent on the system length, but only on the bond dimension. Thus, given an MPS, it is possible to check efficiently if it possesses LRLE, which is a non-local hidden order not always detectable by other known quantities. This order can be relevant for revealing certain kinds of topological quantum phase transitions, an example being the case of the ``deformed'' AKLT model presented in Ref.~\cite{LRLE_QPT}, which has a quantum phase transition at the AKLT point where LRLE occurs.

While our results provide a criterion to check efficiently whether a given MPS has LRLE, it would be desirable to find a figure of merit which indicates if an MPS is close to having LRLE. This figure of merit could be used in numerical simulations done with MPS to show that they cross a quantum phase transition, as numerically obtained MPS will never \textit{exactly} possess LRLE. Our work might facilitate further research in this direction.

The mayor part of this thesis is devoted to chiral PEPS: We provided several families of chiral Gaussian fermionic PEPS (GFPEPS), i.e., they do have a non-zero Chern number. This result defies arguments which have been proposed that seemed to rule out the existence of chiral PEPS, at least in the free fermionic limit. The reason why those families can withstand the arguments is that they all silently assumed the PEPS to be injective (and thus to have local, gapped parent Hamiltonians). They thus do not rule out chiral PEPS in general and are consistent with our rigorous proof, which demonstrates that chiral GFPEPS necessarily have to have gapless parent Hamiltonians. Their band touching points correspond to singularities in the Gaussian map (in reciprocal space), which give rise to algebraically decaying correlations in real space. Nevertheless, these GFPEPS are still topological in the sense that they are unique ground states of gapped long range Hamiltonians with non-zero Chern numbers. The latter are robust against the addition of random local disorder and translationally invariant perturbations.

We demonstrated that the families of chiral GFPEPS found possess a symmetry that resides only on the virtual level of the PEPS tensor. It can be concatenated to arbitrarily large regions and gives rise to string operators that can be inserted into the PEPS without changing it locally. We found that on the torus, the two ground states of the local gapless parent Hamiltonians can be obtained by inserting either one string along any non-contractible loop or two strings along both directions (the PEPS without any strings vanishes on the torus). Furthermore, we defined two kinds of Hamiltonians on the unpaired virtual particles if one contracts the PEPS on a cylinder with unpaired boundary modes, the boundary Hamiltonian and the edge Hamiltonian. The boundary Hamiltonian reproduces the single particle entanglement spectrum, whereas the edge Hamiltonian contains the edge spectrum of the flat band Hamiltonian with open boundary conditions. The physical edge modes they describe are exponentially localized at the boundary of the system. The only exception are $k_y$-points corresponding to maximally entangled Majorana modes between the two edges of the cylinder, which arise due to the symmetry of the PEPS tensor. 

We completely characterized all chiral GFPEPS with one Majorana bond by providing a necessary and sufficient condition on the symmetry that the corresponding PEPS tensor has to possess. This way, it is possible to check locally  if the GFPEPS has a non-vanishing Chern number.

Finally, we demonstrated the existence of chiral (interacting) fermionic PEPS with topological order by Gutzwiller projecting two topological superconductors described by chiral GFPEPS. The resulting conformal field theory is the same as the one of the Gutzwiller projected $p + i p$ model (which is in the same topological phase as the GFPEPS): We found the chiral edge to be described by an $SO(2)_1$ CFT [equivalently $U(1)_4$], which is a chiral Luttinger liquid with central charge $c = 1$ and Abelian anyonic excitations. We verified this by contracting the PEPS on a cylinder and restricting to the symmetric sectors that correspond to minimally entangled states. The entanglement spectra of the latter correspond to the primary fields, whose conformal dimensions were extracted from the two lowest entanglement energies. The presented chiral fermionic PEPS has two symmetries, which are inherited from the two free fermionic layers and a third symmetry arising due to the Gutzwiller projection. Our numerical results indicate furthermore that also this chiral PEPS has polynomially decaying correlations. The local Hamiltonian we calculated has five degenerate ground states on the torus, where four are topologically degenerate, while the fifth appears due to the gapless spectrum above the ground state energy. Those five ground states can again be constructed by inserting string operators (and ``fluxes'' arising from the third symmetry) in the PEPS construction along non-contractible loops of the torus. We believe that there exists also a long range Hamiltonian, which has a gap and only four topologically degenerate ground states (described by the constructed PEPS). So far, we were unable to determine the corresponding Hamiltonian and leave this problem for future work.

Many directions of research on chiral PEPS appear to be relevant: First, it would be interesting to construct an interacting chiral PEPS with a gapped local Hamiltonian. For that it might be necessary to generalize the symmetry~\eqref{eq:d-sym-def} to an operator that is, e.g., of second order in the Majorana operators, but nevertheless can be concatenated. Note that the general requirement for topological PEPS has been given in Ref.~\cite{Top_PEPS}, which might be used for that extension. A challenge would be to show (rigorously) that the corresponding symmetry gives rise to chiral edge modes and does not correspond to a non-chiral topological state. A remote goal could also be to characterize all symmetries which give rise to chiral PEPS, which might help to classify chiral topological phases. 

On the other hand, even if there are no chiral PEPS with exponentially decaying correlations (and gapped parent Hamiltonians), PEPS are very useful for the numerical description of realistic chiral models. E.g., by introducing a small, but finite temperature, PEPS no longer need to be non-injective (and thus fine-tuned) in order to have a non-vanishing Chern number. This is what has been demonstrated in subsection~\ref{sec:numerics}. It would be interesting to use the same approach to numerically approximate Fractional Quantum Hall systems. Moreover, given our results, PEPS are ideally suited to describe (chiral) topological quantum phase transitions, even at zero temperature. Hence, many numerical and analytic questions arise following the discovery of free fermionic and interacting chiral PEPS and we anticipate new findings in the near future.

%% file: A.tex
{
\pagenumbering{Roman}
\setcounter {chapter}{0}
\appendix
\addappheadtotoc
\chaptermark{}

\newpage
\thispagestyle{plain}

\vspace*{\fill}

\centering{\bfseries \Huge{Appendices}}

\vspace*{\fill}
\mbox{}

\newpage
}


\chapter{Proof that LRLE for $D' > 2$ implies LRLE for $D' = 2$}\label{app_show_qubits}

Assume the localizable entanglement (LE) of the state~\eqref{eq:initial_MPS} is non-vanishing in the limit $N \rightarrow \infty$ for some $\mathbb{P}, \mathbb{Q}$.
Then, we would like to show that isometries $\tilde{\P}, \tilde{\mathbb{Q}} \in \mathcal{P} := \{\tilde{\P}: \mathbb{C}^2 \rightarrow \mathbb{C}^D \ \mathrm{s. t.} \ \tilde{\P}^\dg \tilde{\P} = \Id_{2 \times 2} \}$ exist such that the quantity
\begin{align}
2 \sup_{\{|i\rangle\}} \sum_{i_1, ..., i_N} \left| \det(\tilde{\P}^\dg A_{i_1} ... A_{i_N} \tilde{\mathbb{Q}}) \right| \label{app_LE_det}
\end{align}
is non-vanishing in the thermodynamic limit (assuming we are using the optimal basis $\{|i\rangle\}$ for the LE).

We define the coefficient matrix $\Psi_{\mb i}$ of $| \psi^\mathcal{M}_{\mb i} \rangle$ [see Eq.~\eqref{eq:MPS_proj}] via $(\Psi_{\mb i})_{kl} = \frac{1}{\sqrt{p_{\mb i}}} (k|\P^\dg A_{i_1} ... A_{i_N} \mathbb{Q}|l)$. Let $\tilde{\P}_{\mb i}$ and $\tilde{\mathbb{Q}}_{\mb i}$ denote the isometries contained in $\mathcal{P}$ projecting $A_{i_1} ... A_{i_N}$ into the subspace of dimension 2 corresponding to the maximum Schmidt coefficients. For the resulting state we obtain the coefficient matrix $\tilde{\Psi}_{\mb i}= \frac{1}{\sqrt{\tilde{p}_{\mb i}}} \tilde{\P}^\dg_{\mb i} A_{i_1} ... A_{i_N} \tilde{\mathbb{Q}}_{\mb i}$, where $\tilde p_{\mb i}$ is the probability, which arises from keeping track only of the projection into the space of the two highest Schmidt coefficients, i.e., $\tilde p_{\mb i} = \tr(\tilde{\P}^\dg_{\mb i} A_{i_1} ... A_{i_N} \tilde{\mathbb{Q}}_{\mb i} \tilde{\mathbb{Q}}_{\mb i}^\dg A_{i_N}^\dg ... A_{i_1}^\dg \tilde{\P}_{\mb i})$. Now, we want to show that
\begin{align}
p_{\mb i} E\left(|\Psi_{\mb i}^\mc{M}\rangle\langle\Psi_\mb{i}^\mc{M}|\right) \leq f(D') \tilde p_{\mb i} C(\tilde{\Psi}_{\mb i}), \label{inequ}
\end{align}
where $p_\mb{i}$ is given by Eq.~\eqref{eq:measurement_prob} and $E(\cdot)$ is the entropy of entanglement~\cite{Entropy_ent}, $C(\cdot)$ the concurrence \cite{concurrence} and $f(D')$ some finite-valued function to be analyzed below. $E(\cdot)$ is given by the von Neumann entropy of the normalized reduced density operator of the bipartite state $\rho_{AB}$, $E(\rho_{AB}) = - \tr(\rho_A \log_2(\rho_A))$ with $\rho_A = \tr_B(\rho_{AB})$.
The concurrence, $C(\cdot)$, is defined for a pure state $| \psi \rangle$ of two qubits as $C(|\psi\rangle) = | \langle \psi^* | \sigma_y \otimes \sigma_y | \psi \rangle | = 2 |\det (\Psi)| \leq 1$, where $\Psi$ is the $2 \times 2$ coefficient matrix of the state.

We drop the index $\mb i$ for the variables introduced in the following and denote by $Q_1 \geq Q_2 \geq ... \geq Q_{D'}$ the eigenvalues of the unnormalized 1-particle reduced density operator, $\rho'_A$, of $\rho'_{AB} = p_{\mb i} |\psi_{\mb i}^\mathcal{M}\rangle \langle \psi_{\mb i}^\mathcal{M}| = p_{\mb i} \, \rho_{AB}$ after the measurement. Thus, we get
\begin{align}
p_{\mb i} = \sum_{k=1}^{D'} Q_k \leq (Q_1 + Q_2) \frac{D'}{2} = \tilde{p}_{\mb i} \frac{D'}{2},
\end{align}
and it is sufficient to show that
\begin{align}
E(\Psi_{\mb i}) \leq g(D') C(\tilde{\Psi}_{\mb i}) \label{show_g}
\end{align}
with $g_i = \tfrac{2 f(D')}{D'} \leq \tfrac{f(D')\tilde p_\mb{i}}{p_\mb{i}}$.
In order to do so, we first derive an upper bound for the left hand side of \eqref{show_g}. The eigenvalues of the normalized 1-particle reduced density operator, $\rho_A$, of $\rho_{AB} = |\psi_{\mb i}^{\mathcal{M}} \rangle \langle \psi_{\mb i}^\mathcal{M} |$ are $P_k = {Q_k}/{\sum_{n=1}^{D'} Q_n}$, $k = 1, \ldots, D'$. Hence, we obtain
\begin{align}
E\left(|\Psi_{\mb i}^\mc{M}\rangle\langle\Psi_\mb{i}^\mc{M}|\right) = - \sum_{k=1}^{D'} P_k \log_2 P_k \leq -P_1 \log_2 P_1 - (1-P_1) \log_2 \frac{1-P_1}{D'-1} := w(P_1, D'),
\end{align}
since the sum from the second term on is maximized for $P_2 = ... = P_{D'} = (1-P_1)/(D'-1)$. 

A lower bound for the right hand side of \eqref{show_g} can be found by using the entropy of entanglement of the qubit system into which has been projected, $E\left(|\Psi_{\mb i}^\mc{M}\rangle\langle\Psi_\mb{i}^\mc{M}|\right) = \mathcal{F}(C(\tilde{\Psi}_{\mb i})) \leq C(\tilde{\Psi}_{\mb i})$ \cite{concurrence}, where $\mathcal{F}(C) = h ((1+ \sqrt{1-C^2})/2)$ and $h(x) = -x \log_2 x - (1-x) \log_2 (1-x)$, and thus $\mathcal{F}(C) \leq C$ for $0 \leq C \leq 1$. In our case, by definition of $\tilde{\P}_{\mb i}$ and $\tilde{\mathbb{Q}}_{\mb i}$, the entropy of entanglement is
\begin{align}
E\left(|\Psi_{\mb i}^\mc{M}\rangle\langle\Psi_\mb{i}^\mc{M}|\right) &= - \frac{P_1}{P_1 + P_2} \log_2 \frac{P_1}{P_1 + P_2} - \frac{P_2}{P_1 + P_2} \log_2 \frac{P_2}{P_1 + P_2} \notag \\
&\geq - \frac{P_1 (D' - 1)}{P_1(D'-1) + 1 - P_1} \log_2 \frac{P_1 (D'-1)}{P_1 (D' - 1) + 1 - P_1} \notag \\
&- \frac{1-P_1}{P_1(D'-1) + 1 - P_1} \log_2 \frac{1-P_1}{P_1(D'-1) + 1 - P_1} := r(P_1, D'),
\end{align}
since for fixed $P_1$ first sum is minimized for $P_2 = \frac{1-P_1}{D'-1}$. Therefore, we have
\begin{align}
\frac{E\left(|\Psi_{\mb i}^\mc{M}\rangle\langle\Psi_\mb{i}^\mc{M}|\right)}{C(\tilde{\Psi}_{\mb i})} \leq \frac{w(P_1, D')}{r(P_1, D')},
\end{align}
which is finite for $\frac{1}{D'} \leq P_1 < 1$. In the limit $P_1 \rightarrow 1$ the ratio takes the value $D' - 1$, thus verifying that there exists a finite-valued function $g(D')$ such that ${E\left(|\Psi_{\mb i}^\mc{M}\rangle\langle\Psi_\mb{i}^\mc{M}|\right)}/{C(\tilde{\Psi}_{\mb i})} \leq g(D')$, which demonstrates~\eqref{inequ}.

Now we apply \eqref{inequ} to bound the LE for arbitrary $D'$
\begin{align}
L(\rho) = \sup_{\{|i\rangle\}} \sum_{\mb i} p_{\mb i} E(\Psi_{\mb i}) \leq f(D') \sup_{\{|i\rangle\}} \sum_{\mb i} \tilde{p}_{\mb i} C(\tilde{\Psi}_{\mb i}),
\end{align}
where $\rho = |\Psi\rangle\langle \Psi|$ denotes the density matrix of the MPS.
Using Eq.~\eqref{eq:MPS_proj} and $C(\Psi) = 2 \det(\Psi)$, we obtain
\begin{align}
L(\rho) \leq 2 f(D') \sup_{\{|i\rangle\}} \sum_{i_1, ..., i_N} |\det(\tilde{\P}_{\mb i}^\dg A_{i_1} ... A_{i_N} \tilde{\mathbb{Q}}_{\mb i})|.
\end{align}
We introduce an $\epsilon$-net ($\epsilon > 0$) $\mathcal{N}_{\epsilon}$ of isometries contained in $\mathcal{P}$, that is, for any such isometry $\tilde{\P}$ there exists a $\mathbb{V} \in \mathcal{N}_{\epsilon}$ with operator norm
$\| \tilde{\P} - \mathbb{V} \| \leq \epsilon$. By choosing $\epsilon$ sufficiently small, we get (all remaining equations of this proof are understood with $\mc{O}(\epsilon)$ corrections)
\begin{align}
L(\rho) &\leq 2 f(D') \sum_{\mathbb{V} \in \mathcal{N}_\epsilon} \sum_{\mathbb{W} \in \mathcal{N}_\epsilon} \sum_{i_1, ..., i_N} | \det(\mathbb{V}^\dg A_{i_1} ... A_{i_N} \mathbb{W})| \delta_{\mathbb{V}, \tilde{\P}_{\mb i}} \delta_{\mathbb{W}, \tilde{\mathbb{Q}}_{\mb i}} \notag \\
&:= 2 f(D') \sum_{\mathbb{V} \in \mathcal{N}_\epsilon} \sum_{\mathbb{W} \in \mathcal{N}_\epsilon} \omega(\mathbb{V}, \mathbb{W}) \label{omega2},
\end{align}
where $\delta_{\mathbb{V},\tilde{\P}}$ is 1 if $\mathbb{V}$ and $\tilde{\P}$ lie in the same $\epsilon$-hypercube and 0 otherwise.


If the LE is non-zero in the thermodynamic limit, the final expression on the right hand side of Eq.~\eqref{omega2} is lower bounded by $L(\rho) > 0$. We use this to show that~\eqref{app_LE_det} is finite in the limit $N \rightarrow \infty$: For all $r > \epsilon$ there must exist isometries $\tilde{\P}$ and $\tilde{\mathbb{Q}}$ such that the regions
\begin{align*}
\mc{K}_r^1 := \{ \mathbb{V} \in \mc{N}_\epsilon \ \mr{s.t.} \ \| \mathbb{V} - \tilde{\P}\|_\mr{F} \leq r\} \ \mr{and} \ \mc{K}_r^2 := \{ \mathbb{W} \in \mc{N}_\epsilon \ \mr{s.t.}\  \| \mathbb{W} - \tilde{\mathbb{Q}}\|_\mr{F} \leq r\}
\end{align*}
($\| \cdot \|_\mr{F}$ denoting the Frobenius norm) fulfill
\be
L(\rho) \leq 2 f(D') \sum_{\mathbb{V} \in \mc{K}_r^1} \sum_{\mathbb{W} \in \mc{K}_r^2} \omega(\mathbb{V},\mathbb{W}) \frac{|\mc N_\epsilon|^2}{|\mc R_r^1| |\mc R_r^2|},
\ee
where $|\mc{K}|$ denotes the cardinality of the set $\mc{K}$. In words, there have to be spherical regions $\mc{K}^1_r$ and $\mc{K}^2_r$ where the points of the $\epsilon$-net have at least average weight. We insert back the definition of $\omega(\mathbb{V},\mathbb{W})$,
\begin{align}
L(\rho) &\leq 2 f(D') \sum_{\mathbb{V} \in \mc{K}_r^1} \sum_{\mathbb{W} \in \mc{K}_r^2} \sum_{i_1, \ldots, i_N} \left| \det(\mathbb{V}^\dg A_{i_1} \ldots A_{i_N} \mathbb{W})\right| \delta_{\mathbb{V},\tilde{\P}_\mb{i}} \delta_{\mathbb{W},\tilde{\mathbb{Q}}_\mb{i}} \frac{|\mc N_\epsilon|^2}{|\mc K_r^1| |\mc K_r^2|} \notag \\
&= \frac{2 f(D') |\mc N_\epsilon|^2}{|\mc K_r^1| |\mc K_r^2|} \sum_{\mathbb{V} \in \mc{K}_r^1} \sum_{\mathbb{W} \in \mc{K}_r^2} \sum_{i_1, \ldots, i_N} \left| \det(\tilde{\P}^\dg A_{i_1} \ldots A_{i_N} \tilde{\mathbb{Q}})\right| \delta_{\mathbb{V},\tilde{\P}_\mb{i}} \delta_{\mathbb{W},\tilde{\mathbb{Q}}_\mb{i}} (1 + \mc{O}(r)) \notag \\
&=  \frac{2 f(D') |\mc N_\epsilon|^2}{|\mc K_r^1| |\mc K_r^2|}  \sum_{i_1, \ldots, i_N} \left| \det(\tilde{\P}^\dg A_{i_1} \ldots A_{i_N} \tilde{\mathbb{Q}})\right| (1 + \mc{O}(r)), \label{thorsten}
\end{align}
where in the expression before we obtained an upper bound by replacing $\mc K_r^1$ and $\mc K_r^2$ by $\mc N_\epsilon$ in the summation ranges. We take $r$ sufficiently small. The prefactor in the final expression of Eq.~\eqref{thorsten} converges to a constant in the limit $\epsilon \rightarrow 0$. Finally, we take $N \rightarrow \infty$, which completes the proof. \qed

\chapter{Properties of free fermionic chiral Projected Entangled Pair States}
\chaptermark{Properties of free fermionic chiral PEPS}

\section{Decay of correlations in real space}\label{app:corr}

In this part of the appendix we show that the correlations of the
GFPEPS defined via Eq.~\eqref{eq:Psi1ex} and therefore also the hoppings
of the corresponding flat band Hamiltonian $\mc H_\mr{fb}$ in
Eq.~\eqref{eq:H-Gout} decay like the inverse of the distance cubed. More
precisely, we will show that the $\hat d_j(\mb k)$ in Eq.~\eqref{eq:G-param}
($j = x,y$) decay at least as fast as $\tfrac{\log(r)}{r^3}$ (with $r = |\mb r|$), but not
faster than $\tfrac{1}{r^3}$ in real space (by analogous arguments it
can be shown that $\hat d_z(\mb k)$ corresponds to a faster decay than the
inverse distance cubed). Crudely speaking, the reason for this decay is
that the $\hat d_j(\mb k)$ have a non-analytic point at $\mb k = (0,0)$, where
they are continuous, but not continuously differentiable. 

An important fact which we will need in the proof is the following
relation between the decay of Fourier coefficients and the smoothness of
the corresponding Fourier series, stated for the relevant case of two
dimensions: Given that the Fourier coefficients decay faster than
$r^{-(2+d)}$ (i.e., they are upper bounded by a constant 
times $r^{-(2+d+\delta)}$ for some
$\delta>0$), it follows that the Fourier series is $d$ times continuously
differentiable (continuous if $d=0$);  see, e.g., Proposition 3.2.12 in
Ref.~\cite{Grafakos}.

Let us start by considering the behavior of $\hat d(\mb k)$ around the
non-analytic point $\mb k = (0,0)$. For simplicity, we again restrict
ourselves to $\lambda = 1/2$, but the arguments for other $\lambda$ are
the same. We expand the
numerators and denominators in Eqs.~\eqref{eq:dx} and \eqref{eq:dy} to
second order and those in Eq.~\eqref{eq:dz} to fourth order around $\mb k =
(0,0)$ to obtain 
\begin{align}
\hat d_x(\mb k) &= \frac{-2 k_x k_y^2}{k_x^2 + k_y^2} + \mc{O} (\mb k^2)  \label{eq:dx-approx},\\
\hat d_y(\mb k) &= \frac{2 k_x^2 k_y}{k_x^2 + k_y^2} + \mc{O} (\mb k^2)  \label{eq:dy-approx},\\
\hat d_z(\mb k) &= -1 + \frac{2 k_x^2 k_y^2}{k_x^2 + k_y^2} + \mc{O} (\mb k^3).  \label{eq:dz-approx}
\end{align}
This shows that the $\hat d_j(\mb k)$ are continuous, but not continuously
differentiable at $\mb k = (0,0)$, whereas $\hat d_z(\mb k)$ is both (and only its
second derivative is non-continuous). This implies that the $\hat d_j$
cannot asymptotically decay faster than $\tfrac{1}{r^3}$ in real space,
since otherwise their Fourier transform would be continuously
differentiable.  This demonstrates the claimed lower bound bound on the
decay of the correlations. 

The upper bound is obtained by formally carrying out the Fourier transform
and bounding the terms obtained after
partial integration: Let us assume
that the site coordinates fulfill $|x| \geq |y|$ ($x \neq 0$); in the
opposite case the line of reasoning is the same. We integrate its Fourier
transform twice with respect to $k_x$ by parts, ($\mb r = (x,y)$)
\begin{align}
d_{j, \mb r} = \int_\mr{BZ} \hat d_j(\mb k) e^{-i \mb k \cdot \mb r} \mr{d} k_x \mr{d} k_y 
= \left(-\frac{1}{-i x}\right)^2 \int_{-\pi}^\pi \mr{d} k_y \int_{-\pi}^\pi \frac{\partial^2 \hat d(\mb k)}{\partial k_x^2} e^{-i \mb k \cdot \mb r} \mr{d} k_x, \label{eq:double-int}
\end{align}
where BZ denotes the first Brillouin zone, that is, $(-\pi,\pi] \times
(-\pi,\pi]$. Let us first show that the last double integral is defined,
although its integrand might diverge at $\mb k = (0,0)$: For that, we will
demonstrate the bounds 
\begin{equation} 
\left| \frac{\partial^2 \hat
d_j(\mb k)}{\partial k_x^2}\right| < \frac{c}{k}, \ \left| \frac{\partial^3
\hat d_j(\mb k)}{\partial k_x^3}\right| < \frac{c'}{k^2} \label{eq:bounds}
\end{equation}
with $k = |\mb k|$ and $c, c' > 0$. In order to show the first bound, we realize that $\frac{\partial^2 \hat d_j(\mb k)}{\partial k_x^2} \sqrt{k_x^2 + k_y^2}$ and $\frac{\partial^3 \hat d_j(k)}{\partial k_x^3} (k_x^2 + k_y^2)$
cannot diverge anywhere but at $k = (0,0)$. We expand them for $j = x$ around this point by setting $\mb k = (k \cos(\phi), k \sin(\phi))$ and obtain
\begin{align}
\frac{\partial^2 \hat d_j(\mb k)}{\partial k_x^2} \sqrt{k_x^2 + k_y^2} &\xrightarrow[k \rightarrow 0]{}  \frac{-4(\cos(3 \phi) \sin^2(\phi)) + \mc{O}(k)}{1 + \mc{O}(k)}, \label{eq:2nd-limit} \\
\frac{\partial^3 \hat d_j(\mb k)}{\partial k_x^3} (k_x^2 + k_y^2) &\xrightarrow[k \rightarrow 0]{} \frac{12 \cos(4 \phi) \sin^2(\phi) + \mc{O}(k)}{1 + \mc{O}(k)}. \label{eq:3rd-limit}
\end{align}
Therefore, the limit $k \rightarrow 0$ exists for all
$\phi$ and is uniformly bounded, and as a result, the expressions on the
left hand side of Eqs.~\eqref{eq:2nd-limit} and~\eqref{eq:3rd-limit} are
bounded for any $\mb k \in \mr{BZ}$. The same thing is encountered for $j = y$.  Since the left hand sides of Eqs.
\eqref{eq:2nd-limit} and
\eqref{eq:3rd-limit} do not diverge for any $\mb k$ and are defined for a
finite region (the first Brillouin zone), the bounds \eqref{eq:bounds} are
correct. The first bound implies that the double integral
\eqref{eq:double-int} is defined (and finite).

It will be convenient to split the integral \eqref{eq:double-int} into two parts, one with range over the full circle $C_\epsilon$ of radius $\epsilon$ centered at $\mb k = (0,0)$ and the rest. The first part is bounded in absolute value by $\int_{C_\epsilon} \frac{c}{k}\mr{d}^2 k = 2 \pi c \epsilon$. Thus, employing another integration by parts, we obtain
\begin{align}
&|d_{j, \mb r}| <  \frac{2 \pi c \epsilon}{x^2} + \frac{1}{x^2} \left| \int_{\mr{BZ} \setminus C_\epsilon}  \frac{\partial^2 \hat d_j(\mb k)}{\partial k_x^2} e^{-i \mb k \cdot \mb r} \mr{d}^2 k \right| \notag \\
&= \frac{2 \pi c \epsilon}{x^2} + \frac{1}{x^2} \left| \left(\frac{1}{-i x}\right) \left(  \int_{-\epsilon}^\epsilon \mr{d} k_y \left[\frac{\partial^2 \hat d_j(\mb k)}{\partial k_x^2} e^{-i \mb k \cdot \mb r}  \right]^{-\sqrt{\epsilon^2-k_y^2}}_{+\sqrt{\epsilon^2 - k_y^2}}  - \int_{\mr{BZ} \setminus C_\epsilon} \frac{\partial^3 \hat d_j(\mb k)}{\partial k_x^3}e^{-i \mb k \cdot \mb r}  \mr{d}^2 k \right) \right|.
\end{align}
We use the bounds on the second and third derivative of $\hat d_j(\mb k)$,
\begin{align}
|d_{j,\mb{r}}| &< \frac{2 \pi c \epsilon}{x^2} + \frac{1}{|x|^3} \left(2 \pi c  + \left| \int_{\mr{BZ} \setminus C_\epsilon} \frac{\partial^3 \hat d_j(\mb k)}{\partial k_x^3} e^{-i \mb k \cdot \mb r}  \mr{d}^2 k \right|\right) \\
&< \frac{2 \pi c \epsilon}{x^2} + \frac{1}{|x|^3} \left(2 \pi c + \int_{\mr{BZ} \setminus C_\epsilon} \frac{c'}{k^2} \mr{d}^2 k \right) \notag \\
&< \frac{2 \pi c \epsilon}{x^2} + \frac{1}{|x|^3} \left(2 \pi c  + 2 \pi c' (\ln(\sqrt{2} \pi) - \ln(\epsilon))\right).
\end{align}  
We now set $\epsilon = \tfrac{1}{|x|}$ to obtain
\be
|d_{j,\mb{r}}| < \frac{2 \pi (2 c  + c' \ln(\sqrt{2} \pi |x|))}{|x|^3}.
\ee
After realizing that $|x| \geq \tfrac{r}{\sqrt{2}}$, this leads to
\be
|d_{j,\mb{r}}| < \frac{a + b \ln(r)}{r^3}
\ee
($a, b > 0$). The decay of $\hat d_z$ in real space is faster, since its derivatives start diverging at a higher order. Hence, the hopping amplitude decays at least as fast as $\tfrac{\ln(r)}{r^3}$ and, therefore, for large $r$ as the inverse distance cubed. \qed

\section{Momentum polarization and topological entanglement entropy}
\label{app:euler-maclaurin}

In this section of the appendix, we derive analytic expressions for two quantities which
probe topological order based on the entanglement spectrum, namely
the momentum polarization and the topological entanglement entropy, for the case of
non-interacting fermions, i.e., Gaussian states.  First, we will prove
that the universal contribution to the momentum
polarization~\cite{mom-pol} is exactly determined by the number of
divergences in the entanglement spectrum ($\hat H^\mathrm{b}_\infty(k_y)$ in
the case of GFPEPS);  and second, we will prove that there is no additive
topological correction to the von Neumann entropy $S_\mathrm{vN}$ of the
entanglement spectrum.  Let us stress that both of these
arguments rely only on few properties of the entanglement spectrum and the
corresponding boundary Hamiltonian and are thus not restricted to the
case of GFPEPS.

Both these proofs
are based on the Euler-Maclaurin formulas, which for our purposes say the
following: Given a function $f:[0,2\pi]\rightarrow \mathbb C$ which is $3$
times continuously differentiable, it holds that
\begin{align}
\sum_{n=0}^N f\left(\frac{2\pi n}{N}\right)  - \frac{f(0)+f(2\pi)}{2} &=
\frac{N}{2\pi}\int\limits_{0}^{2\pi}f(x)\,\mathrm{d}x +
\frac{2\pi\,(f'(2\pi)-f'(0))}{12\,N} + \mc O(1/N^3),
\\
\label{eq:euler-maclaurin-2}
\sum_{n=0}^{N-1} f\left(\frac{\pi(2n-1)}{N}\right) & =
\frac{N-1}{2\pi}\int\limits_{0}^{2\pi}f(x)\,\mathrm{d}x -
\frac{2\pi\,(f'(2\pi)-f'(0))}{24\,N} + \mc O(1/N^3)\ .
\end{align}

Let us now first discuss how to compute the momentum polarization; for
clarity, we will focus on two copies of the superconductor defined in
Sec.~\ref{sec:Intro-Examples}, but the arguments can be readily
adapted. 
For a state $|\Psi_\mu\rangle$ defined on a long cylinder, which is partitioned into two halves $A$ and $B$, the momentum polarization $\tau_\mu$ is defined via $\lambda(N_v) = \langle \Psi_\mu | T_L | \Psi_\mu\rangle = \exp\left(\frac{2 \pi i}{N_v} \tau_\mu - \alpha' N_v\right)$, cf. Eqs.~\eqref{eq:twist_half} and~\eqref{eq:mom-pol}, where $T_L$ translates part $A$ boy one site around the cylinder axis. The momentum polarization is related to the conformal dimension $h_\mu$ of the primary field $\mu$ via $\tau_\mu = h_\mu - \tfrac{c}{24}$ (assuming that $|\Psi_\mu\rangle$ is a minimally entangled state, see subsection~\ref{sec:chiral_physical}). $c$ is the central charge of the CFT and $\alpha'$ is non-universal. In our free fermionic system, there is only one sector $\mu$, so this index will be dropped below.

It is immediate to see that this definition is equivalent to evaluating
$\lambda(N_v)=\sum_\ell \varepsilon_\ell e^{i K_\ell}$, where $\varepsilon_\ell$ is the many-body
entanglement spectrum of $A$, i.e., $\ket\Psi = \sum_\ell \sqrt{\varepsilon_\ell}
\ket{\Psi^A_\ell}\ket{\Psi^B_\ell}$, and $K_\ell$ is the many-body momentum of
$\ket{\Psi^A_\ell}$. For PEPS, the entanglement spectrum corresponds to
a state on the boundary degrees of freedom, and therefore this expression
can be evaluated directly at the boundary.  Concretely, in the case of two independent physical
states with one full fermion per bond (i.e., $\chi=2$), such as two copies of
the superconductor of Sec.~\ref{sec:Intro-Examples} (i.e, the energy levels for positive and negative $\mb k$ can be occupied independently), the entanglement
spectrum corresponds to the thermal state of the non-interacting
Hamiltonian $\mc H^\mr{b}_{N}$. That is, $K_\ell = \sum_i k_{i}$ (setting $k = k_y$) and 
\begin{align*}
\varepsilon_\ell = \frac{1}{Z_\ell} \sum_{\substack{\{k_{i}\}\\ \sum_i k_i = K_\ell}}  \prod_{k \in \{ k_i\}} e^{\omega_k} \prod_{k \notin \{ k_i \}} e^{-\omega_k}
\end{align*}
with $Z_\ell = \prod_k \left(e^{-\omega_k} + e^{\omega_k}\right)$, where $\omega_k$ is the energy of the boundary mode with momentum $k$ as shown in Fig.~\ref{Gamma_R1}.
Therefore, the momentum polarization is given by
\begin{equation}
    \label{eq:mompol-sum}
\ln[\lambda(N_v)] = \sum_k
\underbrace{
\ln\frac{e^{-\omega_k}+e^{ik+\omega_k}}{e^{-\omega_k}+e^{\omega_k}}}_{=:f(k)}.
\end{equation}
To evaluate the sum
(\ref{eq:mompol-sum}), we use the Euler-Maclaurin formulas, where $f(k)$
is defined via the summand in (\ref{eq:mompol-sum}) on
the open interval $(0,2\pi)$, and continuously extended to $[0,2\pi]$. In
order to ensure continuity of $f$, we follow the different branches of the
logarithm (i.e., we add $2\pi i$ as appropriate). Moreover, for cases
with a gapless mode at $k=\pi$ (such as the family of subsection~\ref{sec:Chern}), $f(k)$ diverges,
which can be fixed by replacing $e^{ik}$ by $e^{2ik}$ above (and
subsequently correcting for the
factor of $2$ obtained in the scaling). For the examples considered, the functions
$f$ obtained this way are indeed $3$ times continuously differentiable.  Which of the two
Euler-Maclaurin equations we use depends on whether the sum in
(\ref{eq:mompol-sum}) runs over $k=2\pi n/N_v$ or $k=2\pi(n+\tfrac12)/N_v$ ($n = 0, \ldots, N_v-1$),
which is connected to the choice of boundary conditions.  We will focus on
the case $k=2\pi(n+\tfrac12)/N_v$, but let us note that the difference in
the relevant subleading terms is merely a factor of $-2$ in the $1/N_v$
term (which in the examples relates to a non-zero conformal dimension $h_\mu$)
and a trivial additive term proportional to $f(2\pi)-f(0)$
which relates to the treatment of the branches of the logarithm.

With this choice of $k$, using
(\ref{eq:euler-maclaurin-2}) we find that
\[
\ln[\lambda(N_v)] = \alpha' N_v - \frac{2\pi i}{N_v} \tau + \mc O(1/N_v^3)\ ,
\]
where $\alpha'=\tfrac{1}{2\pi}\int_0^{2\pi} f(x)\,\mathrm{d}x$ is non-universal, and
$\tau = \frac{1}{24i} (f'(2\pi)-f'(0))$.  It is now easy to check that
for $k_0=0,2\pi$,
\[
f'(k_0) = \lim_{k\rightarrow k_0} \frac{i\,e^{2 \omega_k}}{1+e^{2\omega_k}} 
\]
and thus a divergence in the entanglement spectrum at $k_0=0$, such as for
the example of Sec.~\ref{sec:Intro-Examples},  implies that $f'(2\pi)-f'(0)=\pm
i$.  We thus find that $\tau$ is universal, with its value only depending
on the presence of a divergence in the entanglement spectrum but not on
the exact form of $\omega_k$. In particular, from $\tau=1/24$ we find a
chiral central charge of $c=1$ for two copies of the superconductor, which
amounts to $c=1/2$ for a single copy of the topological superconductor.
Note that the Euler-MacLaurin formulas can be easily adapted to deal with
more discontinuities and with different values of $k$, by expanding $f(k)$ 
in terms of Bernoulli polynomials; thus, the outlined approach allows
for the analytic calculation of the momentum polarization for general
free fermionic systems with several boundary modes and arbitrary
fluxes through the torus.

Let us conclude by discussing the scaling of the topological entanglement entropy,
which is given by $S_{\mathrm{vN}}(N_v) = \sum_k g(k)$, $g(k)=-p_k\,\log p_k -
(1-p_k)\log(1-p_k)$, $p_k=e^{-\omega_k}/(e^{-\omega_k}+e^{\omega_k})$ (in
particular, $g(k)\rightarrow0$ for $k\rightarrow0,2\pi$). For the cases
discussed in the paper, $g'(k)$ is continuous and periodic, but its second
derivative diverges; thus, the error term in the corresponding Euler-Maclaurin formula applied to $S_\mr{vN}(N_v)$
can be of order $\mc{O}(1/N_v)$. Yet, this is sufficient as we are only
interested in \emph{constant} corrections to the entanglement entropy, and
one immediately finds that both for periodic and anti-periodic boundary
conditions, $S_{\mathrm{vN}}(N_v)=a N_v + \mc{O}(1/N_v)$, with a non-universal
$a=\tfrac{1}{2\pi}\int_0^{2\pi}g(k)\,\mathrm{d}k$, and no constant
topological correction.

\chapter{Numerical implementation of fermionic Projected Entangled Pair States on a cylinder}\label{app:numerical}
\chaptermark{Numerical implementation of fermionic PEPS}

\begin{figure}[tbp]
\centering\includegraphics[width=0.6\columnwidth]{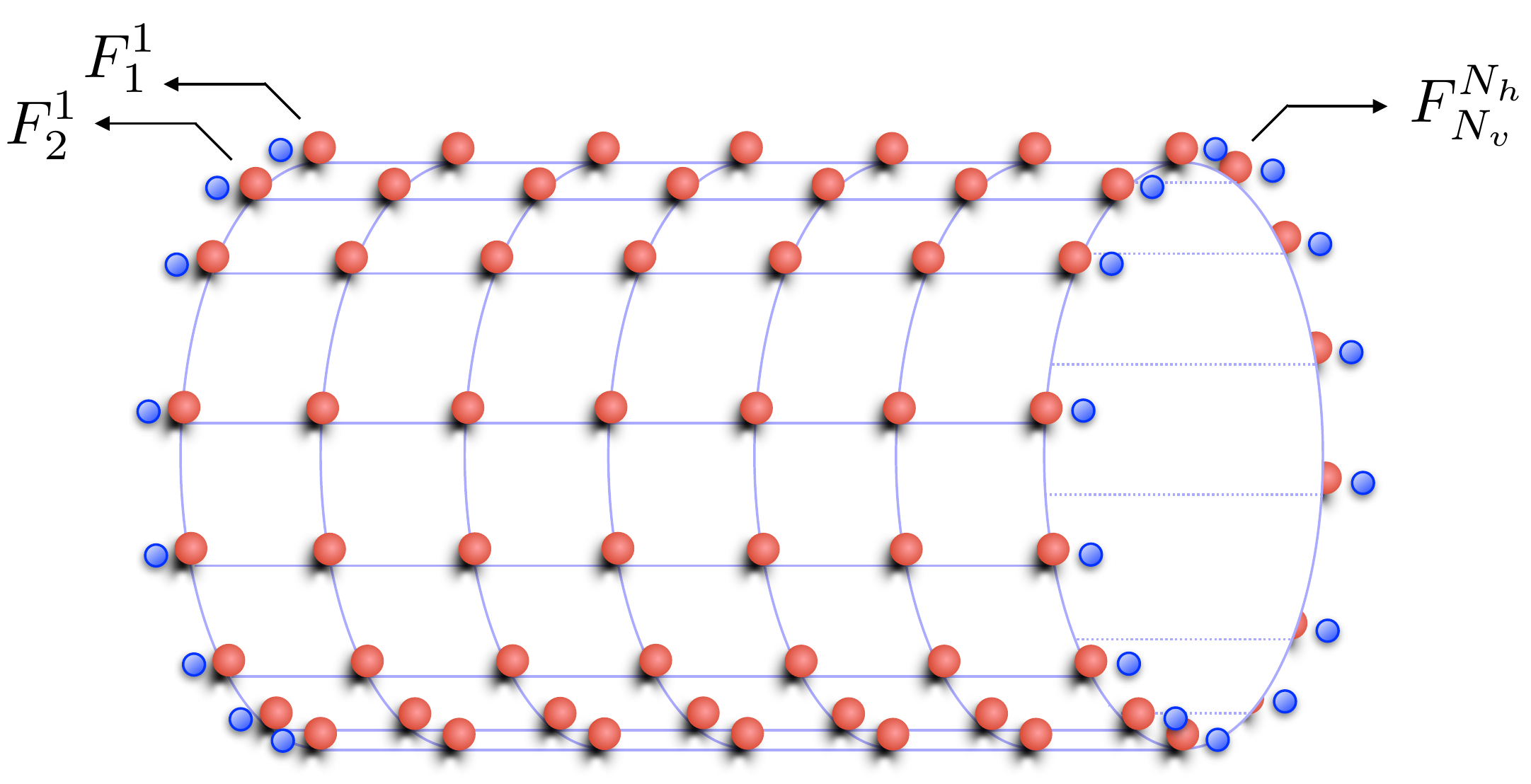}
\caption{Construction of the PEPS on an $N_{v} \times N_{h}$ cylinder. This figure has been reproduced from Ref.~\cite{interacting_cPEPS}.}
\label{appendix:cylinder}
\end{figure}

In the following, we give a description of the numerical method used to
construct the boundary density operator for our chiral PEPS. We restrict ourselves to an $N_{v} \times N_{h}$ cylinder with periodic boundary conditions along the
vertical direction [Fig.~\ref{appendix:cylinder}]. Finally we take the limit of infinite cylinders with $N_h \rightarrow \infty$.


For the double-layer PEPS introduced in the main text, we define in each lattice
site a plaquette containing two physical fermionic modes and four virtual
fermionic modes. The annihilation operators of the virtual modes are denoted
as $L$ (left), $R$ (right), $U$ (up), and $D$ (down), which are constructed
by combining the Majorana modes of the first and second layer in the
same direction. That is, $w = \left(c_{w,1}-i c_{w,2}\right)/2$ for $w=L$, $R$, $U$
and $D$.

We write the fiducial state as $\left| \Psi_1 \right\rangle=F\left| \Omega
\right\rangle$, where $F$ is an operator consisting of the creation and
annihilation operators acting on a plaquette, and $\left| \Omega
\right\rangle$ is the vacuum. We label the rows from up to down and the
columns from left to right. Each plaquette $F_{i}^{j}$ is labelled by its
row $i$ and column $j$ [Fig.~\ref{appendix:cylinder}]. We will omit the
indices whenever there is no ambiguity. Since we have periodic boundary
conditions along the vertical direction, the row index is understood as mod $%
N_v$.

We denote $\omega_{i,i+1}^{j}$ as the maximally entangled operator acting on
two neighboring virtual fermions $D_{i}^{j}$ and $U_{i+1}^{j}$,
\begin{align}
\omega &=(\Id+ic_{D,1} c_{U,1})(\Id+ic_{D,2} c_{U,2})/4  \notag \\
&=(D^{\dagger} D U U^{\dagger}+D D^{\dagger} U^{\dagger} U+i D^{\dagger} U
+i D U^{\dagger})/2.
\label{omega}
\end{align}
Similarly, we denote $\eta_{i}^{j,j+1}$ as the maximally entangled operator
acting on $R_{i}^{j}$ and $L_{i}^{j+1}$,
\begin{align}
\eta =(R^{\dagger} R L L^{\dagger}+R R^{\dagger} L^{\dagger} L +i
R^{\dagger} L +i R L^{\dagger})/2.
\end{align}
We note that
\begin{align}
\omega^{2}=\omega =\omega^{\dagger},  \notag \\
\eta^{2}=\eta =\eta^{\dagger}.  \label{omegaeta}
\end{align}
Also, all the $\omega$'s and $\eta$'s commute among themselves, since they
act on different fermionic modes and are even in the number of fermionic
mode operators.

We denote by $F^{j}=\prod_{i} F_{i}^{j}$, $\omega^{j}=\prod_{i}
\omega_{i,i+1}^{j}$, and $\eta^{j,j+1}=\prod_{i} \eta_{i}^{j,j+1}$. With all
that, we define the wave function
\begin{align}
\left| \Phi_{N_{h}} \right\rangle=\left[ \prod_{j=1}^{N_h-1} \eta^{j,j+1} %
\right] \left[ \prod_{j=1}^{N_{h}} \omega^{j} \right] \left[
\prod_{j=1}^{N_{h}} F^{j} \right] \left| \Omega \right\rangle.
\end{align}
If we ignore all the modes that are projected by the $\omega$'s and $\eta$%
's, this is the state for (i) $N_{v} \times N_{h}$ physical modes $%
p_{i}^{j} $, (ii) $N_{v}$ virtual modes living at the left of the cylinder, $%
L_{i}^{1}$, (iii) $N_{v}$ virtual modes living at the right, $R_{i}^{N_{h}}$
[Fig.~\ref{appendix:cylinder}].

We are interested in the boundary theory of the right virtual modes, that is
the density operator $\sigma_{R}$ acting on the modes $R_{i}^{N_{h}}$ alone.
In order to obtain that, we specify an initial density operator $x^{0}$
onto which we project the modes $L_{i}^{1}$. The operator $\sigma_{R}$ is
defined such that for any operator $G$ acting on the $R_{i}^{N_{h}}$,
\begin{align}
\text{tr}\left( \sigma_{R} G \right)=\left\langle \Phi_{N_{h}} \right| G
x^{0} \left| \Phi_{N_{h}} \right\rangle.
\end{align}
To write this expression in a more suitable form, we define
\begin{align}
x^{n} &=\left\langle \Omega^{n} \right| (F^{n})^{\dagger}
(\omega^{n})^{\dagger} (\eta^{n,n+1})^{\dagger} x^{n-1} \eta^{n,n+1}
\omega^{n} F^{n} \left| \Omega^{n} \right\rangle  \notag \\
&=\mr{tr}_{p^n L^n R^n U^n D^n} \left[ x^{n-1} \eta^{n,n+1} \omega^{n} F^{n} \left| \Omega^{n}
\right\rangle \left\langle \Omega^{n} \right| \left( F^{n} \right)^{\dagger} %
\right],  \label{makerow}
\end{align}
where $\Omega^{n}$ represents the vacuum of all the modes in column $n$. We
have used that certain operators commute and Eq.~(\ref{omegaeta}). Since $%
\omega$ does not depend on the $R^{N_h}_i$'s and commutes with $x^{N_{h}-1}$, we can
now write
\begin{align}
\sigma_{R}=\text{tr}_{p^{N_h} L^{N_h} U^{N_h} D^{N_h}}\left[ x^{N_{h}-1} \omega^{N_{h}} F^{N_{h}} \left|
\Omega^{N_{h}} \right\rangle \left\langle \Omega^{N_{h}} \right|
(F^{N_{h}})^{\dagger} \right].  \label{makelastrow}
\end{align}
The trace is with respect to all the operators in the last column except for the $R$'s. 
For an input $x_{0}$, we determine $x^{n}$ successively
using Eq.~(\ref{makerow}), and in the end obtain $\sigma_{R}$ via Eq.~(\ref%
{makelastrow}).

To be specific, we start out with $x^{n-1}$ and add plaquettes one by one to
obtain $x^{n}$ [Eq.~(\ref{makerow})]. Adding plaquette $F_{1}^{n}$ gives
\begin{align}
y_{1}^{n} &=\mathrm{tr}_{L_{1}^{n}} \left[ x^{n-1} G_{1} \right],
\end{align}
where $G_{1} =\mathrm{tr}_{p_{1}^{n},U_{1}^{n},D_{1}^{n},R_{1}^{n}} $ $[
\eta_{1}^{n,n+1} \omega_{N_{v},1}^{n} \omega_{1,2}^{n} $ $F_{1}^{n} $ $%
\left| \Omega_{1}^{n} \right\rangle $ $\left\langle \Omega_{1}^{n} \right| $
$(F_{1}^{n})^{\dagger} ]$. This is an operator acting on $L_{1}^{n}$, $%
U_{2}^{n}$, $D_{N_{v}}^{n}$, and $L_{1}^{n+1}$. We can do the same for $2
\leq m \leq N_{v}-1$,
\begin{align}
y_{m}^{n} =\mathrm{tr}_{U_{m}^{n},L_{m}^{n}} \left[ y_{m-1}^{n} G_{m}\right].
\end{align}
Here, $G_{m} =\mathrm{tr}_{p_{m}^{n},D_{m}^{n},R_{m}^{n}} $ $[
\eta_{m}^{n,n+1} \omega_{m,m+1}^{n} $ $F_{m}^{n} \left| \Omega_{m}^{n}
\right\rangle $ $\left\langle \Omega_{m}^{n} \right| $ $(F_{m}^{n})^{%
\dagger} ] $. For the last plaquette in column $n$, we have
\begin{align}
x^{n} =\mathrm{tr}_{U_{N_{v}}^{n},D_{N_{v}}^{n},L_{N_{v}}^{n}} \left[
y_{N_{v}-1}^{n} G_{N_{v}} \right]
\end{align}
with $G_{N_{v}} =\mathrm{tr}_{p_{N_{v}}^{n},R_{N_{v}}^{n}} $ $[
\eta_{N_{v}}^{n,n+1} F_{N_{v}}^{n} \left| \Omega_{N_{v}}^{n} \right\rangle $
$\left\langle \Omega_{N_{v}}^{n} \right| (F_{N_{v}}^{n})^{\dagger} ]$. We
note that $G_{2}= \cdots = G_{N_{v}-1}$ up to relabeling of operators and
that $G_{1}$, $G_{2}$ and $G_{N_{v}}$ are the same for all columns, so that
they have to be calculated only once.

Once we have obtained $x^{N_{h}-1}$, the plaquettes of the last column are
added one by one to get $\sigma_{R}$ [Eq.~(\ref{makelastrow})]. Adding
plaquette $F_{1}^{N_{h}}$ gives
\begin{align}
z_{1} =\mathrm{tr}_{L_{1}^{N_{h}}} \left[ x^{N_{h}-1} H_{1} \right],
\end{align}
where $H_{1} =\mathrm{tr}_{p_{1}^{N_{h}},U_{1}^{N_{h}},D_{1}^{N_{h}}} $ $[
\omega_{N_{v},1}^{N_{h}} \omega_{1,2}^{N_{h}} F_{1}^{N_{h}} $ $|
\Omega_{1}^{N_{h}} \rangle $ $\langle \Omega_{1}^{N_{h}} | $ $%
(F_{1}^{N_{h}})^{\dagger} ]$. Adding plaquette $F_{m}^{N_{h}}$ with $2 \leq
m \leq N_{v}-1$ gives
\begin{align}
z_{m} &=\mathrm{tr}_{U_{m}^{N_{h}},L_{m}^{N_{h}}} \left[ z_{m-1} H_{m} %
\right],
\end{align}
where $H_{m} =\mathrm{tr}_{p_{m}^{N_{h}},D_{m}^{N_{h}}} $ $[
\omega_{m,m+1}^{N_{h}} F_{m}^{N_{h}} $ $| \Omega_{m}^{N_{h}} \rangle $ $%
\langle \Omega_{m}^{N_{h}} | $ $(F_{m}^{N_{h}})^{\dagger} ]$. Finally, by
adding the last plaquette $F_{N_{v}}^{N_{h}}$ we have
\begin{align}
\sigma_{R}=\mathrm{tr}%
_{U_{N_{v}}^{N_{h}},D_{N_{v}}^{N_{h}},L_{N_{v}}^{N_{h}}} \left[ z_{N_{v}-1}
H_{N_{v}} \right],
\end{align}
where $H_{N_{v}}=\mathrm{tr}_{p_{N_{v}}^{N_{h}}} $ $[ F_{N_{v}}^{N_{h}} |
\Omega_{N_{v}}^{N_{h}} \rangle $ $\langle \Omega_{N_{v}}^{N_{h}} | $ $%
(F_{N_{v}}^{N_{h}})^{\dagger} ]$. Again, we just have to calculate $H_{1}$, $%
H_{2}$, and $H_{N_{v}}$, since $H_{m}=H_{2}$ up to relabeling for $%
m=3,\cdots,N_{v}-1$.

It is important to remark that one has to be extremely careful with the
definition of the trace and the vacuum when one deals with fermions. The
natural Hilbert space for the fermionic modes does not possess a tensor
product structure, so that one cannot simply write $\left| \Omega
\right\rangle=\left| \Omega^{1} \right\rangle \otimes \cdots \otimes \left|
\Omega^{N_{h}} \right\rangle$. Similarly, the trace is defined in terms of
an orthonormal basis, which will be built in terms of creation operators. It
cannot be moved inside an expression since those operators do not commute
with each other. The appropriate way of doing that is using a Jordan-Wigner
transformation (JWT), so that we can deal with spins. In particular, when
calculating the trace, we define the JWT such that the corresponding
operators are in the right order. We ensure that the operators we do not
trace do not have any strings corresponding to the ones we do trace. Then we
can trace the desired spins as usual. Alternatively, one may
explicitly calculate the signs caused by fermion anti-commutation relations
and absorb them into local tensors. In this way, the memory and CPU
requirements are largely reduced. Finally, the complexity of the calculation remains the same as for the spin models in Refs.~\cite{bulk_boundary,TE_RVB,transfer,edge_theories}.

%% file: B.tex
\chapter*{Acknowledgments}
\setcounter {chapter}{0}
\chaptermark{Acknowledgments}

Gratitude makes us aware that things cannot be taken for granted but are the result of other people's hard work and commitment. It can be a major impetus, as it increases our joy in what has been given to us, strengthens our relationships with others and enhances our desire to do something in return. In that sense, 
I am delighted to thank those people who supported me during my PhD and regret that the following list is neither complete nor my acknowledgments sufficient given all the moral and practical support I received.

In the first place, I would like to thank my supervisor Ignacio Cirac for everything I could learn from him, his guidance and inspiration. I particularly appreciate that he repeatedly took time to help working out many of the purely technical results presented in this thesis. Along the same lines, I am indebted to my collaborators who contributed to the research projects presented: I would like to thank David P\'{e}rez-Garc\'{i}a for his brilliant ideas, easygoingness and understanding. The same applies to Norbert Schuch, who contributed many crucial ideas to our research on chiral PEPS. I am especially grateful to Hong-Hao Tu for everything he taught and explained to me, for his constant readiness to help and, in particular, for his support regarding my PostDoc applications. I also thank Shuo Yang for her kindness and all the hard work she invested into our research. 

I gratefully acknowledge all other people who contributed to the success of this work through explanation, inspiration and advice: Stefan K\"{u}hn, who provided invaluable assistance in using the Garching high performance computing cluster, Benedikt Herwerth and Anna Hackenbroich for their guidance in Conformal Field Theory and Tao Shi for discussions on topological systems. I would also like to thank Toby Cubitt, who gave me important hints regarding localizable entanglement. 

Furthermore, I owe special thanks to Michael Lubasch for his advice on the formalities regarding my PhD from the very beginning and to my former office mates, Christine Muschik, Juan Bermejo-Vega, Xiaotong Ni, Anika Pflanzer and Leonardo Mazza for all their help and for providing such a welcoming and casual atmosphere. Overall, I would like to thank the theory group as a whole for all the fond memories I will take away from the MPQ, especially Yue Chang and Erez Zohar apart from those already mentioned. Moreover, I also have to acknowledge the warm greetings by \mbox{Mr Weser} in the morning, which on many occasions made a substantial difference for me as to how I approached the beginning day.
I am grateful to the secretaries of our group, Veronika Lechner and Andrea Kluth for all their loving support.

Finally, I would like to thank my entire family. In particular, I am grateful to my father, who aroused my first curiosity in physics, and to both of my parents for their love and devotion. I owe special thanks to my grandparents for their financial and moral support and for their earnest caring. 

This PhD project has been partially funded by the Quantum Computing, Control and Communication (QCCC) project, founded by Thomas Schulte-Herbr\"{u}ggen, to whom I am indebted.